\documentclass[final,3p,times,sort&compress]{elsarticle}

\usepackage{notoccite}
\usepackage{graphics,epsfig}
\usepackage{color,epsfig,bm}
\usepackage{fancybox,amsmath,amssymb,rotating,pifont}
\usepackage{epic,eepic,calc,array,shadow}
\usepackage{npsfont}
\usepackage{graphicx}
\usepackage{dcolumn}  
\usepackage{subfigure}
\usepackage[section]{placeins}
\usepackage{multirow}
\usepackage[section]{placeins}

\makeatletter
\newlength{\earraycolsep}
\def\eqnarray{\stepcounter{equation}\let\@currentlabel%
\theequation \global\@eqnswtrue\m@th
\global\@eqcnt\z@\tabskip\@centering\let\\\@eqncr
$$\halign to\displaywidth\bgroup\@eqnsel\hskip\@centering
$\displaystyle\tabskip\z@{##}$&\global\@eqcnt\@ne \hskip
2\earraycolsep \hfil$\displaystyle{##}$\hfil &\global\@eqcnt\tw@
\hskip 2\earraycolsep $\displaystyle\tabskip\z@{##}$\hfil
\tabskip\@centering&\llap{##}\tabskip\z@\cr} \makeatother

\newcommand{\DZero}{\rm D\O}
\newcommand{\Dze}   {\ensuremath{ D^0 }}

\newcommand{\Dstn} {\ensuremath{ D^{*0} }}

\newcommand{\DstnDn}  {\ensuremath{\Dstn \bar{\Dze}}}

\newcommand{\jpsi}{\ensuremath{J/\psi}}
\def\beq{\begin{equation}}
\def\enq{\end{equation}}
\def\beqa{\begin{eqnarray}}
\def\enqa{\end{eqnarray}}
\def\nnb{\nonumber}

\def\MeV{\nobreak\,\mbox{MeV}}
\def\GeV{\nobreak\,\mbox{GeV}}

\def\comq{\lag\bar{q}q\rag}
\def\uu{\lag\bar{u}u\rag}
\def\dd{\lag\bar{d}d\rag}

\def\mix{\lag\bar{q}g\si.Gq\rag}

\def\G3{\lag g^3G^3\rag}
\def\pli{p^\prime}

\def\la{\lambda}

\def\ga{\gamma}
\def\Ga{\Gamma}

\def\rh{\rho}
\def\si{\sigma}

\def\al{\alpha}

\def\lb{\label}
\def\nn{\nonumber}

\def\xsla{x\kern-.5em\slash}
\def\psla{p\kern-.5em\slash}
\newcommand{\rag}{\rangle}
\newcommand{\lag}{\langle}

\newcommand{\qq[1]}{\langle \bar{#1} #1 \rangle}
\newcommand{\GG}{\langle g_s^2 G^2 \rangle}
\newcommand{\qGq[1]}{\langle \bar{#1} G #1 \rangle}

\newcommand{\qqqq[1]}{\langle \bar{#1} #1 \bar{#1} #1 \rangle}
\newcommand{\GGG}{\langle g_s^3 G^3 \rangle}

\journal{Journal of Physics G}

\begin{document}
\begin{frontmatter}

\title{QCD Sum Rules Approach to the $X,~Y$ and $Z$ States }

\author[UERJ]{Raphael M. Albuquerque}
\ead{raphael.albuquerque@uerj.br}

\author[IF]{Jorgivan M. Dias}
\ead{dias@if.usp.br}

\author[UNIFESP]{K.~P.~Khemchandani}
\ead{kanchan.khemchandani@unifesp.br}

\author[IF]{A.~Mart\'inez~Torres}
\ead{amartine@if.usp.br}
  \author[IF]{Fernando S. Navarra}
\ead{navarra@if.usp.br}

\author[IF]{Marina Nielsen\corref{cor1}}
\cortext[cor1]{Corresponding author} \ead{mnielsen@if.usp.br}

\author[UERJ]{Carina M. Zanetti}
\ead{carina.zanetti@gmail.com}

\address[UERJ]{Faculdade de Tecnologia, Universidade do Rio de Janeiro (FAT, UERJ), Rod. Presidente Dutra Km 298, 27537-000, Resende, RJ, Brazil}
\address[IF]{Instituto de F\'isica, Universidade de S\~ao Paulo, Rua do Mat\~ao, Travessa R187, 05508-090 S\~ao Paulo, S\~ao Paulo, Brazil}
\address[UNIFESP]{Universidade Federal de S\~ao Paulo, C.P. 01302-907, Diadema, S\~ao Paulo, Brazil}



\begin{abstract}

  In the past decade, due to the experimental observation of many charmonium-like states, there has been a revival of  hadron spectroscopy. In particular, the experimental observation of charged charmonium-like, $Z_c$ states, and bottomonium-like, $Z_b$ states, represents a challenge since they can not be accommodated
within the naive quark model. These charged states are good candidates of either  tetraquark or molecular states and their observation motivated a vigorous theoretical activity.  This is a rapidly evolving field with enormous amount 
of new experimental information.  In this work, we review the current experimental progress and investigate various theoretical interpretations of these candidates of the multiquark states. The present review is written from the perspective of the QCD sum rules approach, where we present the main steps of concrete calculations and compare the results with other approaches and with experimental data. 

\end{abstract}

\date{\today}


\end{frontmatter}

\clearpage\tableofcontents
\vspace{1cm}
\section{\label{Introduction}Introduction}

If asked, most of the physicists today will say that protons and neutrons are made 
of quarks. Indeed, the constituent quark model is still widely used to 
represent all known hadrons and it has been valid for more than half century. 
On the other hand, more sophisticated  QCD-inspired models based on quarks 
and gluons have predicted the existence of more complex structures than 
simple mesons (quark-antiquark bound states) 
or baryons (3 quark bound states), which are called exotic states. 
The idea of unconventional quark structures is quite old and due to decades  
of investigation, the existence of exotic meson has been recently confirmed.

In the charmonium sector, below the open-charm threshold the $c \bar{c}$  charmonium states are  
successfully described using the quark model supplemented with quark potentials. 
All the predicted states have been observed 
with the expected properties below this  threshold and excellent 
agreement is achieved between theory and experiment. Indeed,  
theoretical models and experiments  achieved  an overall agreement of $2-3$~MeV/$c^2$ 
precision in the mass measurements of charmonium states. Above the open-charm 
threshold, however, there are still many predicted states that have not yet 
been discovered, and, surprisingly, several unexpected states have been 
observed since 2003. Interesting examples of these so - called (exotic) 
charmonium-like XYZ states are the axial-vector  $X(3872)$, the 
vector states $Y(4260)$, $Y(4360)$ and $Y(4660)$  and the charged state 
$Z_c(3900)^{\pm}$. This latter is a   manifestly exotic state. 
It became evident that charmonium-like states with more than a quark and an antiquark exist  
and several new models and possible interpretations have been advanced. These interpretations are 
still open,  
mostly due to the poor available statistics in the past experiments to  
investigate them,  i.e. to perform a full amplitude analysis.

The number of the exotic charmonium states has increased year by year. 
Up to now there are more than  twenty of these $X,~Y,~Z$ states. 
In Table~\ref{tab_summary} we give a list of these charmonium states. These states have been studied 
using different kind of models and there are several reviews about these studies 
 \cite{Jaffe:2004ph,Swanson:2006st,Zhu:2007wz,Klempt:2007cp,Voloshin:2007dx,Godfrey:2008nc,Nielsen:2009uh,
Brambilla:2010cs,Druzhinin:2011qd,Li:2012me,Liu:2013waa,Brambilla:2014jmp,Olsen:2014qna,Nielsen:2014mva,Esposito:2014rxa,Briceno:2015rlt,Hosaka:2016pey,Chen:2016qju,Esposito:2016noz,Guo:2017jvc}. An experimental review can 
be found in \cite{Yuan:2018inv}.

The study of spectroscopy and the decay properties of the heavy flavor 
mesonic states provides us with useful information about the dynamics of     
quarks and gluons at the hadronic scale.  One interesting question about the  
QCD dynamics refers to the existence of diquarks. Whether or not quarks form diquark clusters 
inside a baryon or in multiquark states, it has implications for the spectrum of the radial 
excitations. If diquarks are relevant, then the number of possible excitations is smaller. 
This fact may be verified experimentally. A systematic scan of the states lying in this energy region is 
now feasible. 
\begin{table*}[tbp]
{\scriptsize
  \caption{The $X,~Y$ and $Z$ states in the $c\bar{c}$ region 
ordered by mass. Masses $m$ and widths $\Gamma$ represent
the weighted averages from the listed sources, or are taken from \cite{pdg}
when available. The citation given in {\color{red} red} is for the first
observation and the citation given in {\color{blue} blue} is for a non
confirmation. The quoted year is the year of the first observation  and the given charge conjugation ($C$) of the isovector states is for the neutral state in the multiplet.
 } 
\begin{center}
\label{tab_summary}
\hspace*{0cm}
\begin{tabular}{lcccccc}
\hline\hline
\rule[10pt]{-1mm}{0mm}
 State & $m$~(MeV) & $\Gamma$~(MeV) & $J^{PC}$ & \ Process~(mode) & 
      \ experiment & Year \\
\hline
\rule[10pt]{-1mm}{0mm}
$X(3872)$& 3871.69$\pm$0.17 & $<1.2$ &
    $1^{++}$
    & $B\to K (\pi^+\pi^-J/\psi)$ &
    {\color{red} Belle} \cite{Choi:2003ue,Adachi:2008te,Choi:2011fc}, 
    BaBar~\cite{Aubert:2008gu} & 2003  \\[0.7mm]
& & & & $p\bar p\to (\pi^+\pi^- J/\psi)~(...)$ &
    CDF~\cite{Acosta:2003zx,Abulencia:2006ma,Aaltonen:2009vj}, \DZero~\cite{Abazov:2004kp} & \\[0.7mm]
& & &   & $B\to K (\omega J/\psi)$ &
    Belle~\cite{Abe:2005ix},
    BaBar~\cite{delAmoSanchez:2010jr} & \\[0.7mm]
& & & & $B\to K (\DstnDn)$ &
    Belle~\cite{Gokhroo:2006bt,Adachi:2008sua}, 
    BaBar~\cite{Aubert:2007rva} & \\[0.7mm]
& & & & $B\to K (\gamma J/\psi)$ &
    Belle~\cite{Abe:2005ix}, BaBar~\cite{Aubert:2006aj,Aubert:2008ae}&\\[0.7mm]
& & & & $B\to K (\gamma \psi(2S)$ & BaBar~\cite{Aubert:2008ae}, LHCb~\cite{Aaij:2014ala} & \\[0.7mm]
& & &   &  $e^+ e^- \to \pi^+ \pi^- J/\psi$ &
    BESIII~\cite{Ablikim:2013dyn}& \\[0.7mm]
& & & & $pp\to (\pi^+\pi^- J/\psi)~(...)$ &
    LHCb~\cite{Aaij:2011sn,Aaij:2013zoa}, CMS~\cite{Chatrchyan:2013cld}& \\[1.89mm]
$Z_c(3900)$ & $3886.6\pm2.4$ & 28.2$\pm$2.6 & $1^{+-}$ &
    $Y(4260) \to (J/\psi~\pi^+) \pi^-$ &
     {\color{red} BESIII}~\cite{Ablikim:2013mio}, Belle~\cite{Liu:2013dau}, CLEO-c \cite{Xiao:2013iha}]& 2013 \\[0.7mm]
     & & & & $Y(4260)\to (D \bar{D}^*)^+ \pi^-$ &
    BESIII~\cite{Ablikim:2013xfr} &  \\[1.89mm]
$Y(3940)$ & $3918.4\pm1.9$ & 20$\pm$5 & $0/2^{++}$ &
      $B\to K~(J/\psi\omega)$ &
    {\color{red} Belle}~\cite{Abe:2004zs},
    BaBar~\cite{Aubert:2007vj,delAmoSanchez:2010jr} & 2004\\[0.7mm] 
&&&&  $e^+e^-\to e^+e^- (\omega J/\psi)$ &
    {Belle}~\cite{Uehara:2009tx}, BaBar~\cite{Lees:2012me}&  \\[1.89mm]
$X(3940)$ & $3942^{+9}_{-8}$ & $37^{+27}_{-17}$ & $?^{?+}$ &
      $e^+e^-\to J/\psi\ (...)$ &
     {\color{red} Belle}~\cite{Abe:2007jna} & 2005\\[0.7mm] 
&&&&  $e^+e^-\to J/\psi\ (D D^*)$ &
      Belle~\cite{Abe:2007sya} & \\[1.89mm] 
$Y(4008)$ & $3891 \pm 42$ & 255$\pm$42 & $1^{--}$ &
     $e^+e^-\to\pi^+\pi^-J/\psi$ &
      {\color{red} Belle}~\cite{Yuan:2007sj,Liu:2013dau},
      {\color{blue} BESIII}~\cite{Ablikim:2016qzw}& 2007  \\[1.89mm]
$Z_c(4020)$ & $4024.1\pm1.9$ & $13\pm5$ & $?^{?-}$&
     $ e^+e^-\to \pi^-(\pi^+h_c)$ &
     {\color{red} BESIII}~\cite{Ablikim:2013wzq} & 2013 \\[0.7mm]
 & & & &
     $ Y(4260)\to \pi^-(D^*\bar{D}^*)^+$ &
     {BESIII}~\cite{Ablikim:2013emm} &   \\[1.89mm]
$Z_1(4050)$ & $4051^{+24}_{-43}$ & $82^{+51}_{-55}$ & $?^{?-}$&
     $ B\to K (\pi^+\chi_{c1}(1P))$ &
     {\color{red} Belle}~\cite{Mizuk:2008me}, {\color{blue} BaBar}~\cite{Lees:2011ik}& 2008 \\[1.89mm]
$Z_c(4055)$& $4054 \pm 3$ & $45 $ &     $(?^{?-})$
     &$e^+e^- \to \pi^- (\pi^+\psi(2S))$& {\color{red} Belle}~\cite{Wang:2014hta} & 2014 \\[1.89mm]
$Z_c(4100)$& $(4096\pm^{+28}_{-32})$& $152^{+70}_{-45})$ &$0^{++}/1^{-+}$&
     $B^0 \to K^+ (\pi^-\eta_c(1S))$& {\color{red} LHCb}~\cite{Aaij:2018bla} & 2018 \\[1.89mm]
$Y(4140)$ & $4146.8\pm2.4 $ & $22^{+8}_{-7}$ & $1^{++}$ &
      $B\to K (\phi J/\psi)$ &
     {\color{red} CDF}~\cite{Aaltonen:2009tz,Aaltonen:2011at}, D0~\cite{Abazov:2013xda}, LHCb~\cite{Aaij:2016iza}, {\color{blue} BESIII}~\cite{Ablikim:2014atq,Ablikim:2017cbv} & 2009  \\[1.89mm]
$X(4160)$ & $4156^{+29}_{-25} $ & $139^{+113}_{-65}$ & $?^{?+}$ &
     { $e^+e^- \to J/\psi (D^* \bar{D}^*)$} &
     {\color{red} Belle}~\cite{Abe:2007sya} & 2007 \\[1.89mm]
$Z_c(4200)$ & $4196^{+35}_{-30}$ & $370^{+99}_{-110}$ & $1^{+-}$&
     $B\to K (\pi^+\jpsi)$ &
     {\color{red} Belle}~\cite{Chilikin:2014bkk}     & 2014 \\[1.89mm]
$Y(4220)$ & $4218^{+5}_{-4}$ & $59^{+12}_{-10}$  & $1^{--}$ &
	$ e^+e^-\to \chi_{c0} \:\omega$ &
	{\color{red}BESIII}~\cite{Ablikim:2014qwy} & 2014 \\[0.7mm]
& &&&     
	$ e^+e^-\to h_c \:\pi^+ \pi^- $ &
	{BESIII~\cite{BESIII:2016adj}} &  \\[0.7mm]
& &&&     
	$ e^+e^-\to \psi(2S) \:\pi^+ \pi^- $ &
	{BESIII~\cite{Ablikim:2017oaf}}&  \\[0.7mm]
& &&&
	$ e^+e^- \to D^0 D^{\ast \:-} \:\pi^+ $ &
	{BESIII~\cite{Ablikim:2018vxx}} &  \\[1.89mm]
$Z_2(4250)$ & $4248^{+185}_{-\ 45}$ &   177$^{+321}_{-\ 72}$ &$?^{?+}$&
     $ B\to K (\pi^+\chi_{c1}(1P))$ &
     {\color{red} Belle}~\cite{Mizuk:2008me}, {\color{blue} BaBar}~\cite{Lees:2011ik}  & 2008 \\[1.89mm]
$Y(4260)$ & $4230\pm8$ & $55\pm$19 & $1^{--}$ &
     $e^+e^-\to\pi^+\pi^- J/\psi$ &
     {\color{red} BaBar}~\cite{Aubert:2005rm,Lees:2012cn}, CLEO-c~\cite{He:2006kg}, {Belle~\cite{Yuan:2007sj,Liu:2013dau}, BESIII~\cite{Ablikim:2016qzw}} & 2005\\[0.7mm] 
& & & & {$e^+e^-\to K^+ K^- J/\psi$} & { CLEO-c~\cite{Coan:2006rv}, BESIII~\cite{Ablikim:2013xfr}} & \\[0.7mm]
& & & & {$e^+e^-\to \pi^0\pi^0 J/\psi$} & { CLEO-c~\cite{Coan:2006rv}} & \\[0.7mm]
& & &&    
	$ e^+e^-\to Z_c(3900)^{\pm} \:\pi^{\mp}$ &
	{Belle~\cite{Liu:2013dau}, BESIII~\cite{Ablikim:2013mio}}
	&  \\[1.89mm]
$X(4350)$ & $4350.6^{+4.6}_{-5.1}$ & $13.3^{+18.4}_{-10.0}$ & ?$^{?+}$ &
     {$e^+e^-\to \phi J/\psi$} &
     {\color{red} Belle}~\cite{Shen:2009vs} & 2009 \\[1.89mm] 
$Y(4360)$ & $4368 \pm 13$ & 96$\pm$7 & $1^{--}$ &
     { $e^+e^-\to\pi^+\pi^- \psi(2S)$} &
     {\color{red} BaBar}~\cite{Aubert:2007zz,Lees:2012pv}, Belle~\cite{Wang:2007ea,Wang:2014hta}, BESIII~\cite{Ablikim:2017oaf} & 2007   \\[0.7mm]
& & & & $e^+e^-\to\pi^+\pi^- J/\psi$ &
    BESIII~\cite{Ablikim:2016qzw} & \\[1.89mm]
$Y(4390)$ & $4391.5^{+7.3}_{-7.8}$  & $139.5^{+16.3}_{-20.7}$ & $1^{--}$ &
	$ e^+e^- \to h_c \:\pi^+ \pi^-$ &
	{BESIII~\cite{BESIII:2016adj}} & 2016 \\[1.89mm]
$Z(4430)$ & $4478^{+15}_{-18}$ & $181\pm31$ & $1^{+-}$&
     $B\to K^- (\pi^+\psi(2S))$ &
     {\color{red} Belle}~\cite{Choi:2007wga,Mizuk:2009da,Chilikin:2013tch}, 
{\color{blue} BaBar}~\cite{Aubert:2008aa}, LHCb~\cite{Aaij:2014jqa}
     & 2007 \\[0.7mm]
& & & & $B\to K^-(\pi^+ J/\psi)$ &
    Belle~\cite{Chilikin:2014bkk}, BaBar~\cite{Aubert:2008aa} & \\[1.89mm]
$X(4630)$ & $4634^{+\ 9}_{-11}$ & $92^{+41}_{-32}$ & $1^{--}$ &
     { $e^+e^- \to \Lambda_c^+ \Lambda_c^-$} &
     {\color{red} Belle}~\cite{Pakhlova:2008vn} & 2008 \\[1.89mm]
$Y(4660)$ & 4643$\pm$9 & 72$\pm$11 & $1^{--}$ &
     {$e^+e^-\to\pi^+\pi^- \psi(2S)$} &
     {\color{red} Belle}~\cite{Wang:2007ea,Wang:2014hta}, 
	BaBar~\cite{Lees:2012pv} & 2007\\[0.7mm]
& & &&    
	$ e^+e^-\to \Lambda_c^+ \Lambda_c^-$ &
	{BESIII~\cite{Pakhlova:2008vn}} &  \\[1.89mm]
\hline\hline
\end{tabular}
\end{center}
}
\end{table*}
In the case of some of the X and Y states,  we can say that there are still 
attempts to interpret them as  $c - \bar{c}$. One can pursue this approach  
introducing corrections in the potential, such as quark pair creation. This 
``screened potential'' changes 
the previous results, obtained with the unscreened potential and allows to 
understand some of the new data in the  $c - \bar{c}$ approach. Departing 
from the $c - \bar{c}$ assignment, the next option  is a system composed by       
four quarks, which can be uncorrelated, forming a kind of bag, or can be grouped  
in diquarks which 
then form a bound system. These configurations are called tetraquarks. 
Alternatively,  these four quarks can  form two mesons which then interact 
and form a bound 
state. If the mesons contain only one charm quark or antiquark, 
this configuration is referred to as a molecule. 
If one of the 
mesons is a charmonium, then the configuration is called hadro-charmonium. 
Another possible configuration is a hybrid charmonium.
In this case, apart from the 
$c - \bar{c}$ pair, the state contains excitations of the gluon field.  
In some 
implementations of the hybrid, the excited gluon field is represented by a 
``string'' or 
flux tube, which can oscillate in normal modes. 

The above mentioned configurations  are quite different 
and are governed by different dynamics. 
In quarkonia states the quarks have a short range 
interaction 
dominated by one gluon exchange and a long range non-perturbative confining 
interaction, 
which is often parametrized by a linear attractive potential. In tetraquarks 
besides these 
two types of interactions, we may have a diquark-antidiquark interaction, 
which is not very well 
known. In molecules and hadro-charmonium the interaction occurs through  
meson exchange. 
Finally, in some models inspired by lattice QCD results, there is a flux 
tube formation 
between color charges and also string junctions. With these building blocks 
one can construct 
very complicated ``stringy'' combinations of quarks and antiquarks and their 
interactions 
follow the rules of string fusion and/or recombination. 
In principle, the knowledge of the interaction should be enough to determine 
the spatial 
configuration of the system. In practice, this is only feasible in simple 
cases, such as the 
charmonium in the non-relativistic approach, where having the potential one 
can solve the 
Schr\"odinger equation and determine the wave function. In other approaches 
the spatial 
configuration must be guessed and it may play a crucial role in the production 
and decay of 
these states.

All recent analyses performed  on exotic states show statistics limitation, 
not allowing a final conclusion. It is definitively necessary to upgrade all 
the experiments in order to have more statistics. One example of upgrade is the 
project Belle II.  In 2018 the first collisions happened, probably marking the 
beginning of a new era for the $e^+ e^-$ colliders, which will last at least  
ten years. With the expected high luminosity, Belle II can improve for sure some 
of the measurements already performed by Belle, and look for new still  
undisclosed forms of exotic matter. It will be possible to search for more rare 
decays, up to now not possible due to the limited statistics. With such high 
statistics amplitude analysis can be performed and the quantum numbers can be 
determined.

A comparison of running and future experiments can be found in 
recent papers and help in understanding the  future opportunities in spectroscopy.

A good reason to  write a report on the subject  is because this is a 
rapidly evolving field with enormous amount  
of new experimental information coming from the analysis of BELLE II, BESIII and  
LHCb accumulated data. In the present review we include and discuss  
data which were not yet available to the previous reviewers. 
This astonishing progress on  the experimental side  has opened up new  
challenges in the understanding of heavy flavor  hadrons and from time to 
time it is necessary to organize the theoretical and experimental advances in 
short review papers. This is the goal of this text.

Any theoretical review is biased and naturally emphasizes the approach 
followed by the   
authors. We focus on the theoretical developments and more specifically on the  
works done with QCD sum rules (QCDSR). The present review is   written from  
the perspective of QCD sum rules, where we present the main steps  
of concrete calculations and compare the results with other approaches and with  
experimental data. 
In what follows we will review and comment the work presented in   
Refs.~\cite{Matheus:2004gx,Navarra:2006nd,Matheus:2006xi,Lee:2007gs,Lee:2008tz,Lee:2008gn,Albuquerque:2008up,Bracco:2008jj,Lee:2008uy,Albuquerque:2009ak,Matheus:2009vq,Albuquerque:2010fm,Nielsen:2010ij,Narison:2010pd,Finazzo:2011he,Zanetti:2011ju,Dias:2011mi,Albuquerque:2011ix,Sun:2012sy,Dias:2012ek,Dias:2013xfa,Torres:2013saa,Dias:2013qga,Khemchandani:2013iwa,Torres:2013lka,Albuquerque:2013owa,Albuquerque:2015nwa,Albuquerque:2015kia,Torres:2016oyz,Albuquerque:2016znh,Albuquerque:2017vfq,Albuquerque:2018jss}.

In the next Section we review the basic concepts of the QCDSR method  and, in Sec. 2.9, we discuss some limitations of the application of QCDSR to the exotic states.
In Section 3 we describe the progress achieved in the study of the $X(3872)$, which is the 
best known exotic charmonium state. Section 4 is dedicated to the vector exotic states $Y$.
In Section 5 we review the electrically charged $Z$ exotic states. In Section 6 we present an 
updated discussion of the controversial $Y$ states, i.e., those which need confirmation.
In all the Sections, we give a brief experimental introduction, a short review of the theoretical 
works on the states and then, in more detail, their interpretation in QCD Sum Rules.  Finally, in
Section 7 we present the recent theoretical developments in QCDSR involving perturbative 
$\alpha_s$ corrections and, in the end, we finish with our concluding remarks in the summary. All the 
sections are to some extent self-contained.  
Somewhat inspired by the PDG, this text allows the fast reader to go directly to the particle of 
his interest and look for recent information.

\section{\label{qcdsr}QCD sum rules technique}

In this section, we  discuss in detail
one of the most active nonperturbative approach in 
Quantum Chromodynamics (QCD), which is also an entirely 
analytical tool for obtaining valuable information about 
hadronic states, called QCD sum rules (QCDSR) or also 
known as SVZ sum rules after Shifman, 
Vainstein and Zakharov, in 1979 \cite{Shifman:1978bx,
Shifman:1978by}. 

The QCD sum rules approach \cite{Shifman:1978bx,Shifman:1978by, Reinders:1984sr, Narison:2007spa, Narison:1989aq, Narison:1980ti} allows us to extract 
properties of  hadronic states from QCD 
parameters like quark masses, QCD coupling, 
and QCD condensates. In contrast to some 
nonperturbative approaches in QCD, for instance, 
potential models, the QCDSR is an analytic method 
and fully relativistic \cite{Narison:2002pw}. 
Over the last decade, this approach has been 
successfully used to describe the new hadronic 
states in the charmonium and bottomonium 
spectrum. It was also employed to investigate 
the exotic structure of some states, recently 
observed by the LHCb collaboration 
\cite{Aaij:2015tga,Aaij:2016phn,Aaij:2016ymb}, 
which are candidates to be of Pentaquark nature 
\cite{Chen:2016qju}. 
It has been largely used in many applications, 
different to testing the exotic nature of the 
hadronic matter. In the beginning, it was applied 
just to investigate mesons and its extension 
to baryons was done afterwards by Ioffe 
\cite{Ioffe:1981kw}. There are some textbooks 
\cite{Narison:2002pw,Shifman:2001ck} 
and articles \cite{Radyushkin:1998du,
Novikov:1977dq,Reinders:1984sr,Shifman:1998rb} 
in which the initial basic aspects of the 
method are discussed in many details, as well as the 
extensions to studies involving the nuclear medium 
\cite{Cohen:1994wm}.

\subsection{Correlation Functions}

The approach focuses on the correlation 
functions of local 
composite operators. A generic two-point
correlation function is given by
\beq
\Pi(q)\equiv i\int d^4 x\, e^{iq\cdot x}
\lag{0}| T [j(x)j^\dagger(0)]|0\rag\ ,
\label{cor}
\enq
where $j(x)$ is  the local composite operator, $\langle ... \rangle$ is the QCD vacuum expectation value, while $T$ is the time 
ordered product between the operators. These operators are 
build up from quarks and gluons fields 
in such a way that they carry the quantum 
numbers of the hadron under investigation. 
We often refer to such operators as 
interpolating fields or currents. The interpolating currents 
for non-exotic states have the following generic expression:
\begin{equation}\label{jnonex}
j_n(x)=\bar{q}_a(x)\Gamma_n q_a(x)\, ,
\end{equation}
where $q(x)$ is the spinor representing the quark 
field,  $a$ stands for the color index and $\Gamma_n$ is any structure made of 
Dirac matrices, $\Gamma_n=1,\,\gamma_{\mu},\,\gamma_{\mu}
\gamma_{\nu},\,\gamma_{\nu}\gamma_5\, .\,.\,.$, which allows us to
characterize the 
tensorial structure of the current. For instance, 
$\Gamma_0=1$ gives us $j(x)=\bar{q}_a(x)q_a(x)$ 
and describes a scalar meson with $J^P=0^+$, while 
$\Gamma_{\mu}=\gamma_{\mu}$ gives
$j_{\mu}(x)=\bar{q}_a(x)\gamma_{\mu}q_a(x)$, 
that is used to describe a vector meson with $J^P=1^-$. Hence 
we can choose the scalar, vectorial or tensor 
character of the interpolator by simply 
choosing the appropriate gamma matrix $\Gamma_n$. 
The quark content is determined by the flavor of the hadron. For instance,
the current $\bar{c}_a(x)
\gamma_{\mu}c_a(x)$ describes a meson with charm-anti-charm quark content
with $J^p=1^-$, and can be used to study, within the QCDSR approach, 
the $J/\psi$ meson.

From the Quark-Hadron 
duality concept \cite{Narison:2002pw,Shifman:2000jv,Shifman:2001qm}, 
which establishes a correspondence 
between two different descriptions of 
the correlation functions, we can match 
a QCD description of a correlation function 
with a phenomenological one. More 
specifically, we can take into account 
the QCD degrees of freedom, quarks and 
gluon fields, and calculate the correlation 
function using Wilson's Operator Product 
Expansion (OPE) \cite{Wilson:1969zs} in 
order to separate the physics of short 
and long-range distances. In this description, 
the correlator is expressed as the sum of 
coefficients (called Wilson's coefficients), 
which are c-numbers, multiplied by the expectation 
values of composite operators, which give 
rise to the condensates, for instance, the quark 
and gluon condensates. It is through those 
QCD vacuum condensates that the nonperturbative 
effects are included into the method. By 
doing this, we are calculating the QCD side 
(also called OPE side) of the QCDSR. 
On the other hand, we can also consider 
the interpolating fields in the correlation 
functions representing the hadronic degrees 
of freedom, in such a way that they are no 
longer the quarks/gluon operators, but 
are associated to the creation/annihilation 
operators for the hadron itself. This description 
is usually called the Phenomenological side of the 
QCDSR. Assuming that there is an interval in momentum 
for which the QCD side and the Phenomenological 
one are equivalent, we can compare both sides 
and extract the hadronic parameter we are interested with.

Writing the correlation functions is also the 
starting point of other nonperturbative 
QCD approaches like Lattice QCD calculations 
(in the Euclidean coordinate space) and, in 
this case, the uncertainties can be 
systematically improved \cite{Cohen:1994wm}. 
In the QCDSR technique 
we make use of some phenomenological inputs, 
limiting the accuracy of the method to be around $10\%-20\%$ 
\cite{Narison:2002pw,Leinweber:1995fn}. 
This estimate can get worst when we 
take into account the expectation values of higher 
dimensional operators on the OPE side, since, as we will 
discuss later, we have to assume some factorization 
hypothesis, i.e., we replace the expectation values 
of higher dimensional operators by the products of the 
lower dimensional ones. This is one of the main source of 
uncertainty of the method.

\subsection{The QCD side}
\label{s-ope}

As we are considering the quarks and gluons fields 
as the building blocks, i.e., the QCD degrees of 
freedom, we have to deal with the effects of 
soft gluons and quarks fields populating the QCD 
vacuum. In other words, we have to take into 
account the complex structure of the QCD vacuum. 
This means that the expectation values of 
the operators associated with those fields are 
non-zero, giving rise to what we call condensates. 
One way to deal with this feature of the QCD vacuum 
is to use the OPE~\cite{Wilson:1969zs}. 

As mentioned before, the correlation 
functions can be written as a sum of Wilson's 
coefficients times the expectation values of the 
composite operators. The perturbative 
part is encoded into those coefficients, 
which are obtained using the perturbative 
QCD. The information on the 
complex structure of the QCD vacuum is  
in the condensates, i. e., the nonperturbative 
effects due to the QCD vacuum is contained into 
the expectation values of the composite operators. 
Therefore, using the OPE we have a clear separation 
of scales, that is, the short distances effects 
in the coefficients, and the long-range 
ones due to the condensates. Therefore, we can write
\begin{equation}\label{opedef}
\Pi(q)= i\int d^4 x\, e^{iq\cdot x}
\lag{0}| T[j(x)j^\dagger(0)|0\rag\ = \sum_d\, C_d(Q^2)\,\langle \hat{O}_d
\rangle \, ,
\end{equation}
where $C_d(Q^2)~(Q^2=-q^2)$ is the Wilson's coefficients, 
and $\langle \hat{O}_d \rangle$ is the 
expectation value of the composite local operators.
This series is ordered by the dimension of the operator, 
denoted by the $d$ index. The lowest-dimension
operator with $d=0$ is the unit operator associated with the perturbative
contribution: $C_0(Q^2)=\Pi^{per}(Q^2)$, $\hat{O}_0=1$.
Since one cannot build gauge invariant composite 
operators for $d=2$ \cite{Dominguez:2013ata}, 
the next term in the expansion of Eq.~\eqref{opedef} 
is for $d=3$.
Considering only the lowest dimension operators in the OPE, one obtains:
\begin{eqnarray}
   \hat{O}_{3} ~~=~~ &:\!\bar{q}(0) q(0)\!:& ~~\equiv~~ \bar{q}q \nonumber \\
   \hat{O}_{4} ~~=~~ &:\!g_s^2 \:G^{N}_{\alpha\beta}(0) G^{N}_{\beta\alpha}(0)\!:& 
   	~~\equiv~~ g_s^2 G^2 \nonumber \\
   \hat{O}_{5} ~~=~~ &:\!\bar{q}(0) \: g_s \:\sigma^{\alpha\beta} G_{\beta\alpha}(0)\: q(0)\!:&
   	~~\equiv~~ \bar{q}Gq \nonumber \\
   \hat{O}^q_{6} ~~=~~ &:\!\bar{q}(0)q(0) \: \bar{q}(0) q(0)\!:& ~~\equiv~~ \bar{q}q\bar{q}q \nonumber \\
   \hat{O}^G_{6} ~~=~~ &:\!f_{_{NMK}} g_s^3 \:G^{N}_{\alpha\beta}(0) 
   	G^{M}_{\beta\gamma}(0) G^{K}_{\gamma\alpha}(0)\!:& ~~\equiv~~ g_s^3 G^3
\end{eqnarray}
where the symbol $: :$ represents the normal ordering of the operators, $q(0)$ is the quark field, 
$G^{N}_{\alpha\beta}(0)$ is the gluon field tensor, $f_{_{NMK}}$ is the structure constant of the SU(3) group 
and $\sigma_{\!\alpha\beta}=\frac{i}{2} ~[ \gamma_{\alpha}, \gamma_{\beta} ]$. 
The vacuum expectation value (VEV) of these local operators
\begin{equation}
  \langle 0| \:\hat{O}_n\: |0 \rangle ~,
\end{equation}
gives the quark condensate $\qq[q]$, gluon condensate $\GG$, mixed condensate 
$\qGq[q]$, four-quark condensate $\qqqq[q]$ and the triple gluon condensate $\GGG$.
In general, the quark condensate and the gluon condensate are enough to
reliably investigate non-exotic mesonic systems 
\cite{Nielsen:2009uh}, for instance, the $J/\psi$ 
meson \cite{Shifman:1978bx,Shifman:1978by,Reinders:1980wy,
Reinders:1981si}. Contrarily, for exotic 
systems like the ones situated in 
the charmonium spectrum, the $X(3872)$ for example, one has to 
go a step further in the expansion, including the
terms of higher dimensions~\cite{Matheus:2006xi}.

\begin{figure}[h]
\begin{center}
\scalebox{0.4}{\includegraphics{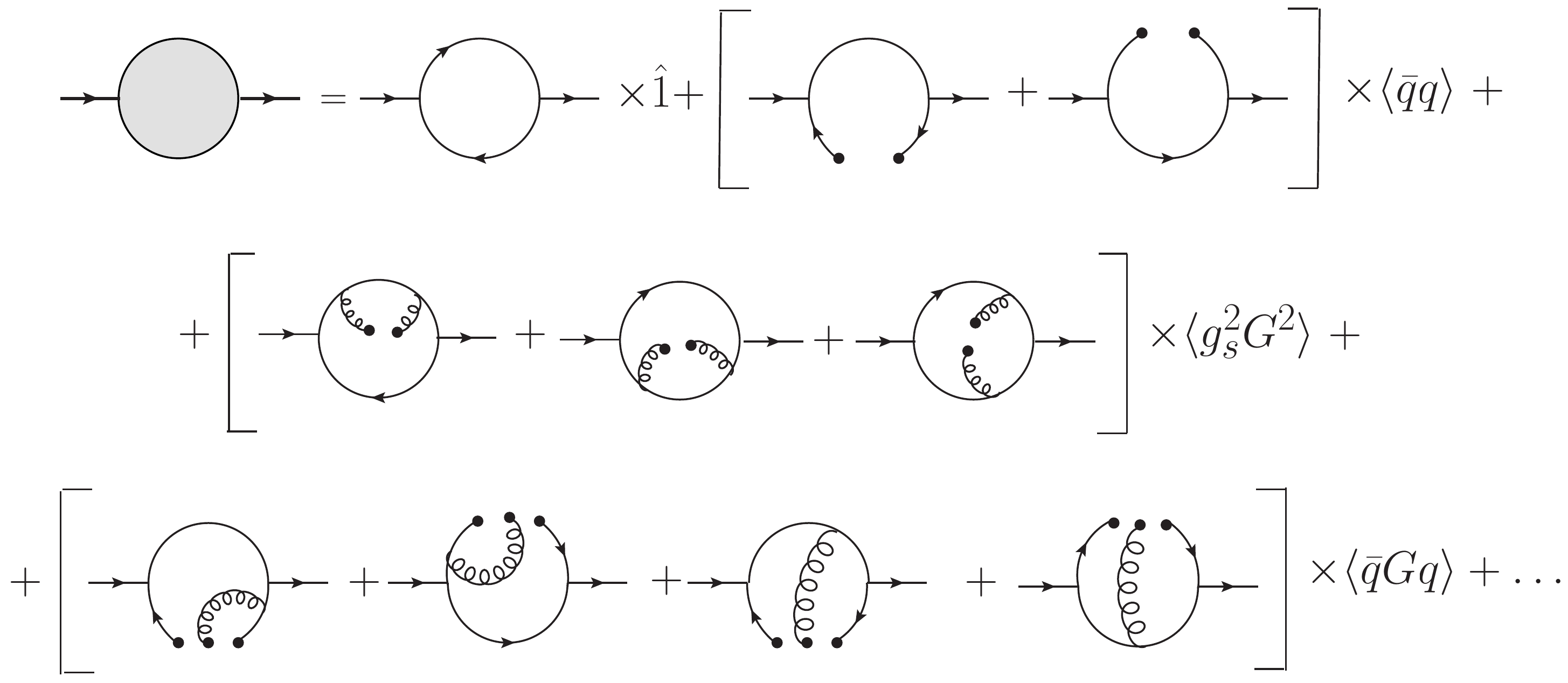}}
\end{center}
\caption{The OPE for the two-point function 
Eq.~\eqref{exp8}.}
\label{OPEexp}
\end{figure}

Taking, as an 
example, the $2$-point function for a scalar state, the OPE for this case will be
(up to dimension-5):
\begin{eqnarray}\label{exp8}
\Pi(Q^2) &=& C_0(Q^2)\, 1 + C_3(q^2)
 \langle \bar{q}q\rangle + C_4(q^2)\langle 
g_s^2 G^2  \rangle 
+ C_5(q^2)\langle \bar{q}\,G\,q\rangle \;.
\end{eqnarray}
The coefficients $C_0,\, C_3,\,C_4,\, C_5$ 
multiplying the condensates in Eq.~\eqref{exp8}, 
are obtained  by calculating the Feynman diagrams, 
whose topology depends on the particular choice 
of the interpolating current $j$. Examples of 
such diagrams can be seen in Fig.~\ref{OPEexp}.

The values for the condensates are not 
determined directly from the experiment 
and, in general, we extract them from Lattice 
QCD calculations or, as in the case of quark 
and gluon condensates, they can be obtained 
from a QCDSR calculation for a given well-known 
state. 
For instance, the gluon condensate was estimated 
in Ref.~\cite{Narison:2010cg}, where the authors  used the 
QCDSR method to investigate the charmonium system. 
As a result, they obtained $\langle g^2\, G^2\rangle= (7.5\pm 2.0) 
\times 10^{-2}$ GeV$^{4}$. The quark condensate 
can be estimated in terms of the pion mass 
$m_{\pi}$ and its decay constant $f_{\pi}$ 
through the Oakes-Renner relation 
\cite{GellMann:1968rz}, 
\begin{equation}
\langle \bar{q}q\rangle = -\frac{f_{\pi}^2m_{\pi}^2}{(m_u+m_d)}\, ,
\end{equation}
where $m_u(m_d)$ is the quark up (down) mass. 
Using the following values for the pion mass 
and its decay constant, $f_{\pi}=93.0$ MeV and 
$m_{\pi}=139.0$ MeV \cite{Tanabashi:2018oca}, 
with $m_u+m_d=14$ MeV, we have 
$\langle\bar{q}q\rangle=-(0.23\pm 0.03)^3$ GeV$^3$. 

On the other hand, for higher dimensional 
condensates, as is the case for $d=6$ and $d=8$, i. e.,  
four-quark condensate, and eight-dimensional 
condensate, respectively, we have to assume 
that they can be factorized and their saturation 
values are: $\langle\bar{q}q\bar{q}q\rangle
=\langle\bar{q}q\rangle^2$, $\langle\bar{q}q\bar{q} g\sigma . 
G q\rangle= \langle\bar{q}q\rangle\langle\bar{q} g\sigma . 
G q\rangle$. In order to give an estimation 
of their precise values, a more involved analysis 
has to be done, once it is related to a non-trivial 
choice of factorization scheme \cite{Bagan:1984zt}. 
In general, one introduces a parameter $\rho$, 
that assumes values from $1$ up to $2.1$, which 
account for deviations of the factorization 
hypothesis, with $\rho=1$ related to vacuum 
saturation values and $\rho=2.1$ to the violation 
of the factorization hypothesis assumption 
\cite{Narison:1989aq,Launer:1983ib,Narison:2009vy,Braghin:2014nva}. 
This approximation also contributes to the uncertainties 
of the sum rule calculation.

In conclusion, on the OPE side, we calculate the correlator in terms of 
the OPE series, and we have to consider the contributions from condensates that are sufficient to guarantee a good OPE convergence. The next step is to write the correlator using the K\"allen-Lehmann representation or the dispersion relation:
\begin{eqnarray}\label{cope}
  \Pi^{OPE}(q^2) &=& \int\limits_{s_{min}}^{\infty} \frac{\rho^{OPE}(s)}{s - q^2} \:ds 
\, ,
\end{eqnarray}
where
\beq
\rho^{OPE}(s)=\frac{1}{\pi}\mbox{Im}[\Pi^{OPE}(s)]\;,
\enq
is the spectral density function and $s_{min}$ stands for a kinematical 
limit in the integral. In the case of interpolating 
currents with two heavy quarks (charm and bottom), $s_{min}=4m^2_{c(b)}$, with 
$m_{c(b)}$ the charm (bottom) quark mass.

\subsection{The phenomenological side}

In this case, the correlation function is 
evaluated considering the hadron itself as the 
degree of freedom, i. e., the operators 
in terms of which the correlation function is 
defined is now represented by the hadronic state we are interested in.
 Again, let us consider the two-point 
function as an example, with the interpolating 
current $j$ associated with the operator that 
creates/annihilates the hadron under investigation.
We have
\begin{equation}\label{2p}
\Pi^{phen}(q^2)= i\int d^4x \, e^{ipx}
\langle 0|T\Big(j(x)j^{\dagger}(0)\Big)|0\rangle \, .
\end{equation}
Using the Kallen-Lehmann representation, Eq.~\eqref{2p} is 
written as
\begin{equation}\label{klr}
\Pi^{phen}(q^2)=\int_{0}^{\infty}\, ds \frac{\rho(s)}{s-q^2+i\epsilon} +\textrm{ subtraction terms}\, ,
\end{equation} 
where $\rho(s)$ is the spectral function. 
It is defined as $\rho(s)=\sum_n |\langle n|j(0)|0\rangle|^2$, 
and it means that all intermediate states 
coupling to the operator $j$  contribute to 
the integral in Eq.~\eqref{klr}. The spectral 
representation is a special case of dispersion 
relations. Within the QCDSR approach, dispersion 
relations are useful since it encodes the 
Quark-Hadron duality, which is the underlying 
concept of the QCDSR method. Therefore, the use 
of dispersion relations will connect the QCD side 
with the Phenomenological side, allowing to extract 
the hadronic observables from the QCD parameters.

In order for the QCDSR technique to be useful, one must parametrize
$\rho(s)$ with a small number of parameters. For this, we recall that in a hadron spectrum, in general, the lowest resonance
is often fairly narrow, whereas higher-mass states are broader.
Therefore, one can parameterize the spectral density as a single sharp
pole representing the lowest resonance of mass $m_H$, plus a smooth continuum
of resonances representing higher mass states:
\beq
\rho(s)=\lambda^2\delta(s-m_H^2) +\rho^{cont}(s)\,.
\label{den}
\enq
Using this equation into Eq.~\eqref{klr}, we 
get the following expression for the Phenomenological 
side
\begin{equation}\label{phen1}
\Pi^{phen}(q^2)=\frac{\lambda^2}{q^2-m_H^2}+\int_{s_{min}}^{\infty}\,
ds\,\frac{\rho^{cont}(s)}{s-q^2} +\textrm{ subtraction terms}\, ,
\end{equation}
where $\lambda$ measures the coupling of 
the low mass hadron state $|H\rangle$ to the interpolating 
current $j$, i. e.
\beq
\lag 0 |j|H\rag =\lambda.
\label{cou}
\enq

For simplicity, one often assumes that the continuum contribution to the
spectral density, $\rho_{cont}(s)$ in Eq.~(\ref{den}), vanishes below a
certain continuum threshold $s_0$. Above this threshold, it is assumed to be
given by the result obtained with the OPE.  Therefore, one uses the ansatz
\beq
\rho_{cont}(s)=\rho^{OPE}(s)\Theta(s-s_0)\;.
\enq
Hence, we finally get
\begin{equation}
\Pi^{phen}(q^2)=\frac{\lambda^2}{q^2-m_H^2}+\int\limits_{s_0}^{\infty}\,
ds\,\frac{\rho^{OPE}(s)}{s-q^2}+\textrm{ subtraction terms}\,
\end{equation}

As we should discuss later, the second term 
on the RHS of Eq.~\eqref{phen1} is suppressed, 
when a Borel transform is applied, then we can 
extract the $\lambda$ as well as the mass of 
the low-lying state coupling to the current $j$. 
In the next subsections, we are going to show 
that we can extract mass of the hadronic state 
under investigation from the two-point functions. 
Three-point functions are useful to get the form 
factors, which provide the coupling we need to 
know in order to determine the decay widths.

\subsection{Borel Transform}

According to the Quark-Hadron duality, 
the sum rule is obtained by equating the 
correlator evaluated on the OPE side with  the one written on the Phenomenological 
side:
\begin{eqnarray}\label{QHdual}
  \Pi^{OPE}(Q^2) &=& \Pi^{phen}(Q^2)\;.
\end{eqnarray}
The validity of Eq. (\ref{QHdual}) gives us information about the properties of hadrons in terms of the QCD variables. However, such a duality is weakened by: i) the presence 
of subtraction terms which appear as unknown polynomials in $Q^2$; ii) dominance of excited state contributions 
in comparison with the lowest hadronic state contribution; and iii) the truncation of the OPE. In order to overcome these problems and to obtain a self-consistent and a reliable match between OPE 
and Phenomenological sides, the authors in Refs.~\cite{Shifman:1978bx,Shifman:1978by} suggested to perform a Borel transformation on both sides 
of the sum rule. The Borel transformation (also known as inverse Laplace transformation) is defined as:
\begin{equation}\label{borel}
\Pi(M^2)=\mathcal{B}[\Pi(Q^2)]\equiv 
\lim\limits_{\tiny \begin{matrix} Q^2, n 
\rightarrow \infty \\ Q^2/n = M^2 \end{matrix}} 
\frac{(Q^2)^{n+1}}{(n)!}\Big(-\frac{d}{dQ^2}\Big)^n\, \Pi(Q^2)\,,
\end{equation}
where the parameter $M^2$ is often called as the
Borel mass. Some typical and useful examples of the Borel transformation are:
\begin{eqnarray}
\mathcal{B}[(Q^2)^k]&=&0\,,\\
\mathcal{B}\Big[\frac{1}{(Q^2)^k}\Big]&=&\frac{(-1)^k}{(k-1)!(M^2)^{k-1}}\,,\\
\mathcal{B}\Big[\Big(\frac{1}{s+Q^2}\Big)^k\Big]&=&\frac{1}
{(k-1)!}\Big(\frac{1}{M^2}\Big)^{k-1}\,e^{-s/M^2}\, .
\end{eqnarray}
Evidently, the Borel transformation kills any eventual subtraction terms in the correlators and suppresses exponentially 
the continuum contribution, improving the convergence of the dispersion integral. Furthermore, it suppresses factorially 
the higher-order operators in OPE, which contain inverse powers of $Q^2$, justifying the truncation of the OPE and favoring a good OPE convergence.

\subsection{QCD sum rules: Two-point function and Mass}
\lb{sr-sta}

After transferring the continuum contribution to the OPE side, and performing a 
Borel transformation on both sides, the sum rule can be written as
\begin{equation}\label{srmass}
\lambda^2\, e^{-m_{H}^2/M^2}=\int_{s_{min}}^{s_0}\,ds\,\rho^{OPE}(s)\, 
e^{-s/M^2}\, .
\end{equation}

By taking the derivative of Eq.~\eqref{srmass} 
with respect to $1/M^2$ and dividing the result 
by Eq.~\eqref{srmass}, we obtain 
\begin{equation}\label{sumrule}
m^2_{H}=\frac{\int\limits_{s_{min}}^{s_0}\,ds\,s\,\rho^{OPE}(s)\,e^{-s/M^2}}
{\int\limits_{s_{min}}^{s_0}\,ds\,\rho^{OPE}(s)\,e^{-s/M^2}}\, .
\end{equation}

To extract reliable results from Eq. (\ref{sumrule}), it is necessary to work in a region with a $M^2$-stability, a 
dominance of the lowest hadronic state (or pole dominance) over the continuum contribution, and a good OPE 
convergence. When it is possible to find a Borel range of $M^2$ where all above mentioned requirements are fulfilled, then we can 
define the so-called Borel window. Notice that the Borel mass, $M^2$, is intrinsically related to the energy scale of the 
hadronic system. Therefore, considering higher values for $M^2$ would correspond to the higher-energy regime 
where the continuum contribution dominates. In order to avoid this region, one usually sets an upper bound on the Borel 
mass value, $M_{{max}}^2$, where the pole dominance is guaranteed. A maximum value is determined by 
imposing the condition that the pole contribution is equal to the continuum contribution. On the other hand, considering smaller 
values for $M^2$, implies working in the low-energy regime, where the truncated OPE no longer provides a 
reasonable information on the lowest hadronic state. The non-perturbative effects become extremely important and 
higher dimension condensates must be included in the OPE series. Then, one naively expects that there is a minimum 
value in the Borel mass, $M_{{min}}^2$, which still provides a good OPE convergence. Typically, one defines the 
$M_{{min}}^2$ value where the contribution of the higher dimension condensates in the OPE is smaller than 
10\% to 25\% of the total contribution. This can be controlled by the $\epsilon_{N}$-parameter
\begin{eqnarray}
  \epsilon_N &\equiv& \left| 1 - \frac{\mbox{OPE}_{N-1}}{\mbox{OPE}_N} \right| ~=~ 0.10 ~\mbox{to}~ 0.25 ~.
\end{eqnarray}
where $\mbox{OPE}_{N}$ is the summation up to the $N$-dimension operator in the OPE series.
Finally, one expects that the hadron mass has a certain stability in the $M^2$ 
parameter inside the Borel window. Then, Borel windows with a large $M^2$-instability could indicate that the 
obtained hadron mass is not reliable, and more improvements must be done for these sum rule calculations. 
Sometimes, the inclusion of more condensates in the OPE improves such a $M^2$-stability. If one can not find a Borel window, then the QCDSR method can not be used to
draw any conclusions.

Another important point is the choice of the continuum threshold, $s_0$. It is a physical parameter that 
should be determined from the hadronic spectrum. Using a harmonic-oscillator potential model, it was shown in 
Ref.~\cite{Pascual:1984zb} that a constant continuum threshold is a very poor approximation. 
The actual accuracy of the parameters extracted from the sum rules improves considerably when using a Borel 
dependent continuum threshold. It also allows to estimate realistic systematic errors \cite{Lucha:2007pz}. 
However, to be able to fix the form of the Borel dependent continuum threshold, one needs to use the experimental 
value of the mass of the hadron. Since in our study we want to determine the mass of the state and not to use the experimental value, it is not possible to fix the Borel dependent continuum 
threshold. For this reason, although aware of the limitations of the values we extract mass from the sum rule, as 
 a first estimate for such states, by using a constant continuum threshold. 
In many cases, a good approximation for the value of the continuum threshold is the value of the mass of the first 
excited state squared. In some known cases, like the $\rho$ and $J/\psi$ mesons, the first excited state has a mass 
approximately $0.5 ~\mbox{GeV}$ above the ground state mass. As we do not know, in principle, the spectrum of the 
hadrons we  study, the range for the continuum threshold is  fixed to be the smallest value which 
provides a valid Borel window. The optimal choice for $s_0$ is taken when there is a $M^2$-stability inside the 
Borel window. Therefore, the sum rule calculation that respects these optimal criteria, can reliably extract the mass of 
hadronic states through Eq. (\ref{sumrule}).

\subsection{QCD Input Parameters}
We consider here the same values for the quark masses and condensates as used
in 
Refs.~\cite{Matheus:2006xi,Lee:2007gs,Lee:2008gn,Lee:2008tz,Bracco:2008jj,Albuquerque:2008up,Lee:2008uy,Narison:2004vz}, listed in Table \ref{QCDParam}.

\begin{table}[h]
\begin{center}
\setlength{\tabcolsep}{1.25pc}
\caption{QCD input parameters.}
\begin{tabular}{ll}
&\\
\hline
Parameters&Values\\
\hline
$m_b$ & $(4.24 \pm 0.05) \GeV$ \\
$m_c$ & $(1.23 \pm 0.05) \GeV$ \\
$m_s$ & $(0.13 \pm 0.03)\GeV$ \\
$\qq[q]$ & $-(0.23 \pm 0.01)^3\GeV^3$\\
$\langle\al_s^2 G^2\rangle$ & $0.88~\GeV^4$\\
$\kappa \equiv \qq[s] / \qq[q]$ & $(0.74 \pm 0.03)$\\
$m_0^2 \equiv \qGq[q] / \qq[q]$ & $(0.8\pm0.2\GeV^2$\\
\hline
\end{tabular}
\label{QCDParam}
\end{center}
\end{table}

\subsection{QCD sum rules: Three-point function and Decay Width}

The use of three-point or vertex 
functions in QCDSR technique 
is related to the decay width, where a coupling constant is 
involved.  Consider, for instance, 
the hadronic decay process $H_1(p)\to H_2(p^{\prime})\,H_3(q)$,
in which a given hadronic state $H_1(p)$, 
with four-momentum $p$, decays into two 
hadrons $H_2(p^{\prime})$ and $H_3(q)$ each with momentum 
$p^{\prime},\, q$, respectively. 
The three-point function associated 
with this vertex is written as
\begin{equation}\label{3pdef}
\Pi(p^2,p^{\prime},q^2)=\int\,d^4x\,\int\ 
d^4y\, e^{ip^{\prime}\cdot x}e^{q\cdot y}
\,\langle 0|T\Big\{j_{H_3}(x)j_{H_2}(y)
j_{H_1}^{\dagger}(0)\Big\}|0\rangle \, ,
\end{equation}
with $j_{H_1},\,j_{H_2}$ and $j_{H_3}$ the interpolating currents 
associated with the hadrons $H_1,\,H_2$ and $H_3$, respectively.

The evaluation of Eq.~\eqref{3pdef} follows 
the same steps  as those used for the two-point case. 
That is, it can be evaluated using 
QCD degrees of freedom and, in this case, the OPE is used 
in order to deal with the complex QCD vacuum 
structure. The coefficients of this operator 
expansion are determined by perturbative 
calculation, while the vacuum expectation value of the operators are
parametrized in terms of the condensates. 
On the other hand, Eq.~\eqref{3pdef} can also be 
evaluated using hadronic degrees of freedom. In this case, the 
three-point function is written in terms of the 
matrix elements of hadronic states. These matrix elements
are, in general, obtained by using 
some effective field theory approach, where an 
effective Lagrangian provides the information 
on the dynamics. 

It is worth mentioning that since the 
analytical structure of the three-point function 
can be different from the two-point case, the 
use of a dispersion integral should be treated 
with care \cite{Kallen:1958ifa, 
Kallen:1959kza, Martin:1999cr, Zwicky:2016lka}. 

\subsubsection{QCD or OPE side}

\begin{figure}[h]
\begin{center}
\scalebox{0.4}{\includegraphics[angle=0]{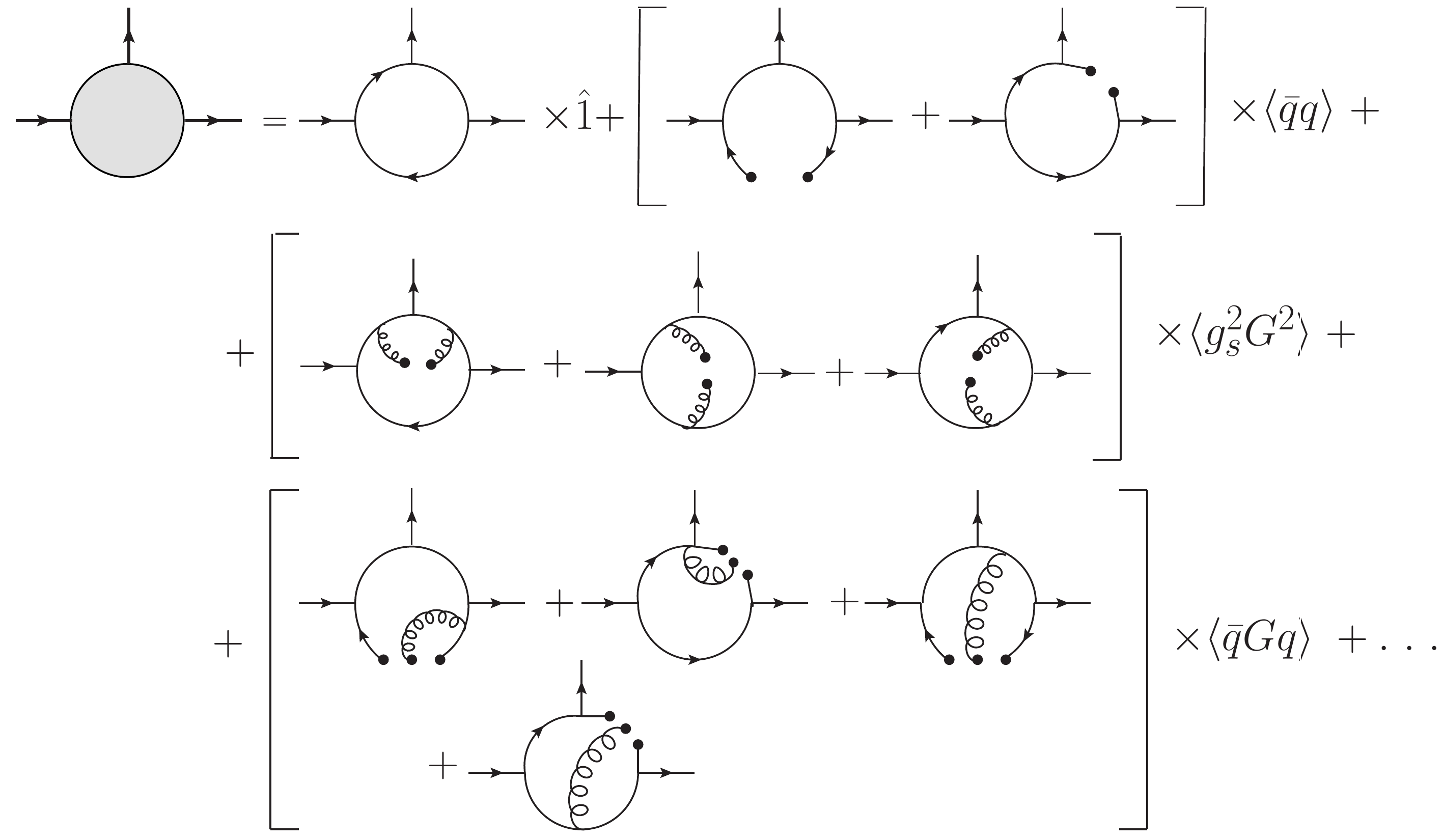}}
\end{center}
\caption{The OPE expansion for the three-point function of non-exotic mesons.
Possible permutations are not shown}
\label{3pOPE}
\end{figure}

Analogously to the two-point function case, 
the OPE for Eq.~\eqref{3pdef} can be written 
as 
\begin{equation}\label{3pOPEdef}
\Pi^{OPE}(P^2,P^{\prime\,2},Q^2)=\sum\limits_{n=0}^{\infty}
\,C_n(P^2,P^{\prime\,2},Q^2)\,\langle \hat{O}_n\rangle\, ,
\end{equation}
with $C_n(P^2,P^{\prime\,2},Q^2)$ being the 
OPE coefficients. The vacuum expectation values of the local 
operators are the same ones already defined in 
Eq.~\eqref{opedef} for the two-point case. 
The coefficients $C_n(P^2,P^{\prime\,2},Q^2)$ are 
obtained by calculating some Feynman diagrams. As 
an illustration, Fig.~\ref{3pOPE} shows 
a typical OPE (for the first lowest dimensions) 
in which some Feynman diagrams (other permutations are possible) are inside the brackets. 
The first one in Fig.~\ref{3pOPE} is associated 
with the first coefficient on the RHS of 
Eq.~\eqref{3pOPEdef}, and it gives the coefficient 
$C_0$ for the unit operator, reflecting the fact 
that at zero order, we have contribution only from 
the perturbative QCD physics domain. The second term in Fig.~\ref{3pOPE} 
has dimension $d=3$ and is used to determine the coefficient 
$C_3$. The third term in the series in Fig.~\ref{3pOPE}, with dimension $d=4$, determines $C_4$ that is multiplied to the gluon condensate 
$\langle g^2_s\,G^2 \rangle$. The last set of diagrams 
contributes to the coefficient $C_5$ multiplying the quark-gluon mixed 
condensate. According to Ref.~\cite{Shifman:1978bx} 
these power corrections are more important than higher 
order perturbative $\alpha_s$ corrections. This will be discussed in the last
section of this manuscript.

Once these diagrams are calculated we arrive at an expression for
$\Pi^{OPE}(p^2,p^{\prime\,2},q^2)$ that has the generic form: 
\beq
\Pi^{OPE}(p^2,p^{\prime\,2},q^2)=\sum_i \Gamma_i(p^2,p^{\prime\,2},q^2)T_i
\enq
where $ \Gamma_i(p^2,p^{\prime\,2},q^2)$ are invariant functions of the momenta and
$T_i$ are the tensorial structures, i.e., products
of Dirac matrices, the metric tensor and the four momenta, carrying Lorentz
indices.  The amount of Lorentz indices in  those structures 
depends on the tensorial nature of the current 
$j_n$ defined in Eq.~\eqref{jnonex}. 
The invariant functions 
$\Gamma_i(p^2,p^{\prime\,2},q^2)$ can be 
written in terms of a double dispersion 
relation over the virtualities 
$p^{\prime\,2}$ and $p^2$ 
\cite{Ioffe:1982qb, Reinders:1983wi, Nesterenko:1982gc, 
Bracco:2011pg}
\begin{equation}\label{doubleDIS}
\Gamma_i(p^2,p^{\prime\,2},q^2)=-\frac{1}{4\pi^2}\int\limits_{s_{min}}^{\infty}\,
ds\,\int\limits_{u_{min}}^{\infty}\, du\,\frac{\rho^{OPE}_i(s,u,q^2)}
{(s-p^2)(u-p^{\prime\,2})}+\,.\,.\,.\, ,
\end{equation}
where $\rho^{OPE}_i(s,u,q^2)$ is the 
double discontinuity that can be calculated
using the Cutkosky$^\prime$s rules \cite{Cutkosky:1960sp}. 
The terms not written explicitly, represented 
by the dots in Eq.~\eqref{doubleDIS}, are 
polynomials in $p^{\prime\,2}$ and $p^2$, and 
they vanish after a double Borel transform 
is applied to Eq.~\eqref{doubleDIS}.
It is just a Borel transform as in Eq.~\eqref{borel} applied twice, with 
$P^2=-p^2\to M^2$ and $P^{\prime\,2}=p^{\prime\,2}\to M^{\prime\, 2}$. A double 
transform applied to Eq.~\eqref{doubleDIS} leads to
\begin{equation}\label{doubBorel}
\mathcal{B}\{\mathcal{B}[\Gamma_i(P^2,P^{\prime\,2},Q^2)]\}=-\frac{1}{4\pi^2}\,
\int\limits_{s_{min}}^{\infty}\,ds\,\int\limits_{u_{min}}^{\infty}\,du\,
\rho_i^{OPE}(s,u,Q^2)\,e^{-s/M^2}\,e^{-u/M^{\prime\,2}}\, .
\end{equation}
 The Borel parameters, $M^2,~M^{\prime\,2}$, are chosen in such a way 
that the contribution 
of the higher states is suppressed at the same 
time we keep the power corrections under control.

\begin{figure}[h]
\begin{center}
\scalebox{0.35}{\includegraphics[angle=0]{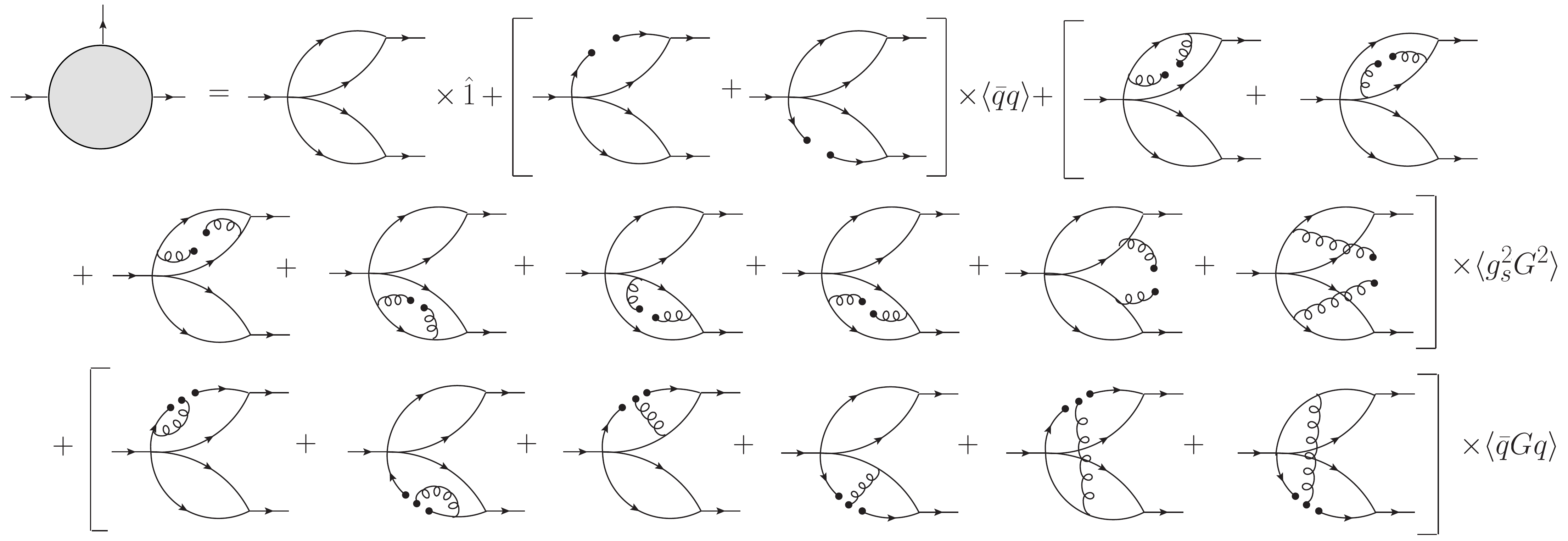}}
\end{center}
\caption{The OPE for the three-point function written in 
terms of exotic interpolating currents. Possible permutations are not shown.
}
\label{exOPE}
\end{figure}

As the interpolating currents are 
written in the form given by Eq.~\eqref{jnonex}, 
the three-point function describes 
the coupling between non-exotic 
states, i. e., a vertex where a non-exotic 
meson decays into two other non-exotic mesons 
\cite{Bracco:2011pg}. 
However, most of the new charmonium states have been described as exotic
states. In order to describe, within QCDSR approach, those states 
as exotic four-quark states, the interpolating currents must have  an 
additional pair of quark-anti-quark as compared with Eq.~\eqref{jnonex}:
\beqa
j_{ij}&=&\epsilon_{abc}\epsilon_{dec}({q}_a^TC\Gamma_i q_b)(\bar{q}_d\Gamma_jC\bar{q}_b^T)~,\lb{j4-tetra}\\
j_{ij}&=&(\bar{q}_a\Gamma_i q_a)(\bar{q}_b\Gamma_j q_b)~.\lb{j4-mol}
\enqa
These four-quark currents are used to interpret the exotic structures  with a tetraquark current, Eq.~(\ref{j4-tetra}), or with a molecular current, Eq.~(\ref{j4-mol}). The new diagrams connected with
the coefficients for the OPE in this case 
are constructed in the same manner as in 
the previous one. However, the topology 
changes, i. e. the Feynman diagrams will 
look like the ones depicted in Fig.~\ref{exOPE}. 
This occurs because the additional pair 
of quark-anti-quark fields on the current 
definition, gives rise to the ``petal'' 
form in the diagrams of Fig.~\ref{exOPE}.

For most exotic-type current calculations 
\cite{Dias:2013xfa,Dias:2013qga,Dias:2012ek,Albuquerque:2013owa,Torres:2016oyz,Dias:2016dme}, 
usually, it is sufficient to write the OPE up to 
dimension five $d=5$. Some of the diagrams of dimension five ($d=5$) have 
an interesting feature. When there is a gluon exchange between the petals they
are called color-connected. 
This implies that they are related to an 
intrinsic tetraquarks structure since the two 
petals cannot be considered as representing 
two separated mesons \cite{Dias:2013xfa,Dias:2013qga,Dias:2016dme,Duraes:2002px}.

Before proceeding it is important to stress that the currents in Eqs.~(\ref{j4-tetra}), (\ref{j4-mol}), and all currents used in this review to describe four-quark states, are local. Therefore, it is not possible to clearly distinguish between a tightly-bound tetraquark structure and a weakly-bound molecular structure. It is just the color combination between the quarks in the currents that is similar to a tetraquark or molecular state. Furthermore, due to Fierz transformation (see for instance Ref.~\cite{Nielsen:2009uh}), these currents are related with each other and, in general, it is easy to reproduce the mass of the state for a variety of currents with the same quantum numbers. However, the relation between the currents are suppressed by typical color and Dirac factors and, as a consequence, the coupling between the current and the state can vary largely~\cite{Narison:2010pd}. Therefore, a current with a large overlap with the physical state can still be used as an indi!
 cation of
  the state inner structure.

\subsubsection{Phenomenological side}

The phenomenological side of the three-point function is defined 
in terms of the hadronic degrees of freedom. 
This means that one has to consider the current operators 
as the creation/annihilation of the hadrons in the vertex.
The three-point function 
is evaluated inserting a complete set of 
intermediate states between these operators. This procedure leads 
to hadronic matrix elements such as 
$\langle 0|j_n(p)|H\rangle$, that represents 
the coupling of the hadronic state $H$ 
with the current $j_n$. These matrix 
elements are parametrized in terms of hadron parameters.
Let us define the parameter  $f_{H_i}$, associated to each
hadron in the vertex, as:
\begin{eqnarray}
\langle 0|j_{H_1}(p)|H_1(p)\rangle &=&f_{H_1}\nonumber\\
\langle 0|j_{H_2}(p^{\prime})|H_2(p^{\prime})\rangle &=&f_{H_2}\nonumber\\
\langle 0|j_{H_3}(q)|H_3(q)\rangle &=&f_{H_3}\, .
\end{eqnarray}
The three-point function can be written as 
\begin{equation}\label{3pophen1}
\Pi^{phen}(p^2,p^{\prime\,2},q^2)=
\frac{f_{H_1}f_{H_2}f_{H_3}}
{(p^2-m^2_{H_1})(p^{\prime\,2}-m^2_{H_2})(q^2-m^2_{H_3})}\,
\langle H_2(p^{\prime})H_3(q)|H_1(p)\rangle\, .
\end{equation}
The  matrix element in Eq.~\eqref{3pophen1}:
$\langle H_2(p^{\prime})H_3(3)|H_1(p)\rangle$, 
is associated with the transition 
$H_1\to H_2\,H_3$ and, in general, is obtained 
from an effective Lagrangian describing the 
vertex we are interested in. A form factor, 
$g_{H_1H_2H_3}(q^2)$, is introduced in 
Eq.~\eqref{3pophen1} when an  effective 
Lagrangian is used to calculate the matrix element 
$\langle H_2(p^{\prime})H_3(q)|H_1(p)\rangle$: 
\begin{equation}
\langle H_2(p^{\prime})H_3(q)|H_1(p)\rangle=g_{H_1H_2H_3}(q^2)\,T_i \, ,
\end{equation}
where $T_i$ is the tensorial structure  discussed previously. 
These tensor structures will be important in the definition of 
the sum rules, since the comparison must be done with 
the same tensorial structure on both sides of the sum rule.
In principle all structures are equivalent and should yield the same result. In practice, due to the truncation of the OPE, some differences appear and one structure may be more reliable than others~\cite{Bracco:1999xe}.
The Phenomenological side of the 
sum rule can be written as
\begin{equation}\label{3pophen2}
\Pi^{phen}(p^2,p^{\prime\,2},q^2)=\sum_i\Gamma^{phen}_i(p^2,p^{\prime\,2},q^2)\,T_i\, ,
\end{equation}
with $\Gamma^{phen}_i(p^2,p^{\prime\,2},q^2)$ 
defined as
\begin{equation}\label{poleterm}
\Gamma^{phen}_i(p^2,p^{\prime\,2},q^2)=\frac{f_{H_1}f_{H_2}f_{H_3}}
{(p^2-m^2_{H_1})(p^{\prime\,2}-m^2_{H_2})(q^2-m^2_{H_3})}\,
g_{H_1H_2H_3}(q^2)\, .
\end{equation}

In Eqs.~\eqref{3pophen1} and \eqref{poleterm} 
we have written explicitly only the pole 
contribution to the three-point function. 
On the other hand, as  done on the 
OPE side, Eq.~\eqref{poleterm} can also 
be written in terms of a double dispersion 
relation, in such a way that the effects of  higher 
states can be taken into account also
on the Phenomenological side. Therefore, we write:
\begin{equation}\label{doubDISphen}
\Gamma^{phen}_i=-\frac{1}{4\pi^2}\,
\int\limits_{s_{min}}^{\infty}\,ds\,
\int\limits_{u_{min}}^{\infty}\,du\,
\frac{\rho^{phen}_i(s,u,Q^2)}{(s-p^2)(u-p^{\prime\,2})}\, ,
\end{equation}
where $\rho^{phen}_i(u,s,Q^2)$ is the double 
discontinuity of the amplitude 
$\Gamma_i(p^2,p^{\prime\,2},Q^2)$. In Ref.~\cite{Bracco:2011pg} this spectral 
density was generically expressed as
\begin{eqnarray}\label{spec}
\rho^{phen}_i(s,u,Q^2)&=&a\delta(s-m^2_{H_1})\,\delta(u-m^2_{H_2})
+b\delta(s-m^2_{H_1})\theta(u-u_0)\nonumber\\
&+&c\delta(u-m^2_{H_2})\theta(s-s_0)+
\rho^{cont}(s,u,Q^2)\theta(s-s_0)\theta(u-u_0)\, ,
\end{eqnarray}
where $s_0,\,u_0$ are the continuum 
thresholds associated with the hadrons 
$H_1$ and $H_2$, respectively. Equation~\eqref{spec} 
has a simple kinematical interpretation. 
The first term describes the kinematical situation 
where the hadrons $H_1$ and $H_2$ are 
on the ground state, while $H_3$ is off-shell, 
with arbitrary Euclidian four momentum $Q^2=-q^2$. 
The second term refers 
to a situation where $H_1$ is still on the 
ground state, but the ground state of the hadron $H_2$
is absent in the vertex, which contains only its excitations starting at
$u_0$. The third term is analogous to the second one with the exchange $H_2\leftrightarrow H_1$. Finally, the last term represents the excitations of $H_1$
and $H_2$, which start at $s_0$ and $u_0$ respectively.
Substituting Eq.~\eqref{spec} 
into Eq.~\eqref{doubDISphen}, with the parameter 
$a$ (in Eq.~(\ref{spec})) identified as the pole term given 
in Eq.~\eqref{poleterm}, we obtain
\begin{eqnarray}\label{3pophen}
\Gamma^{phen}_i&=&
\frac{f_{H_1}\,f_{H_2}\,f_{H_3}
g_{H_1H_2H_3}(q^2)}{(p^{ 2}-m^2_{H_1})
(p^{\prime 2}-m^2_{H_2})
(q^2-m^2_{H_3})} 
-\frac{1}{4\,\pi^2}\Bigg[\frac{1}{m^2_{H_1}-p^2}
\int\limits^{\infty}_{u_0}du\,\frac{b(u,q^2)}
{(u-p^{\prime 2})}\nonumber\\
&+&\frac{1}{m^2_{H_2}-p^{\prime 2}}\int\limits^{\infty}_{s_0}
ds\,\frac{c(s,q^2)}
{(s-p^2)}\Bigg]+\int\limits_{s_0}^{\infty}ds\int
\limits_{u_0}^{\infty}du\,\frac{\rho^{cont}_n(s,u,Q^2)}{(s-p^2)
(u-p^{\prime 2})}\, . \nonumber\\
\end{eqnarray} 
In Eq.~(\ref{3pophen}), $b(s,q^2)$ and 
$c(s,q^2)$ are unknown functions contributing 
to the pole-continuum transitions~\cite{Ioffe:1983ju, Matheus:2009vq}
of the hadrons $H_1$ and $H_2$, respectively. 
These functions can be determined adopting the model discussed in 
Ref.~\cite{Eidemuller:2005jm}.

\subsubsection{Three-point function sum rule}
\label{s-3p}

Analogously to the two-point function 
case, the three-point function sum rule is obtained by matching 
the OPE and the Phenomenological sides. In order to do this a given tensorial 
structure, $T_i$,  must be present 
on both sides. Hence, for a given $T_i$ structure, after doing a Borel transform
in both $P^2\to M^2$ and $P^{\prime 2}\to M^{\prime 2}$ one has
\begin{equation}
\Gamma^{phen}_i(M^2,M^{\prime\, 2},Q^2)=\Gamma^{OPE}_i(M^2,M^{\prime\, 2},Q^2)\, .
\end{equation}

Since this review is dedicated to the exotic charmonium states,
all of the applications discussed here are related to 
hadronic states that are described by exotic
four-quark interpolating currents. As a 
consequence, the invariant function, 
$\Gamma^{OPE}_i$, depends only on $P^{2}$ and $Q^2$ or on
$P^{\prime\,2}$ and $Q^2$ four momenta 
and, in this case, a double Borel transform eliminates the OPE side.
To overcome this problem we first notice that 
in the calculations involving systems with heavy flavors the approximation 
$P^2\approx P^{\prime\,2}$  is very good \cite{Reinders:1983wi}. Therefore,
taking $P^2= P^{\prime\,2}$ we do  a single Borel transform to
$P^2=P^{\prime 2}\to M^{2}$ on both sides of the sum rules.
However, as first noticed by Ioffe and Smilga \cite{Ioffe:1983ju} the
pole-continuum transitions are not exponentially suppressed, as compared to
the pole contribution, when only one Borel transformation is done in both
$P^2$ and $ P^{\prime\,2}$. Following Ref.~\cite{Ioffe:1983ju} here
we introduce an unknown function $B(Q^2)$ parametrizing the 
two integrals in the brackets in Eq.~(\ref{3pophen}). Therefore, the sum rule
can be written as:
\begin{eqnarray}\label{g1}
\frac{f_{H_1}f_{H_2}f_{H_3}\,g_{H_1H_2H_3}(Q^2)}{(Q^2+m^2_{H_3})}
\left({e^{-m^2_{H_1}/M^2}-e^{-m^2_{H_2}/M^2}\over m_{H_2}^2-m_{H_1}^2}\right) + B(Q^2)e^{-s_0/M^2}&=&
\Gamma^{OPE}(M^2,Q^2)\, .
\end{eqnarray}
The second term on the LHS of Eq.~\eqref{g1} 
accounts for the pole-continuum 
transitions. An expression for the 
form factor is obtained by taking 
the derivative of Eq.~\eqref{g1} 
with respect to $1/M^2$ and using 
the result to eliminate $B(Q^2)$ 
from the equations.

\subsection{Extrapolation of the form factor and the coupling constant}
\lb{exproc}

The coupling constant is obtained 
from the form factor, $g_{H_1H_2H_3}(Q^2)$, in Eq.~(\ref{g1}).
In fact, it is defined as the value assumed 
by the form factor at the hadron pole mass, that is 
$g_{H_1H_2H_3}=g_{H_1H_2H_3}(Q^2=-m^2_{H_3})$. 
However, we cannot simply consider 
$Q^2=-m^2_{H_3}$ in Eq.~\eqref{g1} since 
the sum rule is valid only for $Q^2>0$. 
Therefore, in order to overcome this problem 
we use a procedure that allows us to extrapolate 
the QCDSR results to the time-like region, 
where the value for the coupling constant can 
be extracted.
To this end we use a function 
that fits the QCDSR results for the form factor 
$g_{H_1H_2H_3}(Q^2)$. 

The form factor  depends not only on $Q^2$, but 
also on the Borel mass $M^2$. 
The QCDSR results would be independent of this parameter 
if one could take into account the whole OPE series, without 
truncating it at some dimension.
Since the OPE series is always truncated at some order, 
we have to look for some interval at $M^2$ in 
which the QCDSR results are as independent of the  
Borel mass as possible. Following this procedure, 
we are able to guarantee that the QCDSR results 
depend only on $Q^2$. Therefore, we usually 
plot $g_{H_1H_2H_3}(Q^2)$  as a function of $Q^2$ 
and $M^2$ in order to look for such an interval. As an example,
Fig.~\ref{gexample}  shows the  plot of the form factor 
of the $Z^{+}_c(3900)\to \eta_c\,\rho^+$ vertex, 
studied in Ref.~\cite{Dias:2013xfa}. As can be seen, from values 
within the interval $4.0\leq M^2\leq 10.0$ 
GeV$^2$, the form factor $g_{Z_c\eta_c\rho}$ has almost no 
$M^2$ dependence and consequently $g_{Z_c\eta_c\rho}$ has 
the same $Q^2$ dependence for every $M^2$ value 
assumed on that interval. 

\begin{figure}
  \subfigure[]{\lb{gexample}\includegraphics[width=0.5\textwidth]{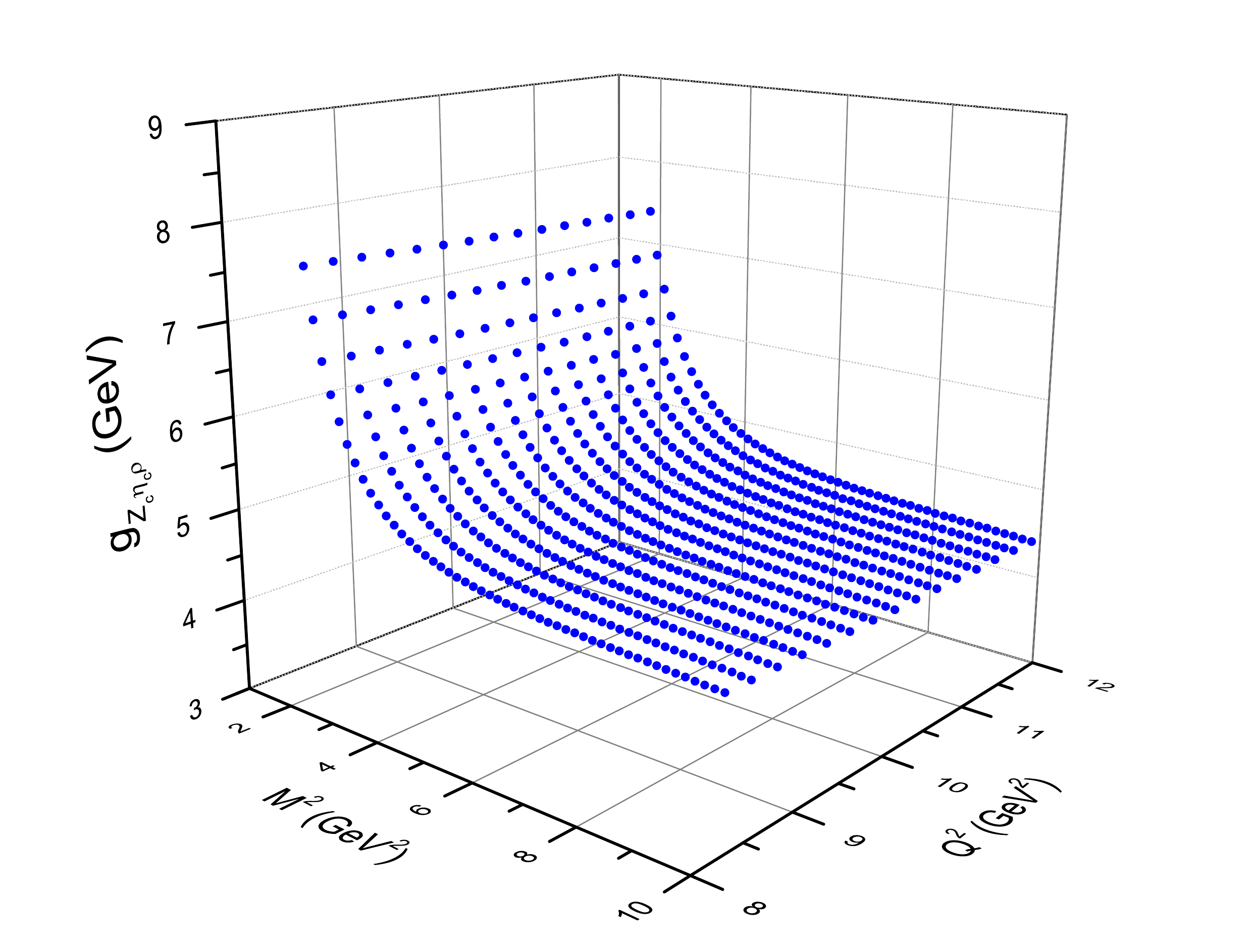}}
  \subfigure[]{\lb{gQ2rho}\includegraphics[width=0.6\textwidth]{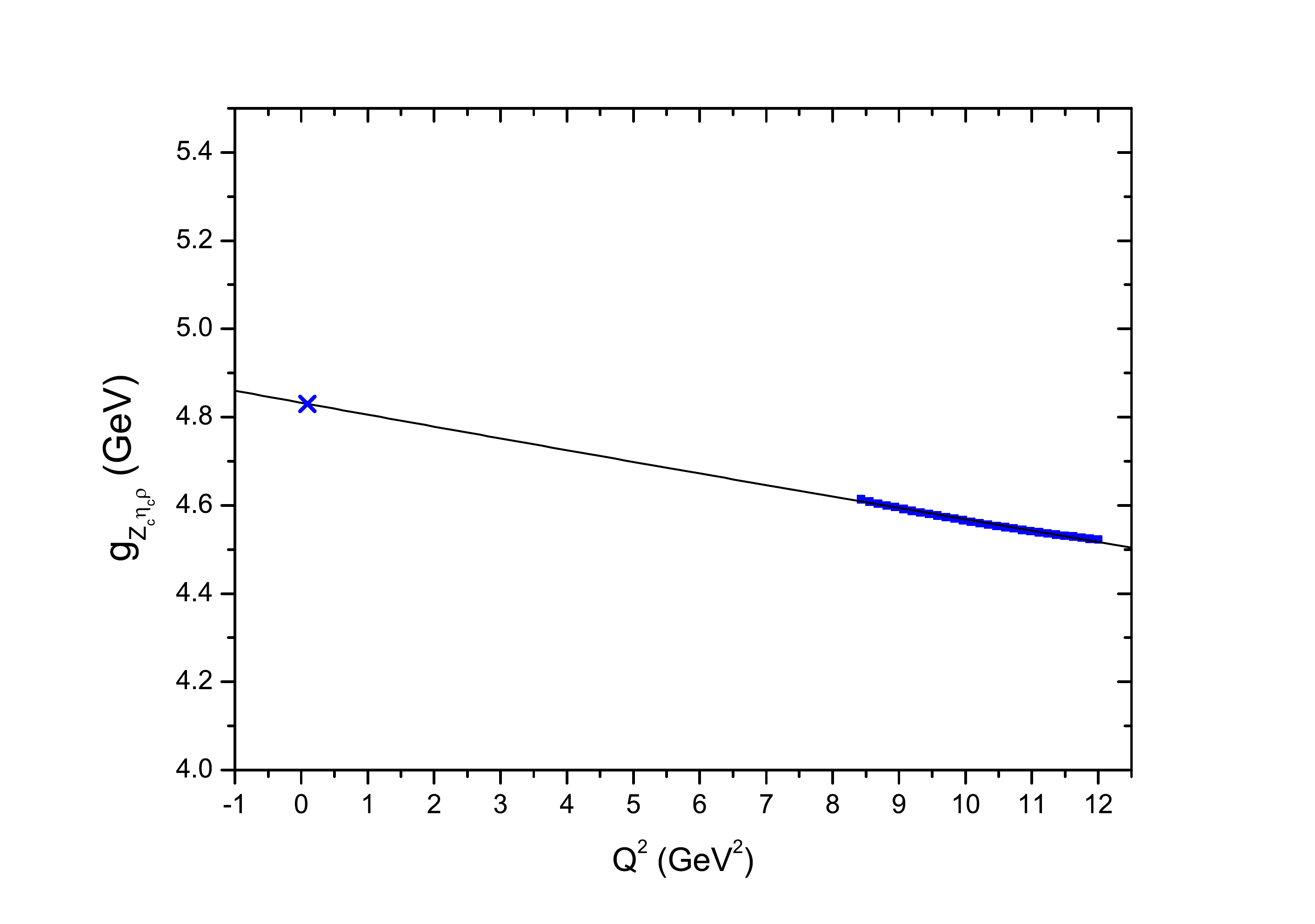}}
    \caption{(a) QCDSR results for the form factor  $g_{Z_c\eta_c \rho}(Q^2)$ as 
a function of $Q^2$ and $M^2$. (b) QCDSR results for $g_{Z_c\eta_c \rho}(Q^2)$, as a function of $Q^2$ for Borel mass values within the interval 
$4.0\leq M^2 \leq 10.0$ GeV$^2$ (squares). 
The solid line gives the parametrization of the QCDSR results  through Eq. 
(\ref{exprho}). The cross gives the value of the coupling constant. Taken from Ref.~\cite{Dias:2013xfa}.}
\end{figure}
Fig.~\ref{gQ2rho} shows $g_{Z_c\eta_c\rho}$ as a 
function of $Q^2$ (represented by the squares) for 
Borel mass values within the interval 
$4.0\leq M^2 \leq 10.0$ GeV$^2$. Note that the QCDSR 
results are evaluated in the interval $8.5\leq Q^2\leq 12.0$ 
GeV$^2$. This is the interval in which 
the results are independent of $M^2$ and, therefore, 
we can say that the QCDSR results have a good degree of reliability. 

In the $g_{Z_c\eta_c\rho}$ case, we extrapolate the 
QCDSR results (given by the squares in Fig.~\ref{gQ2rho}) 
by using an exponential function
\begin{equation}\lb{exprho}
g_{Z_c\eta_c\rho}(Q^2)= g_1\,e^{-g_2Q^2}\, ,
\end{equation}
that fits the QCDSR results (the solid line in 
Fig.~\ref{gQ2rho}) for $g_1=4.83$ GeV and 
$g_2=5.6\times 10^{-3}$ GeV$^2$. The QCDSR results 
in Fig.~\ref{gQ2rho} were obtained by considering the $\rho$ 
meson as being off-shell. The coupling constant 
$g_{Z_c\eta_c\rho}$ is then given by 
$g_{Z_c\eta_c\rho}(Q^2=-m^2_{\rho})$, leading to  
the following result~\cite{Dias:2013xfa}:
\begin{equation}\label{coupetarho}
g_{Z_c\eta_c\rho}(Q^2=-m^2_{\rho})=(4.85\pm 0.81)\,\, \textrm{GeV}\,.
\end{equation}

\begin{figure}[h]
\begin{center}
\scalebox{0.6}{\includegraphics[angle=0]{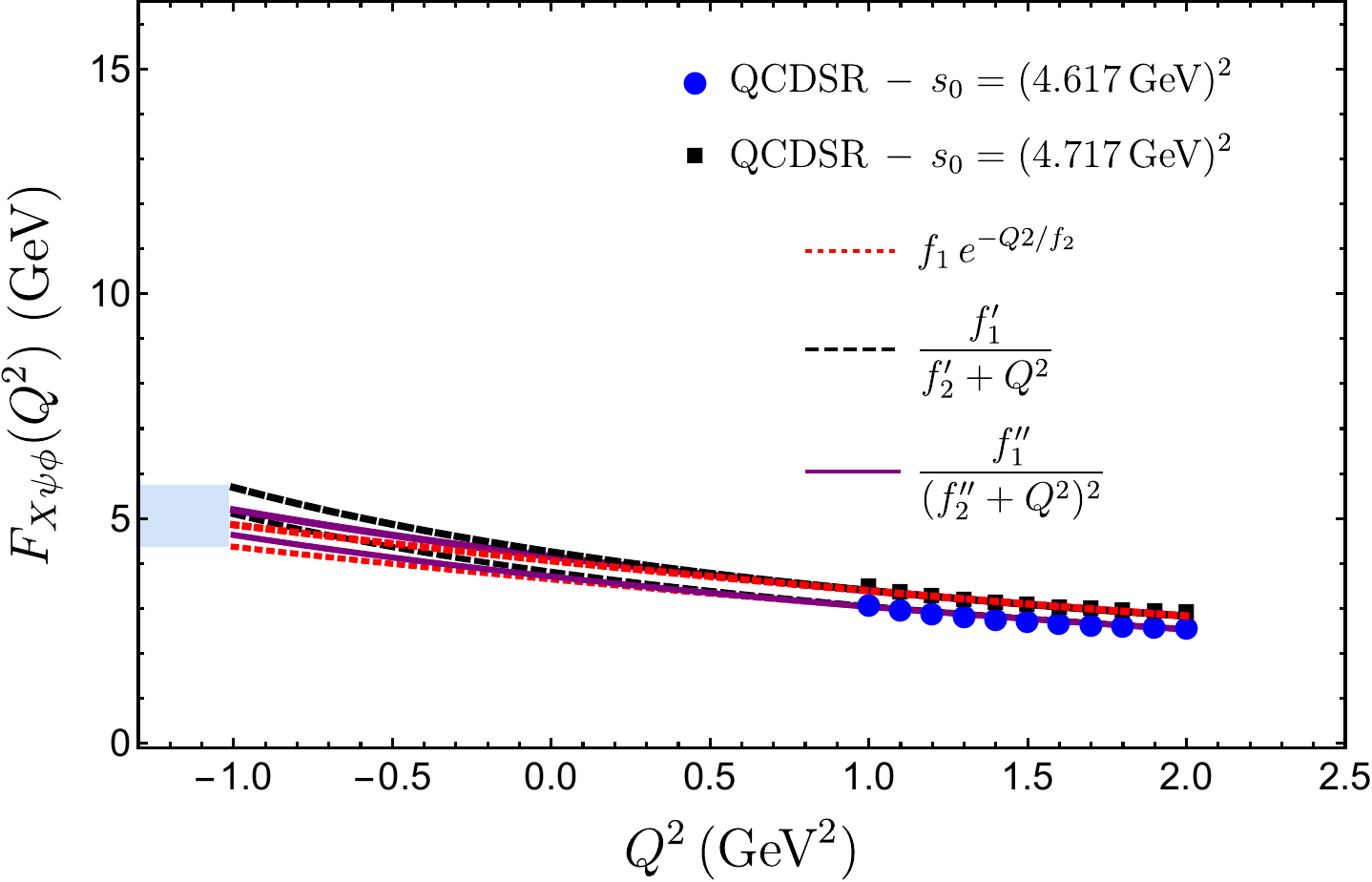}}
\end{center}
\caption{QCDSR results (filled circles) for $YJ/\psi\phi$ 
vertex function. The dotted, 
dashed and solid lines, correspond to the fits using different 
expressions for the form factor $F_{Y\psi\phi}$, as can be 
seen on the plot. The shaded area near the vertical 
axis indicates the range of the values for the coupling 
constant $F_{Y\psi\phi}$ due to the different choices 
for the form factor. Taken from Ref.~\cite{Torres:2016oyz}.}
\label{manyFQ2}
\end{figure}

The value of the coupling 
constant extracted from the QCDSR results 
depends on the choice of the function $g(Q^2)$. 
More concretely, if we would choose another 
function to fit the QCDSR results, shown in Fig.~\ref{gQ2rho},
we would  get a different value for the coupling 
constant, $g_{Z_c\eta_c\rho}$, in Eq.~(\ref{coupetarho}), since two different
functions that fit the QCDSR results might have a different 
behavior at the time-like region. This will give rise to some uncertainty in the
extracted value of the coupling constant.
In order to illustrate this uncertainty we show, 
in Fig.~\ref{manyFQ2}, the 
calculation of the  $Y\to J/\psi\phi$ coupling constant 
using different parametrizations to extract the coupling constant~\cite{Torres:2016oyz}.

The QCDSR results in Fig.~\ref{manyFQ2} are 
represented by the circles and squares. In order to fit these 
points three different functions, denoted by $F(Q^2)$, 
were used. As it was expected, each one gives a 
different value for the coupling constant $F_{Y\psi\phi}$, 
as indicated by  the shaded area projected into the 
vertical axis. 

In Table~\ref{tabFF}, we list some few examples of the 
most common analytical functions used to describe form 
factors in QCDSR calculations: the monopolar, exponential 
and Gaussian forms. All of them have in common the fact that, 
they vanish as $Q^2\to \infty$.
\begin{table}[h]
\caption{Some examples of analytical functions for the 
form factors used in QCDSR calculations.}
\centering
\begin{tabular}{c c }
\hline\hline
 ~& \textrm{Analytical functions for the form factor}\\
\hline
\\
\vspace{0.3cm}
Monopolar ~&~ ~~~$\frac{f_1}{f_2+Q^2}$\\
\vspace{0.3cm}
Exponential  ~&~ ~~~$f_1\,e^{-f_2Q^2}$\\
\vspace{0.3cm}
Gaussian ~&~  ~~~ $f_1 e^{-(f_2+Q^2)^2/f_3}$\\
\hline\hline
\end{tabular}
\label{tabFF}
\end{table}

In general, as it is discussed in \cite{Bracco:2011pg}, 
these systematic uncertainties are reduced 
by performing a double QCDSR analysis. For a given 
$H_1H_2H_3$ vertex, we take  $H_1$ to be off-shell,
calculate the QCDSR points and extrapolate them to the 
time-like region. At the same time we take $H_2$ to be 
off-shell and repeat the procedure. When doing the 
extrapolations we impose the two form factors to yield
the same coupling constant. This matching condition
significantly reduces the freedom in the choice of the
parameters $f_1$, $f_2$, ...etc. This procedure 
is enough to reduce the uncertainties and provides 
results consistent with the experimental values, 
as it was the case for the $D^*D\pi$ and $J/\psi D^*D^*$ 
studied in Refs.~\cite{Navarra:2001ju,Bracco:2004rx}.

Even after the above mentioned improvement one might 
argue that the choice of the functional form, $g_{H_1H_2H_3}(Q^2)$,
remains arbitrary and this implies a systematic uncertainty.
In order to reduce this freedom of choice one may try to 
match the QCDSR results with meson loops calculated via effective 
field theory approach. This idea was advanced in 
Ref.~\cite{Duraes:2004uc}, and can be used to 
impose constraints that allow us to reduce the 
freedom of choice of the function parametrizing the 
QCDSR results in the deep Euclidean region. 

As it was shown in Ref.~\cite{Duraes:2004uc}, in a certain $Q^2$ 
region the $D^*D\pi$ vertex is described by an effective Lagrangian 
obeying  chiral symmetry. Beyond the tree level one can compute 
meson loop diagrams. These computations bring up some unknown parameters 
arising from the renormalization of the loops. Fortunately, 
these unknown loop parameters can be isolated into some basic constants, 
and by matching the calculation of the form factor to the QCDSR results, 
they can be determined, providing a knowledge of the form factor. 

This match between chiral and QCDSR results is justified. 
In the region in which the QCDSR approach does not provide reliable results  
(at low or negative $Q^2$ values) the effective field theory approach is valid. On the other hand, for intermediate and higher $Q^2$ regions the QCDSR is more reliable than the effective theories.

\subsection{Limitations of the applications of the QCDSR to the exotic states}

  One of the limitations, as already discussed in Sec. 2.7.1, is due to the use of local currents. One can not distinguish between  tetraquarks or molecular states, considered as spatially large objects.
 The only thing one can say is that if a molecular kind of current couples strongly with the state, this gives us a hint about the color organization of the quarks in the state. In order to consider the size of the states, a possible alternative  would be to use non-local currents. To our knowledge, non-local currents have never been used in the QCDSR approach, since quantum corrections to these non-local operators are awful~\cite{Dosch:1994wj}. However, non-local currents for the exotic states, in particular $Z_c(3900)$ and $Z(4430)$, have already been used in a covariant quark model~\cite{Goerke:2016hxf}. It is a challenge but it would be important to consider them also in the QCDSR approach.

The lack of $\alpha_s$ corrections, in most of the previous calculations, can be also a limitation. However, as shown in Sec.~7,   $\alpha_s$ corrections have already been considered in the QCDSR for the exotic states, leading to small corrections to their masses.

Another drawback of the method is the fact that most of the $X,~Y,~Z$ states are found close to the two particle $S$-wave thresholds to which they seem to have quite strong couplings. In studies of non-exotic hadrons, the ground state of the spectrum is a zero width state and is well separated from the continuum. In this case, a reasonable Ansatz is adopted for the phenomenological description of the spectral function, which is taken to be  a sharp pole separated from a continuum, see Eq.(\ref{den}).  
The same Ansatz has been extended to the case of $X,~Y,~Z$ states, assuming that each $X$, $Y$ or $Z$ state is the ground state of the related spectrum and that the corresponding higher excited states lie in the high energy region. Such a description can be questioned in the case of exotic hadrons, where two meson thresholds often lie close to their masses and  which can couple to them in $S$-wave. In fact, such hadron channels may even lie below the mass of the exotic states, contributing to their widths and  smearing up a continuum in the background of the spectral function.

There have been several attempts to consider the contribution of such two particle intermediate states to the QCDSR of  exotic states~\cite{Kondo:2004cr,Lee:2004xk,Lee:2007mva,Chen:2009gs}. In all cases these two-hadron-reducible contributions~\cite{Kondo:2004cr} can be included by adding a term in the phenomenological side of the sum rule. In general, the $H_1-H_2$ continuum contribution to the exotic state $X$, that couples in $S$-wave with these two hadrons, can be included by modifying the phenomenological side of the QCDSR in Eq.~(\ref{den}) as:
\beq
\rho(s)=\lambda^2\delta(s-m_X^2) +\rho^{cont}(s) +\rho_{H_1H_2}(s)\,.
\label{swave}
\enq
To find an expression for $\rho_{H_1H_2}(s)$ one needs to introduce a coupling between the current, representing the exotic state, and the two particles (considering here, for simplicity, as spin 0 states):
\beq
\lambda_{H_1H_2}=\langle0|j_X|H_1~H_2\rangle\,.
\enq
The correlation function of the $H_1-H_2$ continuum is then given by~\cite{Chen:2009gs}:
\beq
\Pi_{H_1H_2}(p^2)=i|\lambda_{H_1H_2}|^2\int~{d^4q\over(2\pi)^4}{i\over q^2-m_{H_1}^2}{i\over(p-q)^2-m_{H_2}^2}\;,
\enq
and, therefore, $\rho_{H_1H_2}(s)$ is given by
\beq
\rho_{H_1H_2}(s)={1\over\pi}{\mbox Im}[\Pi_{H_1H_2}(s)]\;.
\enq
The problem in such approach is how to evaluate $\lambda_{H_1H_2}$. 
It could be evaluated by using the method of current algebra, if the properties of the resonance state are known. It could be evaluated by a new QCDSR, as in Refs.~\cite{Kondo:2004cr,Lee:2004xk,Lee:2007mva}, with the consequence of introducing more uncertainties in the calculation. In Ref.~\cite{Chen:2009gs} $\rho_{\pi\pi}(s)$ was parameterized as:
\beq
\rho_{\pi\pi}(s)=as^2\sqrt{1-{4m_\pi^2\over s}},
\enq
where $a$ is a new parameter. In Ref.~\cite{Chen:2009gs} such $\pi-\pi$ continuum contribution to the QCDSR of the light scalar $\sigma(600)$  (considered as a tetraquark) was included. The sum rule  shows a much better stability of the obtained resonance mass with respect to the continuum threshold parameter $s_0$. 

In all quoted cases~\cite{Kondo:2004cr,Lee:2004xk,Lee:2007mva,Chen:2009gs}, the inclusion of the two-hadron-reducible contribution to the QCDSR improves the stability of the results, but it does not change drastically the results for the mass of the exotic state. Therefore one can say that, although an effort  should be made to include such contribution to the QCDSR of the $X,~Y,~Z$ states, the results obtained so far can still be trusted.

The presence of two-hadron thresholds near the mass of exotic states can give rise to yet other type of complications.
Sometimes the opening of these thresholds can lead to cusp like structures in cross-sections/invariant mass spectra \cite{Swanson:2014tra, Bugg:2011jr}. The
cusps usually show up with line shapes different to a Breit-Wigner state, though it may not be straightforward to distinguish
between the two. However, it is argued in Ref.~\cite{Guo:2014iya}, that a narrow peak must correspond to a bound state/resonance (pole in the complex plane,
in the amplitude). Other difficulty is that sometimes the pole may not be present in the right Riemann sheet  and the state may have an alternative
interpretation, such as, a virtual state (see the review  \cite{Guo:2017jvc} for more discussions). Such
states can also lead to an enhancement in the experimental data. It is, thus, important to investigate if a possible state found in experiments
is a genuine state or a threshold effect or a virtual state, etc. Effects, like, cusps, virtual states, etc., cannot be directly identified in QCD sum rules.
One would reach a negative result (based on the conditions to be satisfied in order for the results to be reliable) concluding the
nonexistence of a state, which would indicate that the enhancement found in  experiments may have a different interpretation.

\section{\label{X(3872)} The X(3872) state}

It has been fifteen  years since the first observation  of the $X(3872)$
state was reported.  It was the  first charmonium state that could not
be accommodated within the usual quark-antiquark meson model.  Ever since then,
the study  of hadron spectroscopy is continuously being revised,
with the observation of several other states with exotic properties in
the heavy quark sector.  The first observation was reported in 2003 by
Belle Collaboration with the measurement of a narrow resonance in
the $B$  meson decay channel $B^\pm\to  K^\pm X(\to J/\psi\pi^+\pi^-)$
\cite{Choi:2003ue}, and  soon it  was confirmed by  BABAR in  the same
channel  \cite{Aubert:2004ns}, and  by  D0 and  CDF Collaborations  in
$p\bar{p}$        collisions       \cite{Acosta:2003zx,Abazov:2004kp}.
Subsequently, the  $X(3872)$ was  observed in many  other experiments
and in several  different channels,  leading to  a vast  amount of
experimental  data,   which  are  collected  and   summarized  by  the PDG
\cite{pdg}.   Here, we  list  some  of  the main  experimental
results and their connection with the main theoretical interpretations of $X(3872)$.

The    current   world    average   mass    of   the    $X(3872)$   is
$3871.69\pm0.17\MeV$ and it is a very narrow state, with an upper limit on the
decay  width  of  $\Gamma<1.2\MeV$  at  90\% confidential level (CL) \cite{pdg}.   The  first
interesting aspect one  can readily  notice  is the proximity
of   the    mass of the state to   the  $D^0\bar{D}^{*0}$     threshold,
$3871.81\pm0.09\MeV$.  The  state quantum  numbers have  been 
completely determined  as $J^{PC}=1^{++}$,  corresponding to  an axial
vector  meson  \cite{Aaij:2015eva}.  The determination of  the charge-conjugation parity was stablished unambiguously  to be $C=+1$, due to the  observation of the radiative   decay   $X\to\gamma   J/\psi$   reported   by   Belle   in Ref.~\cite{Abe:2005ix}.  In Ref.~\cite{Abulencia:2006ma}, the CDF
Collaboration performed  an analysis of the  angular distributions for
the decay channel $X(3872)\to J/\psi\pi^+\pi^-$, $J/\psi\to\mu^+\mu^-$
comparing the  obtained outcome with the  theoretical predictions, and
of  all possible  assignments, only  the quantum numbers $1^{++}$  and
$2^{-+}$ were consistent  with the data, while the other quantum numbers were excluded with 99.7\%  CL.  Finally, the LHCb Collaboration performed a  full amplitude analysis of  the angular correlations, in five dimensions, between the products of the decay mode $B^\pm\to K^\pm X(3872)$, establishing the
quantum numbers of the state as  $J^{PC}=1^{++}$, definitely  ruling out  the
$2^{-+}$ possibility \cite{Aaij:2015eva}.

The  discussions on the puzzling nature  of  the  $X(3872)$  started immediately after its
observation.  A  possible candidate, within  the  quark-antiquark
conventional  model with  proper quantum numbers, would be the $2^3P_1$ state,
also  known  as  $\chi_{c1}(2P)$ or  $\chi_{c1}^\prime$.
However,  the masses obtained for this state, from constituent quark models
or lattice QCD, are not compatible  with  the $X(3872)$ mass
\cite{Barnes:2003vb,Barnes:2005pb,Okamoto:2001jb}.  The fact that the  $X(3872)$  could
not  be accommodated within the  conventional quark model and that its mass is very  close to the  $D^0\bar{D}^{*0}$
 threshold,  strongly  suggested a possible molecular structure with small  biding energy. In fact,  the existence of a  molecule, with
$1^{++}$ (and also with $0^{-+}$) quantum numbers, near the $D^0\bar{D}^{*0}$
threshold was  predicted   by Tornqvist,  using the  potential
model, many  years before the discovery of $X(3872)$ \cite{Tornqvist:1993ng}.

The measurement of  the branching ratio between final  states with two
and three pions  was definitive to establish  the unconventional nature
of  $X(3872)$.   In Ref.~\cite{Abe:2005ix},  Belle  reported  the
branching  ratio   for  the   channels  $X\to   J/\psi\pi^+\pi^-$  and
$J/\psi\pi^+\pi^-\pi^0$:
\beq
{{\cal B}(X \to J/\psi\,\pi^+\pi^-\pi^0)\over {\cal B}(X\to\!J/\psi\pi^+
\pi^-)}=1.0\pm0.4\pm0.3.
\label{rate}
\enq
The BaBar Collaboration~\cite{delAmoSanchez:2010jr} also  observed the decay
$X \to  J/\psi\omega$ at a rate  compatible with  Eq.~({\ref{rate}}):
\beq {{\cal
    B}(X  \to  J/\psi\pi^+\pi^-\pi^0)\over {\cal  B}(X\to\!J/\psi\pi^+
  \pi^-)}=0.8\pm0.3.
\label{barate}
\enq
The  decay mode  $ J/\psi\pi^+\pi^-$  occurs via  $\rho J/\psi$,  with
isospin $I=1$,  and $J/\psi\pi^+\pi^-\pi^0$ via $\omega  J/\psi$, with
isospin $I=0$, indicating  a strong isospin and  $G$ parity violation.
The  isospin violating  modes  are strongly  suppressed for  $c\bar{c}$
states,  while  should  be  a  common  feature  for  molecular  states
\cite{Swanson:2004pp}.  Besides, this result  was predicted by Swanson
in   Ref.~\cite{Swanson:2003tb},  considering   the  $X(3872)$   as  a
$D^0\bar{D}^{*0}$ molecule with a small admixture of $\rho J/\psi$ and
$\omega J/\psi$ components.

The radiative decays  are other important processes that lead
to  distinguishable  results  between  the  charmonium  and  molecular
states, as pointed out by Swanson in Ref.~\cite{Swanson:2004pp}.
The  Belle  Collaboration  reported   the  first  observation  of  the
radiative decay channel \cite{Abe:2005ix}, $X\to\gamma J/\psi$, with a branching ratio of
\beq
\frac{\Gamma(X\to\gamma J/\psi)}{\Gamma(X\to\pi^+\pi^- J/\psi)}=0.14\pm0.05.
\label{rad}
\enq
This  result  is  incompatible  with  the preferred  $c\bar{c}$  
candidate  $\chi_{c1}$, but it is in  agreement with  the molecular  description for $X(3872)$
 \cite{Swanson:2004pp,Aceti:2012cb}. Although the  topic is
still not settled, this result has contributed to the general consensus about
the unconventional nature of  $X(3872)$.

Another  possibility for  the nature of $X(3872)$ as a four-quark state is a
tetraquark structure,   as   proposed  by Maiani {\it et. al.}~\cite{Maiani:2004vq}.   In this  model,  the state  is  formed by  the
binding  of  a  diquark  and an antidiquark  pair,  with  a symmetric  spin
distribution.   The  mixing  of   pure  isospin  states  provides the
possibility of  isospin violating modes.

An additional radiative decay mode, $X\to \gamma\psi(2S)$, was reported by Babar
\cite{Aubert:2008ae}, with  a large  branching ratio in  comparison to
the $\gamma J/\psi$ mode:
\beq
R_{\psi\gamma} = {{\cal B}(X \to\gamma \psi(2S))\over {\cal B}(X\to\gamma\psi)}=3.4\pm1.4.
\label{rategaexp}
\enq
The   Babar   result  was   confirmed   by   the  LHCb   Collaboration
\cite{Aaij:2014ala},   where   the   branching  ratio   measured   was
$R_{\psi\gamma}=2.46\pm0.64\pm0.29$.   This  ratio  varies  widely  in
different theoretical  models, and  can be  used as  a distinguishable
feature    between     different    models.     For     example,    in
Ref.~\cite{Swanson:2004pp},  the   molecular  model  gives rise to a 
suppression  of  the $\gamma\,\psi(2S)$  channel, providing:
${\Gamma(X             \to
  \psi(2S)\,\gamma)\over\Gamma(X\to\psi\gamma)}\sim4\times10^{-3}$.
In fact, neither pure charmonium or pure molecule can accommodate this
experimental result (see  \cite{Aaij:2014ala} and references therein),
and the  predicted ratio in Eq.~(\ref{rategaexp})  favors a different
scenario, in which  $X(3872)$ is an admixture of charmonium and four-quark
states. The necessity for
such type  of admixture  was anticipated in  several studies,  see for
example
Refs.~\cite{Barnes:2003vb,Eichten:2005ga,Suzuki:2005ha,Meng:2005er}.

Besides the four-quark and mixed  models, there are other theoretical
descriptions,      like, a   cusp      \cite{Bugg:2004rk},   a   hybrid structure
\cite{Li:2004sta,Close:2003mb},  and a glueball,  \cite{Seth:2004zb} that
were also presented as alternative interpretations for the nature of
  $X(3872)$.   The debate  regarding  the  puzzling nature  of  the
$X(3872)$ is  still not  settled, although the  molecule/tetraquark and
admixtures  of   charmonium  and  molecule  scenarios   are  the  most
promising ones,  being widely  studied and tested  so far  in several
theoretical frameworks,  including the  QCDSR, which  is the  focus of
this review.

\subsection{QCDSR calculations for $X(3872)$}

The $X(3872)$ state has been studied  by various authors using  the QCDSR
technique.  The  first QCDSR  calculations using a  tetraquark current
were presented in Refs.~\cite{Matheus:2006xi,Navarra:2006nd}, where the
mass and decay  width were obtained.  Afterwards,  several other QCDSR
calculations  for  the  mass of  $X(3872)$ were  done,  considering  different
hypothesis      for its     quark      structure:    a  molecule
\cite{Lee:2008uy,Zhang:2009em,Wang:2013daa,Mutuk:2018zxs}, a tetraquark
\cite{Chen:2010ze,Wang:2013vex},  a hybrid
\cite{Harnett:2012gs,Chen:2013zia},    and  a  mixed    hybrid-molecule
\cite{Chen:2013pya,Palameta:2018yce}. In  particular, in all  works considering a four-quark structure, tetraquark or molecule,
the   mass obtained  for    $X(3872)$ is  compatible  with  the
experimental one, considering the uncertainties.  Thus, from a QCDSR point
of view, the mass of
tetraquark and molecule structures are indistinguishable and cannot be
used as the  sole test to determine the state  configuration. This finding is
not surprising since the currents  describing these two structures are
related   by  a   Fierz   transformation  \cite{Nielsen:2009uh}.    In
Refs.~\cite{Lee:2008tz,Narison:2010pd},   the    equivalence   between
the tetraquark and molecule  results for the mass  is addressed, asserting
the indistinguishability of both results.  The only feature that could
set apart the  structures is that in the molecular configuration a better
Borel stability is obtained \cite{Lee:2008tz}.
Regarding the  hybrid structure,  from the  works
quoted  above,  the   mass  obtained  from  a QCDSR calculation for a state with
$J^{PC}=1^{++}$  is   $5.13\GeV$,  which  is  much   bigger  than  the
experimental mass,  while   the  mixed  hybrid-molecule   provides  compatible
results.   More  recently,  the    properties   of $X(3872)$ in  a  dense   medium
\cite{Azizi:2017ubq} were also studied in the QCDSR framework.

Most of the works quoted above do not address the decay width of the state,
which can be crucial to determine  the structure of a state.  The only
work that deals with the decay properties of  $X(3872)$, considered as a
pure four-quark state,  was done in Ref.~\cite{Navarra:2006nd}.
In Ref.~\cite{Navarra:2006nd}  it was
shown that the decay width of  the  tetraquark  state  is   too  large  in
comparison  with  the experimental decay width of the  $X(3872)$.
A further attempt, using a mixed molecule and  charmonium currents, has been
successful in explaining
the   mass,  decay   and  production   properties  of   the  $X(3872)$
\cite{Matheus:2009vq,Nielsen:2010ij,Zanetti:2011ju}.      This  
possibility is also  in agreement  with the  current radiative  decay data.

In the remaining of this section, we review briefly the QCDSR
studies  of the   $X(3872)$  based on a 
four-quark picture, a tetraquark or  a molecular current,    to conclude that a pure four-quark state is incompatible with the decay
properties of   $X(3872)$. Next, we  present  the QCDSR
analysis considering an admixture of  charmonium and $D\bar{D}^*$ currents.
We show that this kind of current can explain all the data, including those related with the
$X(3872)$  decay  processes. We also present the studies of the $X(3872)$
radiative  decay  and of the $X(3872)$ production in $B$ decays,
providing  the most  complete  and consistent  QCDSR  analysis of  the
$X(3872)$ at the present.

\subsection{QCDSR for pure four-quark structures}

In   Ref.~\cite{Matheus:2006xi}, for the first time a tetraquark structure for   $X(3872)$   was  tested   in   the  framework   of   QCDSR.  A
diquark-antidiquark  current, previously  proposed by  Maiani {\it et. al.}
\cite{Maiani:2004vq} was used.  The  current was constructed for  the state with
quantum  numbers $J^{PC}=1^{++}$  with a  symmetric spin  distribution
$[cq]_{S=1}[\bar{c}\bar{q}]_{S=0}+[cq]_{S=0}[\bar{c}\bar{q}]_{S=1}$,
with the corresponding interpolating current given by
\beqa
j_\mu={i\epsilon_{abc}\epsilon_{dec}\over\sqrt{2}}[(q_a^TC
  \gamma_5c_b)(\bar{q}_d\gamma_\mu     C\bar{c}_e^T)+(q_a^TC\gamma_\mu
  c_b) (\bar{q}_d\gamma_5C\bar{c}_e^T)]\;,
\label{cur-di}
\enqa
where $a,~b,~c,~...$ are color indices, $C$ is the charge conjugation
matrix and $q$ denotes a $u$ or $d$ quark.

\subsubsection{Two-point correlation function}
The  SR for  the mass  of  the state  is obtained  from the  two-point
correlation function:
\beqa
\Pi_{\mu\nu}(q)&=&i\int d^4x ~e^{iq.x}\lag 0
|T[j_\mu^{(q)}(x)j^{(q)\dagger}_\nu(0)]
|0\rag
\nn\\
&=&-\Pi_1(q^2)\left(g_{\mu\nu}-{q_\mu q_\nu\over q^2}\right)+\Pi_0(q^2){q_\mu
q_\nu\over q^2},
\lb{2po}
\enqa
The non-conservation of the axial-vector  current implies that the two
functions $\Pi_1$  and $\Pi_0$ (with  spin 1 and 0,  respectively) are
independent.

The phenomenological  side of  the sum rule, computed  by inserting
intermediate states for $X$, is given by
\beq
\Pi^{phen}_{\mu\nu}(q^2)=\frac{2f_X^2M_X^8}{M_X^2-q^2}\left(-g_{\mu\nu}+\frac{q_\mu q_\nu}{M_X^2}\right)
\enq
where the decay constant $f_X$ is  used to parametrize the coupling of
the axial-vector meson $1^{++}$ to the current $j_\mu$ as
\beq
\lag0\vert j_\mu\vert X \rag = \sqrt{2}f_XM_X^4\epsilon_\mu.\lb{coupjmeson}
\enq

As discussed in Sec.~{\ref{s-ope}, in  the OPE  side,  the  correlation function  $\Pi_1$  is  given by  a
dispersion relation:
\beq
\Pi_1^{OPE}(q^2)=\int_{4m_c^2}^\infty ds {\rho^{OPE}(s)\over s-q^2}=\frac{1}{\pi}
\int_{4m_c^2}^\infty ds~{\rm{Im}[\Pi_1^{OPE}(s)]\over s-q^2} \;.
\lb{ope}
\enq

\begin{figure*}\centering
    \subfigure[]{\lb{fig1a}\includegraphics[angle=0,scale=0.8]{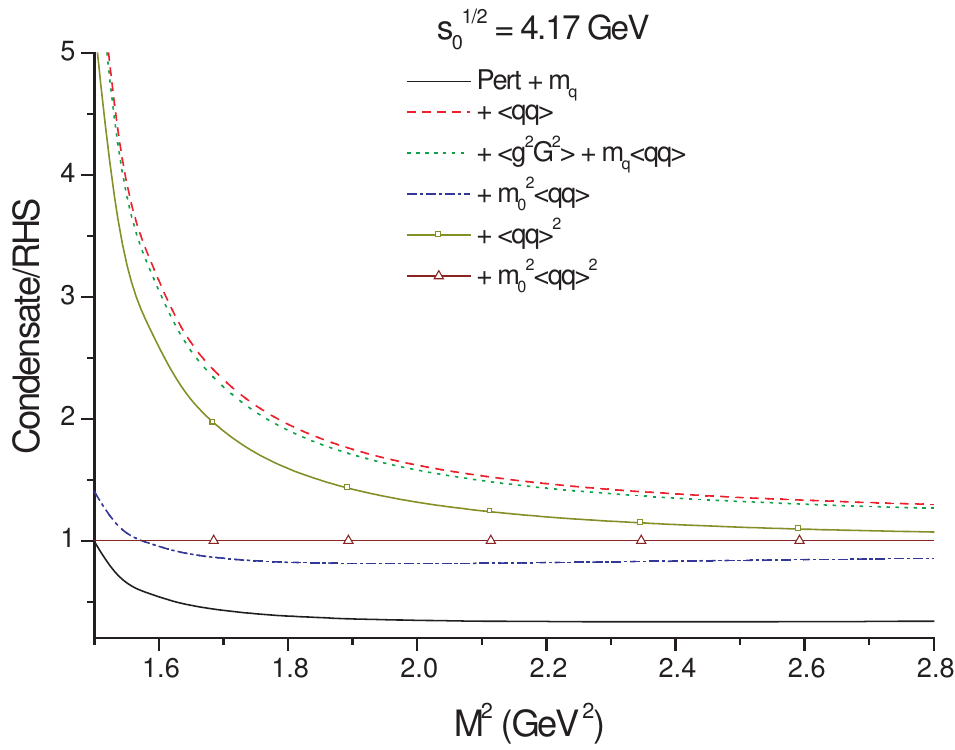}}
    \subfigure[]{\lb{fig1b}\includegraphics[angle=0,scale=0.8]{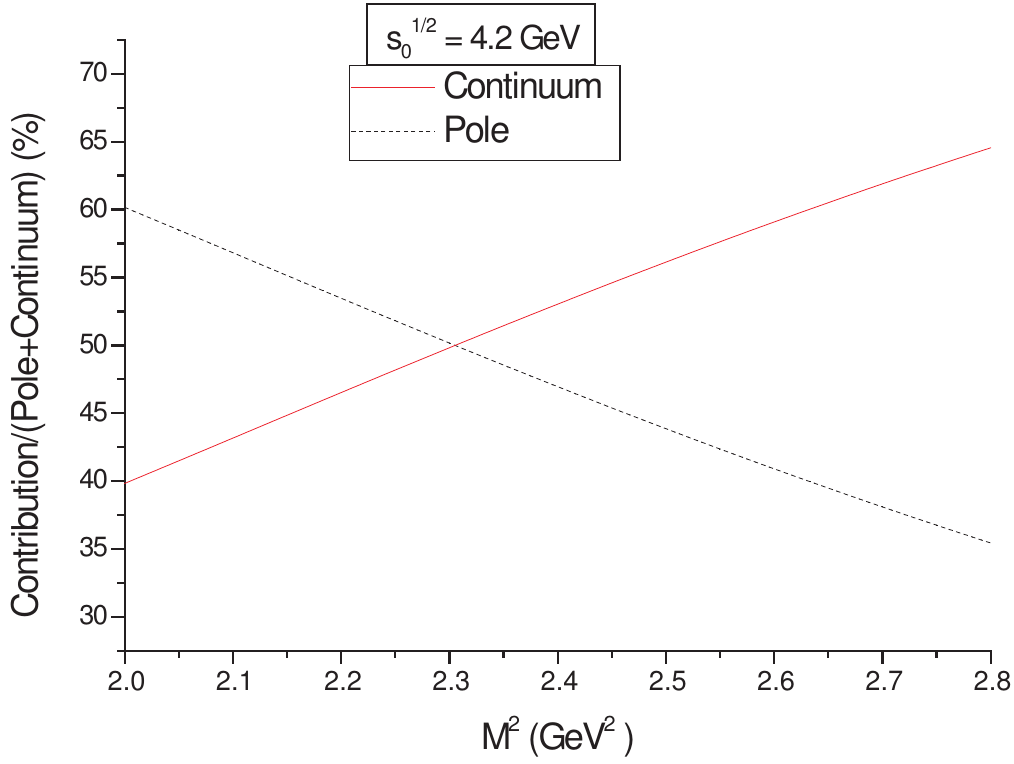}}
  \caption{(a) The  OPE  convergence  in  the region  $1.6 \leq  M^2 \leq  2.8
    \GeV^2$ for $\sqrt{s_0}=4.17\GeV$.  Starting with the perturbative
    contribution  (plus  a  very  small  $m_q$  contribution).  Each
    subsequent line  represents the  addition of the contribution associated with the next condensate
     in the expansion. (b) The dashed line shows the relative pole
contribution (i.e., the pole contribution divided by the total, pole plus
continuum, contribution) and the solid line shows the relative
continuum contribution. Figures taken from~Ref.~\cite{Matheus:2006xi}.}
\end{figure*}

The matching of  both sides of the  sum rule is done  by applying a
Borel  transform  on  both  sides. After  transferring  the  continuum
contribution  to the  OPE side,  the  sum rule  up to  dimension-eight
condensates, is written as
\beq
2f_X^2M_X^8e^{-M_X^2/M^2}=\int_{4m_c^2}^{s_0}ds\,e^{-s/M^2}\rho(s)+\Pi_1^{\rm{mix}\comq}(M^2)\lb{sr-1}
\enq
with
\beqa
\rho(s)=\rho^{pert}(s)+\rh^{m_q}(s)+\rh^{\comq}(s)+\rh^{\lag G^2\rag}
(s)+\rh^{mix}(s)+\rh^{\comq^2}(s)\;,
\lb{rhoeq}
\enqa
and  the  function  $\Pi_1^{\rm{mix}\comq}(M^2)$   is  a  part  of  the
dimension-eight condensate that does not depend on $s$.

The convergence  of the OPE  and the  comparison between the  pole and
continuum    contribution   are    shown   in    Figs.~\ref{fig1a}   and
\ref{fig1b}. The lower limit of the Borel window is determined from the
convergence of the OPE for higher values of $M^2$, and the upper limit
from the constraint  that the pole contribution must  dominate over the
continuum contribution.  This procedure leads to  the following Borel
window: $2.0\leq M^2\leq 2.2\GeV^2$.

The mass  of the $X$ state  as a function  of the Borel mass  $M^2$ is
presented  in Fig.~\ref{Xf3},  showing  that the  sum  rule is  stable
within the  Borel mass window. The  result for the mass  $M_X$, taking
into account the  uncertainties of the parameters and  within the range
of the Borel mass, is
\beq
M_X=(3.92\pm0.13)~\GeV\;,
\enq
a value  compatible with the  experimental data on $X(3872)$.

\begin{figure}
\centering
    \includegraphics[angle=0,scale=0.8]{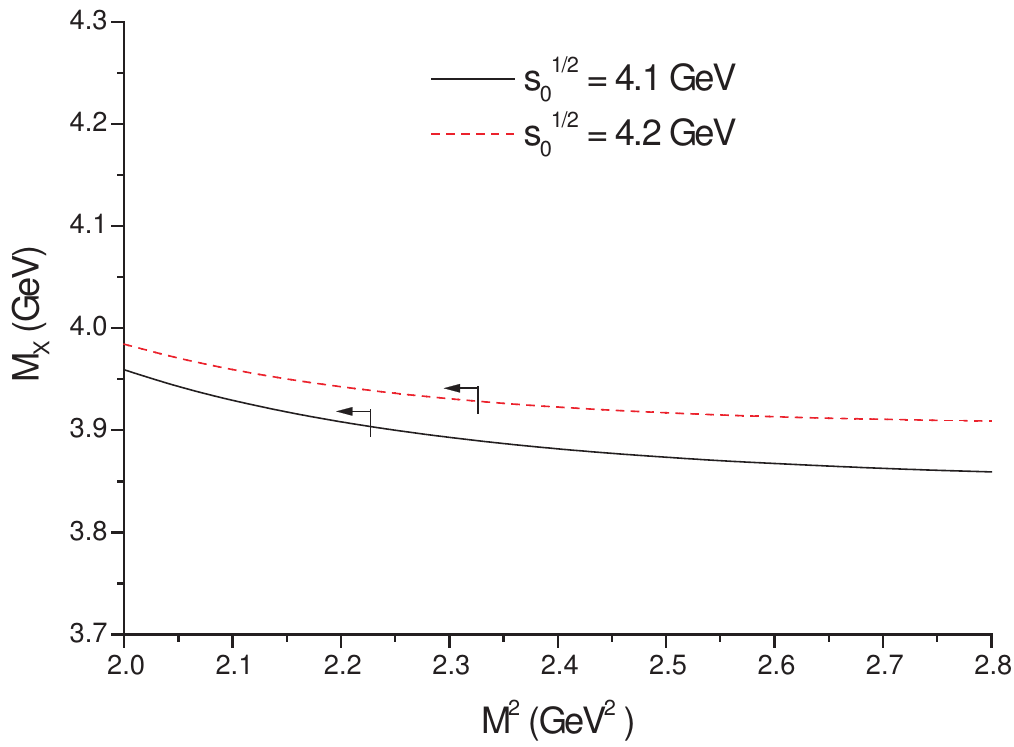}
\caption{The $X$ meson  mass as a function of the  Borel mass $M^2$ for
  different values of the continuum  threshold: $ \sqrt{s_0}= 41 \GeV$
  (solid line)  and $\sqrt{s_0}=4.2  \GeV$ (dashed line).   The arrows
  indicate the region allowed for the  sum rules: the lower limit (cut
  below $2.0 \GeV^2$) is given  by the OPE convergence requirement and the
  upper    limit    by    the    dominance    of    the    QCD    pole
  contribution. Figures taken from~Ref.~\cite{Matheus:2006xi}.}
  \label{Xf3}
  \end{figure}

The  above results were obtained using the tetraquark current in Eq.~(\ref{cur-di}), although other four-quark operators with $1^{++}$ quantum numbers are possible.  For example, a different tetraquark current with $J^{PC}=1^{++}$ can be
constructed by combining the pseudoscalar $0^-$ and vector $1^-$ diquarks,
instead of the  scalar $0^+$ and isovector $1^+$ diquarks, as done in Eq.~(\ref{cur-di}).  Equivalently, other operators can be constructed with meson type currents.  The number of currents increases further if one allows for additional color states; color sextet for the diquark and color octet for the
molecular states.  An extensive study has been carried out for  $X(3872)$
in Ref.~\cite{Narison:2010pd}. As shown in Ref.~\cite{Narison:2010pd,Dias:2011mi}, the choice
of the current does not matter 
much for the determination of the mass of the state, provided that one can work
with quantities less affected by radiative corrections and where the OPE converges quite well. Besides the current in the $\bar{3}-3$ color configuration in 
Eq.~(\ref{cur-di}), the other  currents considered were:

\noindent
 a diquark-antidiquark in the color sextet ($\bar 6-6$) configuration,
\beqa
j^\mu_{6}&=&{i\over\sqrt{2}}[(q_a^TC\gamma_5\lambda^S_{ab}c_b)
(\bar{q}_d\gamma^\mu C\lambda^S_{de}\bar{c}_e^T)+(q_a^TC\gamma^\mu
\lambda^S_{ab} c_b)(\bar{q}_d\gamma_5C\lambda^S_{de}\bar{c}_e^T)]\;,
\label{field}
\enqa
where $\lambda^S$ stands for the six symmetric Gell-Mann matrices:
$\lambda^S=\left(\lambda_0,~\lambda_1,~\lambda_3,~\lambda_4,~\lambda_6,
~\lambda_8\right)$; a $D^*-D$ molecular current,
\beqa
j^{(mol)}_{\mu}(x) & = & {1 \over \sqrt{2}}
\bigg[
\left(\bar{q}_a(x) \gamma_{5} c_a(x)
\bar{c}_b(x) \gamma_{\mu}  q_b(x)\right) - 
\left(\bar{q}_a(x) \gamma_{\mu} c_a(x)
\bar{c}_b(x) \gamma_{5}  q_b(x)\right)
\bigg];
\lb{cur-mol}
\enqa
and  a $\lambda-J/\psi$-like molecular current,
\beqa
j^{\mu}_{\lambda}  =  \left[(\bar{c}\lambda^a\gamma^{\mu}c)
(\ga\bar{q}\lambda_a  \gamma_5 q)\right],
\lb{eq:curr5}
\enqa
where $\lambda_a$ is the color matrix. \\

In particular, it was shown that the three substructure assignments for the $X$-meson ($\bar 3-3$ and $\bar 6-6$ tetraquarks and $D-D^{(*)}$ molecule) lead to (almost) the same mass predictions within the accuracy of the approach. Therefore, a priori, a study of the $X$-mass alone cannot reveal
its  nature and identify if it is mainly composed of these substructures. On the other hand,
the analysis of the $\lambda- J/\psi$-like molecular current in Eq.~(\ref{eq:curr5}) showed that a  lower mass as compared to the previous configurations can be obtained, with the ratio between the masses  obtained with the currents in Eqs.~(\ref{eq:curr5}) and (\ref{cur-di}) being $0.96\pm0.03$.

From the analysis of all these currents, it was found that the distance between the continuum threshold (about 4 GeV) and the resonance mass
is relatively small, which indicates that the separation between the resonance and the continuum may be difficult to achieve. 

The molecular $D\bar{D}^*$ current in Eq.~(\ref{cur-mol}),
together with the tetraquark current, Eq.~(\ref{cur-di}), were also used in Ref.~\cite{Lee:2008tz} to study the importance of including the width of the state,
in a QCDSR calculation. This can be done by replacing the delta function in Eq.~(\ref{den})
by the relativistic Breit-Wigner function:
\begin{equation}
 \delta(s-m^2) \to \frac{1}{\pi}\frac{\Gamma
  \sqrt{s}}{(s-m^2)^2+s\Gamma^2}.\label{eq:BW}
\end{equation}
The mass and width were determined by looking at the stability of the 
 mass results against  the  varying   Borel parameter $M^2$, as usual.

Although the effect of the width was not found to be large, it was possible
to fit the experimental mass, 3872 MeV, and the width,
$\Gamma < 1.2$ MeV, simultaneously for both currents. However, the 
molecular current, Eq.~(\ref{cur-mol}),  gave a better stability as compared with the tetraquark current of Eq.~(\ref{cur-di}).

\subsubsection{Three-point correlation function}
\label{s-3po-tetra}

As discussed in the previous subsection, the  mass of  the  $X(3872)$  can be well reproduced  from  QCDSR calculations for  a variety of four-quark structures. However, it is important to test
if these currents can also explain other properties of the state, like, the
decay widths.

With such a motivation, in Ref.~\cite{Navarra:2006nd}, the  tetraquark current of Eq.~(\ref{cur-di})
was  tested, within  the QCDSR approach, to  the calculation   the  decay  width for  the isospin   violating   channels,   $X(3872)\to   J/\psi\pi^+\pi^-$   and $X(3872)\to   J/\psi\pi^+\pi^-\pi^0$. It was shown in Ref.~\cite{Navarra:2006nd},
 that the  ratio between these decays  can be obtained
from
\beq
{\Gamma(X\to J/\psi\,\pi^+\pi^-\pi^0)\over \Gamma(X\to J/\psi\,\pi^+\pi^-)}
=0.118\left({g_{X\psi\omega}\over g_{X\psi\rho}}\right)^2,
\label{rationum1}
\enq
where the coupling  constants $g_{X\psi V}$, with  $V=\omega,\rho$, can
be evaluated  from the study of  the vertex $X\psi V$  in QCDSR,
through the three point correlation function:
\beq
\Pi^V_{\mu\nu\al}(p,\pli,q)=\int d^4x d^4y ~e^{i\pli.x}~e^{iq.y}
\Pi^V_{\mu\nu\al}(x,y),
\enq
with
\beq
\Pi^V_{\mu\nu\al}(x,y)=\lag 0 |T[j_\mu^{\psi}(x)j_{\nu}^{V}(y){j_\al^{X}}^
\dagger(0)]|0\rag,
\lb{3po}
\enq
where $p=\pli+q$,  and the interpolating fields are given by:
\beqa
j_{\mu}^{\psi}&=&\bar{c}_a\gamma_\mu c_a,\nn\\
j_{\nu}^{\rho}&=&{1\over2}(\bar{u}_a\gamma_\nu u_a-\bar{d}_a\gamma_\nu
d_a),\nn\\
j_{\nu}^{\omega}&=&{1\over6}(\bar{u}_a\gamma_\nu u_a+\bar{d}_a\gamma_\nu
d_a).
\lb{cur-3po}
\enqa
To be  able to reproduce the  experimental data, as was shown in Ref.~\cite{Navarra:2006nd}, it is necessary  to take
the current for the $X$ state as a mixture
\beq
j_{\mu}^{X}=\cos\al j_\mu^{(u)}+\sin\al j_\mu^{(d)},
\label{mixrate}
\enq
where $j_\mu^{(u)}$ and $j_\mu^{(d)}$ are tetraquark currents given by
Eq.~(\ref{cur-di}) with $q=u,d$.

Considering the  light quarks as degenerated, we have:
\beqa
\Pi^\rho_{\mu\nu\al}(x,y)&=&{-i\over2\sqrt{2}}\left(\cos{\al}~
-\sin{\al}\right)~\Pi^q_{\mu\nu\al}(x,y),\nn\\
\Pi^\omega_{\mu\nu\al}(x,y)&=&{-i\over6\sqrt{2}}\left(\cos{\al}~
+\sin{\al}\right)~\Pi^q_{\mu\nu\al}(x,y),
\label{piAI}
\enqa

On the  phenomenological side, the  intermediate states of  the three
mesons are inserted in the vertex, and the following definitions are applied:
\beqa
\lag 0 | j_\mu^\psi|J/\psi(\pli)\rag &=&m_\psi f_{\psi}\epsilon_\mu(\pli),
\nn\\
\lag 0 | j_\nu^V|V(q)\rag &=&m_{V}f_{V}\epsilon_\nu(q),
\nn\\
\lag X(p) | j_\al^X|0\rag &=&\la_X \epsilon_\al^*(p),
\lb{fp}
\enqa
\beq
\lag J/\psi(\pli) V(q)|X(p)\rag=g_{X\psi V}\epsilon^{\si\al\mu\nu}p_\si
\epsilon_\al(p)\epsilon_\mu^*(\pli)\epsilon_\nu^*(q),
\label{coup}
\enq
where  $\la_X=(\cos{\al}+\sin{\al})\la^q$   (with  the  coupling
between  the current  and the  meson as  $\lambda_q=\sqrt{2}f_XM_X^4$,
from Eq.~(\ref{coupjmeson})). The coupling  constant $g_{X\psi V}$ is
extracted from the effective Lagrangian that describes the coupling
between two vector mesons and one axial vector meson:
\beq
{\cal{L}}=ig_{X\psi V}\epsilon^{\mu\nu\al\si}(\partial_\mu X_\nu)\Psi_\al
V_\si.
\enq
The phenomenological side was, thus, computed in Ref.~\cite{Navarra:2006nd}, as
\beqa
\Pi_{\mu\nu\al}^{phen} (p,\pli,q)={i(\cos\al+\sin\al)\lambda^q m_{\psi}
f_{\psi}m_Vf_{V}~g_{X\psi V}
\over(p^2-m_{X}^2)({\pli}^2-m_{\psi}^2)(q^2-m_V^2)}
\times\bigg(-\epsilon^{\al
\mu\nu\si}(\pli_\si+q_\si)-\epsilon^{\al\mu\si\ga}{\pli_\si q_\ga q_\nu
\over m_V^2}
-\epsilon^{\al\nu\si\ga}{\pli_\si q_\ga\pli_\mu\over m_\psi^2}
\bigg)+\cdots\;,
\lb{phen3}
\enqa

The OPE side was computed with condensates up to dimension-five, written in
terms of a dispersion relation for each Dirac structure $i$:
\beq
\Pi_i^{OPE}(M^2,Q^2)=\int_{4m_c^2}^\infty\rho_i^{OPE}(u,Q^2)e^{-u/M^2}du.
\enq
There  are   four  different  Dirac
structures contributing to the correlation function. We chose to
work with the structures that have  more  condensates
contributing to  the OPE. They are:  $\epsilon^{\alpha\mu\nu\sigma}q_\sigma$ and
$\epsilon^{\alpha\nu\sigma\gamma}p^{\prime}_\sigma            q_\gamma
p^{\prime}_\mu$.

The sum rule  for each structure was obtained by  matching both sides,
and performing  a single Borel  transformation to $P^2= P^{\prime2}\to  M^2$.
Transferring the pure continuum contribution  to the OPE side, the
general expression for the sum rule for each structure, $i=1,2$, was obtained:
\beq
C_i^{XV}(Q^2)\left(e^{-m_\psi^2/M^2}-e^{-m_X^2/M^2}\right)+B_ie^{-s_0/M^2}=-i\frac{Q^2+m_V^2}{2\sqrt{2}}\Pi^{\rm{OPE}}_i(M^2,Q^2),\lb{sumrl}
\enq
where, $B_i$ takes into account the pole-continuum contribution, as discussed in Sec.~\ref{s-3p}, and 
\beqa
C_1^{XV}(Q^2)&=&\frac{f_\psi}{m_\psi}\frac{\lambda_q}{m_X^2-m_\psi^2}A_{XV}(Q^2)\nn\\
C_2^{XV}(Q^2)&=&f_\psi m_\psi\frac{\lambda_q}{m_X^2-m_\psi^2}A_{XV}(Q^2)
\enqa
with
\beqa
A_{X\rho}(Q^2)&=&- m_\rho f_\rho\frac{\cos\al+\sin\al}{\cos\al-\sin\al}g_{X\psi\rho}(Q^2)\nn\\
A_{X\omega}&=&3m_\omega f_\omega g_{X\psi\omega}(Q^2).\lb{axlh}
\enqa
The OPE side of the sum  rule in Eq.~(\ref{sumrl}) determines a unique  value  for  $C^{XV}$  for  each structure,  thus, the
following ratio between the form factors was found:
\beqa
\frac{ g_{X\psi\omega}(Q^2)}{ g_{X\psi\rho}(Q^2)}=\frac{m_\rh f_\rho}{3m_\omega f_\omega}\frac{\cos\al+\sin\al}{\cos\al-\sin\al}.
\lb{ff-ro}
\enqa
Using Eq.~(\ref{ff-ro}) into Eq.~(\ref{rationum1}) we finally got
\beq          {\Gamma(X\to         J/\psi\,\pi^+\pi^-\pi^0)\over
  \Gamma(X\to
  J/\psi\,\pi^+\pi^-)}=0.153\left(\frac{\cos\al+\sin\al}{\cos\al-\sin\al}\right)^2.
\lb{rate-mix-al}
\enq
The experimental value in Eq.~(\ref{rate}) was used to  determine the mixing
angle: $\al\sim20^\circ$ \cite{Navarra:2006nd}.

The  decay  width was  finally  computed  using the  coupling  constant
determined in Ref.~\cite{Navarra:2006nd}: $g_{X\psi\omega}=13.8\pm2.0$,
leading to the following result for the partial width:
\beq
\Gamma(X\to J/\psi\,(n\pi))=(50\pm15)\MeV
\enq
which is much  bigger than the experimental lower limit  for the total
width  $\Gamma<1.2\MeV$.   In
Ref.~\cite{Matheus:2009vq} it was  also shown that a  similar result is
obtained for a molecule. Therefore, a  pure four-quark state can not explain
both mass and  width within the QCDSR framework.

\subsection{Mixing of two- and four-quarks structure from QCDSR}
\lb{sec-mixing}

In the  QCDSR studies, the  $X(3872)$ mass is obtained
in  agreement  with  the  experimental data  in  both,  tetraquark  and
molecular,  pictures.  However,  the  QCDSR results for the  decay width
were not found to be in agreement with the available data.  A new attempt to obtain
both, mass and decay width, compatible with experiment
was  made in  Ref.~\cite{Matheus:2009vq}, where 
$X(3872)$ was considered to be a mixture of two-  and four-quarks
states. In particular,  a mixing between charmonium and
molecular $\bar{D}^{*}D$ states was done at the level of  the
currents.  This kind  of   mixture  was  previously considered, in the QCDSR
approach,  in Ref.~\cite{Sugiyama:2007sg} for  the light quark sector,  and it was
also implemented, using  QCD   factorization,   to  study   the   $X(3872)$   
\cite{Suzuki:2005ha}.  In order to keep consistency with the available data
on the isospin breaking decay modes, Eq.(\ref{rate}), a  further
mixing was found to be necessary. Therefore,  $X(3872)$ was  considered  to be
a mixing between charmonium,
$(D^{*0}\bar{D}^0-\bar{D}^{*0}D^0)$                     and
$(D^{*+}\bar{D}^--\bar{D}^{*-}D^+)$ states.  In the next subsections we
discuss the calculations of the mass, decay widths and the $B$ meson production channel of 
$X(3872)$ in the QCDSR approach.

\subsubsection{\lb{2pointsec}Two-point correlation function}

The interpolating current, in Ref.~\cite{Matheus:2009vq}, for the two-quarks and four-quarks mixed states has a
two-quark conventional charmonium axial current part:
\beq
j'^{(2)}_{\mu}(x) = \bar{c}_a(x) \gamma_{\mu} \gamma_5 c_a(x),
\lb{curr2}
\enq
and a four-quark part given by a $D^0\bar{D}^{*0}$ molecular current:
\beqa
j^{(4q)}_{\mu}(x) & = & {1 \over \sqrt{2}}
\bigg[
\left(\bar{q}_a(x) \gamma_{5} c_a(x)
\bar{c}_b(x) \gamma_{\mu}  q_b(x)\right)  - \left(\bar{q}_a(x) \gamma_{\mu} c_a(x)
\bar{c}_b(x) \gamma_{5}  q_b(x)\right)
\bigg].
\lb{curr4}
\enqa
Since these two currents have different dimensions, we follow Ref.~\cite{Sugiyama:2007sg} and to write the two-quark part of the current as:
\beq
j^{(2q)}_{\mu} = {1 \over 6 \sqrt{2}} \comq j'^{(2)}_{\mu}.
\enq
The mixing of the two currents was considered as:
\beq
J_{\mu}^q(x)= \sin(\theta) j^{(4q)}_{\mu}(x) + \cos(\theta) j^{(2q)}_{\mu}(x). 
\lb{field0}
\enq
The two-point correlation  function in Ref.~\cite{Matheus:2009vq} was obtained, as  usual, by inserting
the corresponding current (Eq.~(\ref{field0})) in the two-point correlation
function, to get
\beqa
\Pi_{\mu\nu}(q)  = 
\left({\comq \over 6 \sqrt{2}}\right)^2 \cos^2(\theta)
\,\Pi^{(2,2)}_{\mu\nu}(q) 
 + 
{\comq \over 6 \sqrt{2}}\left(\sin(2\theta)\right) 
\,\Pi^{(2,4)}_{\mu\nu}(q)
+ 
\sin^2(\theta)\, \Pi^{(4,4)}_{\mu\nu}(q),
\enqa  
with
\beq
\Pi^{(i,j)}_{\mu\nu}(q) 
 = 
i\int d^4x ~e^{iq.x}\lag 0
|T[j^{(i)}_\mu(x)j^{(j)\dagger}_\nu(0)]
|0\rag.
\enq

The phenomenological side was obtained by parametrizing  the  hadron-current
coupling  in   terms  of  a  single  parameter
$\lambda^q$:
\beq\label{eq:decay}
\lag 0 |
J_\mu^q|X\rag =\lambda^q\epsilon_\mu~, 
\enq
and by inserting  intermediate states in the correlation function, giving
\beq
\Pi_{\mu\nu}^{phen}(q)={(\lambda^q)^2\over
m_X^2-q^2}\left(-g_{\mu\nu}+ {q_\mu q_\nu\over m_X^2}\right)
+\cdots\;, \lb{phe} \enq
where the chosen Lorentz structure projects out the $1^{++}$ state.  The dots
denote higher mass axial-vector resonances.

The QCDSR was obtained by applying the Borel transformation  on both sides,
and transferring  the continuum contribution to the OPE side:
\beqa (\lambda^q)^2e^{-m_X^2/M^2} 
 =
\left({\comq \over 6 \sqrt{2}}\right)^2 \cos^2(\theta)
\,\Pi^{(2,2)}_{1}(M^2) 
 +
{\comq \over 6 \sqrt{2}}\left(\sin(2\theta)\right) 
\,\Pi^{(2,4)}_{1}(M^2)
+ 
\sin^2(\theta)\, \Pi^{(4,4)}_{1}(M^2). 
\lb{sr} 
\enqa
In Ref.~\cite{Matheus:2009vq} the functions $\Pi^{i,j}(q)$ were  computed in
leading order in $\alpha_s$ up to the dimension-eight condensates. 
\begin{figure*}[t]
  \centering
    \subfigure[]{\lb{fig6a}\includegraphics[scale=0.7,angle=0]{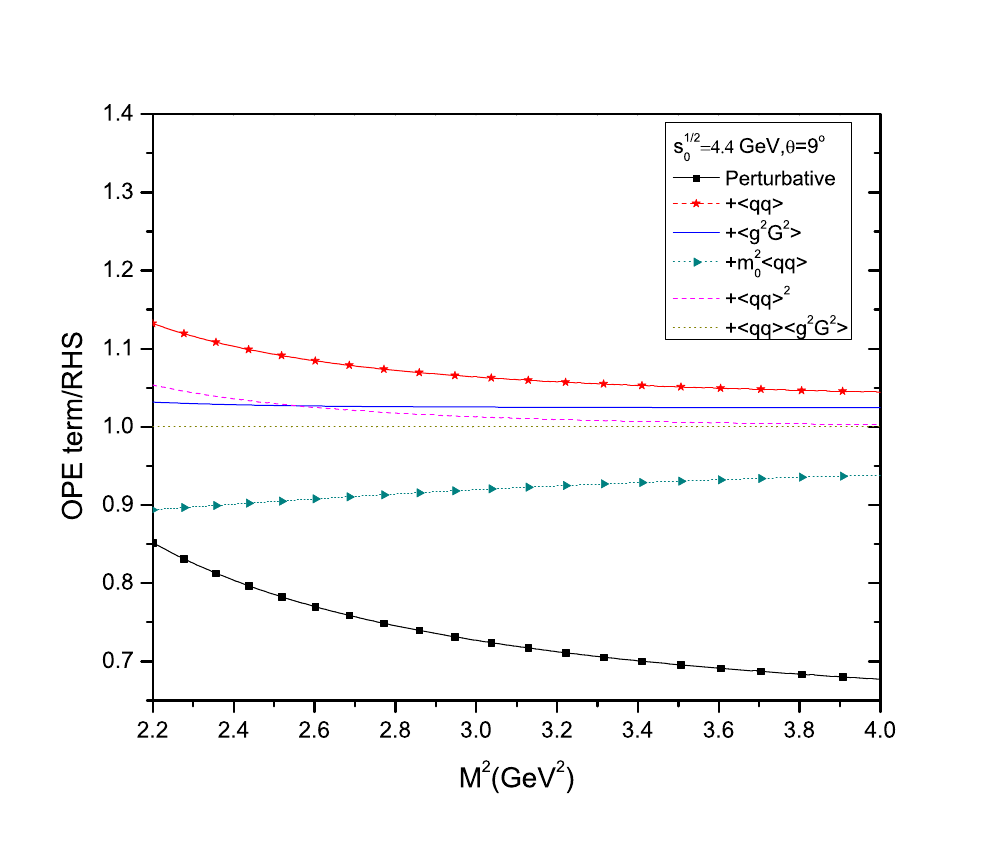}}
    \subfigure[]{\lb{fig6b}\includegraphics[scale=0.7,angle=0]{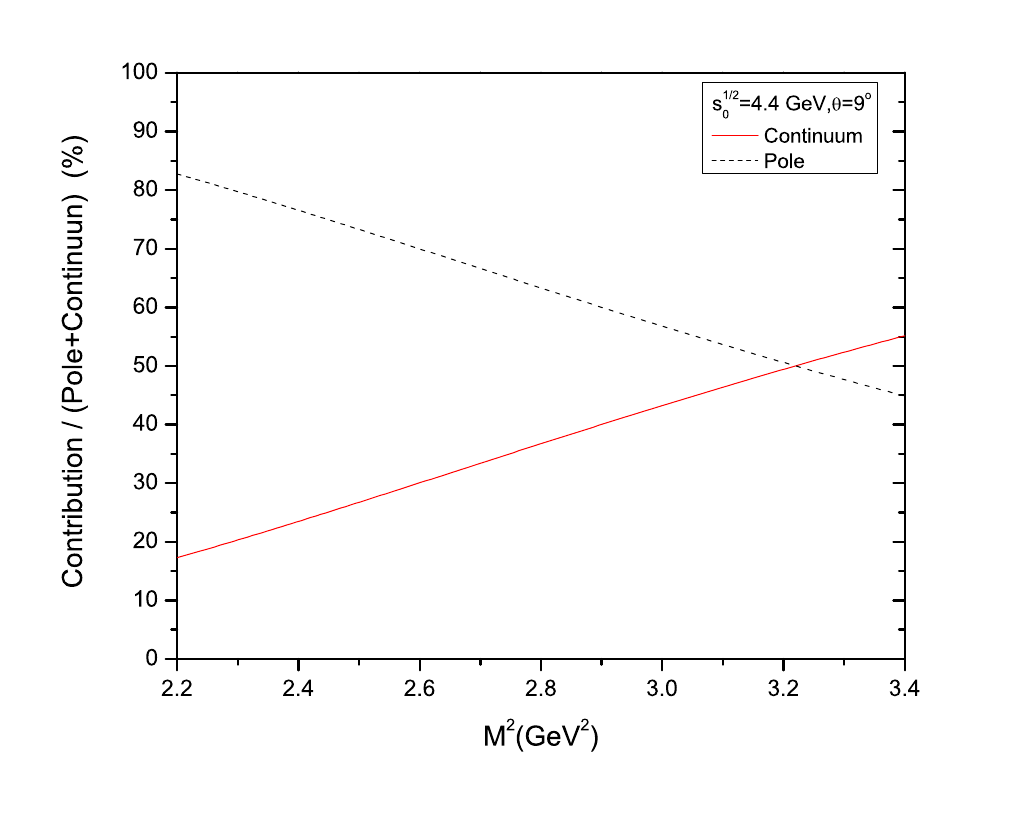}}
     \caption{Figures showing the OPE  convergence for the correlation
       function (left) and the pole-continuum dominance (right). Taken from \cite{Matheus:2009vq}}
\end{figure*}
In Figs.~\ref{fig6a} and \ref{fig6b}, we show the results for the OPE convergence
and for the pole-continuum  contributions as obtained in Ref.~\cite{Matheus:2009vq}.
These results were obtained using $\sqrt{s_0}=4.4 \GeV$ and $\theta=9^\circ$.
From Fig.~\ref{fig6a} we see that there is a good
OPE  convergence for  values of $M^2\geq2.6\GeV^2$. The  upper limit  for the
Borel  mass is determined from the pole-continuum analysis, fixing the Borel
window as:
\beq2.6\GeV^2\leq M^2\leq 3.0\GeV^2,\enq
The   result   from  the   QCDSR   for   the $X(3872)$  mass is presented   in
Fig.~\ref{fig7},  showing  a  good   stability  within  the  Borel
window. It was found that there is no problem in reproducing the experimental mass of 
$X(3872)$
with the mixed current for a large range of the mixing angle $\theta$.
Considering $\theta$ in the region $5^\circ \leq \theta \leq 13^\circ$ one
gets \cite{Matheus:2009vq}:
\beq\label{sr.mx}
m_X = (3.77 \pm 0.18) \GeV,
\enq
which is in good agreement with the experimental data.  The value of
the meson-current coupling parameter, $\lambda^q$, was also obtained:
\beq\label{sr.lamb}
\lambda^q = (3.6 \pm 0.9).10^{-3} \GeV^5.
\enq

\begin{figure}[t]
\centering
    {\includegraphics[angle=0]{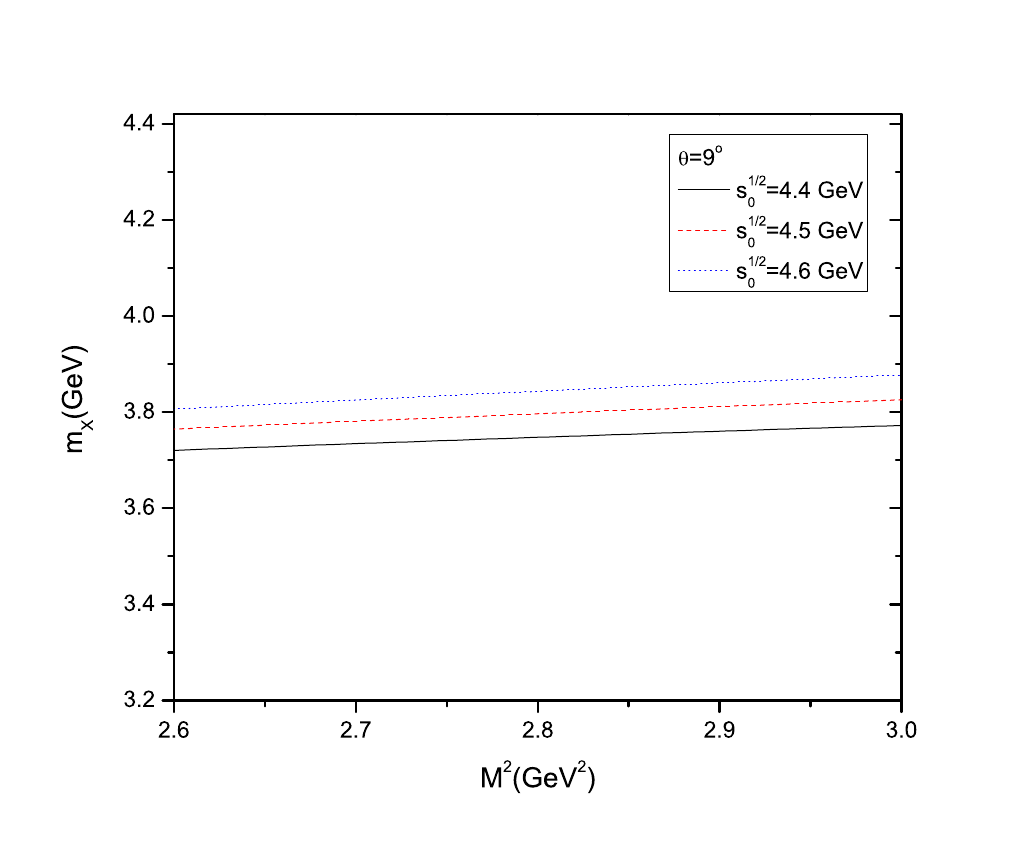}}
    \caption{The $X$  meson mass as a  function of the sum  rule parameter
      ($M^2$) within the Borel window range, for  different  values of  the
      continuum threshold $s_0$. Taken from  \cite{Matheus:2009vq}.}
  \label{fig7}
  \end{figure}

\subsubsection{Three-point correlation function:  $X(3872)\to J/\psi(n\pi)$ decay}

As discussed earlier, to be able to explain the ratio between the decay rates in
Eq.~(\ref{rate})
one needs to consider also a mixture between the $D^+D^{*-}$ and $D^-D^{*+}$ components. Therefore,  the most general current is given by
  \beq    j_{\mu}^X(x)=    \cos\alpha    J_{\mu}^u(x)+\sin\alpha
J_{\mu}^d(x),
\label{4mix}
\enq
with $J_{\mu}^u(x)$ and $J_{\mu}^d(x)$  given by Eq.(\ref{field0}).
The interpolating fields for the other mesons in the vertex 
are given by Eqs.~(\ref{cur-3po}). The
three  point function  for the  vertex $X(3872)J/\psi  V$ is  obtained
inserting the currents (Eq.~(\ref{4mix}) and (\ref{cur-3po})) into the three-point
function of Eq.~(\ref{3po}). 
Taking the quarks  $u$ and $d$ as degenerated, one arrives at Eqs.~(\ref{piAI})
multiplied by $\sin\theta$, showing that it is only the four-quark part of the
current in Eq.~(\ref{field0}) that contributes to these decays.

The phenomenological side of the sum rule is given by Eq.~(\ref{phen3}).
As discussed in Sec.~\ref{s-3po-tetra}, the mixing  angle $\alpha$  is determined in  order to  reproduce the experimental  ratio in Eq.~(\ref{rate}):
$\al=20^\circ$. This is the same result obtained in
Refs.~\cite{Maiani:2004vq,Navarra:2006nd}.
In Ref.~\cite{Matheus:2009vq} the OPE side was evaluated up to dimension five
at leading order in $\alpha_s$.  Taking  the limit $p^2={\pli}^2=-P^2$ and
doing a single Borel  transform to  $P^2\rightarrow  M^2$, we  get  in the
structure $\epsilon^{\al\nu\si\ga}\pli_\si  q_\ga\pli_\mu$
\beqa
C(Q^2)\left(e^{-m_\psi^2/M^2}-e^{-m_X^2/M^2}\right)+B~e^{-s_0/M^2}=
(Q^2+m_\omega^2)\Pi^{OPE}(M^2,Q^2),
\label{3sr2}
\enqa
where
\beq
C(Q^2)={6\over\sin(\theta)}m_\omega f_\omega{f_\psi\lambda^q\over m_\psi
(m_X^2-m_\psi^2)}g_{X\psi\omega}(Q^2),
\label{CXV}
\enq
and    $B$   gives    the    pole-continuum transitions contribution
(see Sec.~\ref{s-3p}). $s_0$ and $u_0$ are  the continuum thresholds for $X$
and $J/\psi$ respectively.  

In Fig.~\ref{fig8} 
we show the function $C(Q^2)$, obtained from Eq.~(\ref{3sr2}),
as a  function    of   $M^2$   and    $Q^2$. As can be seen from this figure,
in the region $3.0\leq M^2\leq3.5\GeV^2$  $C(Q^2)$ is a very stable function
of $M^2$. Therefore, we choose this Borel window to extract $C(Q^2)$, as it is
shown by the dots in Fig.~\ref{fig9}. The QCDSR results for $C(Q^2)$
was fitted using  a  monopole   parametrization:
\beq C(Q^2)= {2.5\times10^{-2}\GeV^7\over Q^2+38\GeV^2},
\label{parC}
\enq
as shown by the solid line in Fig.~\ref{fig9}.   
\begin{figure}
  \subfigure[]{\lb{fig8}\includegraphics[width=0.5\textwidth]{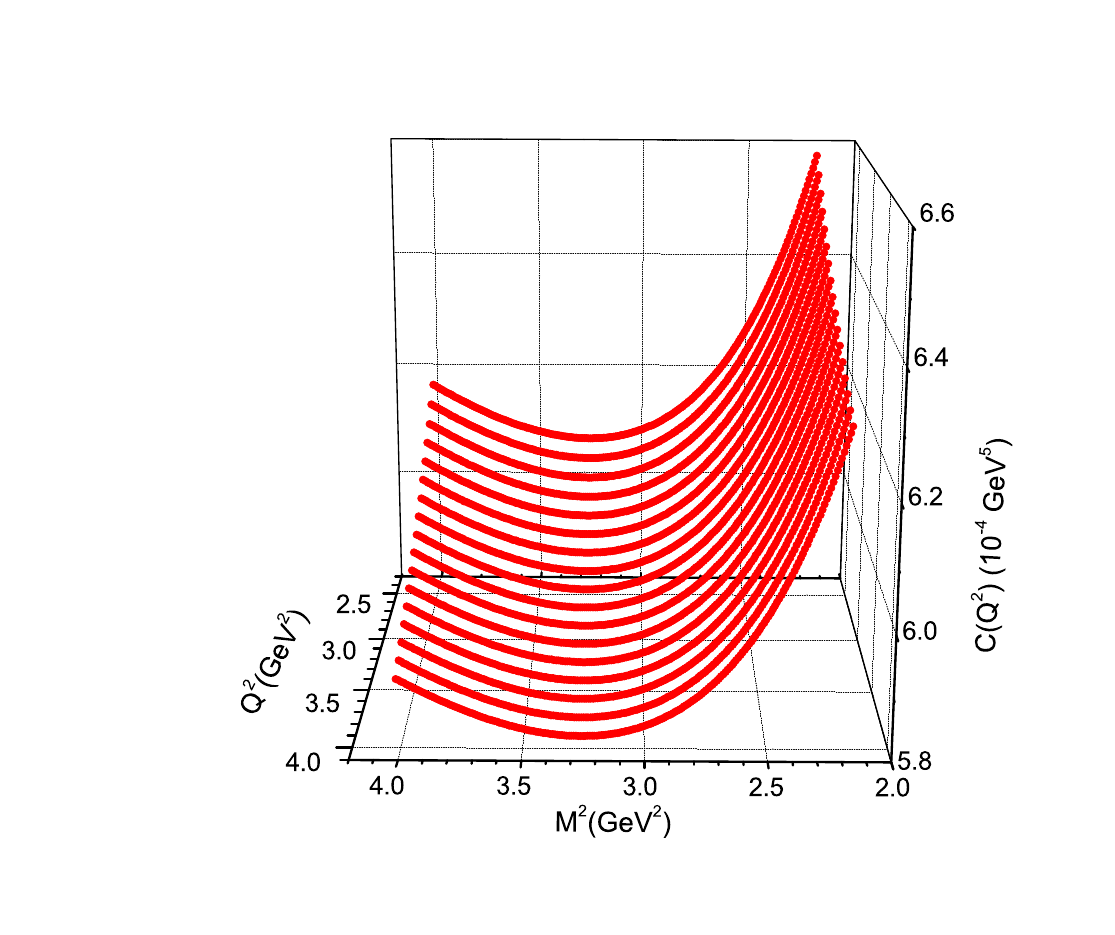}}\subfigure[]{\lb{fig9}\includegraphics[width=0.5\textwidth]{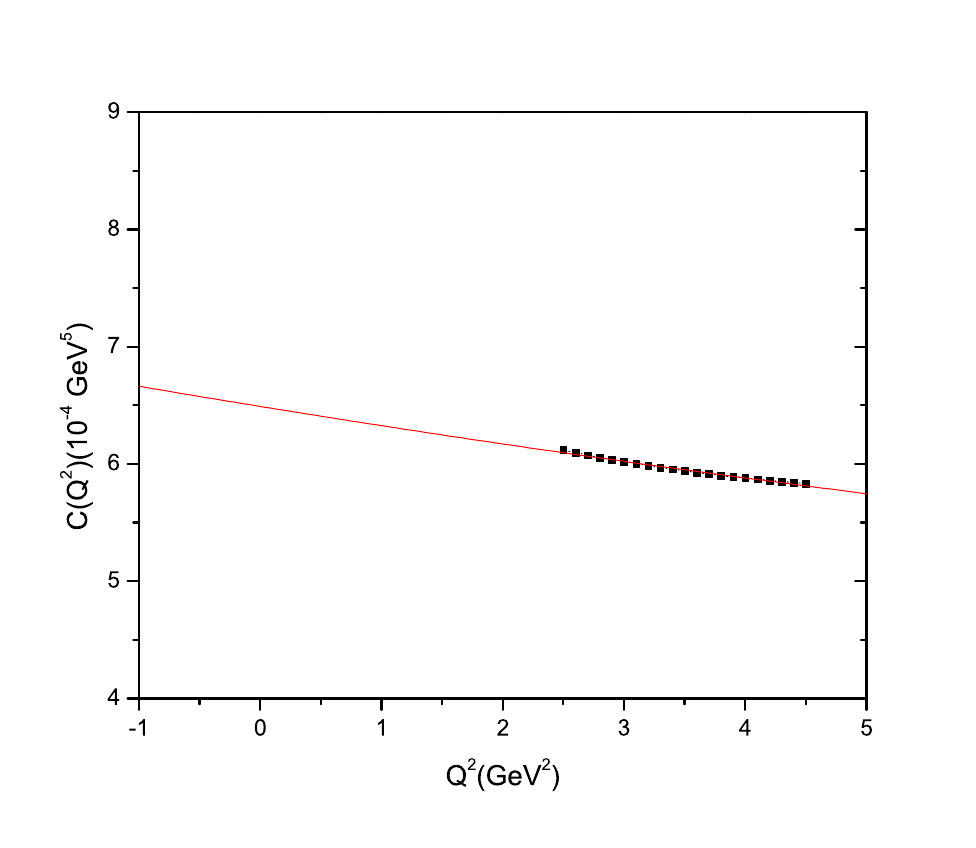}}
    \caption{(a) Values of $C(Q^2)$ obtained by varying both $Q^2$ and
      $M^2$ in  Eq.~(\ref{3sr2}). (b)  Momentum dependence  of $C(Q^2)$
      for  $s_0^{1/2}   =  4.4  \GeV$.   The  solid  line   gives  the
      parametrization   of   the    QCDSR   results   (dots)   through
      Eq.~(\ref{parC}). Taken from \cite{Matheus:2009vq}.}
\end{figure}
The  form  factor  $g_{X\psi\omega}(Q^2)$  can be  obtained  by  using
Eqs.~(\ref{CXV}) and  (\ref{parC}). The  constant coupling  was
calculated  by  extrapolating  the  form  factor  at  the  meson  pole
$Q^2=-m_\omega^2$ as 
\begin{equation}
g_{X \psi \omega} = g_{X\psi\omega}(-m_\omega^2) = 5.4 \pm 2.4,
\label{sr.coupl}
\end{equation}
where the value $5^\circ  \leq \theta \leq 13^\circ$ was used for the
mixing angle. The  decay width  was determined to be:
\beq
\Gamma\left(X\rightarrow J/\psi \pi^+ \pi^- \pi^0 \right) = (9.3 \pm 6.9)~\MeV,
\label{regama}
\enq
which is in agreement with  the  experimental  upper  limit.
Since the QCDSR results for the width and the mass grow with the mixing angle
$\theta$, there is only a small range for the values
of this angle that can provide simultaneously good agreement with the
experimental values of the mass and the decay width, and this range is 
$5^\circ \leq \theta \leq 13^\circ$.

\subsubsection{ $X(3872)\to\gamma J/\psi$ radiative decay }

Since it is possible to explain both, the mass and the total width of the
$X(3872)$, using a  mixed charmonium-molecular current in a QCDSR calculation,
in Ref.~\cite{Nielsen:2010ij} the authors used the  same
current to study the vertex of the radiative decay mode $X(3872)\to \gamma
J/\psi$.  The calculation was done using the same previously determined mixing
angles: $\alpha=20^\circ$ and $\theta=(9\pm4)^\circ$ for the mixings in
Eqs.~(\ref{4mix}) and (\ref{field0}) respectively. To study the $X(3872)$
radiative decay one considers the three-point function:
\beq 
\Pi_{\mu\nu\al}(p,\pli,q)=\int d^4x d^4y ~e^{i\pli.x}~e^{iq.y} 
\lag 0 |T[j_\mu^{\psi}(x)j_{\nu}^{\gamma}(y) 
{j_\al^X}^\dagger(0)]|0\rag,\lb{3po2} 
\enq 
where the $J/\psi$ current is given in Eq.~(\ref{cur-3po}) and the
electromagnetic current is given by:
\beq
j_{\nu}^{\gamma}=\sum_{q=u,d,c} e_q\,\bar{q}\gamma_\nu q\,,\,\,\, e_{u,c}=+\frac{2}{3}\,,\,\,e_{d}=-\frac{1}{3}\;. 
\lb{gamma}
\enq

Parameterizing the coupling of the currents with the states as:
\beqa
&&\langle0\vert j_\mu^\psi\vert\psi(\pli)\rangle=m_\psi f_\psi
\epsilon_\mu(\pli)\,;\nn\\
&&\langle X(p)\vert j_\al^X\vert0\rangle=(\cos\alpha+\sin\alpha)
\la_q\epsilon_\al^*(p)\,,\nn\\
&&\langle\psi(\pli)\vert
j_\nu^\gamma(q)\vert X(p)\rangle=i\,\epsilon_\nu^\gamma(q)\,
\mathcal{M}(X(p)\to\gamma(q)J/\psi(\pli))\,,\enqa
the  phenomenological side is then given by~\cite{Nielsen:2010ij}:
 \beqa  \Pi_{\mu\nu\al}^{\mathrm{phen}}   (p,\pli,q)&=&\frac{i  e 
(\cos\alpha+\sin\alpha)  \lambda_q m_{\psi}f_{\psi}}{m_X^2(p^2-m_{X}^2) 
({\pli}^2-m_{\psi})}\bigg(\epsilon^{\al\mu\nu\si}q_\si\,p\cdot q 
A+\epsilon^{\mu\nu\la\si}\pli_\la        q_\si       q_\alpha       B- 
\epsilon^{\al\nu\la\si}q_\mu q_\si\pli_\la C 
\nn\\ 
&+&\epsilon^{\al\nu\la\si}\pli_\la\pli_\mu      q_\si(C-A)\frac{p\cdot 
  q}{m_\psi^2}                                 -\epsilon^{\mu\nu\la\si}\pli_\la 
q_\si(q_\al+\pli_\al)(A+B)\frac{p\cdot  q}{m_X^2}\bigg)\,.  \lb{phenmix} 
\enqa 

In deriving Eq.~(\ref{phenmix}) we  used the decay amplitude,
$\mathcal{M}(X(p)\to\gamma(q)J/\psi(\pli))$, given in Ref.~\cite{Dong:2008gb}:
\beqa
\mathcal{M}(X(p)\to\gamma(q)J/\psi(\pli))= e\,\varepsilon^{\kappa
\lambda\rho\sigma}\epsilon_X^\alpha(p)\epsilon^\mu_\psi(p^\prime)
\epsilon^\rho_\gamma(q)\frac{q_\sigma}{m_X^2}\left(A\,
g_{\mu\lambda}g_{\al\kappa}p\cdot q+B g_{\mu\lambda}p_\kappa q_\al+C 
g_{\al\kappa}p_\lambda q_\mu\right).\lb{matrix}
\enqa
Notice that  there are  three dimensionless couplings,  $A$, $B$  and $C$,
to be  determined  by  the  QCDSR.  There are  five independent Lorentz
structures in Eq.~(\ref{phenmix}).   The authors in Ref.~\cite{Nielsen:2010ij}
worked with the structures: $\epsilon^{\alpha\mu\nu\sigma}q_\sigma$,
$\epsilon^{\mu\nu\sigma\lambda}p^{\prime}_\sigma       p^\prime_\alpha
q_\lambda$    and   $\epsilon^{\alpha\nu\lambda\sigma}p^\prime_\lambda
q_\sigma q_\mu$, to determine the couplings  $A$, $A+B$, and $C$ respectively.

The OPE  side is defined considering  degenerated quarks $u$ and $d$,
{\it i.e.}, $m_u=m_d$ and $\uu=\dd=\comq$. Using the mixed  current, Eq.~(\ref{4mix}), in  Eq.~(\ref{3po2}), we arrive  at:
\beqa 
\Pi_{\mu\nu\al}(x,y)&=&\frac{\sin\theta}{3}(2\cos\alpha-\sin\alpha) 
\Pi_{\mu\nu\al}^{mol}(x,y)+{\comq\over6\sqrt{2}}\cos\theta( 
\cos\al+\sin\al)\Pi^{c\bar{c}}_{\mu\nu\al}(x,y)\,.
\label{sepi} 
\enqa 
where the charmonium and molecule terms are written as:
\beq 
\Pi^{c\bar{c}}_{\mu\nu\al}(x,y)=\lag 0 |T[j_\mu^{\psi}(x)j_{\nu}^{\gamma}(y) 
{j_\al^{'(2)}}^\dagger(0)]|0\rag,\lb{corrcc} 
\enq 
\beq 
\Pi^{mol}_{\mu\nu\al}(x,y)=\lag 0 |T[j_\mu^{\psi}(x)j_{\nu}^{\gamma}(y) 
{j_\al^{(4q)}}^\dagger(0)]|0\rag,\lb{corrmol} 
\enq 
with ${j_\al^{'(2)}}$ and ${j_\al^{(4q)}}$ given by Eqs.~(\ref{curr2})
and  (\ref{curr4}) respectively.  In Ref.~\cite{Nielsen:2010ij} the  OPE side was evaluated  in leading  order in  $\alpha_s$ up  to dimension-five condensates.

Doing a single Borel transform to $P^2\to  M^2$, in the
limit $p^2=p^{\prime2}=-P^2$, one gets for each structure $i$:
\beqa 
G_i(Q^2)\left(e^{-m_\psi^2/M^2}-e^{-m_X^2/M^2}\right)+B_i(Q^2)~ 
e^{-s_0/M^2}=\bar{\Pi}_i^{OPE}(M^2,Q^2), 
\label{3sr-1} 
\enqa 
where  $Q^2=-q^2$   and  $B_i(Q^2)$  is  the  pole-continuum transitions contribution (see Sec.~\ref{s-3p}). $G_i(Q^2)$ is  related to  the three form factors:
\beqa        
G_1(Q^2)&=&\frac{3\sqrt{2}\pi^2(\cos\al+\sin\al)\la_qm_\psi 
  f_\psi}{m_X^2(m_X^2-m_\psi^2)}A(Q^2)\,\nn\\
G_2(Q^2)&=&\frac{3^22^4\sqrt{2}\pi^2(\cos\al+\sin\al)\la_qm_\psi  
f_\psi(A(Q^2)+B(Q^2))}{\sin\theta(2\cos\al-\sin\al)m_X^4(m_X^2-m_\psi^2)}\,\nn\\
G_3(Q^2)&=&\frac{6\sqrt{2}\pi^2\la_qm_\psi f_\psi}{\cos\theta m_X^2 
(m_X^2-m_\psi^2)}C(Q^2)
\label{Aq2} 
\enqa 

\begin{figure*}  
 \subfigure[]{\includegraphics[width=0.5\textwidth]{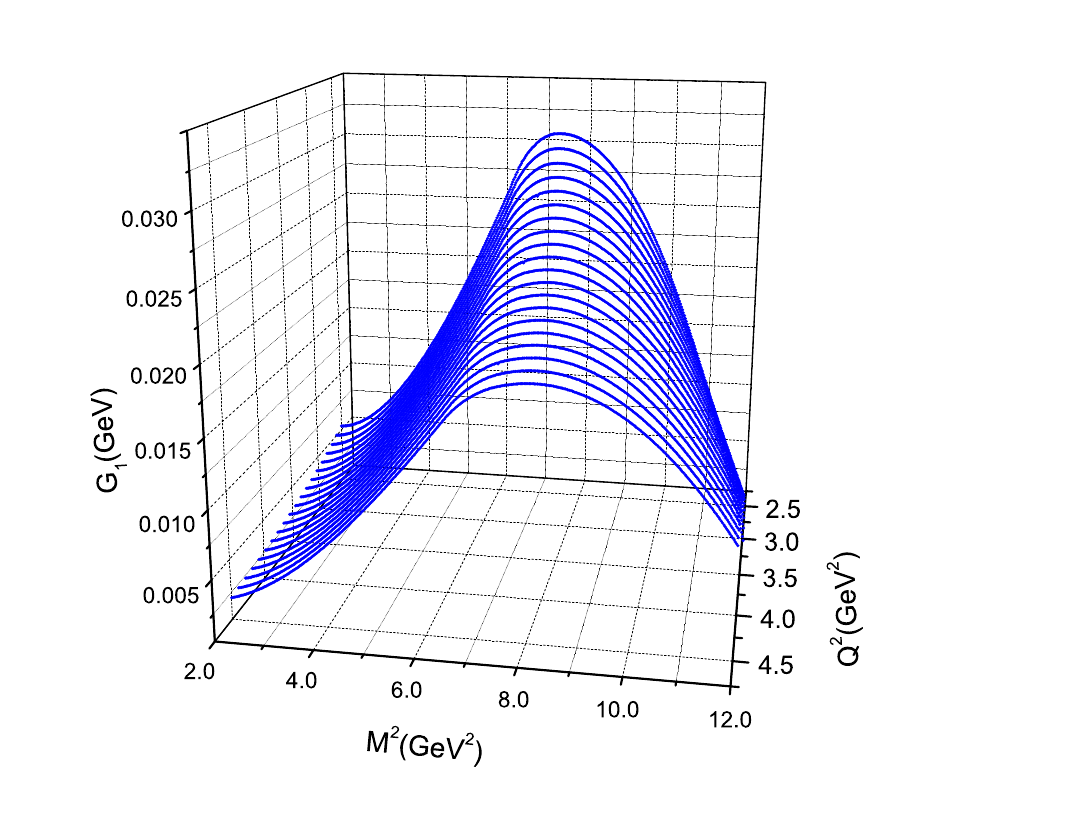}}
 \subfigure[]{\includegraphics[width=0.5\textwidth]{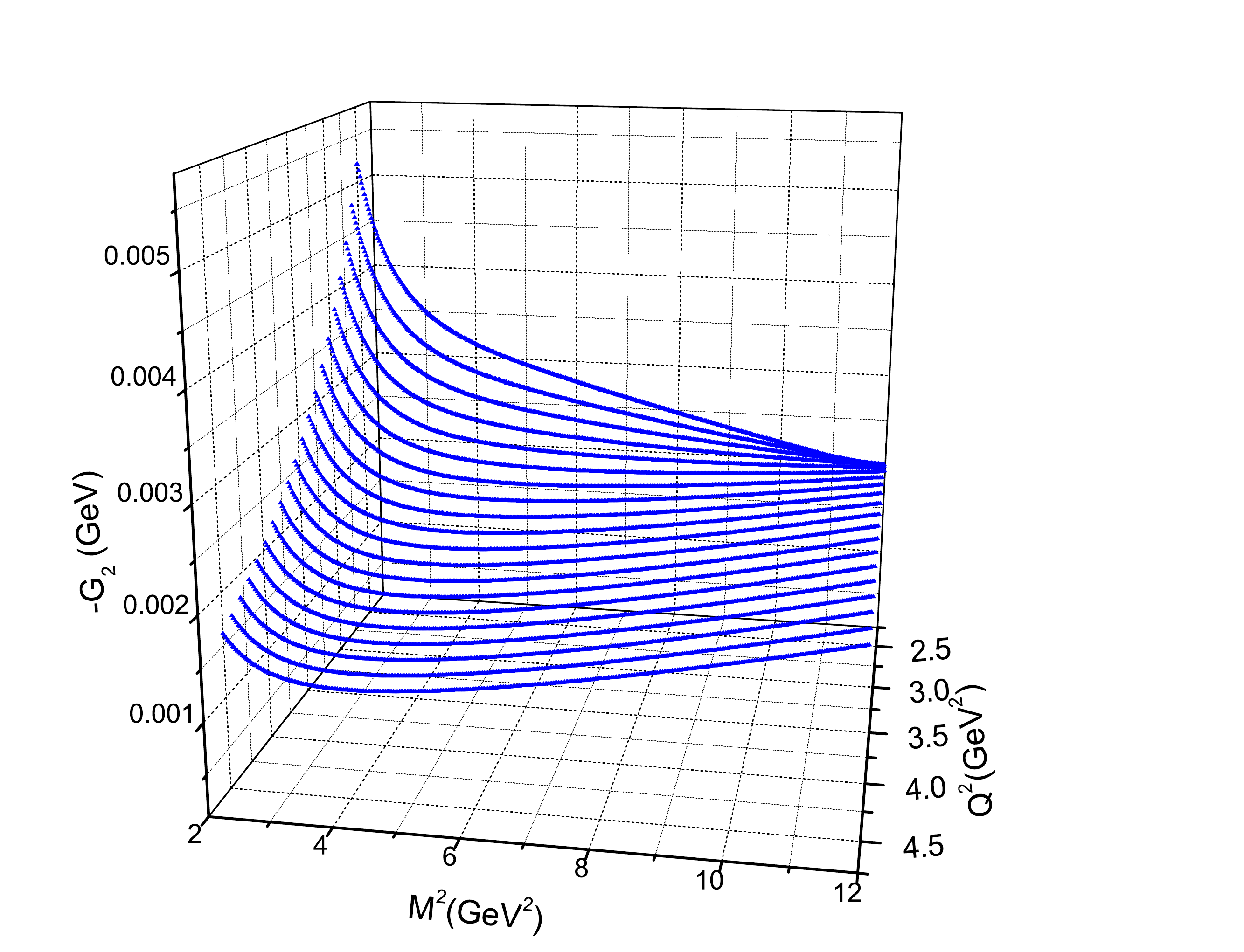}} 
  \subfigure[]{\includegraphics[width=0.5\textwidth]{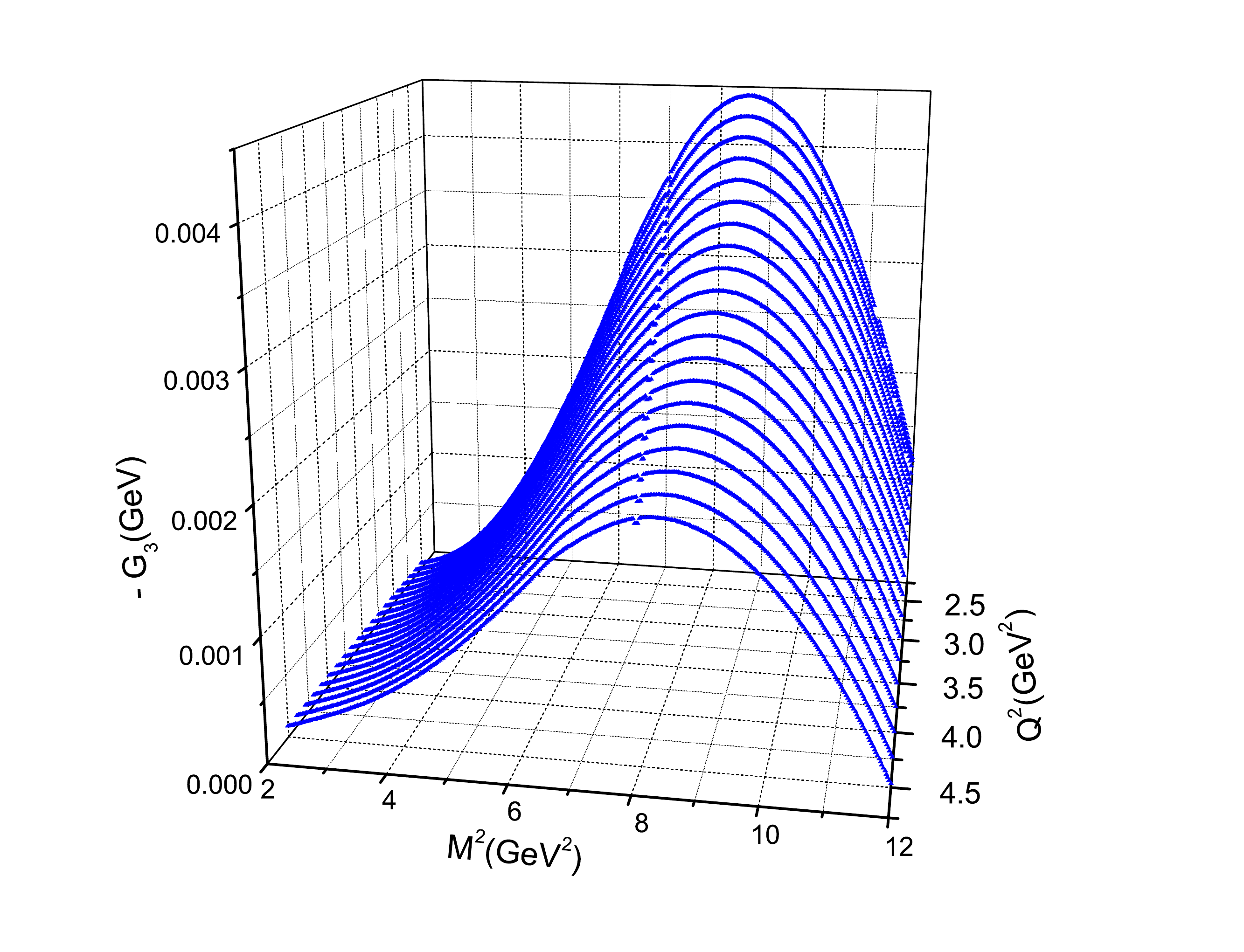}} 
\caption{ \lb{fig7g} Values of the functions obtained by varying both $Q^2$  
and $M^2$: a) $G_1(Q^2)$, b) $G_2(Q^2)$ and c) $G_3(Q^2)$. Taken from \cite{Nielsen:2010ij}} 
\end{figure*} 

The   unknown  functions   $G_i$  and   $B_i$  in   the  LHS   of  
Eq.~(\ref{3sr-1}) are determined by fitting  both sides of the sum rule.
These functions should not depend on  the Borel parameter $M^2$. Therefore,
  the Borel region is fixed by imposing that these functions are
stable as a function of  $M^2$.  Fig.~\ref{fig7g} shows  the QCDSR
results for  the $G_i$'s as  functions of  $Q^2$ and $M^2$.  From these figures it is
possible to determine  the corresponding Borel region for each one of these functions: $7.0\GeV^2\leq M^2\leq8.5\GeV^2$ for $G_1(Q^2)$, $6.5\GeV^2\leq M^2\leq7.5\GeV^2$ for $G_2(Q^2)$, and $8.0\GeV^2\leq
M^2\leq9.0\GeV^2$ for $G_3(Q^2)$. The QCDSR results for the three form factors are shown in Fig.~\ref{fig11}.

The $Q^2$-dependence for all the three form factors can be fitted
by using an  exponential parametrization:
\beq  G_i(Q^2)=g_1 e^{-g_2  Q^2}\,,\lb{giq2}
\enq
The fitted functions  are also  shown in Fig.~\ref{fig11} through the
solid lines. The fitted parameters are given in the Table \ref{tab1}.

\begin{figure*} 
  \subfigure[]{\includegraphics[width=0.4\textwidth]{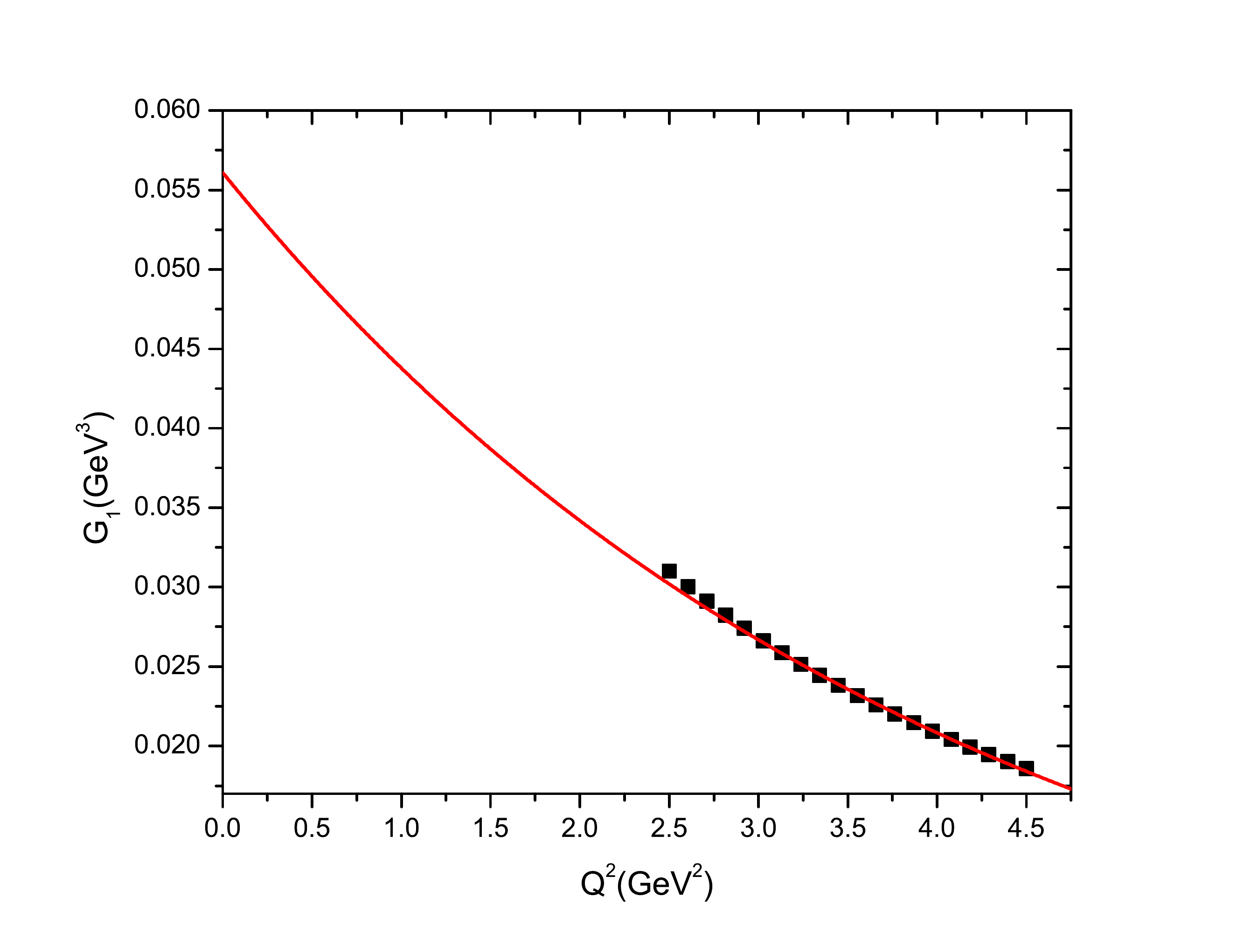}}
  \subfigure[]{\includegraphics[width=0.4\textwidth]{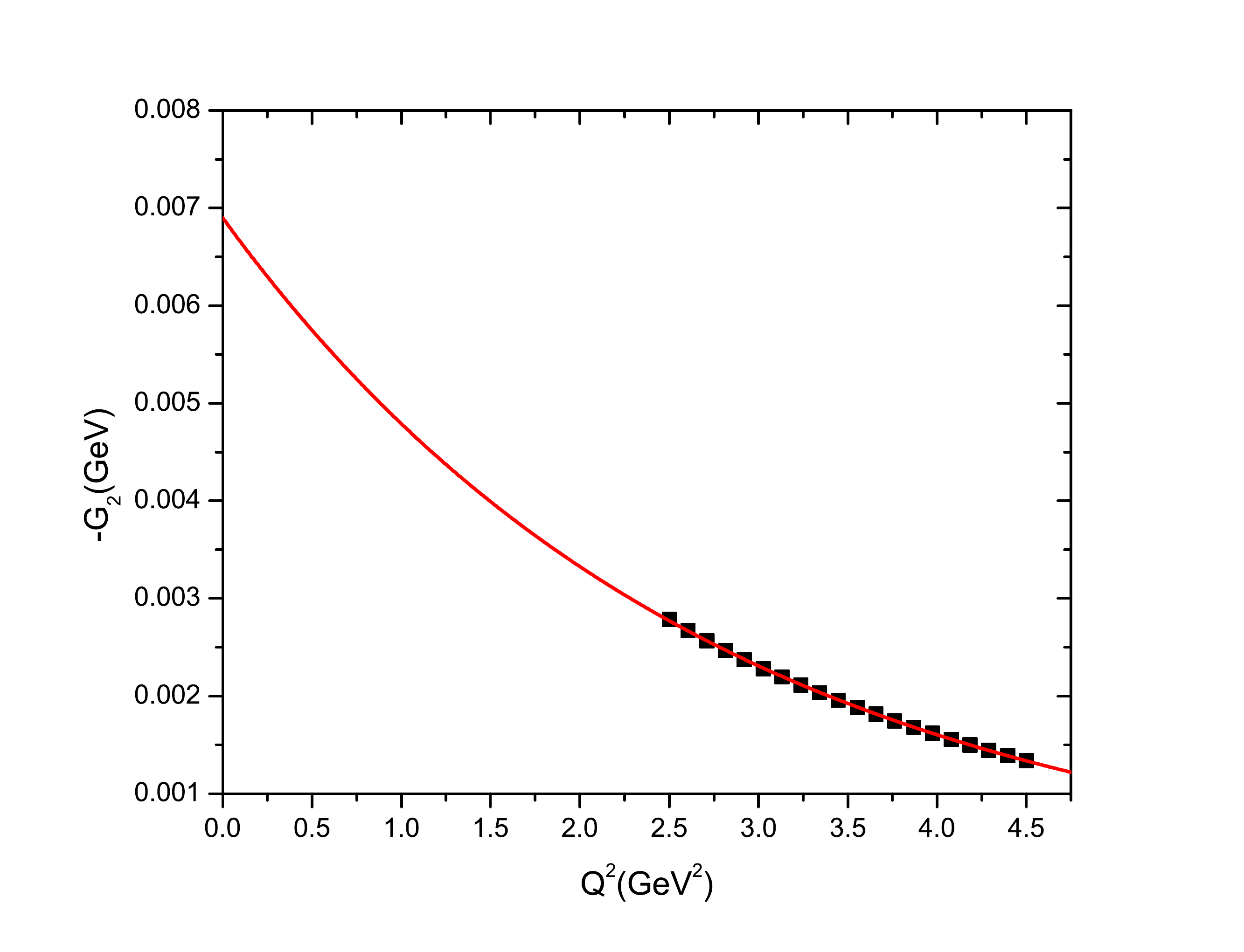}} 
\subfigure[]{\includegraphics[width=0.4\textwidth]{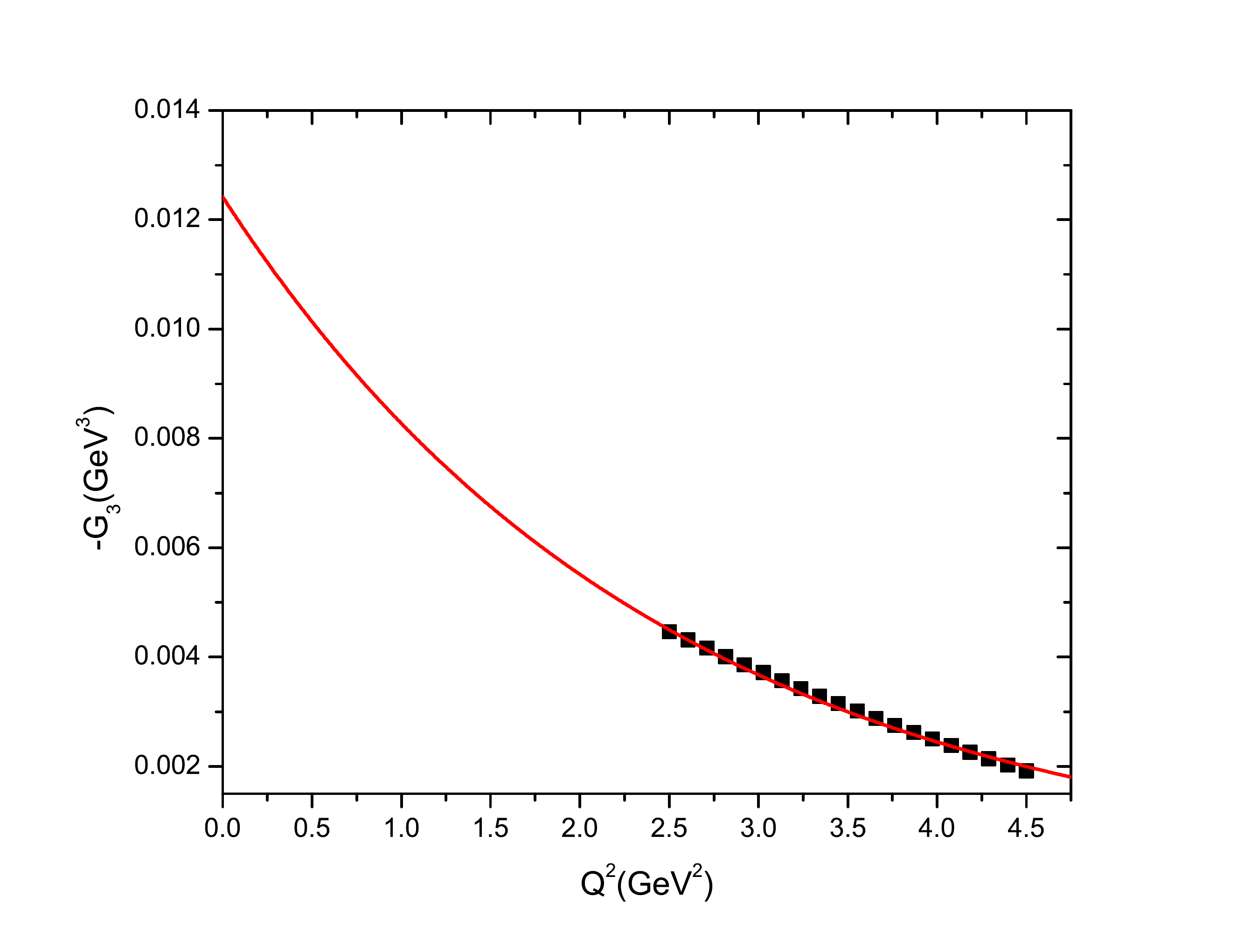}} 
  \caption{  \lb{fig11} Momentum    dependence    of    the    functions     
for  $s_0^{1/2}=4.4\GeV$  and $u_0^{1/2}=3.6\GeV$: (a)  $G_1$, (b)   
$G_2$ and  (c) $G_3$. The solid line  gives the parametrization of the  
QCDSR results (dots) through Eq.~(\ref{giq2}) and the results in  
Table \ref{tab1}. Taken from \cite{Nielsen:2010ij}.}
\end{figure*}

\begin{table}[h] 
\begin{center}
\begin{tabular}{|c||c|c|c|}\hline
      &   $G_1$              &   $G_2$   & $G_3$ \\\hline\hline
$g_1$   & $0.056\GeV^3$     &  $-0.0069\GeV$   &  $-0.013\GeV^3$\\ 
$g_2$   & $0.25\GeV^{-2}$    & $0.365\GeV^{-2}$ &  $0.41\GeV^{-2}$\\\hline
\end{tabular}
\end{center}\caption{\lb{tab1} Results for the fitting parameters in Eq.~(\ref{giq2}).} 
\end{table}

The results  from Table \ref{tab1} can  be used to determine the form factors
$A(Q^2)$, $B(Q^2)$ and $C(Q^2)$  written in terms of
the $G_i(Q^2)$ in Eqs.~(\ref{Aq2}).  The coupling constants are obtained from the form factors  at $Q^2=0$. The couplings are:
\beqa
A&=&A(Q^2=0)=18.65\pm0.94\,;\nn\\
A+B&=&(A+B)(Q^2=0)=-0.24\pm0.11\,;\nn\\
C&=&C(Q^2=0)=-0.843\pm0.008\,.\lb{resabc}
\enqa

The radiative  decay width is obtained from these  couplings through
the expression \cite{Nielsen:2010ij}:
\beq
\Gamma(X\to J/\psi~\gamma)=\frac{\alpha_e}{3}\frac{p^{*5}}{m_X^4}
\bigg((A+B)^2+\frac{m_X^2}{m_\psi^2}(A+C)^2\bigg)\,,\lb{width}\nn\\
\enq  
where $p^*=(m_X^2-m_\psi^2)/(2m_X)$  is the momentum of  the photon in
the rest frame, and $\alpha_e=\frac{1}{137}$ is the fine structure
constant.    Using   the   previous   result      $\Gamma(X\to
J/\psi~\pi\pi)=(9.3\pm6.9)\MeV$, we get the ratio
\beq
\frac{\Gamma(X\to J/\psi~\gamma)}{\Gamma(X\to J/\psi~\pi^+\pi^-)}=0.19
\pm0.13\,,\lb{brfinal}
\enq
which is in excellent  agreement with the experimental result in Eq.~({\ref{rad}}). 

\subsubsection{$X(3872)$ production in $B$ decays}
\lb{sec-pro}

The next application testing the mixed charmonium-molecule current, was
done in Ref.~\cite{Zanetti:2011ju}, which is related to the production of
 $X(3872)$  in  $B$ meson  decays, through  the
channel   $B^\pm\to  K^\pm   X(3872)$.    BaBar Collaboration has reported
the upper limit for the branching ratio of the $X(3872)$
production in $B$ meson decays as \cite{Aubert:2005vi}:
\begin{equation} 
  \label{branchingX} 
  {\mathcal B}(B^\pm\to K^\pm X(3872))<3.2\times10^{-4}. 
\end{equation}

The $B$  meson decay process occurs  via weak decay of  the $b$-quark,
with  the light  quark as  the  spectator.  In  effective theory,  the
Hamiltonian describing  the weak interaction at  the scale $\mu=m_b\ll
m_W$ can be  written in terms of a four-quark  interaction vertex with
an effective  four quark  operator ${\mathcal{O}}_2=(\bar{c}\Gamma_\mu
c)(\bar{s}\Gamma^\mu     b)$,      with     a      $V-A$     structure
$\Gamma_\mu=\gamma_\mu(1-\gamma_5)$. The interaction can be factorized
into two matrix elements, giving the following decay amplitude for the
process:
\begin{eqnarray}\lb{amp} 
{\mathcal M} 
=i\frac{G_F}{\sqrt{2}}V_{cb}V_{cs}^*\left(C_2+\frac{C_1}{3}\right)\langle B(p)\vert J_{\mu}^W\vert K(p^\prime)\rangle\langle X(q) 
\vert J^{\mu(\bar{c}c)}\vert0\rangle, 
\end{eqnarray}
where $V_{ik}$ are CKM matrix  elements, $C_1(\mu)$ and $C_2(\mu)$ are
short  distance Wilson  coefficients computed  at the  renormalization
scale $\mu\sim{\mathcal O}(m_b)$, $p=p^\prime+q$, and the currents are
\begin{equation}\lb{wcurrents}    J_{\mu}^W=\bar{s}\Gamma_\mu   b\,,\quad 
  J_{\mu}^{(\bar{c}c)}=\bar{c}\Gamma_\mu  c\,.\end{equation}
Here  $X(3872)$, considered as a mixture of molecule and  charmonium currents,
interacts through  the $\bar{c}c$  component of  the weak  current.

The  matrix  elements in  Eq.~({\ref{amp}})  are  parametrized, in Ref.~\cite{Zanetti:2011ju}, as
\begin{equation}\lb{2pmatrix} 
\langle X(q)\vert J_{\mu}^{(\bar{c}c)}\vert0\rangle=\lambda_W 
\epsilon_\mu(q)\,, 
\end{equation} 
\begin{equation}\lb{3pmatrix} 
\langle B(p)\vert J_{\mu}^W\vert K(p^\prime)\rangle=f_+(q^2)(p_\mu+ 
p_\mu^\prime)+f_-(q^2)(p_\mu-p_\mu^\prime)\,. 
\end{equation} 
The  coupling between  the  current $J_\mu^{(\bar{c}c)}$  and the  $X(3872)$
is provided  by the  parameter $\lambda_W$  in Eq.~(\ref{2pmatrix}).
The form factors,  $f_\pm(q^2)$, describe the weak  transition process $B\to K$.

The decay width for the process $B^\pm\to X(3872)K^\pm$ is computed from  
\begin{equation}\lb{eqwidth} 
\Gamma(B\to XK)=\frac{1}{16\pi m_B^3}\lambda^{1/2}(m_B^2,m_K^2,m_X^2) 
\vert{\mathcal{M}}\vert^2, 
\end{equation} 
where  $\lambda(x,y,z)=x^2+y^2+z^2-2xy-2xz-2yz$,   and  the  invariant
amplitude squared can be obtained from Eq.~(\ref{amp}):
 
\begin{eqnarray} 
\vert\mathcal{M}\vert^2=\frac{G_F^2}{2}\vert V_{cb}V_{cs}\vert^2\left(C_2 
+\frac{C_1}{3}\right)^2 \lambda(m_B^2,m_K^2,m_X^2)\lambda_W^2f_+^2(q^2)\vert_{q^2\to-m_X^2} 
\,. 
\end{eqnarray}
The unknown parameters, to be determined from   the QCDSR approach, are the
weak coupling, $\lambda_W$, and the value of the form factor $f_+(q^2)$
at the $X(3872)$ pole.

The factorization of the amplitude  allows the four-quark vertex to be
analysed as two separated sub-processes: the creation of  $X(3872)$ 
and the transition $B\to K$, as in Eqs. (\ref{2pmatrix}) and (\ref{3pmatrix}).
Let us start  with  the  two-point  correlator,  describing  the  coupling
between the current $J_\nu^{(\bar{c}c)}$ and  $X(3872)$:
\begin{eqnarray} 
\Pi^{\mathrm{OPE}}_{\mu\nu}(q)=(\cos\alpha+\sin\alpha)~\biggl(\sin\theta\, 
\Pi^{4,2}_{\mu\nu}(q)+\frac{\comq}{6\sqrt{2}}\cos\theta\, 
\Pi^{2,2}_{\mu\nu}(q)\biggr)\,, 
\end{eqnarray} 
where 
\begin{eqnarray} 
\Pi^{4,2}_{\mu\nu}(q)&=&i\int d^4y ~e^{iq\cdot y}\langle0\vert T\{ 
J_\mu^{4q}(y)J_{\nu}^{(\bar{c}c)}(0)\}\vert0\rangle\nn\\ 
\Pi^{2,2}_{\mu\nu}(q)&=&i\int d^4y ~e^{iq\cdot y}\langle0\vert T\{ 
J_\mu^{2q}(y)J_{\nu}^{(\bar{c}c)}(0)\}\vert0\rangle\,. 
\end{eqnarray} 
The   contribution    from   the   vector   part    of   the   current
$J_\nu^{(\bar{c}c)}$ vanishes after the integration is performed, thus
these correlators  are equal  (except for  a minus  sign) to  the ones
calculated  previously  in   Ref.~\cite{Matheus:2009vq} (sub-section \ref{2pointsec})  for  the  two-point
correlator of  $X(3872)$.

On the  phenomenological side  the correlator is  determined inserting 
the intermediate state of  $X(3872)$: 
\begin{eqnarray} 
  \Pi_{\mu\nu}^{phen}(q)&=&\frac{i}{q^2-m_X^2}\langle0\vert J^X_\mu\vert  
X(q)\rangle\langle X(q)\vert J^{(cc)}_\nu\vert0\rangle\,,\nn\\ 
&=&\frac{i\lambda_X\lambda_W}{Q^2+m_X^2}\left(g_{\mu\nu}-\frac{q_\mu q_\nu} 
{m_X^2}\right) 
\end{eqnarray} 
where we  have used the  definition in Eq.~(\ref{2pmatrix}) 
and 
\begin{equation} 
  \langle0\vert J^X_\mu\vert X(q)\rangle=\lambda_X\epsilon_\mu(q)\,. 
\end{equation} 
The parameter defining the  coupling between the current $J^X_\mu$ and 
the $X$ meson has been  calculated in Ref.~\cite{Matheus:2009vq}, and its value 
is given in Eq.~(\ref{sr.lamb}).

The QCDSR, in the structure $g_{\mu\nu}$, obtained after the Borel transformation
 is:
\begin{eqnarray}\lb{2psumrule} 
  \lambda_W\lambda_Xe^{-\frac{m_X^2}{M^2}}=-(\cos\alpha+\sin\alpha) 
\biggl(\sin\theta\,\Pi^{4,2}(M^2)+\frac{\comq}{6\sqrt{2}}\cos\theta\, 
\Pi^{2,2}(M^2)\biggr)\,. 
\end{eqnarray} 
This  expression  is  evaluated  numerically to  obtain  the  coupling
parameter  $\lambda_W$.   To keep consistency  through  the
different  analysis,   the  same   parameters  are  used   as  before:
$\sqrt{s_0}=4.4\GeV$, $2.6 \GeV^2  \leq M^2 \leq 3.0  \GeV^2$, and the
mixing     angles     $\theta=(9\pm4)^\circ$,
and $\alpha=20^\circ  $.

The result  for the  parameter $\lambda_W$, obtained within  the
given ranges  of the  Borel mass  and  the variation in the
mixing angle $\theta$, is~\cite{Zanetti:2011ju}:
\begin{equation}\lb{lambdaW} 
  \lambda_W=(1.29\pm0.51)\GeV^2\,. 
\end{equation} 

The three-point function describing the weak transition
$B\to K$ is written as:
\begin{equation} 
\Pi_{\mu}(p,p^\prime)=\!\!\int d^4x \,d^4y \,e^{i(p^\prime\cdot x-\,p 
\cdot y)}\langle0\vert T\{J_\mu^W(0)J_K(x)J^\dagger_B(y)\}\vert0\rangle, 
\end{equation}  
where the weak  current $J^W_\mu$ is defined  in (\ref{wcurrents}) and
the interpolating currents of the $B$ and $K$ pseudoscalar mesons are:
\begin{eqnarray} 
  J_K(x)=i\,\bar{u}_a(x)\gamma_5 s_a(x)\,,\quad J_B=i\,\bar{u}_a(x) 
\gamma_5 b_a(x)\,. 
\end{eqnarray}

On  the phenomenological  side,  the insertion  of the  intermediate
$B$ and $K$ mesons  gives
\begin{eqnarray} 
  \Pi^{phen}_\mu=-\frac{f_Bf_Km_K^2m_B^2 }{m_b(m_s+m_u)}\frac{(f_+(q^2)(p+ 
p^\prime)_\mu+f_-(q^2)q_\mu)}{(p^2-m_B^2)(p^{\prime2}-m_K^2)}\,,\nn\\ 
\end{eqnarray} 
where we have used Eq.~(\ref{3pmatrix}) and the following  
definitions: 
\begin{eqnarray} 
  \langle0\vert J_K\vert K(p^{\prime})\rangle=f_K\frac{m_K^2}{m_s+m_u}\,,\quad
  \langle0\vert J_B\vert B(p)\rangle=f_B\frac{m_B^2}{m_b}\,. 
\end{eqnarray} 

After a double Borel transform is applied on both sides of the sum rule, $P^2\to M^2$ and $P^{\prime2}\to M^{\prime2}$, we get the sum rule in the structure
$(p_\mu + p^{\prime}_\mu)$:

\begin{eqnarray}\lb{sr3p} 
-\frac{f_Bf_Km_K^2m_B^2 f_+(Q^2)}{m_b(m_s+m_u)}e^{-\frac{m_B^2}{M^2}- 
\frac{m_K^2}{M^{\prime2}}}=\Pi^{OPE}(M^2,M^{\prime2})
\end{eqnarray}
In Ref.~\cite{Zanetti:2011ju}  the  OPE  side was evaluated  at leading  order  in
$\alpha_s$, considering condensates  up to  dimension-five  and terms
linear in  the $s$ quark mass.  The following ansatz,
relating the Borel masses $M^2$ and $M^{\prime2}$, was applied in the numerical analysis:
$  M^{\prime2}=\frac{0.64\GeV^2}{m_B^2-m_b^2}M^2$.

The  QCDSR results for the form factor  $f_+$ is plotted  in
Fig.~\ref{fig2a}  as a  function of  $Q^2$ and  $M^2$, showing  a good
stability  for  $M^2>20.0\GeV^2$. The  obtained $Q^2$  dependence  of the  form
factor, in the Borel region  $26 \GeV^2\leq M^2\leq30\GeV^2$ (compatible with
the $B$ mass) is shown in Fig.~\ref{fig2b}.

Using a  monopolar parametrization for the form factor,  $f_+(Q^2)$,:  
\begin{equation}\lb{fplus} 
  f_+(Q^2)=\frac{(17.55\pm0.04) \GeV^2}{(105.0\pm1.76)\GeV^2+Q^2}\,,
\end{equation} 
the  QCDSR results in Eq.~(\ref{sr3p}) can be well represented, as shown in Fig.~\ref{fig2b}. 
\begin{figure}[t] 
\centerline{ 
\hspace{-0.4cm} 
 \subfigure[]{\label{fig2a}\includegraphics[width=0.45\textwidth]{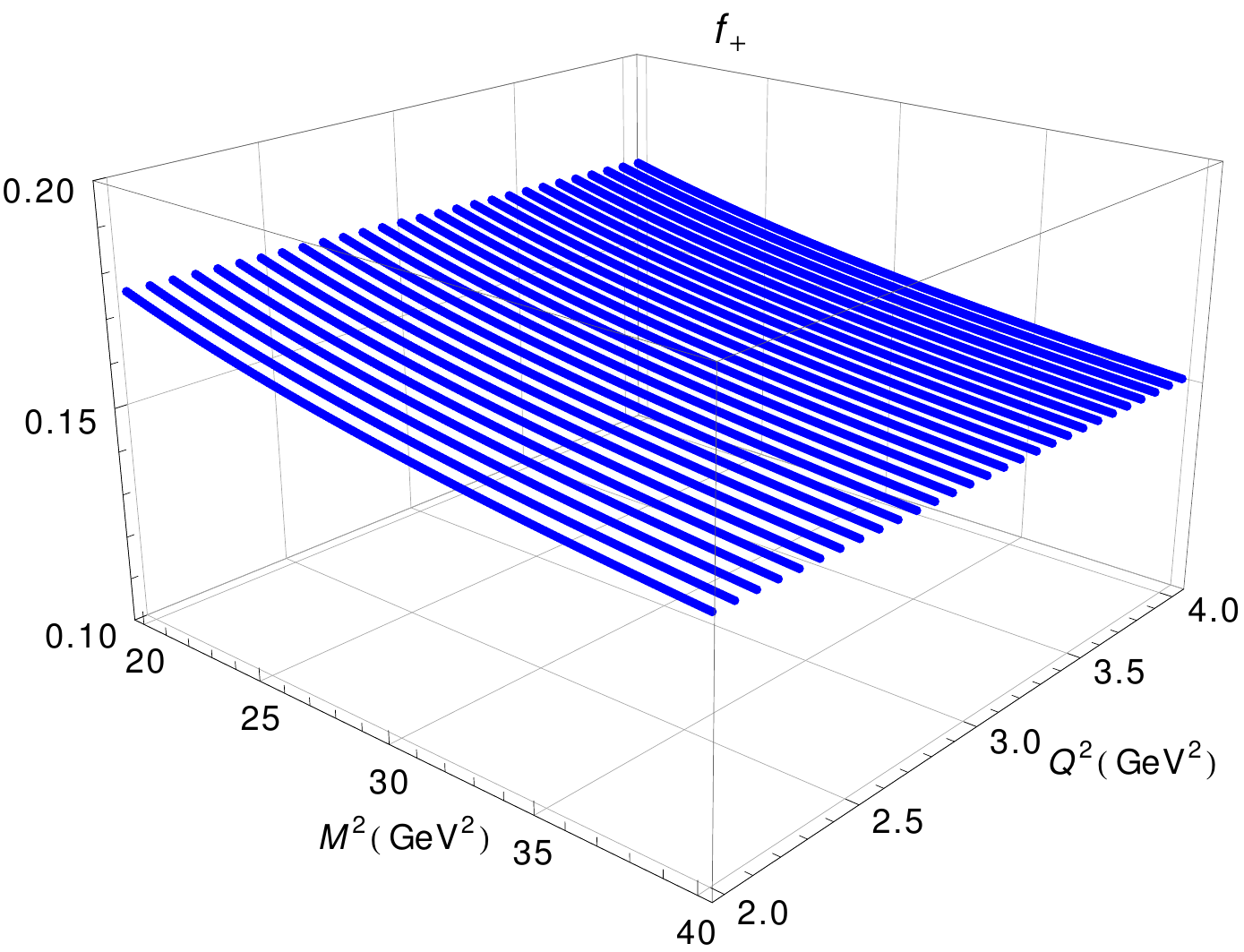}}
 \subfigure[]{\label{fig2b}\includegraphics[width=0.4\textwidth]{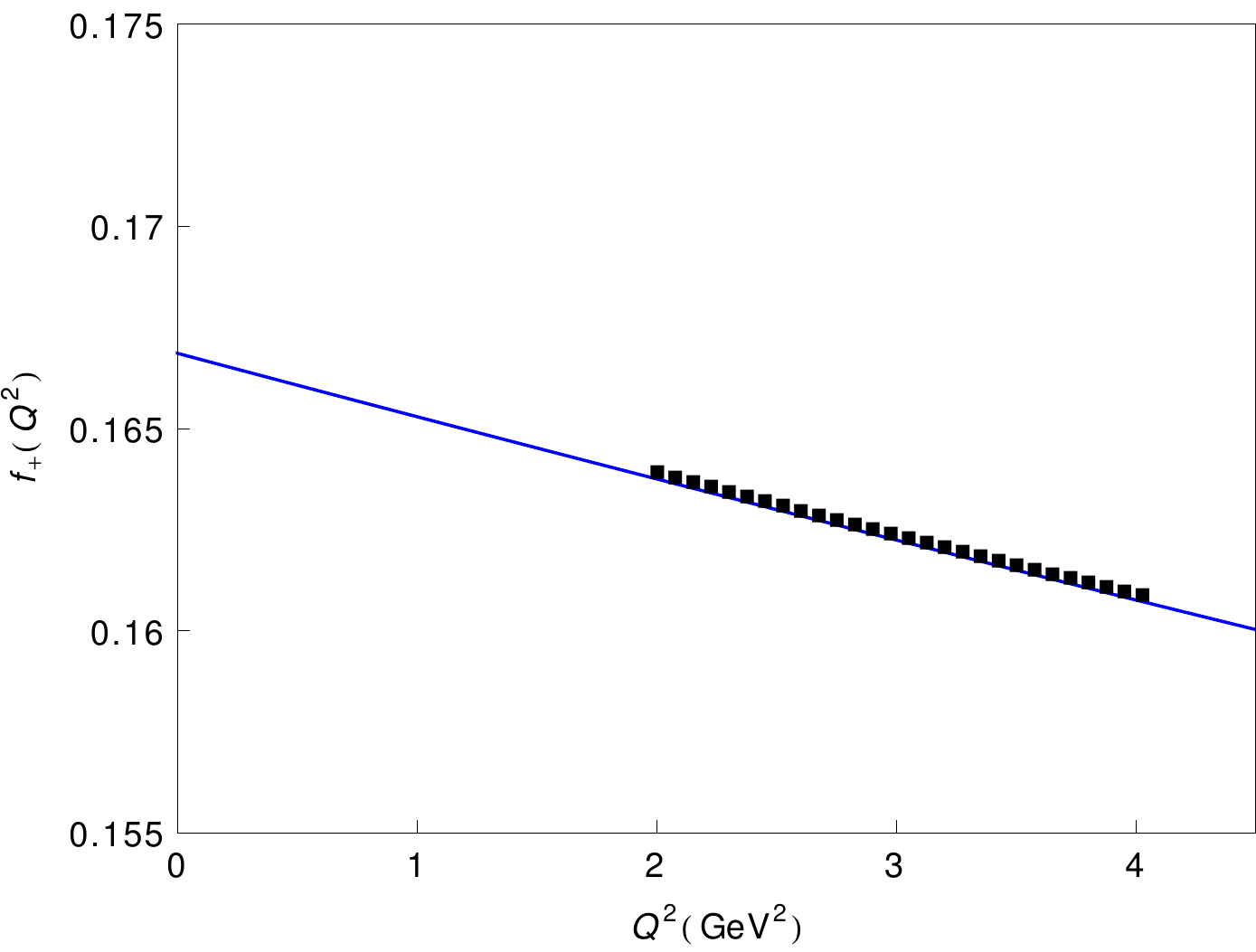}}}
\caption{\label{fit.form} (a)   QCDSR results for the   form  factor
  $f_+$ as a function of $Q^2$ and $M^2$. (b) Momentum dependence of
  $f_+(Q^2)$. The solid line gives the  parametrization of the QCDSR
  results (dots) through Eq.~(\ref{fplus}). Taken from ~\cite{Zanetti:2011ju}.} 
\end{figure}
Using Eq.~(\ref{fplus}),  the value of the form factor  at the $X(3872)$
pole, $Q^2=-m_X^2$, is given by \cite{Zanetti:2011ju}:
\begin{equation}\lb{fpluspolo} 
  f_+(Q^2=-m_X^2)=0.195\pm0.003\,. 
\end{equation} 

The  decay  width is  determined  using  the  value of  the  parameter
$\lambda_W$  obtained  from  the  two-point sum  rule  calculation, Eq.~(\ref{lambdaW}), and the value of the form factor, $f_+$, in Eq.~(\ref{fpluspolo}).
The branching  ratio is  obtained by dividing  this result  by the
total width of the $B$ meson \cite{pdg}: $\Gamma_{\mathrm{tot}}=\hbar/\tau_B$. It
gives \cite{Zanetti:2011ju}:
\begin{equation}\lb{resultX} 
  \mathcal{B}(B\to X(3872)K)=(1.00\pm0.68)\times10^{-5}\,, 
\end{equation}

This value  for the branching ratio  is smaller but compatible  with the
experimental  upper limit in Eq.~(\ref{branchingX}). Since  the factorization
hypothesis  was employed,  this  result  suggests that  non-factorizable
contributions, that were not taken  into account in Ref.~\cite{Zanetti:2011ju},
can be significant. Therefore, the result in Eq.~(\ref{resultX}) should be
taken  as a lower limit for the branching ratio.

\subsection{Summary for $X(3872)$}

The     mixed     charmonium-molecule      current     proposed     in
Ref.~\cite{Matheus:2009vq}  within  the  QCDSR framework,  provides  a
consistent description  of various  properties of   $X(3872)$ state
that are neglected in most  studies, which are usually concerned only with
the mass  of the  state. This type  of mixed state  is favored  by the
$\gamma\psi(2S)$  decay experimental  data,  as discussed  previously.
The consistency of  such an approach is guaranteed  by applying the
same  mixing  parameters in  different  analysis.   The mixing  angles
$\alpha=20^\circ$ and  $\theta=(9\pm4)^\circ$ were fixed in  the first
study of  the mass and decay  width of the channel  $X(3872)\to J/\psi
(n\pi)$,   and   then   applied   to   study   the   radiative   decay
$X(3872)\to\gamma J/\psi$ and the production channel $B\to K~
X(3872)$.   Eq.~(\ref{field0})  may  suggest  that  the  state  is
dominated by  the charmonium component (($\sim$97\%)),  as was pointed out
in the conclusions of Ref.~\cite{Matheus:2009vq}.  However, this is not
so straightforward, since  the $c\bar{c}$ component of  the current is
multiplied by  a dimensional  parameter and hence  the percentage  of each
component is not  provided solely by the mixing  angle.  Despite being
unable to  determine the  percentages of  the charmonium  and molecule
components,  it  is possible  to  establish  that both  components  are
mandatory,  given that  the properties  of   $X(3872)$  can not  be
properly explained with only one of the components.

\section{\label{}Vector Charmonium $Y$ States}
Many of the charmonium-like states observed in the  Initial State Radiation (ISR) process $e^+e^- \to J/\psi \:\pi^+\pi^-$ by BaBar and Belle
collaborations do not fit the quarkonia interpretation, and have stimulated an extensive discussion about exotic 
hadron configurations. In particular,  $Y(4260)$, which was reported in Ref.~\cite{Aubert:2005rm} with a 
mass around 4.26~GeV/$c^2$. This observation was immediately confirmed by CLEO-c \cite{He:2006kg} and 
Belle \cite{Yuan:2007sj}, and more recently by the BESIII Collaboration \cite{Ablikim:2016qzw}. 
Historically, the label $Y$ was used for all states with $I^G(J^{PC})=0^-(1^{--})$ quantum numbers which were produced in $e^+e^-$ 
annihilation. In this review we use the same label, though it is different from the recent 
naming scheme of PDG \cite{pdg}. 

The  conventional neutral $1^{--}$ charmonium states in the mass range (3.8 to 5.0 GeV), such as 
$\psi(4040)$, $\psi(4160)$, and $\psi(4415)$ \cite{pdg}, decay predominantly into open charm final states 
(e.g., $D$ mesons), while the $Y$ states decay  to hidden-charm final states \cite{Mo:2006ss}. 
Furthermore, the observation of the states $Y(4360)$ and $Y(4660)$ in $e^+e^- \to \psi(2S) \:\pi^+\pi^-$ 
\cite{Wang:2014hta, Aubert:2007zz, Wang:2007ea, Lees:2012pv}, together with more resonant structures observed 
in $e^+e^- \to \omega \:\chi_{c0}$ \cite{Ablikim:2014qwy} and $e^+e^- \to h_c \:\pi^+\pi^-$ \cite{BESIII:2016adj}, 
overpopulate the vector charmonium spectrum predicted by potential models \cite{Eichten:1978tg, Eichten:1979ms, 
Godfrey:1985xj}. These facts indicate that the neutral $Y$ states may not be conventional charmonium states, 
and they are good candidates for new types of exotic particles, such as hybrids, tetraquarks, or meson molecules 
\cite{Swanson:2006st, Zhu:2007wz,Nielsen:2009uh, Brambilla:2010cs, Chen:2016qju, Olsen:2009gi,Guo:2017jvc}. 

The list of the most recent experimental results for the $1^{--}$ family is shown in Table~\ref{ylist}. However, one has to be very careful with the information collected from different experimental analysis. One example is the $Y(4220)$ and $Y(4260)$ states. Whether they are two different states (as considered in Ref.~\cite{pdg} and quoted in Table~\ref{ylist}), or whether, as stated in Ref.~\cite{Ablikim:2018vxx}, the structure around 4260 MeV can be interpreted as a superposition of  the two resonances  observed in Ref.~\cite{Ablikim:2016qzw}, the so called $Y(4220)$ and $Y(4320)$, is still an open question.
\begin{table*}[h]
  \caption{A list of the currently known neutral $I^G(J^{PC})=0^-(1^{--})$ charmonium $Y$ states. 
  The current naming scheme used by PDG \cite{pdg} is included in the table. The quoted year is the year of the first observation in each channel.}
  \label{ylist}
\setlength{\tabcolsep}{0.23pc}
\begin{center}
 MeV\begin{tabular}{lcllc}
\hline\hline
\rule[10pt]{-1mm}{0mm}
 State & Name in PDG  &  \ \ \ \ Decay channel & Experiment & Year  \\[0.7mm]
\hline
\rule[10pt]{-1mm}{0mm}
$Y(4220)$ & $\psi(4230)$  &
	$ Y(4220)\to \chi_{c0} \:\omega$ &
	{BESIII~\cite{Ablikim:2014qwy}} & 2015 \\[1.89mm]
& &     
	$ Y(4220)\to h_c \:\pi^+ \pi^- $ &
	{BESIII~\cite{BESIII:2016adj}} & 2017 \\[1.89mm]
& &     
	$ Y(4220)\to \psi(2S) \:\pi^+ \pi^- $ &
	{BESIII~\cite{Ablikim:2017oaf}}& 2017 \\[1.89mm]
& &
	$ Y(4220) \to D^0 D^{\ast \:-} \:\pi^+ $ &
	{BESIII~\cite{Ablikim:2018vxx}} & 2018 \\[4.89mm]
$Y(4260)$ & $\psi(4260)$  &
	$ Y(4260)\to J/\psi \:\pi^+ \pi^-$ &
	{BaBar~\cite{Aubert:2005rm, Lees:2012cn}; 
	  CLEO-c~\cite{He:2006kg}; Belle~\cite{Liu:2013dau, Yuan:2007sj}}
	& 2005 \\[1.89mm]
& &    
	$ Y(4260)\to J/\psi \:\pi^0 \pi^0$ &
	{CLEO-c~\cite{Coan:2006rv}} & 2006 \\[1.89mm]
& &    
	$ Y(4260)\to J/\psi \:K^+ K^-$ &
	{CLEO-c~\cite{Coan:2006rv}; 
	Belle~\cite{Yuan:2007bt, Shen:2014gdm}} & 2006 \\[1.89mm]
& &     
	$ Y(4260)\to J/\psi \:f_0(980)$ &
	{BaBar~\cite{Lees:2012cn}}
	& 2012 \\[1.89mm]
& &     
	$ Y(4260)\to Z_c(3900)^{\pm} \:\pi^{\mp}$ &
	{Belle~\cite{Liu:2013dau}; BESIII~\cite{Ablikim:2013mio}}
	& 2013 \\[1.89mm]
& &     
	$ Y(4260) \to J/\psi \:\pi^+\pi^-$&
        BESIII~\cite{Ablikim:2016qzw}
	& 2017 \\[4.89mm]
$Y(4360)$ & $\psi(4360)$  &
	$ Y(4360) \to \psi(2S) \:\pi^+ \pi^-$ &
	{Belle~\cite{Wang:2007ea, Wang:2014hta}; 
	BaBar~\cite{Aubert:2007zz, Lees:2012pv}; 
	BESIII~\cite{Ablikim:2017oaf}} & 2007 \\[1.89mm]
& &     
	$ Y(4360)\to J/\psi \:\pi^+ \pi^-$ &
	{BESIII~\cite{Ablikim:2016qzw}} & 2017 \\[4.89mm]
$Y(4390)$ & $\psi(4390)$  &
	$ Y(4390) \to h_c \:\pi^+ \pi^-$ &
	{BESIII~\cite{BESIII:2016adj}} & 2017 \\[1.89mm]
& &     
	$ Y(4390)\to \psi(2S) \:\pi^+ \pi^- $ &
	{BESIII~\cite{Ablikim:2017oaf}}& 2017 \\[4.89mm]
$Y(4660)$ & $\psi(4660)$  &
	$ Y(4660) \to \psi(2S) \:\pi^+ \pi^-$ &
	{Belle~\cite{Wang:2007ea, Wang:2014hta}; 
	BaBar~\cite{Lees:2012pv}} & 2007 \\[1.89mm]
& &     
	$ Y(4660)\to \Lambda_c^+ \Lambda_c^-$ &
	{BESIII~\cite{Pakhlova:2008vn}} & 2008 \\[1.89mm]
\hline\hline
\end{tabular}
\end{center}
\end{table*}
In the following  we discuss these $Y$ states.

\subsection{$Y(4260)$}

The $Y(4260)$ state is particularly interesting. It was first observed by the BaBar collaboration in 
the process $e^+e^- \to J/\psi \:\pi^+ \pi^-$ through ISR \cite{Aubert:2005rm}, and it was confirmed 
by CLEO-c \cite{He:2006kg} and Belle \cite{Yuan:2007sj}. The $Y(4260)$ was also observed in the 
$B^-\to Y(4260)K^-\to J/\psi \:\pi^+\pi^- K^-$ decay \cite{Lees:2012cn}, and CLEO-c reported two additional decay 
channels: $J/\psi\pi^0\pi^0$ and $J/\psi K^+K^-$ \cite{Coan:2006rv}. More recently, the BESIII collaboration has 
announced new precise measurements of the $e^+~e^-\to J/\psi \:\pi^+ \pi^-$ cross sections \cite{Ablikim:2016qzw}, reporting not only updated values 
for the mass and width of the $Y(4260)$, but also the presence of a second resonance in the $J/\psi \:\pi^+\pi^-$ mass spectrum. The mass and width of the two observed resonances are, respectively:  $(4222.0\pm3.1\pm1.4)$ MeV and $(44.1\pm4.3\pm2.0)$ MeV for the first one and $(4320.0\pm10.4\pm7)$ MeV
and   $(101.4^{+25.3}_{-19.7}\pm10.2)$ MeV for the second one.
Although in Ref.~\cite{Ablikim:2016qzw}
it is stated that the mass and width of the two observed resonances are in
agreement with 
the those of  $Y(4260)$ and $Y(4360)$  respectively,
in Ref.~\cite{Ablikim:2018vxx} it is said that the structure around 4260 MeV can be interpreted as a superposition of  the two
resonances  observed in Ref.~\cite{Ablikim:2016qzw}. Since this discussion is not yet settled, here we will consider the $Y(4260)$ as one unique state {\it i.e.}, the lowest mass state observed in \cite{Ablikim:2016qzw}.

Since the mass of the $Y(4260)$ is higher than the $D^{(*)}\bar{D}^{(*)}$ threshold, if it was a normal $c\bar{c}$ 
charmonium state, it would decay mainly to $D^{(*)}\bar{D}^{(*)}$. However, the observed $Y$ state does not match 
the peaks in $e^+e^-\to D^{(*)\pm}D^{(*)\mp}$ cross sections measured by Belle \cite{Abe:2006fj} and BaBar 
\cite{Aubert:2009aq, Pakhlova:2008zza}. Besides, the $\psi(3S), \psi(2D)$ and $\psi(4S)$ $c\bar{c}$ states have 
been assigned to the well established $\psi(4040)$, $\psi(4160)$, and $\psi(4415)$ mesons respectively, and the 
prediction from quark models for the $\psi(3D)$ state is 4.52 GeV. Therefore, the mass of the $Y(4260)$ is not 
consistent with any of the $1^{--}$ $c\bar{c}$ states \cite{Zhu:2007wz,Klempt:2007cp,Nielsen:2009uh}.

\subsubsection{Theoretical explanations for $Y(4260)$}
There are many theoretical interpretations for the $Y(4260)$: tetraquark state \cite{Esposito:2014rxa}, hadronic 
molecule of $D_{1} D$  or $D_{0} D^*$ \cite{Ding:2008gr,Wang:2013cya}, $\chi_{c1} \omega$ \cite{Yuan:2005dr}, $\chi_{c1} \rho$ 
\cite{Liu:2005ay}, $J/\psi f_0(980)$ \cite{MartinezTorres:2009xb}, a hybrid charmonium state~\cite{Zhu:2005hp}, a 
charm baryonium \cite{Qiao:2005av}, a cusp \cite{vanBeveren:2006ih,vanBeveren:2009fb,vanBeveren:2009jk}, etc. 
Within the available experimental information, none of these suggestions can be completely ruled out. However, most of the QCDSR calculations  
can not explain the mass of the $Y(4260)$ supposing it to be a tetraquark state \cite{Albuquerque:2008up}, or a 
$D_{1} D$ or $D_{0} D^*$ hadronic molecule \cite{Albuquerque:2008up}, or a $J/\psi f_0(980)$ molecular state 
\cite{Albuquerque:2011ix}. There is only one exception where the mass of the $Y(4260)$ can be explained as a tetraquark state in a QCDSR calculation~\cite{Wang:2018ntv}.

In the next subsections we will show that it is possible to explain not only the mass, but also the decay width of the $Y4260)$, in a QCDSR calculation, if one uses a  mixture of a $J/\psi$  and a $[cq\bar{c}\bar{q}]$ 
tetraquark currents~\cite{Dias:2012ek}, in the same way as discussed in Sec.\ref{sec-mixing} for the $X(3872)$.

\subsubsection{QCDSR calculations for the $Y(4260)$ mass}
\lb{sec-Ymass}

In Ref.~\cite{Dias:2012ek} the $Y(4260)$ was considered as a mixed charmonium-tetraquark state and the QCDSR method was used to study both its 
mass and decay width.
For the charmonium part, the conventional 
vector current was used:
\begin{equation}
j_\mu^{'(2)}=\bar{c}_a(x)\ga_\mu c_a(x),
\label{jcc}
\end{equation}
while the tetraquark part is implemented as~\cite{Albuquerque:2008up}:
\begin{eqnarray}
j_\mu^{(4)} &=& \frac{\epsilon_{abc} \epsilon_{dec}}{\sqrt{2}}
\Big[ \Big( q_a^T(x)C\ga_5 c_b(x) \Big) \: \Big(\bar{q}_d(x)\ga_\mu\ga_5 C\bar{c}_e^T(x) \Big) ~+~
\Big( q_a^T(x)C\ga_5\ga_\mu c_b(x) \Big) \: \Big(\bar{q}_d(x)\ga_5 C\bar{c}_e^T(x) \Big) \Big].
\label{j4q}  
\end{eqnarray}
As in Refs.~\cite{Matheus:2009vq, Sugiyama:2007sg}, we define the normalized two-quark current as
\begin{equation}
j_\mu^{(2)}=\frac{1}{\sqrt{2}}\qq[q] ~j_\mu^{'(2)},
\end{equation}
and from these two currents we build the mixed charmonium-tetraquark $J^{PC}=1^{--}$ current for 
the $Y(4260)$ state:
\begin{equation}
j_\mu(x)=\sin(\theta) \:j_\mu^{(4)}(x)+\cos(\theta) \:j_\mu^{(2)}(x).
\label{jmix}
\end{equation}
As usual, the phenomenological side is evaluated by inserting, in the two-point correlator, a complete set of 
intermediate states with $1^{--}$ quantum numbers. In such a case, the coupling of the 
vector state $Y$ to the current in Eq.~(\ref{jmix}) is parametrized  through the coupling parameter $\lambda_Y$:
\begin{equation}
\langle 0| j_{\mu}(x)|Y\rangle = \lambda_Y \epsilon_{\mu},
\label{Ycoupling}
\end{equation}
where $\epsilon_\mu$ is the polarization vector of $Y(4260)$. Using Eq.~(\ref{Ycoupling}), we can write the 
phenomenological side as
\begin{equation}
\Pi^{phen}_{\mu \nu}(q) = \frac{\lambda_Y^{2}}{m_{Y}^{2} - q^{2}}
\Big(g_{\mu \nu} - \frac{q_{\mu}q_{\nu}}{q^{2}} \Big) + \:.\:.\:.\:
\label{Yphenoside}
\end{equation}
where $m_Y$ is the mass of the $Y$ state and the dots, in the second term in the RHS of 
Eq.~(\ref{Yphenoside}), denotes the higher resonance contributions which will be parametrized, 
as usual, through the introduction of the continuum threshold parameter $s_{0}$ \cite{Ioffe:1981kw}.
In the OPE side, we work at leading order in $\alpha_{s}$ in the operators and we consider the contributions 
from the condensates up to dimension eight. Although we  consider only a part of the  dimension 8 
condensates (related to the quark condensate times the mixed condensate), in Ref.~\cite{Finazzo:2011he} it 
was shown that this is the most important dimension 8 condensate contribution.
Considering the current in Eq.~(\ref{jmix}), the correlator in the OPE side can be written as
\begin{eqnarray}
\Pi_{\mu \nu}(q) &=& \frac{1}{2} \qq[q]^2 \cos^{2}\theta ~~\Pi^{22}_{\mu \nu}(q) ~+~ 
	\sin^{2}\theta ~~\Pi^{44}_{\mu \nu}(q) ~+~ \frac{1}{\sqrt{2}} \:\langle \bar{q}q\rangle \:\sin \theta
	 \:\cos \theta ~~\bigg[ \Pi^{24}_{\mu \nu}(q) + \Pi^{42}_{\mu \nu}(q) \bigg] ,
\label{opeside}
\end{eqnarray}
with
\begin{equation}
\Pi_{\mu \nu}^{ij}(q) = i \int d^{4}x ~e^{i q\cdot x}
\langle 0|T [j_{\mu}^{i}(x)j_{\nu}^{j\dagger}(0)
]|0\rangle .
\end{equation}
In this way, $\Pi_{\mu \nu}^{22}(q)$ and $\Pi_{\mu \nu}^{44}(q)$ are the correlation functions of the 
$J/\psi$ meson and $[cq\bar{c}\bar{q}]$ tetraquark state, respectively.
After making a Borel transform on both sides of the sum rule, and transferring the continuum contributions to the OPE side, the 
sum rule, in the $g_{\mu \nu}$ structure, can be written as:
\begin{eqnarray}
  \lambda_Y^{2} ~e^{- m_{Y}^{2}/M^{2}} &=& \frac{1}{2} \qq[q]^2 \cos^{2}\theta
  	~~\Pi^{22}(M^{2}) ~+~ \sin^{2}\theta ~~\Pi^{44}(M^{2}) ~+~ 
	\frac{1}{\sqrt{2}} \:\qq[q] \:\sin \theta \:\cos \theta ~\bigg[ 
	\Pi^{24}(M^{2}) + \Pi^{42}(M^{2}) \bigg]. ~~
\label{srY4260}
\end{eqnarray}
The expressions for the invariant functions, $\Pi^{ij}(M^2)$, in Eq.~(\ref{srY4260})  are given in Ref.~\cite{Dias:2012ek}.
In Fig.~\ref{figY4260}a, we plot the relative contributions of all the terms in the OPE side. We have 
used $\sqrt{s_{0}} = 4.70$~GeV and $\theta = 53^\circ$. For others $\theta$ values outside the range 
$52.5^\circ \leq\theta\leq53.5^\circ$, we do not have a good OPE convergence. From this figure we see 
that the contribution of the dimension-8 condensates is smaller than 15\% of the total contribution for 
values of $M^{2} \geq 2.40$~GeV$^{2}$, indicating a good OPE convergence. Therefore, we fix 
the lower value of $M^{2}$ in the sum rule window as: $M_{min}^{2} = 2.40$~GeV$^{2}$.
\begin{figure}[t]
\begin{tabular}[b]{p{0.45\textwidth}}
    \includegraphics[width=0.43\textwidth]{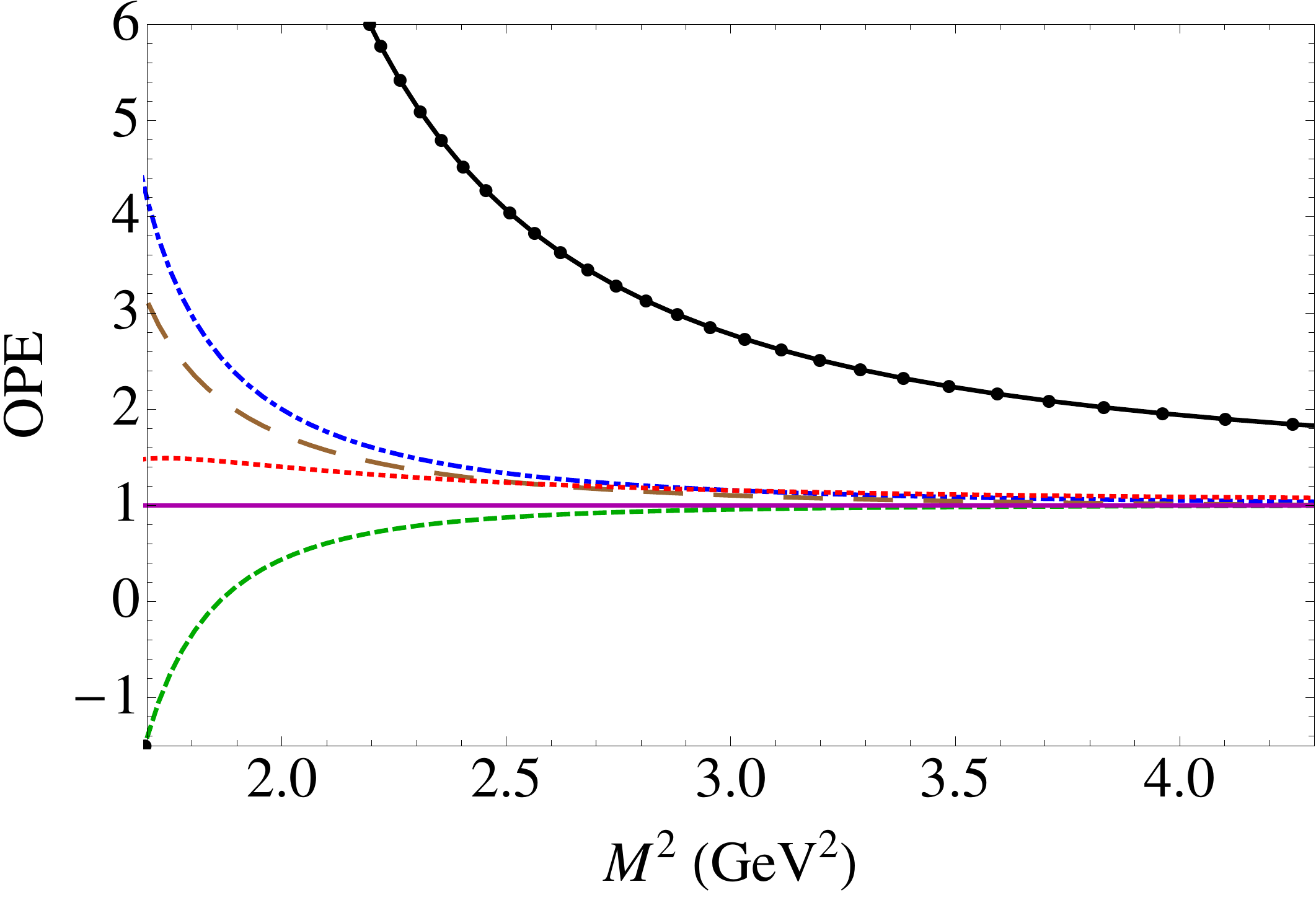}\\ 
    \centerline{(a)}\\ 
    \includegraphics[width=0.43\textwidth]{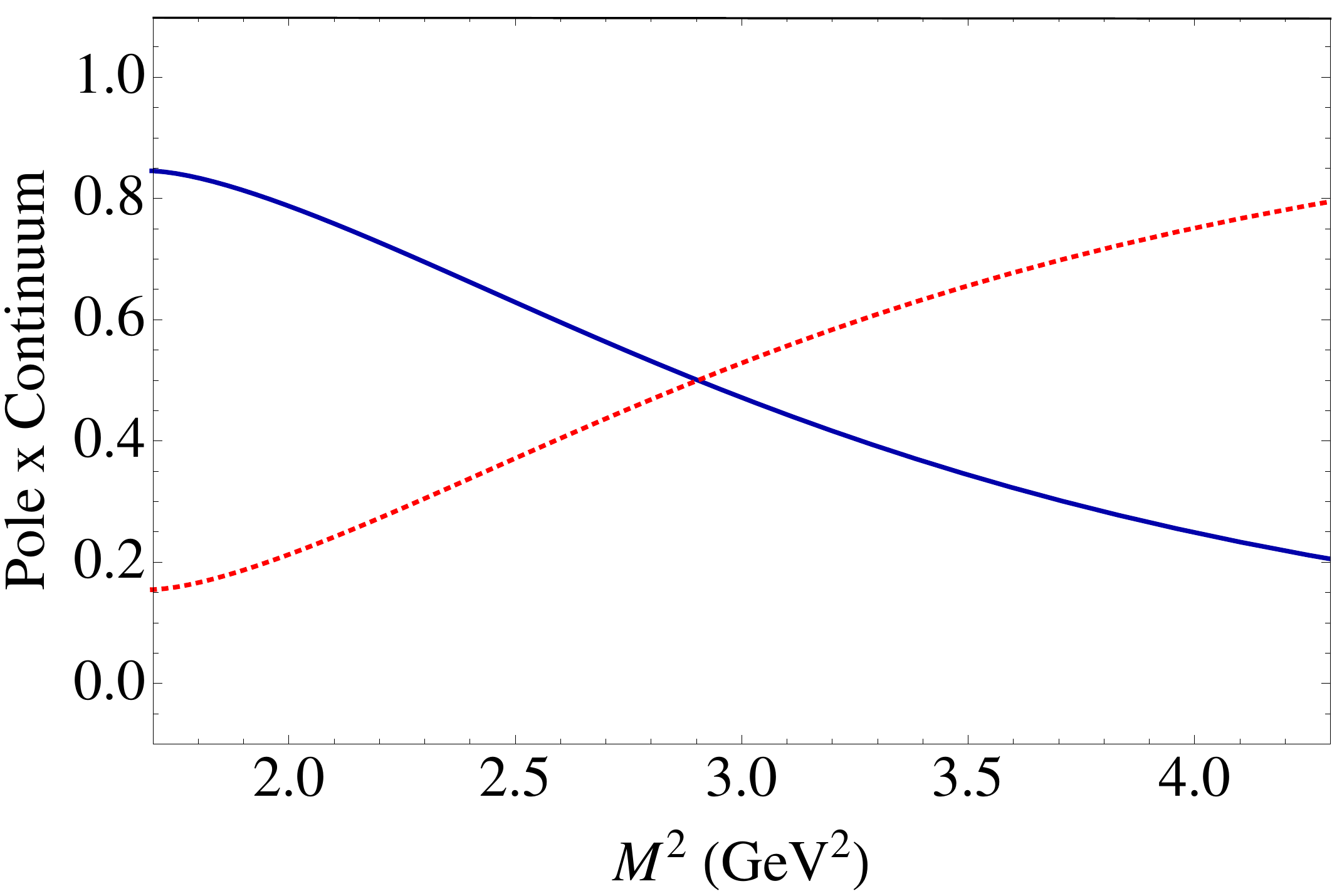}
    \centerline{(b)}\\ 
\end{tabular}
\begin{tabular}[b]{p{0.5\textwidth}}
    \vspace{-7.5cm}
    \includegraphics[width=0.5\textwidth]{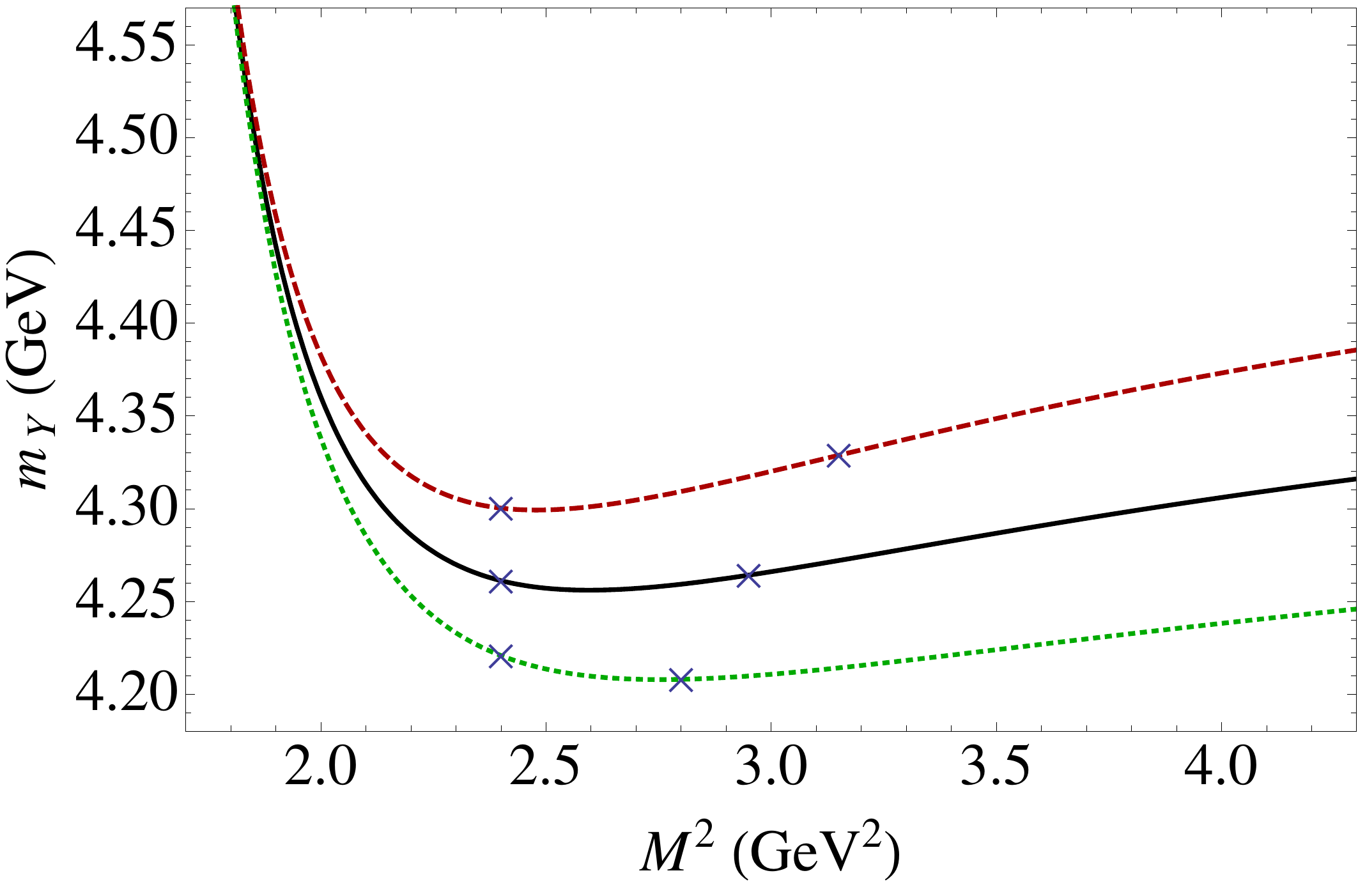}
    \centerline{(c)}\\ 
\end{tabular}
\caption{Sum rule calculation for the $Y(4260)$ state. a) The OPE convergence in the region 
$2.0 \leq M^{2}_{B} \leq 6.0$ GeV$^{2}$ for $\sqrt{s_{0}} =4.70$ GeV. We plot the relative contributions, 
starting with the perturbative contribution (line with circles), and the other lines represent the relative 
contribution after adding the next term  in the expansion: $+ \qq[q]$ (dot-dashed line), $+ \GG$ 
(long-dashed line), $+ \qGq[q]$ (dotted line), $+\qq[q]^2$ (dashed line) and $\qq[q]\qGq[q]$ (solid line). 
b) The pole contribution (solid line) and the continuum contribution (dotted line), for  
$\sqrt{s_{0}} =4.70$~GeV. c) The mass as a function of the sum rule parameter $M^{2}_{B}$ for 
$\sqrt{s_{0}} = 4.60$~GeV (dotted line), $\sqrt{s_{0}} = 4.70$~GeV (solid line), $\sqrt{s_{0}} = 4.80$~GeV 
(long-dashed line). The crosses indicate the valid Borel window. Figures taken from Ref.~\cite{Dias:2012ek}.}
\label{figY4260} 
\end{figure}  
 In Fig.~\ref{figY4260}b, we show a comparison between the pole and 
continuum contributions. It is clear that the pole contribution is equal to the continuum contribution 
for $M^{2}= 2.90$~GeV$^{2}$. Therefore, for $\sqrt{s_{0}} = 4.70$~GeV$^{2}$ and 
$\theta = 53^\circ$ the Borel window is: $2.40 \leq M^{2} \leq 2.90$~GeV$^{2}$. 
The ground state mass is 
shown, as a function of $M^{2}$, in Fig.~\ref{figY4260}c. From this figure we see that there 
is a very good Borel stability in the determined Borel window, which is
represented as crosses in Fig.~\ref{figY4260}c. 
Varying the value of the continuum threshold in the range $\sqrt{s_{0}} = 4.70 \pm 0.10$~GeV, the 
mixing angle in the range $\theta=(53.0\pm0.5)^\circ$, and the other parameters as indicated in 
Table~\ref{QCDParam}, the mass obtained in Ref.~\cite{Dias:2012ek} is:
\begin{equation} \label{ymass}
m_{Y} = (4.26 \pm 0.13) ~ \mbox{GeV},
\end{equation}
which is in excellent agreement with the experimental $Y(4260)$ mass. Once
the mass is determined, its value can be used in Eq.~(\ref{srY4260}) to estimate the meson-current 
coupling parameter, defined in Eq.~(\ref{Ycoupling}). Using the same values of $s_0$, 
$\theta$ and Borel window one  gets~\cite{Dias:2012ek}:
\beq
\lambda_Y = (2.00 \pm 0.23) \times 10^{-2} ~ \mbox{GeV}^5.
\label{lay}
\enq

\subsubsection{$Y(4260)\to J/\psi\pi\pi$ decay width}

To estimate the decay width of the process 
$Y(4260) \rightarrow J/\psi \:\pi^+ \pi^-$, it was assumed in Ref.~\cite{Dias:2012ek} that the two pions in the final state come from 
the $\si$ meson. The coupling constant, associated with the vertex $Y J/\psi \sigma$, is evaluated using the three-point correlator
\beq
\Pi_{\mu \nu}(p,\pli, q) = \int d^4x \:d^4y 
	~e^{i\pli \cdot x} ~e^{iq\cdot y} ~\lag 0|T\{j_{\mu}^{\psi}(x)
j^{\sigma}(y)j_{\nu}^{Y\dagger}(0)\}|0\rag.
\label{3poy4260}
\enq
with $p=\pli+q$. The interpolating fields appearing in Eq.~(\ref{3poy4260}) are the currents for 
$J/\psi$, $\sigma$ and $Y(4260)$, respectively. The currents for $J/\psi$ and $Y(4260)$ are 
defined by Eqs.~(\ref{jcc}) and (\ref{jmix}). For the $\sigma$ meson, we have
\beq
j^{\sigma}=\frac{1}{\sqrt{2}}\Big(\bar{u}_a(x)u_a(x)
+ \bar{d}_a(x)d_a(x)\Big).
\enq  
In order to evaluate the phenomenological side of the three-point correlator we insert, in 
Eq.~(\ref{3poy4260}), intermediate states for $Y$, $J/\psi$ and $\sigma$. Using the definitions: 
\begin{eqnarray} \label{si}
  \lag 0 | \:j_\mu^\psi\: |J/\psi(\pli)\rag &=& m_\psi f_{\psi}\epsilon_\mu(\pli), \nn\\
  \lag 0 | \:j^\sigma\: |\sigma(q)\rag &=& A_{\si},\\
  \lag Y(p) | \:j_\nu^Y\: |0\rag &=& \la_Y \epsilon_\nu^*(p), \nn
\end{eqnarray}
we obtain the following relation:
\beqa
\Pi_{\mu\nu}^{phen} (p,\pli,q) &=& {\la_Y m_{\psi} f_{\psi} A_\sigma~ 
g_{Y\psi \sigma}(q^2) \over (p^2-m_{Y}^2)({\pli}^2-m_{\psi}^2)(q^2-m_\sigma^2)}
\Big[ (\pli \cdot p) g_{\mu\nu} - \pli_\nu q_\mu - \pli_\nu \pli_\mu \Big] ~~+~~ \cdots\;,
\lb{phen}
\enqa
where the dots stand for the contribution of all possible excited states. The form factor, 
$g_{Y\psi \si}(q^2)$, is defined by the generalization of the on-mass-shell matrix element, 
$\lag J/\psi \:\sigma|Y\rag$, for an off-shell $\sigma$ meson~\cite{Dias:2012ek}: 
\beq
\lag J/\psi \:\si |Y\rag=g_{Y\psi \sigma}(q^2)
(\pli \cdot p ~\epsilon^*(\pli)\cdot \epsilon(p) -
\pli \cdot \epsilon(p)~p\cdot \epsilon^*(\pli)).
\label{coupY}
\enq
 In the OPE side, one works at leading order in 
$\al_s$ and considers the condensates up to dimension five. In Ref.~\cite{Dias:2012ek} the authors have chosen to work with the 
${\pli}_{\nu} q_\mu$ structure, since it has more terms contributing to the OPE. Taking the 
limit $p^2 = {\pli}^2=-P^2$ and doing the Borel transform such as $P^2 \rightarrow M^2$, one gets the 
following expression for the sum rule in the structure ${\pli}_{\nu} q_\mu$:
\beqa
\frac{\lambda_Y A_{\si} m_\psi f_\psi}
{(m_Y^2 -m_\psi^2)}~g_{Y\psi \sigma}(Q^2)\left(e^{-m_\psi^2/M^2}
-e^{-m_Y^2/M^2}\right) ~+~ B(Q^2)~e^{-s_0/M^2} &=& 
(Q^2+m_\sigma^2) ~\Pi^{OPE}(M^2,Q^2)
\label{3sr}
\enqa
where $Q^2=-q^2$, and $B(Q^2)$ gives the contribution to the pole-continuum transitions, as discussed in Sec.~\ref{s-3p}. The correlator $\Pi^{OPE}(M^2,Q^2)$ is 
given by~\cite{Dias:2012ek}:
\beqa
  \Pi^{OPE}(M^2,Q^2)&=&\frac{\sin \theta}{3\cdot2^4\sqrt{2}\pi^2}
  \int\limits_{0}^{1} \!\!d\al ~e^{\frac{-m_c^2}{\al (1-\al)M^2}}
  \:\bigg[ \frac{m_c\qGq[q]}{Q^2} \bigg( \frac{1-2\al(1-\al)}{\al(1-\al)} \bigg) - 
  \frac{\GG}{2^5\pi^4} \bigg] ~~.
\label{opeside}
\enqa
The $\sin\theta$ in Eq.~(\ref{opeside}) indicates that only the tetraquark part of the current in 
Eq.~(\ref{jmix}) contributes to the OPE side. In fact, the charmonium part of the current gives 
only disconnected diagrams that are discarded in the calculations. In Eq.~(\ref{3sr}), the values of the mass and 
decay constant of $J/\psi$ and $\si$ mesons are: $m_{\psi}=3.1$~GeV, $f_{\psi}=0.405$~GeV \cite{pdg}, 
and $m_{\si}=0.478$~GeV \cite{Aitala:2000xu}. The parameters $\la_{Y}$ and $A_{\si}$ represent, 
respectively, the coupling of the $Y$ and $\si$ states to the currents defined in Eq.~(\ref{Ycoupling}) 
and (\ref{si}). The value of $\la_{Y}$ is given in Eq.~(\ref{lay}), while $A_{\si}$ was determined in 
Ref.~\cite{Dosch:2002rh} and its value is $A_{\si}=0.197$~GeV$^2$.
To obtain $g_{Y\psi \sigma}(Q^2)$ one uses Eq.~(\ref{3sr}) and 
its derivative, with respect to $M^2$, to eliminate $B(Q^2)$ from these equations. In Fig.~\ref{fig3D}a  $g_{Y\psi\si}(Q^2)$ is shown as a function of both 
$M^2$ and $Q^2$. From Fig.~\ref{fig3D}a we see that, in the region $7.0 \leq M^2 \leq 10.0$~GeV$^2$, the form factor is very stable as a function of $M^2$, for all values of $Q^2$. 
The squares in Fig.~\ref{fig3D}b show the $Q^2$ dependence of $g_{Y\psi\si}(Q^2)$, obtained for 
$M^2=8.0$~GeV$^2$. For other values of the Borel mass, in the range 
$7.0 \leq M^2 \leq 10.0$~GeV$^2$, the results are equivalent. It is possible to fit the QCDSR results for $g_{Y\psi\si}(Q^2)$ using a monopole form:
\beq
g_{Y\psi\si}(Q^2) = \frac{g_1}{g_2 + Q^2},
\label{mono}
\enq
with
\beq
g_1 = (0.58~\pm ~0.04)~\mbox{GeV}; ~~~ g_2=(4.71~\pm ~0.06)~\mbox{GeV}^2,  
\label{c1c2}
\enq
as shown by the solid line in Fig.~\ref{fig3D}b.
\begin{figure}[t]
\begin{tabular}[b]{p{0.48\textwidth}}
    \includegraphics[width=0.45\textwidth]{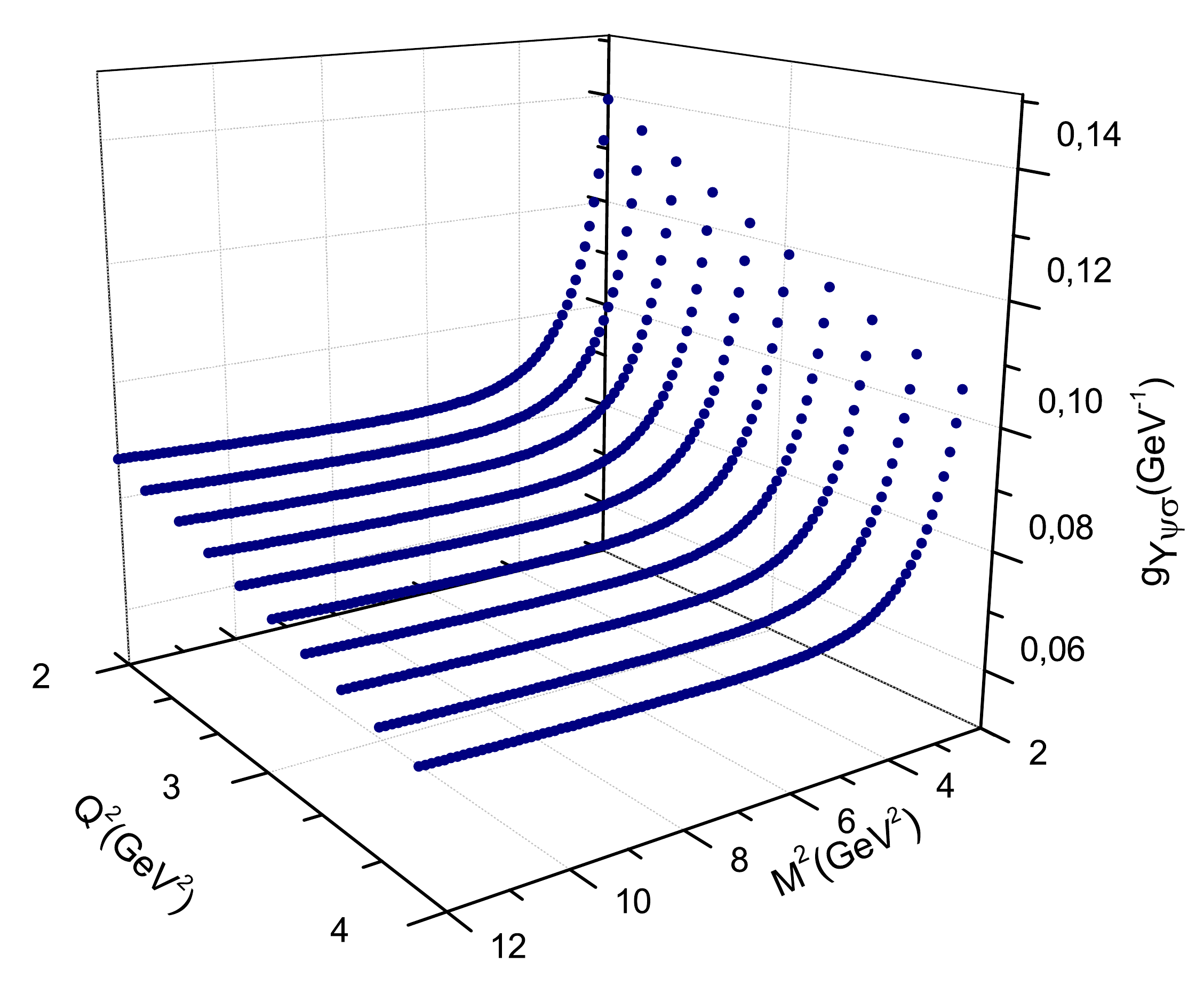}\\ 
    \centerline{(a)}\\ 
\end{tabular}
\begin{tabular}[b]{p{0.48\textwidth}}
    \includegraphics[width=0.48\textwidth]{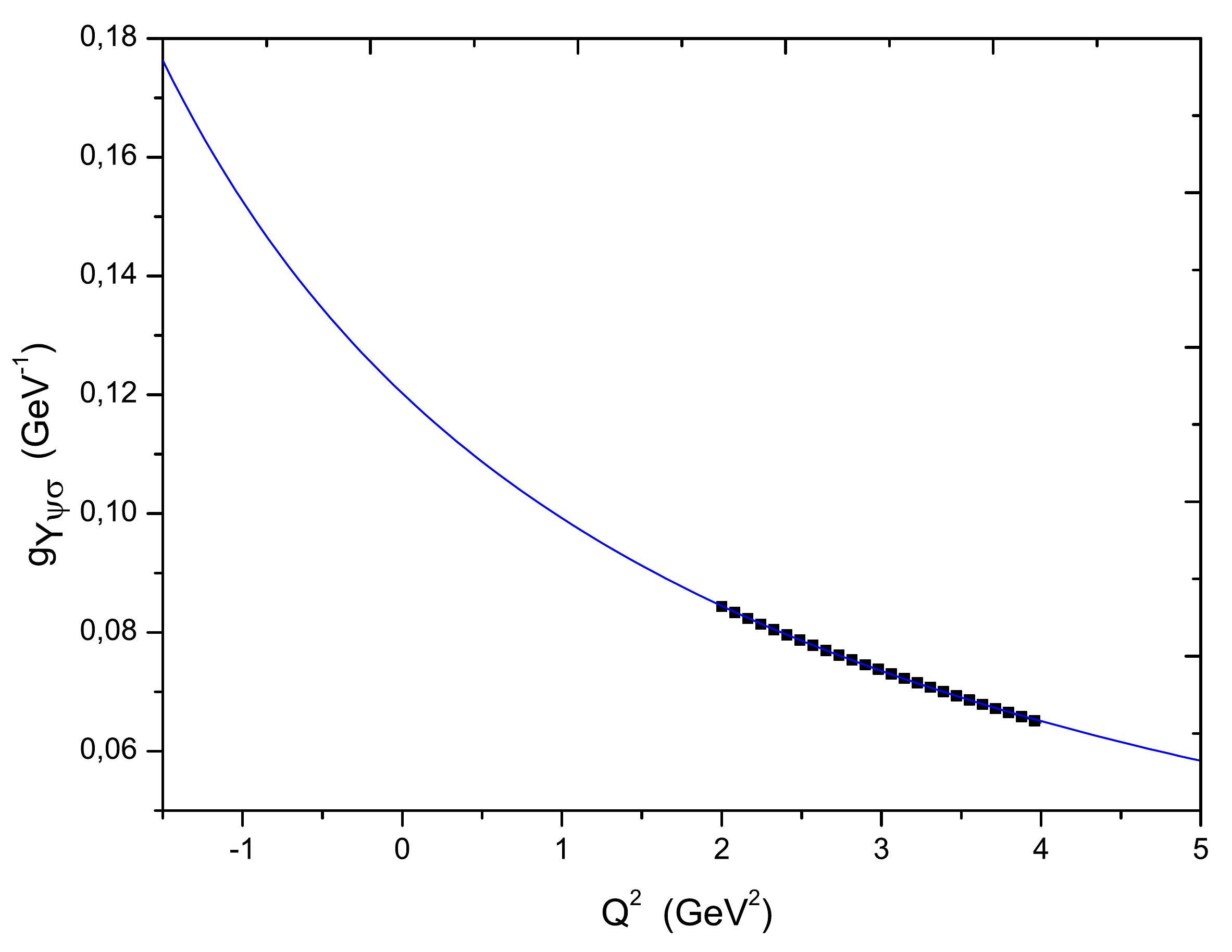}
    \centerline{(b)}\\ 
\end{tabular}
\caption{a) $g_{Y\psi\si}(Q^2)$ as a function of  both $Q^2$ and $M^2$.
b) QCDSR results for $g_{Y\psi\si}(Q^2)$, as a function of $Q^2$, for $\sqrt{s_0}=4.76$~GeV and $M^2=8$ Gev$^2$
(squares). The solid line gives the parametrization of the QCDSR results (see  
Eq.~(\ref{mono})). Figures taken from Ref.~\cite{Dias:2012ek}.}
\label{fig3D} 
\end{figure} 
The coupling constant, $g_{Y\psi\si}$, is given by using $Q^2=-m^2_{\si}$ in 
Eq.~(\ref{mono}). Then one gets~\cite{Dias:2012ek}:
\beq
g_{Y\psi\sigma}=g_{Y\psi\sigma}(-m^2_\sigma)=(0.13 \pm 0.01)~~\mbox{GeV}^{-1}.
\label{coupvalue}
\enq
The error in the coupling constant given above comes from variations in $s_0$ in the range 
$4.6\leq s_0 \leq 4.8$~GeV$^2$, and in the mixing angle $52.5^\circ \leq \theta \leq 53.5^\circ$.

The decay width for the process $Y(4260) \rightarrow J/\psi \:\sigma \rightarrow J/\psi \:\pi \pi$
in the narrow width approximation is given by~\cite{Dias:2012ek}:
\beqa
{d\Gamma\over ds}(Y\to J/\psi \:\pi\pi) &=&\frac{|{\cal{M}}|^2}{8\pi \:m_Y^2}
\left( \frac{m_Y^2 -m^2_{\psi}+s}{2m_Y^2} \right)
\left( \frac{\Ga_{\si}(s) \:m_{\si}}{2m_Y\pi} \right)
\frac{\sqrt{\la(m_Y^2,m_\psi^2,s)}}{(s-m_{\si}^2)^2+(m_{\si}\Ga_{\si}(s))^2},
\label{de1}
\enqa
where $\la(a,b,c)=a^2+b^2+c^2-2ab-2ac-2bc$, and $\Ga_{\si}(s)$ 
is the s-dependent width of an off-shell $\sigma$ meson \cite{Aitala:2000xu}:
\beq \label{gams}
\Ga_{\si}(s)=\Ga_{0\si}\sqrt{\frac{\la(s,m_{\pi}^2, m_{\pi}^2)}
{\la(m_Y^2,m_{\pi}^2, m_{\pi}^2)}}\frac{m_Y^2}{s}.
\enq
Notice that $\Ga_{0\si}$ in Eq.~(\ref{gams}) is the experimental value for the decay of the $\si$ meson into 
two pions. Its value is $\Ga_{0\si}=(0.324\pm 0.042\pm 0.021)$ GeV \cite{Aitala:2000xu}.
The squared invariant amplitude  can be obtained from the matrix element in Eq.~(\ref{coupY}):
\beq
|{\cal M}|^2=g_{Y\psi \sigma}^2(s) \:f(m_Y,m_\psi,s),
\enq
where $g_{Y\psi \si}(s)$ is the form factor in the vertex $YJ/\psi \:\si$, given in Eq.~(\ref{mono}) 
using $s=-Q^2$, and
\beqa
f(m_Y,m_\psi,s)={1\over3}\left(m_Y^2m_\psi^2 + 
{1\over2}(m_Y^2 + m_\psi^2 - s)^2\right).
\label{m2}
\enqa
Therefore, the decay width for the process $Y(4260)\rightarrow J/\psi \pi\pi$ is given by
\beq
\Ga = \frac{m_\si}{32\pi^2 m^5_Y} \int\limits_{(2m_\pi)^2}^{(m_Y-m_\psi)^2} \!\!ds
~g_{Y\psi\si}^2(s) \:\Ga_\si(s) \:(m^2_Y-m^2_\psi +s) \sqrt{\la(m_Y^2,m_\psi^2,s)}
\:\left( \frac{f(m_Y,m_\psi, s)}{(s-m^2_\si)^2+(m_\si \Ga_\si(s))^2} \right) ~.
\label{y4260width}
\enq

Taking variations in $s_0$ and $\theta$ in the same intervals given above, we obtain, 
from Eqs.~(\ref{coupvalue})-(\ref{y4260width}), the following value for the decay width~\cite{Dias:2012ek}: 
\beq
\Ga_\si(Y\rightarrow J/\psi\:\pi\pi) = (1.0\pm 0.2 ) ~\mbox{MeV}.
\enq

The decay channel $Y(4260)\rightarrow J/\psi \:\pi\pi$ can also occur through
the formation of $f_0(980)$ in the
intermediate state. Therefore, we have to determine also the coupling constant associated with the vertex 
$Y \to J/\psi \:f_0(980)$. For this purpose, we consider the $f_0(980)$ meson as a quark-antiquark
state with a mixture of strange and light components. In this case, the interpolating current for 
$f_0(980)$ is given by
\begin{eqnarray}
  j^{f_0} &=& \cos(\alpha) \:\bar{s}s ~+~ \frac{1}{\sqrt{2}} \sin(\alpha) \: (\bar{u}u + \bar{d}d) ~.
\end{eqnarray}
Using  Eq.~(\ref{y4260width}) with the $f_0(980)$ meson parameters instead of the ones for $\si$, {\it i.e.}, \cite{pdg}: 
$m_{f_0} ~=~ (990 \pm 20) ~\mbox{MeV}$,  $\Ga_{0 \:f_0} ~=~ (40 - 100) ~\mbox{MeV}$ 
and taking the variations  $4.6 \leq \sqrt{s_0} \leq 4.8$~GeV and 
$52.3^\circ \leq \theta \leq 53.5^\circ$, we obtain:
\begin{eqnarray}
  \Ga_{f_0} (Y \to J/\psi\:\pi\pi) &=& (3.1 \pm 0.2) ~\mbox{MeV}
\end{eqnarray}
leading to the following decay width into this channel:
\begin{eqnarray}
  \Ga (Y \to J/\psi\:\pi\pi) &=& (4.1 \pm 0.6) ~\mbox{MeV} \:,
\end{eqnarray}
which is consistent with the lower bound given in Ref.~\cite{Brambilla:2010cs}: 
$\Ga (Y \to J/\psi \:\pi\pi) \:>\: 508 ~\mbox{keV}$ at $90\%$ CL. 

Assuming that the two pions in the final state of the decay $Y \to J/\psi \:\pi\pi$ come only
from the $\sigma$ and $f_0(980)$ scalar mesons intermediate states, we obtain a value for the width
$\Ga_{Y \to J/\psi \,\pi\pi} \approx (4.1 \pm 0.6)$~MeV, which is much smaller than the total experimental width:  $\Ga_{exp}\approx (55~\pm~19)$~MeV \cite{pdg}. 
This can be interpreted as an indication that the main decay channel of the $Y(4260)$ should be  into two $D$ mesons. The possibility that the main decay mode
of the $Y(4260)$ is into two $D$ mesons corroborates the interpretation that
the $Y(4260)$ consists of two resonances, as suggested in Ref.~\cite{Ablikim:2018vxx}. If the main component of the $Y(4260)$ is the lower mass resonance,
that is called $Y(4220)$, this component indeed decays into $\pi^+D^0D^{*-}$
\cite{Ablikim:2018vxx}.
Therefore, we conclude that it is possible to explain the $Y(4260)$ state as 
a mixed charmonium-tetraquark state.

\subsubsection{$Y(4260)$ production in $B$ decays}

The same mixed current between the  $J/\psi$ charmonium and a tetraquark state,
proposed in Ref.~\cite{Dias:2012ek}, was used in Ref.~\cite{Albuquerque:2015nwa} to estimate the $Y(4260)$ production in the process $B^- \to Y(4260) \:K^-$. 
The experimental upper limit on the branching fraction for such a production from $B$ meson decay has 
been reported by the BaBar Collaboration \cite{Aubert:2005zh}, with $95\%$ C.L.:  
\begin{equation} 
  \label{branching} 
	{\mathcal B}_{_Y} <\! 2.9\times10^{-5} 
\end{equation} 
where ${\mathcal B}_{_Y} \equiv {\mathcal B}(B^- \!\!\to\! K^- Y(4260),Y(4260) \!\to\! J/\psi\pi^+\pi^-)$.

The process $B \to Y(4260) \:K$ occurs via weak decay of the $b$ quark, while the $u$ quark 
is a spectator. The $Y(4260)$ state, as a mixed state of tetraquark and charmonium, interacts via the
$\bar{c}c$ component of the weak current. As discussed in Sec.~\ref{sec-pro}, in an effective theory the
Hamiltonian describing  the weak interaction can be  written in terms of a four-quark  interaction vertex with
an effective  four quark  operator ${\mathcal{O}}_2=(\bar{c}\Gamma_\mu
c)(\bar{s}\Gamma^\mu     b)$,      with     a      $V-A$     structure
$\Gamma_\mu=\gamma_\mu(1-\gamma_5)$. The interaction can be factorized
into two matrix elements,  giving the following decay amplitude for the
process:
\begin{eqnarray}\lb{ampY} 
  {\mathcal M} &=& i\frac{G_F}{\sqrt{2}}V_{cb}V_{cs}^*\left(C_2+\frac{C_1}{3}\right)
  \:\langle B(p)\vert J_{\mu}^W\vert K(p^\prime)\rangle\langle Y(q) 
  \vert J^{\mu(\bar{c}c)}\vert0\rangle, 
\end{eqnarray} 
where $p = p^\prime+q$ and $J_{\mu}^W$, $ J^{\mu(\bar{c}c)}$ are given in Eq.~(\ref{wcurrents}).
Following Ref.~\cite{Zanetti:2011ju}, the matrix elements in 
Eq.~({\ref{ampY}) are parametrized as:
\begin{equation}\lb{2pmatrix} 
  \langle Y(q)\vert J_{\mu}^{(\bar{c}c)}\vert0\rangle=\lambda_W \epsilon^\ast_\mu(q)\,, 
\end{equation} 
and 
\begin{equation}\lb{3pmatrix} 
  \langle B(p)\vert J_{\mu}^W\vert K(p^\prime)\rangle = 
  f_+(q^2)(p_\mu + p_\mu^\prime)+f_-(q^2)(p_\mu-p_\mu^\prime)\,. 
\end{equation} 

The parameter $\lambda_W$ in Eq.~(\ref{2pmatrix}) gives the coupling between the current 
$J_\mu^{(\bar{c}c)}$ and the $Y$ state. The form factors $f_\pm(q^2)$ describe the weak 
transition $B \to K$. Hence we can see that the factorization of the matrix element describes 
the decay as two separated sub-processes. The decay width for the process $B^-\to Y(4260)K^-$ 
is given by  
\begin{equation}\lb{eqwidth} 
  \Gamma(B \to YK) = 
  \frac{\vert{\mathcal{M}}\vert^2}{16\pi m_B^3}\sqrt{\lambda(m_B^2,m_K^2,m_Y^2)}, 
\end{equation} 
with $\lambda(x,y,z)=x^2+y^2+z^2-2xy-2xz-2yz$. The squared invariant amplitude  can be 
obtained from Eq.~(\ref{ampY}), using Eqs.~(\ref{2pmatrix}) and (\ref{3pmatrix}): 
\begin{eqnarray} 
  \vert\mathcal{M}\vert^2 &=& \frac{G_F^2}{2 m_Y^2} \vert V_{cb}V_{cs}\vert^2 
  \left(C_2 +\frac{C_1}{3}\right)^2 \lambda(m_B^2,m_K^2,m_Y^2)\lambda_W^2f_+^2
\,. 
\end{eqnarray} 
The form factor $f_ +(Q^2)$ was determined in Sec.~\ref{sec-pro} and is given
by Eq.~(\ref{fplus}):
\begin{equation}
  f_+(Q^2)=\frac{(17.55\pm0.04) \GeV^2}{(105.0\pm1.8)\GeV^2+Q^2}\,. \nn
\end{equation} 
For the decay width calculation, we  need the value of the form factor at $Q^2=-m_Y^2$, 
where $m_Y$ is the mass of the $Y(4260)$ meson. Using $m_Y=(4251\pm9)\MeV$ \cite{pdg} 
we get: 
\begin{equation}\lb{fpluspolo} 
  f_+(Q^2=-m_Y^2)=0.206\pm0.004\,. 
\end{equation} 
The parameter $\lambda_W$ can also be determined using the QCDSR approach for the two-point correlator as done in Sec.~\ref{sec-pro}
\begin{equation} 
\Pi_{\mu\nu}(q)=i\int d^4y~e^{iq\cdot y}\langle0 \vert\: T \big[ J_\mu^Y(y) 
\:J_\nu^{(\bar{c}c)}(0) \big] \:\vert 0\rangle\,, 
\label{2point}
\end{equation} 
where the current $J_\nu^{(\bar{c}c)}$ is defined in Eq.~(\ref{wcurrents}).
For the $Y(4260)$ state we consider the mixed charmonium-tetraquark current given in Eq.~(\ref{jmix}). 
 The mixing angle, $\theta$, was determined in Ref.~\cite{Dias:2012ek} 
 to be: $\theta=(53.0\pm0.5)^\circ$.

The phenomenological side of the SR is obtained by considering intermediate $Y$ states: 
\begin{eqnarray} 
  \Pi_{\mu\nu}^{phen}(q) &=& \frac{i}{q^2-m_Y^2} 
  \langle0\vert J^Y_\mu\vert Y(q)\rangle \:\langle Y(q)\vert J^{(\bar{c}c)}_\nu\vert0\rangle 
  ~=~ \frac{i\lambda_Y\lambda_W}{Q^2+m_Y^2}\left(g_{\mu\nu}-\frac{q_\mu q_\nu} {m_Y^2}\right) 
\end{eqnarray} 
where the  definition in Eq.~(\ref{2pmatrix}) was used and 
\begin{equation} 
  \langle0\vert J^Y_\mu\vert Y(q)\rangle=\lambda_Y\epsilon_\mu(q)\,. 
\end{equation} 
The parameter $\lambda_Y$, that defines the coupling between the current $J^Y_\mu$ and the $Y$ 
meson, was determined in Sec.~\ref{sec-Ymass}: 
$\lambda_Y = (2.00 \pm 0.23) \times 10^{-2} ~ \GeV^5$. 

After performing the Borel transform in both sides of the 
sum rule one  gets from the $g_{\mu\nu}$ structure: 
\begin{eqnarray}\lb{2psumrule} 
  \lambda_W\lambda_Ye^{-m_Y^2/M^2} = \frac{1}{\sqrt{2}}\, \sin \theta ~\Pi^{4,2}(M^2)
  + \frac{1}{\sqrt{2}}\, \qq[q] \cos \theta ~\Pi^{2,2}(M^2)
\end{eqnarray} 
where the invariant functions, $\Pi^{2,2}(M^2)$ and $\Pi^{4,2}(M^2)$, are given in 
Ref.~\cite{Albuquerque:2015nwa}. The calculation of the coupling parameter $\lambda_W$  was done
using the same values for the masses and QCD condensates as in Ref.~\cite{Dias:2012ek}, values which are 
listed in Table~\ref{QCDParam}. To be consistent with the calculation of $\lambda_Y$ we also use the 
same region in the threshold parameter $s_0$ as in Ref.~\cite{Dias:2012ek}: 
$\sqrt{s_{0}} = (4.70 \pm 0.10)$ GeV. As one can see in Fig.~\ref{figLW}, the region of 
$M^2$-stability is given by $(8.0 \leq M^2 \leq 25.0) \GeV^2$.  
Taking into account the variation in the Borel mass parameter, in the continuum threshold, in the quark 
condensate, in the coupling constant $\lambda_Y$ and in the mixing angle $\theta$, the result for the 
$\lambda_W$ parameter is: 
\begin{equation}\lb{lambdaW} 
  \lambda_W=(0.90\pm0.32)\GeV^2\,. 
\end{equation} 

The decay width in Eq.~(\ref{eqwidth}) can be evaluated using the values of 
$f_+(-M_Y^2)$ and $\lambda_W$, determined in Eqs.~(\ref{fpluspolo}) and (\ref{lambdaW}) respectively. The 
branching ratio is evaluated dividing the result by the total width of the $B$ meson,
$\Gamma_{\mathrm{tot}}=4.280 \times 10^{-4} \:\mbox{eV}$~\cite{Albuquerque:2015nwa}: 
\begin{equation}\lb{result} 
  \mathcal{B}(B\to Y(4260)K)=(1.34\pm0.47)\times10^{-6}\,, 
\end{equation} 
where we have used the CKM parameters $V_{cs}=1.023$, $V_{cb}=40.6\times10^{-3}$ \cite{pdg}, 
and the Wilson coefficients $C_1(\mu)=1.082$, $C_2(\mu)=-0.185$, computed at $\mu=m_b$ and 
$\bar{\Lambda}_{\mathrm{MS}}=225\MeV$ \cite{Buchalla:1995vs}. 

\begin{figure}[t]
\begin{center}
\includegraphics[width=0.45\textwidth]{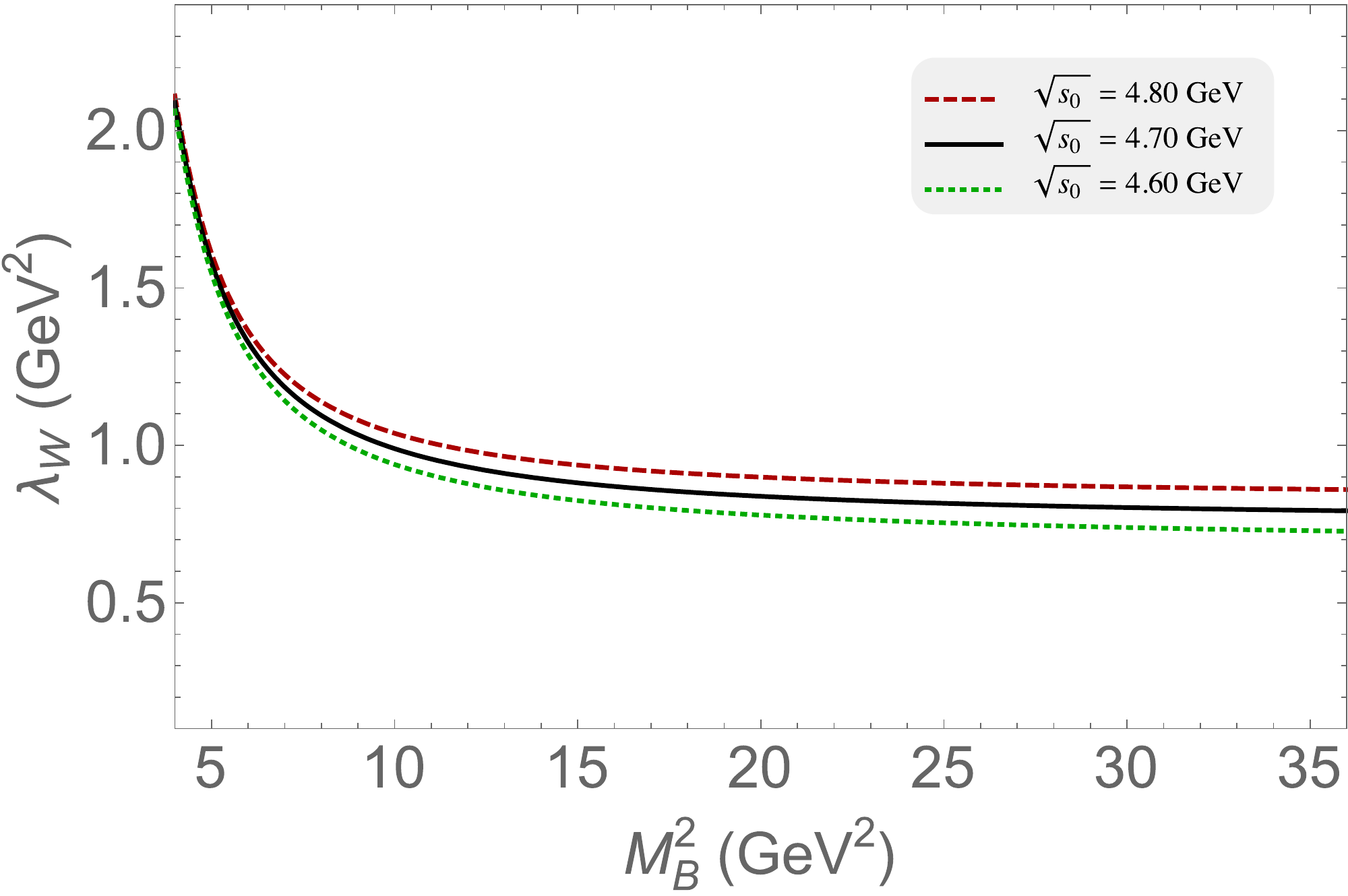}
\caption{The coupling parameter $\lambda_W$ as a function of $M^2$ for 
different values of the continuum threshold. Figure taken from Ref.~\cite{Albuquerque:2015nwa}.}
\label{figLW}
\end{center}
\end{figure}

In order to compare the branching ratio in Eq.~(\ref{result}) with the branching fraction 
obtained experimentally in Eq.~(\ref{branching}), we  use the results found in 
Ref.~\cite{Dias:2012ek}:
\begin{equation}
  \mathcal{B}(Y(4260) \to J/\psi \:\pi^+\pi^-) = (4.3 \pm 0.9)\times10^{-2}\,,
  \label{brY}
\end{equation}
and then, considering the uncertainties, we  estimate
${\mathcal B}_{_Y} >\! 3.0 \times 10^{-8}$.
However, it is important to notice that the authors in Ref.~\cite{Dias:2012ek} have considered 
that the two pions in the final state come only from the intermediate  $\sigma$ and $f_0(980)$ states, which could indicate that the result in Eq.~(\ref{brY}) might be underestimated. 
In this sense, considering that the main decay channel observed for the $Y(4260)$ state is 
into $J/\psi \:\pi^+\pi^-$, we would naively expect that the branching ratio into this channel could 
 be $\mathcal{B}(Y(4260) \to J/\psi \:\pi^+\pi^-) \sim 1.0$, which  leads to the  
result, ${\mathcal B}_{_Y} <\! 1.8 \times 10^{-6}$.
Therefore, we obtain an interval for the branching fraction 
\begin{equation}\lb{rangeBY}
  3.0 \times 10^{-8} < {\mathcal B}_{_Y} < 1.8 \times 10^{-6},
\end{equation}
which is in agreement with the experimental upper limit reported by Babar Collaboration 
given in Eq.~(\ref{branching}). In general the experimental evaluation of the branching fraction 
takes into account additional factors related to the number of reconstructed events for the 
final state ($J/\psi \:\pi^+\pi^- \:K$), for the reference process ($B \to Y(4260) \:K$), and for the 
respective reconstruction efficiencies. However, since such information has not been provided in 
Ref.~\cite{Aubert:2005zh}, these factors were neglected in the calculation of 
the branching fraction ${\mathcal B}_Y$. Therefore, the comparison of the
result in Eq.~(\ref{rangeBY}) with the 
experimental result could be affected by these differences.

Remember that the above result was obtained by considering the mixing angle in Eq.~(\ref{jmix}) in 
the range $\theta=(53.0\pm0.5)^\circ$. Since there is no new free parameter analysis presented above,
the result shown here strengthens the interpretation of  $Y(4260)$ as a mixture between a 
$J/\psi$ charmonium and a tetraquark state. 

As discussed in Ref.~\cite{Zanetti:2011ju}, it is not simple to determine the charmonium and the tetraquark 
contribution to the state described by the current in Eq.~(\ref{jmix}). From Eq.~(\ref{jmix}) one can 
see that, besides the $\sin\theta$, the $c\bar{c}$ component of the current is multiplied by a dimensional 
parameter, the quark condensate, in order to have the same dimension as the tetraquark part of the 
current. Therefore, it is not clear that only the angle in Eq.~(\ref{jmix}) determines the percentage of 
each component. One possible way to evaluate the importance of each part of the current is 
to analyze what one would get for the production rate with each component, {\it i.e.}, using $\theta=0$ and 
$90^\circ$ in Eq.~(\ref{jmix}). Doing this we get respectively for the pure tetraquark and pure charmonium: 
\begin{eqnarray}
  {\mathcal{B}}(B\to Y_{\mathrm{tetra}}K) &=& (1.25\pm0.23)\times10^{-6}\,, \\
  {\mathcal{B}}(B\to Y_{\bar{c}c}K) &=&(1.14\pm0.20)\times10^{-5}\,.   
\end{eqnarray}
Comparing the results for the pure states with the one for the mixed state in Eq.~(\ref{result}), we can see that 
the branching ratio for the pure tetraquark is one order of magnitude smaller, while for the pure charmonium it is larger.  
From these results we see that the $c\bar{c}$ part of the state plays a very important role in the 
determination of the branching ratio. On the other hand, in the decay $Y\to J/\psi\pi^+\pi^-$, the width  
obtained in our approach for a pure $c\bar{c}$ state is \cite{Dias:2012ek}: 
\begin{equation}\lb{xppcc} 
  \Gamma(Y_{\bar{c}c}\to J/\psi\pi\pi)=0\,, 
\end{equation} 
and, therefore, the tetraquark part of the state is the only one that contributes to this decay, playing 
an essential role in the determination of this decay width. Therefore, although we can not determine 
the percentages of the $c\bar{c}$ and the tetraquark components in the $Y(4260)$, we may say that 
both components are extremely important, and that, in our approach, it is not possible to explain all the 
experimental data about the $Y(4260)$ with only one component. 

\subsection{Remarks on $Y(4220)$}

As discussed above, recently, the BESIII Collaboration has announced  the precise 
measurement of the production cross section for $e^{+} e^- \to J/\psi \:\pi^+ \pi^-$~\cite{Ablikim:2016qzw}. The results show
two resonances with masses around $4220$ MeV and $4320$ MeV. The mass 
of the lower resonance is  consistent with the prediction 
of the $D \bar{D}_1(2420)$ molecular model~\cite{Cleven:2013mka} and is also consistent with the $Y(4260)$ mass ~\cite{Ablikim:2016qzw}. Furthermore, a $Y(4220)$ resonance has also been 
reported by the BESIII collaboration in the cross-section measurements of $e^{+} e^- \to \omega \:\chi_{c0}$
\cite{Ablikim:2014qwy}, $e^{+} e^- \to h_c \:\pi^+ \pi^-$ \cite{BESIII:2016adj}, $e^{+} e^- \to \psi(2S) \:\pi^+\pi^-$ 
\cite{Ablikim:2017oaf}, and $e^{+} e^- \to \pi^+ D^0 D^{\ast \:-}$ \cite{Ablikim:2018vxx}. It is important to 
notice that the cross section of $e^{+} e^- \to \pi^+ D^0 D^{\ast \:-}$ was first measured by the Belle 
Collaboration using ISR events \cite{Pakhlova:2009jv}, with no evidence for the presence of charmonium-like
states in this channel. On the other hand, the results found by the BESIII Collaboration in 
Ref.~\cite{Ablikim:2018vxx}  can be the first experimental evidence for open-charm production associated 
with the $Y$ states. The authors in Ref.~\cite{Ablikim:2014qwy} argue that the parameters found for the $Y(4220)$ are inconsistent with those obtained for the $Y(4260)$ state.  However the authors in Ref.~\cite{Ablikim:2016qzw} found the parameters for the state with mass around 4220 MeV consistent with those for the $Y(4260)$.

In the previous sections we have evaluated a sum rule with a mixed charmonium-tetraquark current to describe the 
$Y(4260)$ state. However, the uncertainty in the mass found in Eq.~(\ref{ymass}) 
shows that our result is also compatible with the $Y(4220)$ mass. Indeed, the quantum numbers are the same for 
the $Y(4220)$ and $Y(4260)$ states. As noted in Ref.~\cite{Ablikim:2018vxx}, the measured cross section 
of $e^{+} e^- \to \pi^+ D^0 D^{\ast \:-}$ at the $Y(4220)$ peak is higher than the sum of the known 
hidden-charm channels. Since no other open-charm production associated with this $Y$ state has
yet been reported, the $\pi^+ D^0 D^{\ast \:-}$ final state might be the dominant decay mode of the
$Y(4220)$. In principle, this conclusion is compatible with the results found for the mixed 
charmonium-tetraquark current, which says that the main decay channel could be into $D$ mesons. 
Therefore, the charmonium-tetraquark current given in Eq.~(\ref{jmix}) can be used to describe either the $Y(4220)$ state or the $Y(4260)$ state, which points in the direction that there is only one state in this mass region. More experiments are needed to settle the question if there are two, as considered in the PDG \cite{pdg}, or just one state, as considered in Ref.~\cite{Guo:2017jvc}, in the $4220\sim4260$ MeV mass region.

\subsection{$Y(4360)$}

Repeating the same kind of analysis that led to the observation of the $Y(4260)$ state, the BaBar collaboration 
has used ISR events to study the channel $e^+e^-\to \psi(2S) \:\pi^+\pi^-$ in the c.m. 
energy range $3.95$ to $5.95$~GeV. Initially, they found  a broad peak at 
a mass around $4.34$~GeV~\cite{Aubert:2007zz}. Soon after, the Belle collaboration not only confirmed the presence of such a state, but also
observed another resonant state in the $\psi(2S) \:\pi^+ \pi^-$ mass spectrum at around 
$4.67$~GeV \cite{Wang:2007ea}. More recently, the BaBar collaboration has announced improvements to their 
analysis  and confirmed the experimental findings from the Belle collaboration of a structure 
near $4.65$~GeV \cite{Lees:2012pv}. Both states were named as $Y(4360)$ and $Y(4660)$, respectively. In order to investigate 
more precisely the properties of these two states, and for a better understanding of their nature, Belle revisited  the process $e^+e^- \to \psi(2S) \:\pi^+\pi^-$ using the ISR technique with a larger data 
sample \cite{Wang:2014hta}. The results improved the experimental measurements of the $Y(4360)$ state \cite{Wang:2014hta}: 
\begin{eqnarray*}
  M_{Y(4360)} ~=~ (4347 \pm 6 \pm 3)~\mbox{MeV} &~~~~\mbox{and}~~~~& 
  \Ga_{Y(4360)} ~=~ (103 \pm 9 \pm 5)~\mbox{MeV}
\end{eqnarray*}
Recent experiments carried out by the BESIII collaboration  confirmed once more the 
existence of such a state and, for the first time, announced its observation in the $J/\psi \:\pi\pi$ mass spectrum \cite{Ablikim:2016qzw}.

More recently the BESIII collaboration measured  the $e^+eˆ-\to \psi(2S)\pi^+\pi^-$ cross section between 4.0 to 4.6 GeV and found two resonances with mass around 4210 MeV and 4380 MeV~\cite{Ablikim:2017oaf}. However, instead of identifying the
higher mass resonance with the $Y(4360)$, they stated that it could be the
same state as that observed by the BESIII collaboration in the process $e^+eˆ-\to \pi^+\pi^-h_c$ \cite{BESIII:2016adj}, the so-called $Y(4390)$. If such identification is confirmed, this measurement could be the first non confirmation for the existence of the $Y(4360)$ state.

\subsubsection{Theoretical explanations for $Y(4360)$}

Some interpretations for this state can be found in Refs.~\cite{Brambilla:2010cs,Nielsen:2009uh,Olsen:2009gi}.  
The absence of open charm decay channels (into $D$ mesons) does not favor the conventional $c\bar{c}$ 
explanation for the $Y(4360)$ state. Although it does not seem to fit the charmonium spectrum \cite{pdg}, 
the authors in Refs.~\cite{Ding:2007rg, Li:2009zu} propose to accommodate it as a conventional $c\bar{c}$ state, 
in particular as a $3^3 D_1$ state. Some possible interpretations are radial excitation of the $Y(4260)$ state 
\cite{Qiao:2007ce}, charmed baryonium \cite{Cotugno:2009ys}, vector hybrid charmonium 
\cite{Kalashnikova:2008qr}, radial excitation of $D^\ast D_1$ molecule \cite{Close:2010wq} and 
$[cq\bar{c}\bar{q}]$ tetraquark state \cite{Albuquerque:2008up}.

\subsubsection{QCDSR calculations for $Y(4360)$}

For this study one considers the lowest dimension tetraquark current, with $J^{PC} = 1^{--}$, and symmetric 
spin distribution $[cq]_{S=0} [\bar{c}\bar{q}]_{S=1} ~+~ [cq]_{S=1} [\bar{c}\bar{q}]_{S=0}$~\cite{Albuquerque:2008up}:
\begin{eqnarray}\label{currY4360}
  j_{\mu} &=& \frac{\epsilon_{abc} \epsilon_{dec}}{\sqrt{2}} \bigg[ 
  (q_a^T C\gamma_5 c_b) (\bar{q}_d \gamma_\mu \gamma_5 C \bar{c}_e^T) + 
  (q_a^T C\gamma_5 \gamma_\mu c_b) (\bar{q}_d \gamma_5 C \bar{c}_e^T) \bigg]\;.
\end{eqnarray}
It is interesting to notice that 
the structure of the current relates the spin of the charm quark with the spin of the light quarks. The sum 
rule calculations are done in Ref.~\cite{Albuquerque:2008up} and the results are shown in Fig.~\ref{figY4360}.
From  Fig.~\ref{figY4360}a, we see that a good OPE convergence is obtained for 
$M^2 \geq 3.2 ~\mbox{GeV}^2$. 
Again the upper limit for $M^2$ is obtained by imposing that the 
pole contribution should be bigger than the continuum contribution. In Fig.~\ref{figY4360}b, we show the 
relative continuum (solid line) and pole (dashed line) contributions, using $\sqrt{s_0} = 4.9$~GeV, from 
where we clearly see that the pole contribution is bigger than the continuum contribution for
$M^2 \leq 3.5 ~\mbox{GeV}^2$. Thus we obtain a certain stability for the $m_Y$ mass, in the allowed 
sum rule window, as can be seen in Fig.~\ref{figY4360}c. Taking into account the variations on $M^2$, 
$s_0$, $\qq[q]$ and $m_c$ we get \cite{Albuquerque:2008up}:
\begin{eqnarray}\label{massY4360}
  m_Y &=& (4.49 \pm 0.11) ~\mbox{GeV},
\end{eqnarray}
which is bigger than the $Y(4360)$ mass, but consistent with it considering the uncertainty. Therefore,
from a sum rule point of view we can describe the $Y(4360)$ state as a $[cq\bar{c}\bar{q}]$ tetraquark 
state. However, it would be better to explore other possibilities for the $Y(4360)$ structure, before 
reaching a definite conclusion about its nature.

\begin{figure}[t]
\begin{tabular}[b]{p{0.45\textwidth}}
    \includegraphics[width=0.43\textwidth]{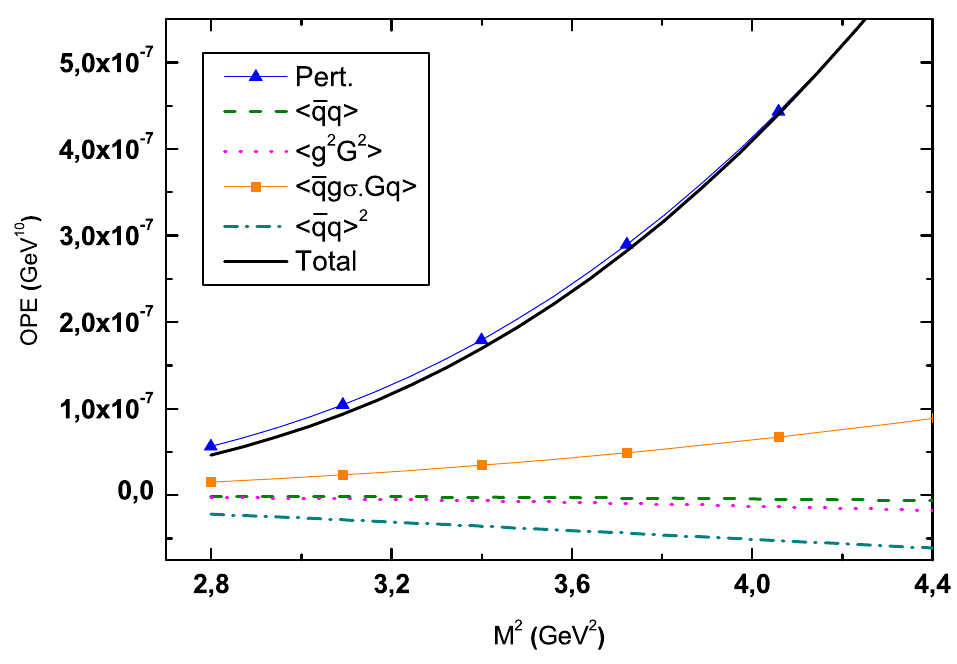}\\ 
    \centerline{(a)}\\ 
    \includegraphics[width=0.43\textwidth]{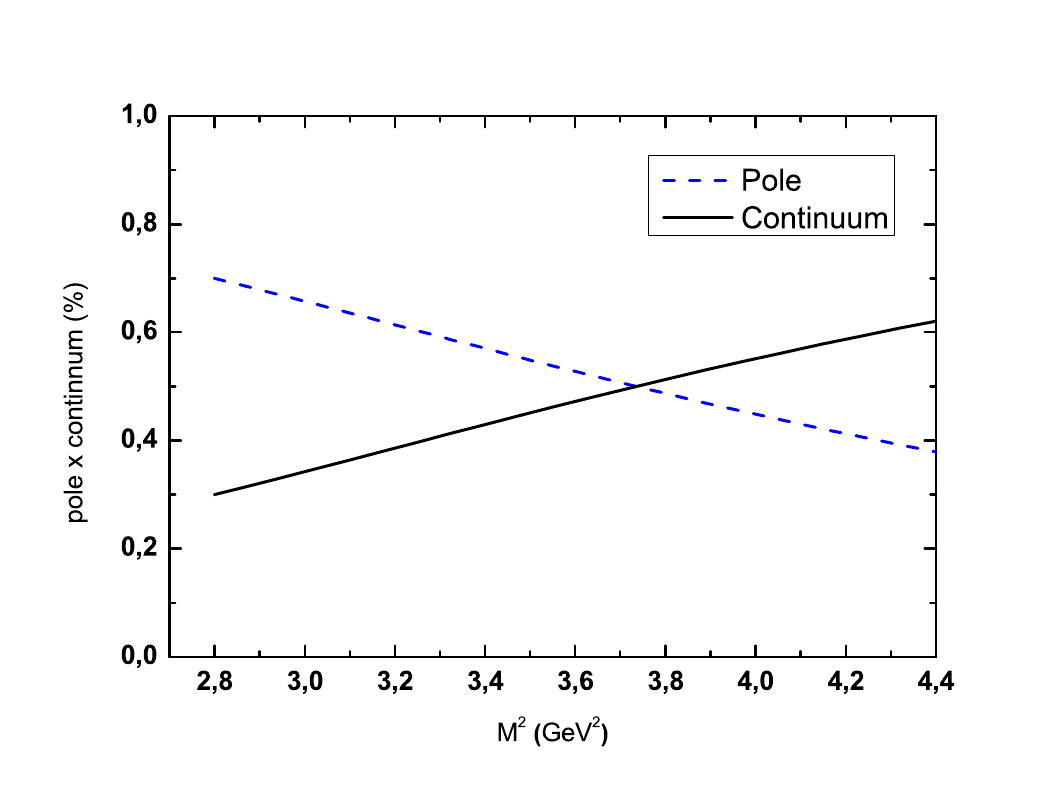}
    \centerline{(b)}\\ 
\end{tabular}
\begin{tabular}[b]{p{0.5\textwidth}}
    \vspace{-7.5cm}
    \includegraphics[width=0.5\textwidth]{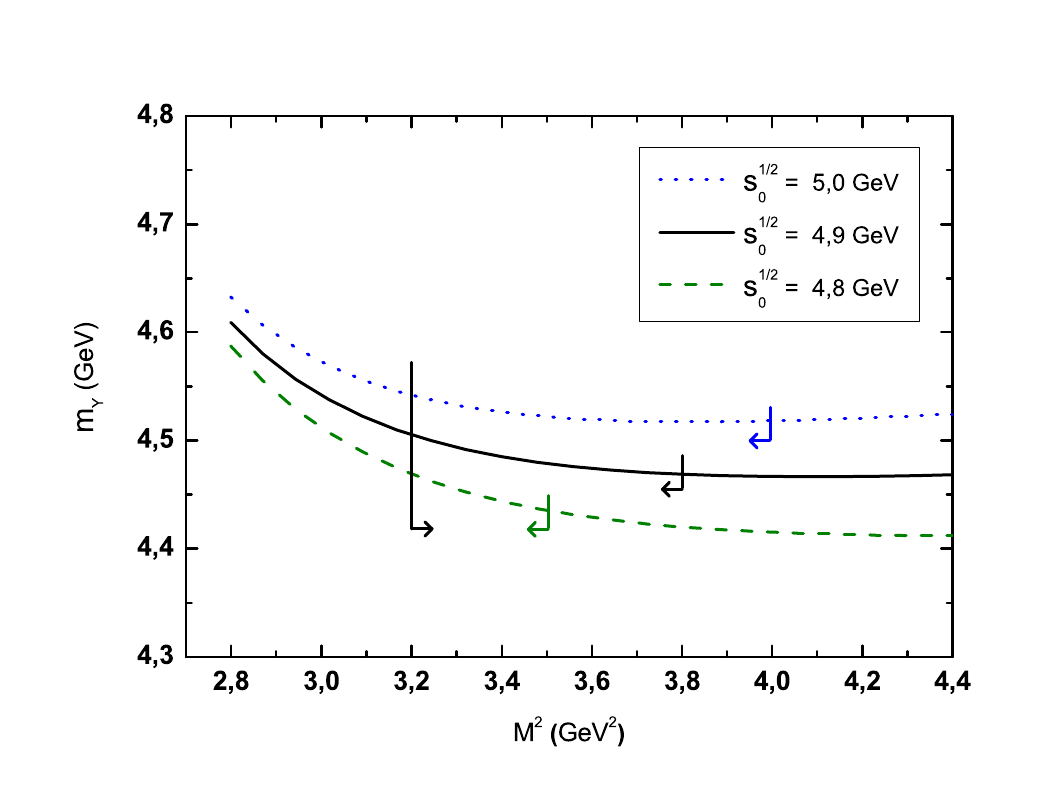}
    \centerline{(c)}\\ 
\end{tabular}
\caption{Sum rule calculation for the $Y(4360)$ state. 
a) The OPE convergence in the region $2.8 \leq M^{2} \leq 4.4$ GeV$^{2}$ for
$\sqrt{s_{0}} =4.9$ GeV. We plot the OPE contributions starting with perturbative 
(line with triangles), quark condensate $\qq[q]$ (dashed line), gluon condensate $\GG$ (dotted line), 
mix condensate $\qGq[q]$ (line with squares), four-quark condensate $\qq[q]^2$ (dot-dashed line) and 
eighth condensate $\qq[q]\qGq[q]$ (solid line). b) The pole contribution (dashed line) and the continuum 
contribution (solid line), for the $\sqrt{s_{0}} =4.90$ GeV. c) The mass as a function of the sum rule 
parameter $M^{2}$ for $\sqrt{s_{0}} = 4.8$~GeV (dashed line), $\sqrt{s_{0}} = 4.9$~GeV (solid line), 
$\sqrt{s_{0}} = 5.0$~GeV (dotted line). The arrows indicate the valid Borel Window.  Figures taken from \cite{Albuquerque:2008up}.}
\label{figY4360} 
\end{figure}  
%

\subsection{Remarks on $Y(4390)$}

Besides the observation of the $Y(4220)$ state, the BESIII Collaboration has also reported another peak 
resonance with a mass around $4390$ MeV, in the processes $e^{+} e^- \to h_c \:\pi^+ \pi^-$ 
\cite{BESIII:2016adj} and $e^{+} e^- \to \psi(2S) \:\pi^+ \pi^-$ \cite{Ablikim:2017oaf}. 
The mass of the $Y(4390)$ state is about 45 MeV greater than that of the $Y(4360)$ state. As pointed out by the   
BESIII Collaboration, the open-charm decay channel for the $Y(4390)$ needs more experimental 
evidence since the resonance parameters for this enhancement are strongly dependent 
on the model assumptions \cite{Ablikim:2018vxx}.

The experimental confirmation of the $Y(4390)$ is very important, since
the tetraquark current in Eq.~(\ref{currY4360}) could be used to explain such a state. Notice that the obtained mass in Eq.~(\ref{massY4360}) is closer to the $Y(4390)$ mass than to the $Y(4360)$ mass. However, results from QCDSR calculations are not so precise to discriminate between these two states. Therefore, 
we need more information from future experiments to determine if there are two, or only one state with mass in the region (4350 -- 4390) MeV. If there are really two states, with more experimental information 
we could be able to understand which of them, $Y(4360)$ or $Y(4390)$, can be
better explained 
as a $[cq\bar{c}\bar{q}]$ tetraquark state, with $J^{PC} = 1^{--}$, and symmetric spin distribution.

\subsection{$Y(4660)$}

The most recent experimental data for the $Y(4660)$ state was reported by BaBar 
and Belle collaborations \cite{Lees:2012pv, Wang:2014hta}. This state was observed in the channel 
$e^+e^-\to \psi(2S) \:\pi^+\pi^-$ with a mass and width given by:
\begin{eqnarray*}
  M_{Y(4660)} ~=~ (4652 \pm 10 \pm 8)~\mbox{MeV} &~~~~\mbox{and}~~~~& 
  \Ga_{Y(4660)} ~=~ (68 \pm 11 \pm 1)~\mbox{MeV}~.
\end{eqnarray*}
A critical information for understanding the structure of the $Y(4660)$ state is whether the pion pair comes 
from a resonance state. Both collaborations state that most of the di-pion candidates are consistent with a 
$f_0(980)$ decay. 

\subsubsection{Theoretical explanations for $Y(4660)$}
From the di-pion invariant mass spectra shown in Ref.~\cite{Wang:2007ea} there is some indication that only 
the $Y(4660)$ has a well-defined intermediate state consistent with $f_0(980)$. Due to this fact and the proximity 
of the mass of the $\psi(2S) \:f_0(980)$ system with the mass of the $Y(4660)$ state, in Ref.~\cite{Guo:2008zg}, 
the $Y(4660)$ was considered as an $\psi(2S) \:f_0(980)$ bound state. The $Y(4660)$ was also suggested to be 
a baryonium state \cite{Qiao:2007ce, Cotugno:2009ys}, a conventional $5^3 S_1$ $c\bar{c}$ state 
\cite{Ding:2007rg}, a hadro-charmonium \cite{Dubynskiy:2008mq} and tetraquark state 
\cite{Ebert:2008kb, Albuquerque:2008up, Zhang:2010mw, Sundu:2018toi}. We still have no evidence for open 
charm decay channels for this state, which does not favor the conventional $c\bar{c}$ explanation for the 
$Y(4660)$ state.

\subsubsection{QCDSR calculations for $Y(4660)$}

In Ref.~\cite{Albuquerque:2008up} the $Y(4660)$ was considered as a $[cs\bar{c}\bar{s}]$ 
tetraquark state, with $J^{PC} = 1^{--}$, and symmetric spin distribution 
$[cs]_{S=0} [\bar{c}\bar{s}]_{S=1} ~+~ [cs]_{S=1} [\bar{c}\bar{s}]_{S=0}$. The  lowest dimension
tetraquark current is given by:
\begin{eqnarray}\lb{cur-y4660}
  j_{\mu} &=& \frac{\epsilon_{abc} \epsilon_{dec}}{\sqrt{2}} \bigg[ 
  (s_a^T C\gamma_5 c_b) (\bar{s}_d \gamma_\mu \gamma_5 C \bar{c}_e^T) + 
  (s_a^T C\gamma_5 \gamma_\mu c_b) (\bar{s}_d \gamma_5 C \bar{c}_e^T) \bigg].
\end{eqnarray}

The QCDSR analysis for this state is the same as the one done for  $Y(4360)$, with
the only difference being the substitution  of  the light quark condensates by those related to the strange quark for the $Y(4660)$. 
The quark content in Eq.~(\ref{cur-y4660}) is consistent with the di-pion invariant mass spectrum \cite{Wang:2007ea}.
\begin{figure}[ht]
\begin{tabular}[b]{p{0.45\textwidth}}
    \includegraphics[width=0.43\textwidth]{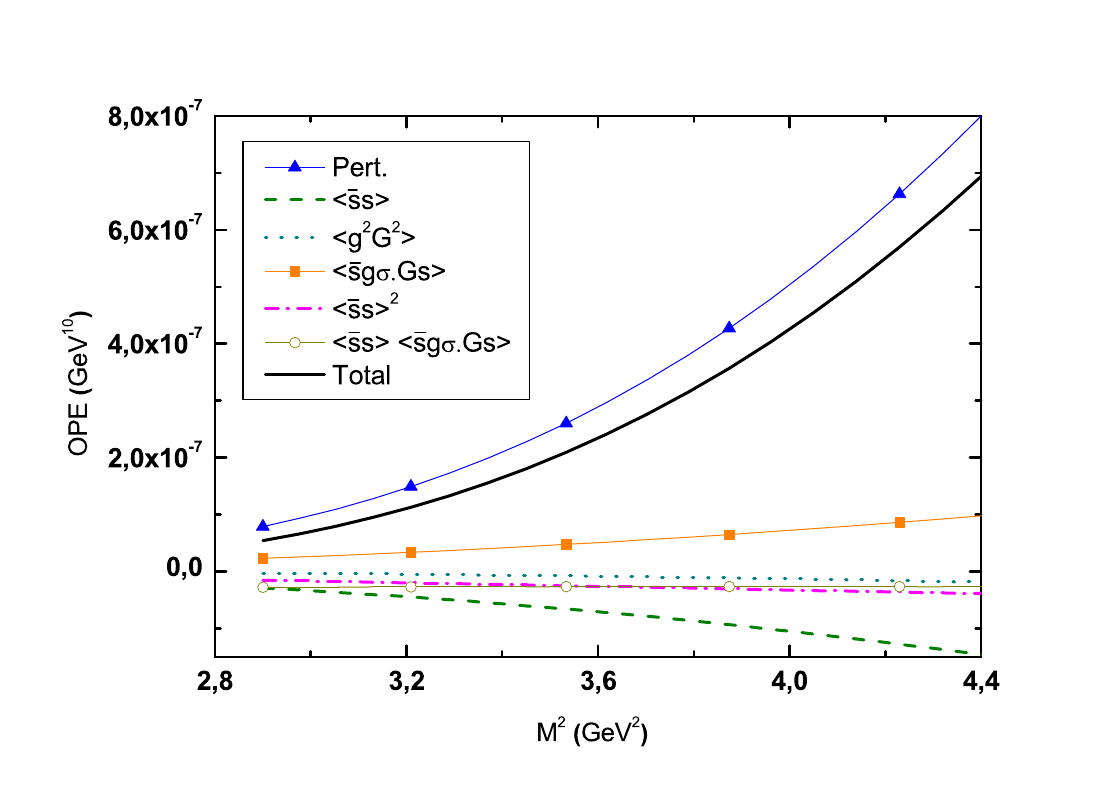}\\ 
    \centerline{(a)}\\ 
    \includegraphics[width=0.43\textwidth]{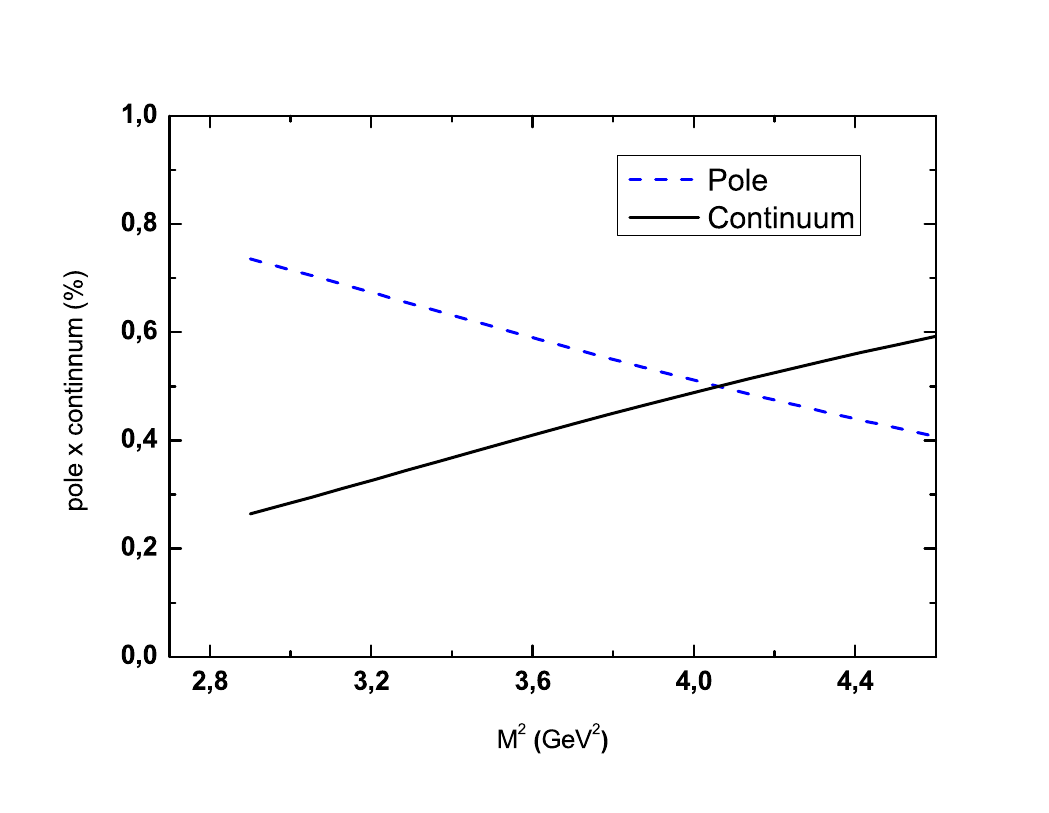}
    \centerline{(b)}\\ 
\end{tabular}
\begin{tabular}[b]{p{0.5\textwidth}}
    \vspace{-7.5cm}
    \includegraphics[width=0.5\textwidth]{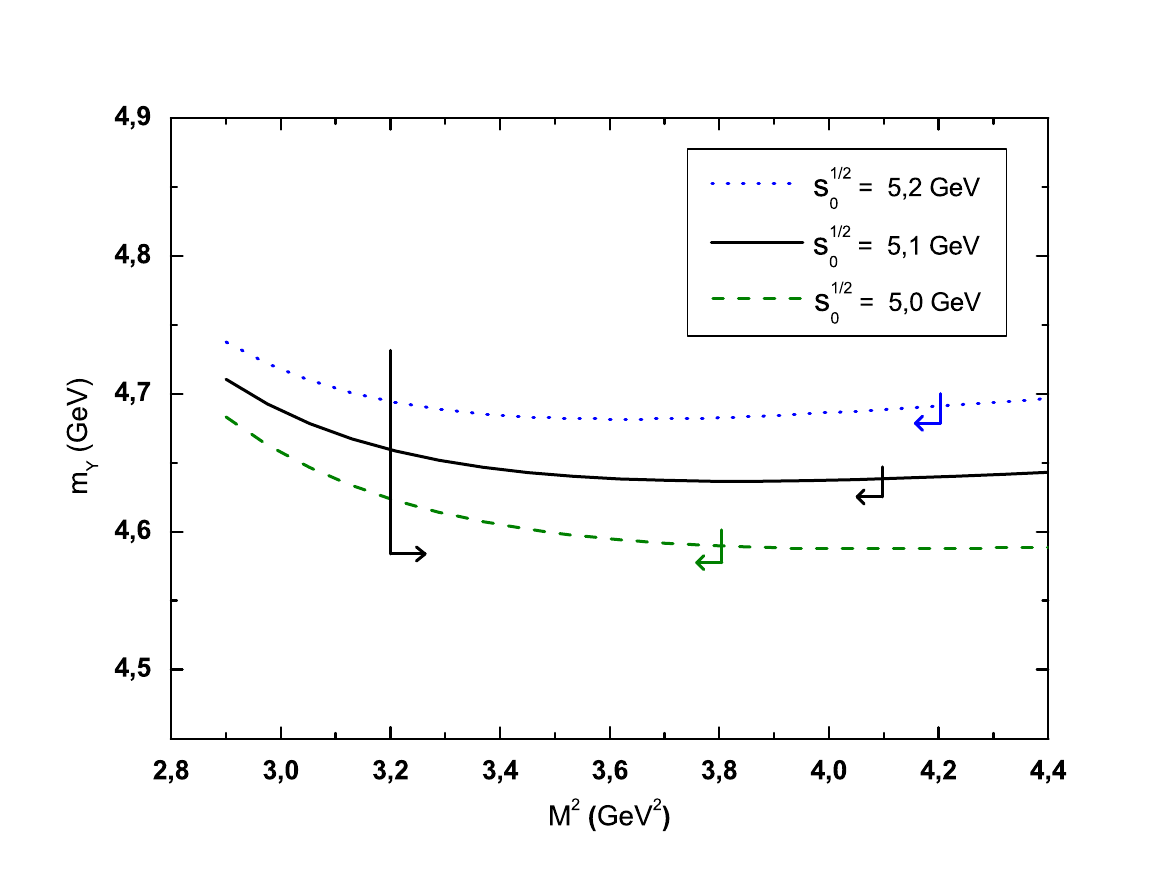}
    \centerline{(c)}\\ 
\end{tabular}
\caption{Sum rule calculation for the $Y(4660)$ state. 
a) The OPE convergence in the region $2.8 \leq M^{2} \leq 4.5$ GeV$^{2}$ for
$\sqrt{s_{0}} =5.1$ GeV. We plot the OPE contributions starting with perturbative 
(line with triangles), quark condensate $\qq[q]$ (dashed line), gluon condensate $\GG$ (dotted line), 
mix condensate $\qGq[q]$ (line with squares), four-quark condensate $\qq[q]^2$ (dot-dashed line) and 
eighth condensate $\qq[q]\qGq[q]$ (solid line). b) The pole contribution (dashed line) and the continuum 
contribution (solid line), for the $\sqrt{s_{0}} =5.1$ GeV. c) The mass as a function of the sum rule 
parameter $M^{2}$ for $\sqrt{s_{0}} = 5.0$~GeV (dashed line), $\sqrt{s_{0}} = 5.1$~GeV (solid line), 
$\sqrt{s_{0}} = 5.2$~GeV (dotted line). The arrows indicate the valid Borel Window. Figures taken from Ref.~\cite{Albuquerque:2008up}.}
\label{figY4660} 
\end{figure}  
The sum rule is evaluated in the Borel range $2.8 \leq M^2 \leq 4.6 ~\mbox{GeV}^2$, and with $s_0$ in the range 
$5.0 \leq \sqrt{s_0} \leq 5.2 ~\mbox{GeV}$.  From Fig.~\ref{figY4660}a, we see that there is a quite good OPE convergence for 
$M^2 \geq 3.2 ~\mbox{GeV}^2$. Therefore, we fix $M_{min}^2 = 3.2 ~\mbox{GeV}^2$. This figure also shows that the dimension-eight condensate contribution is 
very small. In Fig.~\ref{figY4660}b, we show the comparison between the pole and continuum contributions for 
$\sqrt{s_0} = 5.1~\mbox{GeV}$, and we see that for $M^2 \leq 4.05 ~\mbox{GeV}^2$, the pole contribution is bigger 
than the continuum contribution. Therefore, we fix $M^2 = 4.05 ~\mbox{GeV}^2$ as the upper limit of the Borel 
window for $\sqrt{s_0} = 5.1 ~\mbox{GeV}$. In Fig.~\ref{figY4660}c, we show the $m_Y$ mass, for different values of 
$s_0$, as a function of $M^2$, with the upper and lower Borel window limits indicated by the arrows. From this 
figure we see that there is a very good Borel stability for $m_Y$. Taking into account the variations on $M^2$, 
$s_0$, $\qq[s]$, $m_s$ and $m_c$ in the regions mentioned above, we get~\cite{Albuquerque:2008up}:
\begin{eqnarray}
  m_Y &=& (4.65 \pm 0.10) ~\mbox{GeV}
\end{eqnarray}
which is in excellent agreement with the mass of the $Y(4660)$ state. Therefore we conclude that the  
$Y(4660)$ can be described by a diquark-antidiquark $[cs\bar{c}\bar{s}]$ tetraquark state with a spin configuration 
given by scalar and vector diquarks. This quark content is consistent with the di-pion invariant mass spectra 
shown in Ref.~\cite{Wang:2007ea}, which shows that there is some indication that the $Y(4660)$ has a well-defined 
di-pion intermediate state consistent with the formation of $f_0(980)$. 

Another possible interpretation for the $Y(4660)$ could be as a $\psi(2S) f_0(980)$ bound state. The decay channel 
into $\psi(2S) \:\pi^+\pi^-$ favors such a molecular interpretation. However, it is very difficult to work with excited 
states in a QCDSR calculation. For this reason, in Ref.~\cite{Albuquerque:2011ix}
the following current, which couples to a $J/\psi \:f_0(980)$ molecular state with quantum numbers $J^{PC}=1^{--}$, was considered:
\beq
  j_\mu = \left( \bar{c}_i \:\gamma_\mu \:c_i \right)\left( \bar{s}_j\:s_j\right).
  \label{singlet}
\enq
Although there 
are conjectures that the $f_0(980)$ itself could be a tetraquark state \cite{Hooft:2008we}, in 
Ref.~\cite{Matheus:2007ta} it was shown that it is difficult to explain the light scalars as tetraquark states from 
a QCDSR calculation. Therefore, in Ref.~\cite{Albuquerque:2011ix} a simple quark-antiquark current describing the $f_0(980)$ meson was used.
The value used for the continuum 
threshold  is in the range: 
$5.0\leq\sqrt{s_0}\leq 5.2~\GeV$.

With this current it is also possible to get a good OPE convergence  and 
to determine a Borel window with good Borel stability for the mass of
the state, as can be seen in Ref.~\cite{Albuquerque:2011ix}.
Varying the value of the continuum 
threshold in the range $\sqrt{s_0} = 5.1 \pm 0.1 \GeV$, and the other parameters as indicated in 
Table~\ref{QCDParam}, one gets \cite{Albuquerque:2011ix}:
\beq
 m_Y = (4.67\pm0.09)~\GeV ~.
 \label{Msinglet}
\enq
This result is in  an excellent agreement with the mass of the $Y(4660)$ state. The obtained mass is far above the 
$J/\psi~f_0(980)$ threshold and, therefore, such a  state cannot be interpreted as a $J/\psi~f_0(980)$ bound state. 
It is important to remember that the current in Eq~(\ref{singlet}) is written in terms of the currents that couples to the $J/\psi$ and 
$f_0(980)$ mesons, but it also couples with all excited states with the  $J/\psi$ and $f_0(980)$ quantum numbers. 
From the QCDSR analysis presented here we can only guarantee that the mass in Eq.~(\ref{Msinglet}) is the mass 
of the ground state of all the states described by the current in Eq~(\ref{singlet}), but not that their constituents, described 
by the $\bar{c}_i\gamma_\mu c_i$ and $\bar{s}_js_j$ currents, are the ground states of these currents: the $J/\psi$ 
and $f_0(980)$ mesons. 

Therefore, it is possible that the mass obtained in Eq~(\ref{Msinglet}) describes a $\psi(2S) \:f_0(980)$ molecular state, 
since the $\psi(2S)\:f_0(980)$ threshold is at $4.66~\GeV$, compatible with a loosely bound state. The interpretation
of the $Y(4660)$ as a  $\psi(2S)\:f_0(980)$ molecular state was first proposed in Ref.~\cite{Guo:2008zg} and is also in 
agreement with the $Y(4660)$ main decay channel: $Y(4660)\to \psi(2S)~\pi^+\pi^-$. It is also important to mention 
that this result indicates that, from a QCDSR point of view, there is no $J/\psi~f_0(980)$ bound state.

It is straightforward to extend the study presented in the above section to the non-strange case. To do that one only 
has to replace $\qq[s] \to \qq[q]$ and to use $m_s=0$ in the spectral density expressions given in Ref.~\cite{Albuquerque:2011ix}. In Ref.~\cite{Albuquerque:2011ix} it was shown that
the OPE convergence is worse in this case as compared to the $J/\psi f_0(980)$ case. This is
due to the fact that the dimension-3 and dimension-5 condensates do not contribute to the sum rule.
Varying the continuum threshold in the range $5.0 \leq \sqrt{s_0} \leq 5.2 ~\GeV$, and the other QCD parameters in 
Table~\ref{QCDParam} one gets \cite{Albuquerque:2011ix}:
\beq
 m_Y = (4.63\pm0.10)~\GeV ~.
 \label{Msigma}
\enq
The result found for the $J/\psi ~\sigma(600)$ molecular current is also in agreement with the results
obtained with the $J/\psi~ f_0(980)$ current. This kind of findings 
is not uncommon in QCDSR calculations for multiquark states \cite{Matheus:2006xi}. Again, since the masses obtained 
are largely above the $J/\psi ~\sigma(600)$ threshold, we conclude that there is no $J/\psi ~\sigma(600)$ bound state. 
In this case, since the mass obtained is also above the $\psi(2S) \:\sigma(600)$ threshold we can not interpret the 
$Y(4660)$ as a $\psi(2S) \sigma(600)$ molecular state, despite the fact that the obtained mass is in agreement with the $Y(4660)$ mass.

\subsection{Summary for the vector $Y$ states}

The     mixed     charmonium-molecule      current     proposed     in
Ref.~\cite{Dias:2012ek}  within  the  QCDRS framework,  provides  a
consistent description  of various  properties of the  $Y(4260)$ state.
Fixing the  mixing  angle fixed as $\theta=(53.0\pm0.5)^\circ$ it was possible to
describe not only the mass of the $Y(4260)$, but also its decay width into
$J/\psi\pi^+\pi^-$, and the branching fraction for its production in the $B$
meson decay channel $B\to K Y(4260)$. The presented results for $Y(4260)$ are
also consistent for a state with a smaller mass around $4220$ MeV. Therefore,
if future experiments confirm the hypothesis presented in Ref.~\cite{Ablikim:2018vxx}, that the $Y(4260)$ is in fact a superposition of two states with masses
around  $(4220)$ MeV and $(4320)$ MeV, the $Y(4220)$ resonance could be explained as such mixed state.

The $Y(4360)$ state can be explained as a normal $[cq\bar{c}\bar{q}]$ tetraquark, although the obtained mass is slightly bigger than the $Y(4360)$ mass.
Therefore, if the state recently observed by the BESIII collaboration in the process $e^+eˆ-\to \pi^+\pi^-h_c$ \cite{BESIII:2016adj}, the  $Y(4390)$ state, is confirmed, the proposed tetraquark current could describe such state.

In the case of the $Y(4660)$ it was show that it is possible to describe this state with a tetraquark current $[cs\bar{c}\bar{s}]$ with a spin configuration 
given by scalar and vector diquarks, or with a molecular current,
$\left( \bar{c}_i \:\gamma_\mu \:c_i \right)\left( \bar{s}_j\:s_j\right)$, representing a $\psi(2S)f_0(980)$ bound state.

\section{\label{}Isovector states with hidden charm}

The recent  years might be recalled in the future as a period of revolution in the field of hadron physics since several manifestly exotic states have been discovered.  Among them, mesons labelled as ``$Z$'' may be considered as specially interesting, since they have a charmonium like mass but are electrically charged at the same time. A description of such properties  unavoidably requires  (at least) four valence quarks in the wave function.  The first among the series of $Z$'s discovered was $Z^\pm(4430)$. After this discovery, several other charged states with hidden charm  (and their neutral partners) have been reported to exist. We list these states in Table.~\ref{zlist}, denoting all of them by $Z_c$ or simply $Z$, although the names are different in the latest naming scheme of the Particle Data Group (PDG)  \cite{pdg}. Within this scheme, the label $Z_c$ is used  to represent isovector states with hidden charm and with well defined quantum numbers among which the pari!
 ty ($P$) 
 and $C$-parity ($C$) is $+-$. The  label $X$ is used for states with not yet defined quantum numbers. We shall  use $Z$ ($Z_c$), throughout this review, to refer to isovector states with hidden charm, independently of their quantum numbers.
\begin{table*}[h]
  \caption{A list of the currently known isovector mesons with a hidden charm content. The current naming scheme, used by PDG \cite{pdg}, is  included in the table. The quoted year is the year of the first observation in each channel and the quoted charge conjugation, $C$, is for the neutral state in the multiplet.}\setlength{\tabcolsep}{0.23pc}
\begin{center}
\begin{tabular}{lcclcc}
\hline\hline
\rule[10pt]{-1mm}{0mm}
 State & name in PDG  & $I^G(J^{PC})$ & \ \ \ \ Decay channel & 
     Experiment & Year  \\[0.7mm]
\hline
\rule[10pt]{-1mm}{0mm}
$Z_c(3900)$ & $Z_c(3900)$  & $1^+(1^{+-})$&
     $ Z_c^+(3900)\to \pi^+\jpsi$ &
     {BESIII}~\cite{Ablikim:2013mio}, Belle~\cite{Liu:2013dau}, CLEO-c \cite{Xiao:2013iha}]& 2013 \\[1.89mm]
& & &     $ Z_c^+(3900)\to D\bar{D}^{*+}$ &
     {BESIII}~\cite{Ablikim:2013xfr}& 2013 \\[1.89mm]
$Z_c(4020)$ & $X(4020)$&$1^+(?^{?-})$ &
     $ Z_c^+(4020)\to \pi^+h_c$ &
     {BESIII}~\cite{Ablikim:2013wzq} & 2013 \\[1.89mm]
& & &
     $ Z_c^+(4020)\to (D^*\bar{D}^*)^+$ &
     {BESIII}~\cite{Ablikim:2013emm} & 2013  \\[1.89mm]
$Z_1^+(4050)$ & $X(4050)$&  $1^-(?^{?+})$&
     $ Z_1^+(4050)\to \pi^+\chi_{c1}$ &
     {Belle}~\cite{Mizuk:2008me} & 2008 \\[1.89mm]
$Z_c(4055)$& $X(4055)$&$1^+(?^{?-})$&$Z_c^+(4055) \to \pi^+\psi(2S)$& {Belle}~\cite{Wang:2014hta} & 2014 \\[1.89mm]
$Z_c(4100)$& --- &$1^-(0^{++})$ or $(1^{-+})$&$Z_c^-(4100)\to \pi^-\eta_c(1S)$& {LHCb}~\cite{Aaij:2018bla} & 2018 \\[1.89mm]
$Z_c(4200)$ & $Z_c(4200)$& $1^+(1^{+-})$ &
     $Z_c^+(4200)\to \pi^+\jpsi$ &
     {Belle}~\cite{Chilikin:2014bkk}
     & 2014 \\[1.89mm]
$Z_2(4250)$ & $X(4250)$&$1^-(?^{?+})$&
     $ Z_2^+(4250)\to \pi^+\chi_{c1}$ &
     {Belle}~\cite{Mizuk:2008me}  & 2008 \\[1.89mm]
$Z(4430)$ & $Z_c(4430)$&$1^+(1^{+-})$&
     $Z^+(4430)\to \pi^+\psi(2S)$ &
     {Belle}~\cite{Choi:2007wga,Mizuk:2009da,Chilikin:2013tch}, LHCb~\cite{Aaij:2014jqa}
     & 2007 \\[0.7mm]
& & & $Z^+(4430)\to \pi^+ J/\psi$ &
    Belle~\cite{Chilikin:2014bkk} & 2014\\[1.89mm]
\hline\hline
\end{tabular}
\end{center}
\end{table*}

\subsection{$Z^+(4430)$}

The real turning point in the discussion regarding the structure of the new charmonium states was the  observation announced by the Belle Collaboration of a charged state decaying into $\psi'\pi^+$, produced in $B^+\to K\psi'\pi^+$ \cite{Choi:2007wga}. After its discovery, the subsequent  progress on the experimental studies of  $Z^+(4430)$ was astonishing. Soon after the Belle observation, the Babar Collaboration  searched for the $Z^-(4430)$ signature in four decay modes and concluded that there was no significant evidence for the presence of a  relevant signal  in any of these processes \cite{Aubert:2008aa}. However, using the same data sample as in ref.~\cite{Choi:2007wga}, Belle performed a full Dalitz plot analysis  and  confirmed the observation of the $Z^+(4430)$ signal with a 6.4$\sigma$ statistical significance \cite{Mizuk:2009da}. It was only after four years of this disagreement that the controversy came to an end. First the Belle Collaboration confirmed the $Z^+!
 (4430)$ o
 bservation and  determined the  preferred assignment of the quantum numbers to be $J^{P} = 1^{+}$ \cite{Chilikin:2013tch}, and soon after that, the LHCb Collaboration  confirmed both, the $Z^+(4430)$ observation and  the  preferred assignment of the quantum numbers \cite{Aaij:2014jqa}. The LHCb Collaboration also did the first attempt to demonstrate the resonant behavior of the $Z^+(4430)$ state. They have performed a fit in which the Breit-Wigner amplitude was replaced by a combination of independent complex amplitudes at six equally spaced points in the $m_{\psi(2S)\pi}$ range covering the $Z^+(4430)$ peak region  ~\cite{Aaij:2014jqa}. The resulting Argand diagram is consistent with a rapid phase transition at the peak of the amplitude, just as expected for a resonance.   Therefore, the confirmation of the observation of $Z^+(4430)$ by the LHCb Collaboration  with the demonstration of its  resonant behavior can be considered as the first experimental proof of  the existen!
 ce of the
  exotic states. Finally, Belle also searched for this state in the $J/\psi\pi^+$ channel and a 4$\sigma$ signal consistent with $Z^+(4430)$ was found ~\cite{Chilikin:2014bkk}. Comparing the measured product of branching fractions~\cite{Chilikin:2013tch},
\beq
    \mathcal{B}\left({B}^0\to K^+Z^-(4300)\right)\times\mathcal{B}\left(Z^-(4430) \to \psi(2S) \pi^-\right)=\left(6.0^{+1.7+2.5}_{-2.0-1.4}\right)\times10^{-5}\;.
\label{z-}
    \enq        
and~\cite{Chilikin:2014bkk}
\beq
    \mathcal{B}\left(\bar{B}^0\to K^-Z^+(4300)\right)\times\mathcal{B}\left(Z^+(4430) \to J\psi \pi^+\right)=\left(5.4^{+4.0+1.1}_{-1.0-0.6}\right)\times10^{-6}\;,
\enq        
the ratio between the two observed decay channels is estimated to be~\cite{Goerke:2016hxf}:
\beq
    {\mathcal{B}(Z^\pm(4430) \to \psi(2S) \pi^\pm)\over\mathcal{B}(Z^\pm(4430) \to J/\psi \pi^\pm)}=11.1^{+18}_{-8.6}\;.
    \label{ratioZ}
    \enq

The averaged mass and width of this state are $M=(4478\pm17)\MeV$ and $\Gamma=(180\pm31)\MeV$ \cite{pdg}.

\subsubsection{History of theoretical studies of $Z^+(4430)$}

Before the quantum numbers of $Z^+(4430)$ were determined, due to the proximity of its mass with the $\bar{D}^*D_1$ threshold, Rosner \cite{Rosner:2007mu} suggested that it was an $S$-wave threshold effect,  Bugg considered it to be a cusp in the $\bar{D}^*D_1$ channel \cite{Bugg:2007vp}, while in Ref.~\cite{Dubynskiy:2008mq} it was considered as a hadro-charmonium. Other authors considered it to be  a natural candidate for a loosely bound $S$-wave $\bar{D}^*D_1$ molecular state with quantum numbers  $J^P=0^-$ \cite{Lee:2007gs,Bracco:2008jj,Meng:2007fu,Liu:2007bf,Liu:2008xz,Ding:2008mp,Zhang:2009vs}. There exists also a quenched lattice QCD calculation which found attractive interaction in the $\bar{D}^*D_1$ system in the $J^P=0^-$ channel \cite{Meng:2009qt}. The authors of ref.~\cite{Meng:2009qt} also find positive scattering length. Based on these findings, they conclude that although the interaction between the two charmed mesons is attractive in this channel, it is unlik!
 ely that 
 they can form a genuine bound state right below the threshold. 

The first theoretical proposition for the correct quantum numbers of $Z^+(4430)$ was made by Maiani, Polosa and Riquer in Ref.~\cite{Maiani:2007wz}, where this state was interpreted as the first radial excitation of the tetraquark  supermultiplet to which  $X(3872)$ belongs \cite{Maiani:2004vq}. In Ref.~\cite{Maiani:2004vq} it was conjectured that  $X(3872)$ must have a charged partner $X^+$  with $J^{PC}= 1^{+-}$ with a similar mass. In Ref.~\cite{Maiani:2007wz} it was pointed out that since the mass difference
\beq
M_{Z^+ (4430)} -  M_{X^+(3872)}\sim 560 \, \MeV
\label{difz}
\enq
is close to the mass difference $M_{\Psi(2S)}-M_{\Psi(1S)}=590~\MeV$, the  $Z^+(4430)$  may well be the first radial excitation of  $X^+$.
Later, after the discovered of the $Z_c^+(3900)$ state, Maiani {\it et al.} identified it as the predicted $X^+$~\cite{Faccini:2013lda}, and $Z^+(4430)$ as the first radial excitation of $Z_c^+(3900)$~\cite{Maiani:2014aja}. In Refs.~\cite{Goerke:2016hxf,Ebert:2008kb,Navarra:2011xa,Patel:2014vua,Hadizadeh:2015cvx,Wang:2014vha,Agaev:2017tzv} $Z^+(4430)$ was also interpreted as the first radial excitation $(2S)$ of a charged diquark-antidiquark $[cu][\bar{c}\bar{d}]$ tetraquark state. However, in Ref.~\cite{Deng:2015lca}, using a color flux-tube model with a four-body confinement potential, the authors could not explain $Z^+(4430)$ as a tetraquark state.

After Belle and LHCb established the quantum numbers of $Z^+(4430)$ to be $J^P=1^+$, it was clear that the S-wave $\bar{D}^*D_1$ molecular state assignment of $Z^+(4430)$ is not possible. Following the latter findings, in Ref.~\cite{Ma:2014zua} the authors proposed three possible molecular configurations for $Z^+(4430)$:  a P-wave excitation of the $D_1\bar{D}^*$ or $D_2\bar{D}^*$ molecule; an S-wave molecule composed of a $D$ or $D^*$ meson and a D-wave vector $D$ meson; (3) a cousin of the molecular state of $Z_c(3900)$  composed of a $D$ or $D^*$ meson and their radial excitations. In Refs.~\cite{Barnes:2014csa,Liu:2014eka} $Z^+(4430)$ is interpreted as a $D \bar D^*(2S)$ state, and the authors of Ref.~\cite{Liu:2014eka} showed that the ratio  measured by Belle~\cite{Chilikin:2014bkk} in Eq.~(\ref{ratioZ}) can be explained by considering $Z^+(4430)$ as a $D \bar D^*(2S)$ molecular state. The ratio in  Eq.~(\ref{ratioZ}) can also be explained by considering $Z^+(4430)$ as !
 the first
  radial excitation $(2S)$ of a tetraquark state \cite{Agaev:2017tzv}.

From the  above discussions we can conclude that the two possible explanations for the $Z^+(4430)$ structure are: 1) the first radial excitation $(2S)$ of the charged diquark-antidiquark $[cu][\bar{c}\bar{d}]$ $Z_c^+(3900)$ tetraquark state;
2) a $D \bar D^*(2S)$ molecular state. It is very interesting to notice that in the Supersymmetric Light Front Holographic QCD \cite{Dosch:2015nwa,Dosch:2015bca,Brodsky:2016yod,Dosch:2016zdv,Nielsen:2018uyn} $Z_c^+(3900)$ and $Z^+(4430)$ are also identified as tetraquark states, but as, respectively, the first and second radial excitation of the state  $\chi_{c1}(3510)$, considered as the tetraquark superpartner of the $\Xi_{cc}$ baryon \cite{Nielsen:2018ytt}, as can be seen in Fig. \ref{charm}.

\vspace{10pt}
\begin{figure}[h] 
\centerline{\includegraphics[width=8.0cm]{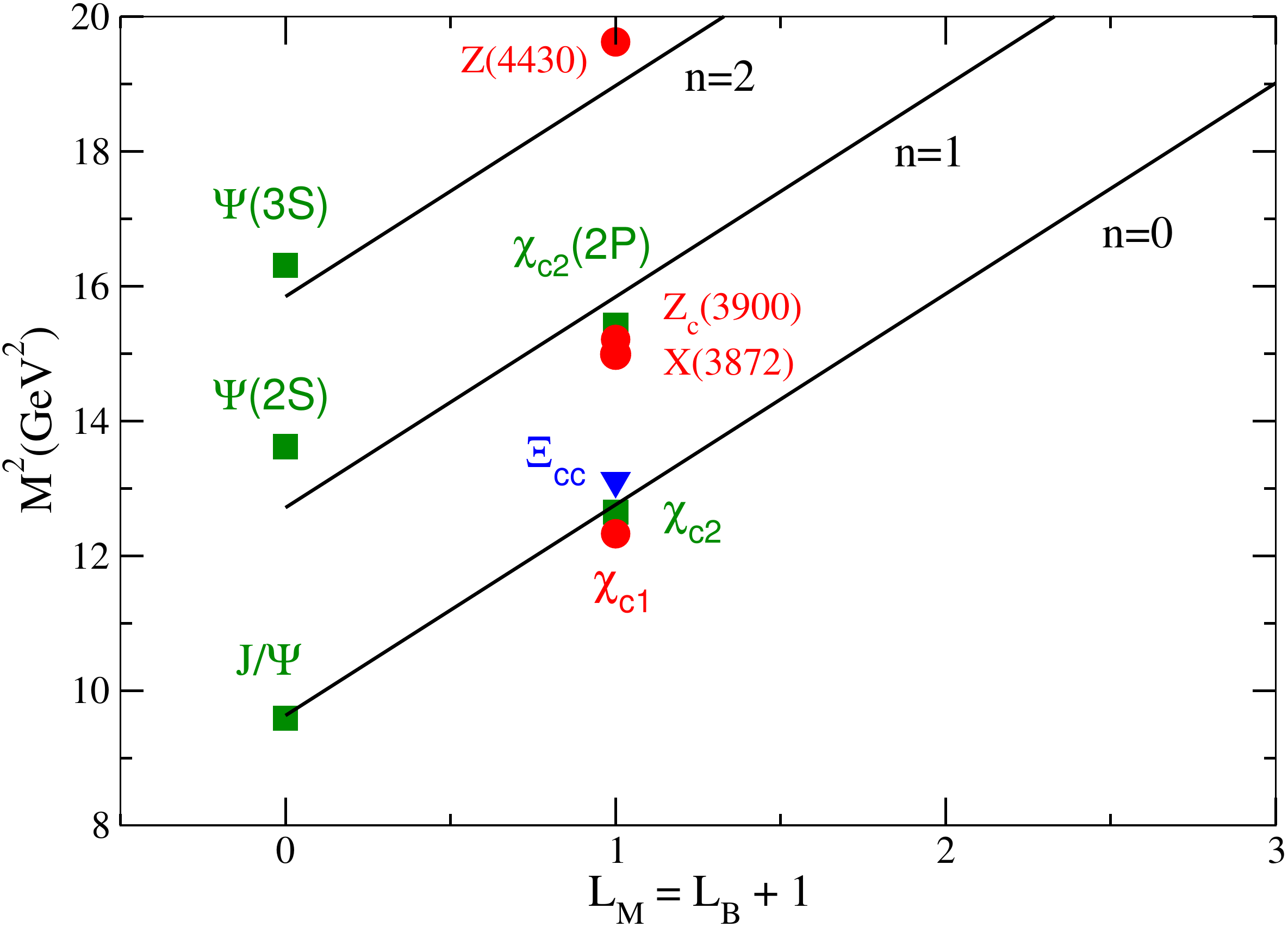}}
\caption{ \label{charm}  Double charm mesons (shown as green squares) baryons (shown as  blue triangles) and tetraquarks (shown as red circles) with different values of angular momentum $L$ and radial excitation $n$. The solid lines are the trajectories fit from \cite{Nielsen:2018ytt}. Hadron masses are taken from PDG~\cite{pdg}.}
\end{figure}

\subsection{$Z_c^+(3900)$}

In 2013, the BESIII Collaboration announced the observation of a charged
charmonium-like state, called $Z_c(3900)$, in the $J/\psi \pi^\pm$ invariant mass distribution of the $e^+e^-\to Y(4260) \to J/\psi \pi^+\pi^-$ process \cite{Ablikim:2013mio}. This structure, was also observed, at the same time, by the  Belle collaboration \cite{Liu:2013dau} and was confirmed by the  authors of Ref. ~\cite{Xiao:2013iha} using the CLEO-c data. 
From these three experiments, assuming the orbital angular momentum between  $J/\psi$ and $\pi$ to be zero, the quantum number of $Z_c(3900)$ was argued to be $I^GJ^P=1^+1^+$. The confirmation of the spin and parity of $Z_c^+(3900)$ as $J^P=1^+$ was done in Ref.~\cite{Collaboration:2017njt} with a statistical significance larger than 7$\sigma$. In Ref.~\cite{Xiao:2013iha} an evidence of the existence of the neutral state $Z_c(3900)^0$ decaying into $\pi^0J/\psi$ was also brought forward and in Ref.~\cite{Ablikim:2015tbp} $Z_c(3900)^0$ was observed by BESIII in the $e^+e^-\to \pi^0Z_c(3900)^0 \to \pi^0\pi^0 J/\psi$ process. The mass and decay width of $Z_c(3900)$ from all these different experiments are consistent with each other. The averaged mass and width are: $M=(3886.6\pm2.4)$ MeV and $\Gamma=(28.2\pm2.6)$ MeV \cite{pdg}.

Soon after the $Z_c^+(3900)$ observation,  the BESIII Collaboration announced the finding of three other charged states: $Z_c(3885)$ \cite{Ablikim:2013xfr,Ablikim:2015swa}, $Z_c(4020)$ \cite{Ablikim:2013wzq} and $Z_c(4025)$ \cite{Ablikim:2013emm}. All these structures were observed in the process $e^+e^-\to Y(4260)\to \pi^-Z_c^+$. 

The $Z_c^+(3885)$ state was found in the process $e^+e^-\to Y(4260) \to(D\bar{D}^*)^\pm\pi^\mp$  with mass $M=(3881.7\pm1.6\pm1.6)$ MeV and width $\Gamma=(26.5\pm1.7\pm2.1)$ MeV \cite{Ablikim:2015swa}. Since its measured mass was slightly lower than that of $Z_c(3900)$ measured in the $J/\psi\pi$ channel by BESIII : $M=(3899.0\pm3.6\pm4.9)$ MeV \cite{Ablikim:2013mio} and by Belle: $M=(3894.5\pm6.6\pm4.5)$ MeV \cite{Liu:2013dau}, BESIII called it $Z_c^+(3885)$. However, the measured mass and width of $Z_c(3885)$ are consistent with those of the $Z_c(3900)$ state obtained by Xiao \textit{et al.}: $M=(3886\pm4\pm2)$ MeV \cite{Xiao:2013iha}.
BESIII also reported the finding of  a neutral state $Z_c(3885)^0$ in the $e^+e^-\to (D\bar D^*)^0\pi^0$ process \cite{Ablikim:2015gda}. The analysis on the angular distribution of the $\pi Z_c(3885)$ system performed by BESIII supports the $J^P=1^+$ assignment  \cite{Ablikim:2013xfr}. With the same spin-parity and similar mass and width, $Z_c(3900)$ and $Z_c(3885)$ are probably the same state and they are considered as the same state in PDG \cite{pdg}. Under this assumption, the ratio of the partial decay width of these two decay modes is \cite{Esposito:2016noz}
\begin{eqnarray}
\frac{\Gamma(Z_c(3900)\to D\bar D^*)}{\Gamma(Z_c(3900)\to
  J/\psi\pi)}=7.7\pm1.3\pm2.8\, .
\label{ratiozc}
\end{eqnarray}

\subsubsection{Theoretical explanations for $Z_c^+(3900)$}

As discussed in the case of $Z^+(4430)$, the two possible explanations for the $Z_c^+(3900)$ structure are:  a charged diquark-antidiquark $[cu][\bar{c}\bar{d}]$   state, or a $D \bar D^*$ molecular state. Concerning the molecular configuration, there are many calculations that could not accommodate $Z_c^+(3900)$ as a $J^{P}=1^+$ $D \bar D^*$ molecule \cite{Zhao:2014gqa,He:2014nya} including lattice QCD calculations \cite{Prelovsek:2013xba,Prelovsek:2014swa,Chen:2014afa}. However, in  Refs.~\cite{Aceti:2014uea,Karliner:2015ina,He:2015mja,Wang:2013daa,Wilbring:2013cha,Dong:2013iqa,Ke:2013gia,Gutsche:2014zda,Esposito:2014hsa,Chen:2015igx,Gong:2016hlt,Ke:2016owt} the authors did find a  $D \bar D^*$ molecular state compatible with  $Z_c^+(3900)$. There are also some QCDSR calculations for the  $Z_c^+(3900)$, done using molecular  type of interpolating currents, for which the obtained mass agrees with the experimental values within
errors \cite{Wang:2013daa,Chen:2015ata,Zhang:2013aoa,Cui:2013yva}. However, it is important to remember that, although the interpolating current is of molecular type, the current is local and, therefore, the four quarks in the current have the same space-time position as in the case of tetraquark currents~\cite{Narison:2010pd}. In the case of tetraquark configuration, many calculations, using different approaches, found a positive signal \cite{Dias:2013xfa,Faccini:2013lda,Wang:2013vex,Deng:2014gqa,Agaev:2016dev}.

In all calculations, which found a positive signal, the mass of the   $Z_c(3900)$  is relatively easily  reproduced. However, the  $Z_c(3900)$ decay width represents a challenge to theorists. While its mass is very close to the $X(3872)$ mass, it has a much larger decay width. Indeed, while the $Z_c(3900)$ decay width is in the range $30$ MeV, the $X(3872)$ width is smaller than  $ 1.2 $ MeV.  This difference can be attributed to  the fact that $X(3872)$ may contain a significant $|c \bar{c} \rangle$ component~\cite{Matheus:2009vq}, which is absent in $Z_c(3900)$.  As pointed out in Ref.~\cite{Wang:2013cya}, this would also 
explain why $Z_c$ has not been observed in $B$ decays. 

According to the experimental observations, $Z_c^+(3900)$ decays into $J/\psi \,  \pi^+$ with a relatively large decay width. If $Z_c$ is a real $D^* - \bar{D}$  molecular state, its decay into  $J/\psi \,  \pi^+$  must involve the exchange of charmed mesons.   When the distance between $D^*$ and the  $ \bar{D}$ is large, as expected for a   $D^* - \bar{D}$  molecular state, it becomes more difficult to exchange mesons, since the exchange  of heavy mesons is a short range process. In Ref.~\cite{Mahajan:2013qja}  it was shown that, in order to reproduce the measured $Z_c(3900)$ width, the effective radius must be $\langle r_{eff} \rangle \simeq 0.4$ fm. This size scale is small and represents a challenge to the molecular picture. In Ref.~\cite{Wilbring:2013cha}, the  $Z_c^+(3900)$ was also treated as a charged $D^* - \bar{D}$ molecule in which the 
interaction between the charm mesons is described by a pionless effective 
field theory. Introducing electromagnetic interactions through the minimal 
substitution in this theory, the authors explored its  electromagnetic structure, arriving at the conclusion that its charge radius is of the order of  $\langle r^2 \rangle \simeq 0.11$ fm$^2$. Taking this radius as a measure of the spatial size of the state, we conclude that it is more compact than $J/\psi$, for which $\langle r^2 \rangle \simeq 0.16$ fm$^2$.  In Ref.~\cite{Dias:2013xfa} the 
combined results of Refs.~\cite{Mahajan:2013qja} and \cite{Wilbring:2013cha} were taken  as an indication that $Z_c$ is a compact object, which may be better understood as a quark cluster, such as a tetraquark.

\subsubsection{QCDSR calculations for the $Z_c^+(3900)$ width}

In Ref.~\cite{Dias:2013xfa} $Z_c^+(3900)$ was interpreted as the isospin 1 partner of $X(3872)$, as in Ref.~\cite{Faccini:2013lda}. The quantum numbers for the neutral state in the isospin multiplet are $I^G(J^{PC})=1^+(1^{+-})$ and,
therefore, the interpolating field for $Z_c^+(3900)$, considered as a tetraquark state, is given by:
\beq
j_\alpha={i\epsilon_{abc}\epsilon_{dec}\over\sqrt{2}}[(u_a^TC\gamma_5c_b)
(\bar{d}_d\gamma_\alpha C\bar{c}_e^T)-(u_a^TC\gamma_\alpha c_b)
(\bar{d}_d\gamma_5C\bar{c}_e^T)]\;,
\label{field}
\enq
where $a,~b,~c,~...$ are color indices, and $C$ is the charge conjugation
matrix. Considering SU(2)  symmetry, the mass obtained in QCDSR for the
$Z_c$ state is exactly the same as that obtained  for $X(3872)$ \cite{Matheus:2006xi,Narison:2010pd}. As discussed above, QCDSR calculations for the $Z_c$ state using a $\bar{D}D^*$ molecular type interpolating current  lead to similar results for the mass of the state \cite{Wang:2013daa,Chen:2015ata,Zhang:2013aoa,Cui:2013yva}. These calculations only confirm the results presented in Refs.~\cite{Matheus:2006xi,Narison:2010pd}.

The evaluation of the $Z_c^+(3900) \to  J/\psi \, \pi^+$ decay width in the 
QCDSR approach  is based on the three-point function:
\beq
\Pi_{\mu\nu\al}(p,\pli,q)=\int d^4x~ d^4y ~e^{i\pli.x}~e^{iq.y}~
\Pi_{\mu\nu\al}(x,y),
\lb{3po}
\enq
with $\Pi_{\mu\nu\al}(x,y)=\lag 0 |T[j_\mu^{\psi}(x)j_{5\nu}^{\pi}(y)
j_\alpha^\dagger(0)]|0\rag$,
where $p=\pli+q$ and $j_\mu^{\psi},~j_{5\nu}^{\pi}$ are the interpolating fields for $J/\psi$ and $\pi$ respectively.

The phenomenological side of the sum rule is obtained by   inserting intermediate states for $Z_c$, $J/\psi$ and $\pi$ into Eq.(\ref{3po}). One arrives at \cite{Dias:2013xfa}:
\beqa
\Pi_{\mu\nu\al}^{(phen)} (p,\pli,q)={\lambda_{Z_c} m_{\psi}f_{\psi}F_{\pi}~
g_{Z_c\psi \pi}(q^2)q_\nu
\over(p^2-m_{Z_c}^2)({\pli}^2-m_{\psi}^2)(q^2-m_\pi^2)}
~\left(-g_{\mu\lambda}+{\pli_\mu \pli_\lambda\over m_{\psi}^2}\right)
\left(-g_\alpha^\lambda+{p_\alpha p^\lambda\over m_{Z_c}^2}\right)
+\cdots\;,
\lb{phen}
\enqa
where the dots stand for the contribution of all possible excited states. 
The form factor, $g_{Z_c\psi \pi}(q^2)$, is defined as the generalization 
of the on-mass-shell matrix element, $\lag J/\psi \,  \pi \,| \, Z_c\rag$,
for an off-shell pion: 
\beq
\lag J/\psi(\pli) \pi(q)|Z_c(p)\rag=g_{Z_c\psi \pi}(q^2)
\varepsilon^*_\lambda(\pli)\varepsilon^\lambda(p),
\label{coup}
\enq
where
$\varepsilon_\alpha(p),~\varepsilon_\mu(\pli)$ are  the polarization
vectors of the $Z_c$ and $J/\psi$ mesons  respectively.

In Ref.~\cite{Dias:2013xfa} the coupling constant, $g_{Z_c\psi \pi}$, was evaluated directly by considering a sum rule at the pion-pole \cite{Bracco:2011pg}, 
valid only at $Q^2=0$, as suggested in \cite{Reinders:1984sr} for the pion-nucleon coupling constant. It consists of neglecting the pion mass in the denominator
of Eq.~(\ref{phen}) and working at $q^2=0$. In the OPE side only terms 
proportional to $1/q^2$ will contribute to the sum rule. Therefore, up to 
dimension five the only diagrams that contribute are the quark condensate
and the mixed condensate. Besides,  only the diagrams with non-trivial color structure, which are called color-connected (CC) diagrams, as shown  in Fig.~\ref{fig1}, were considered on the OPE side.  Possible permutations (not shown) of the diagram in Fig.~\ref{fig1} also contribute.

\begin{figure}[h] 
\centerline{\epsfig{figure=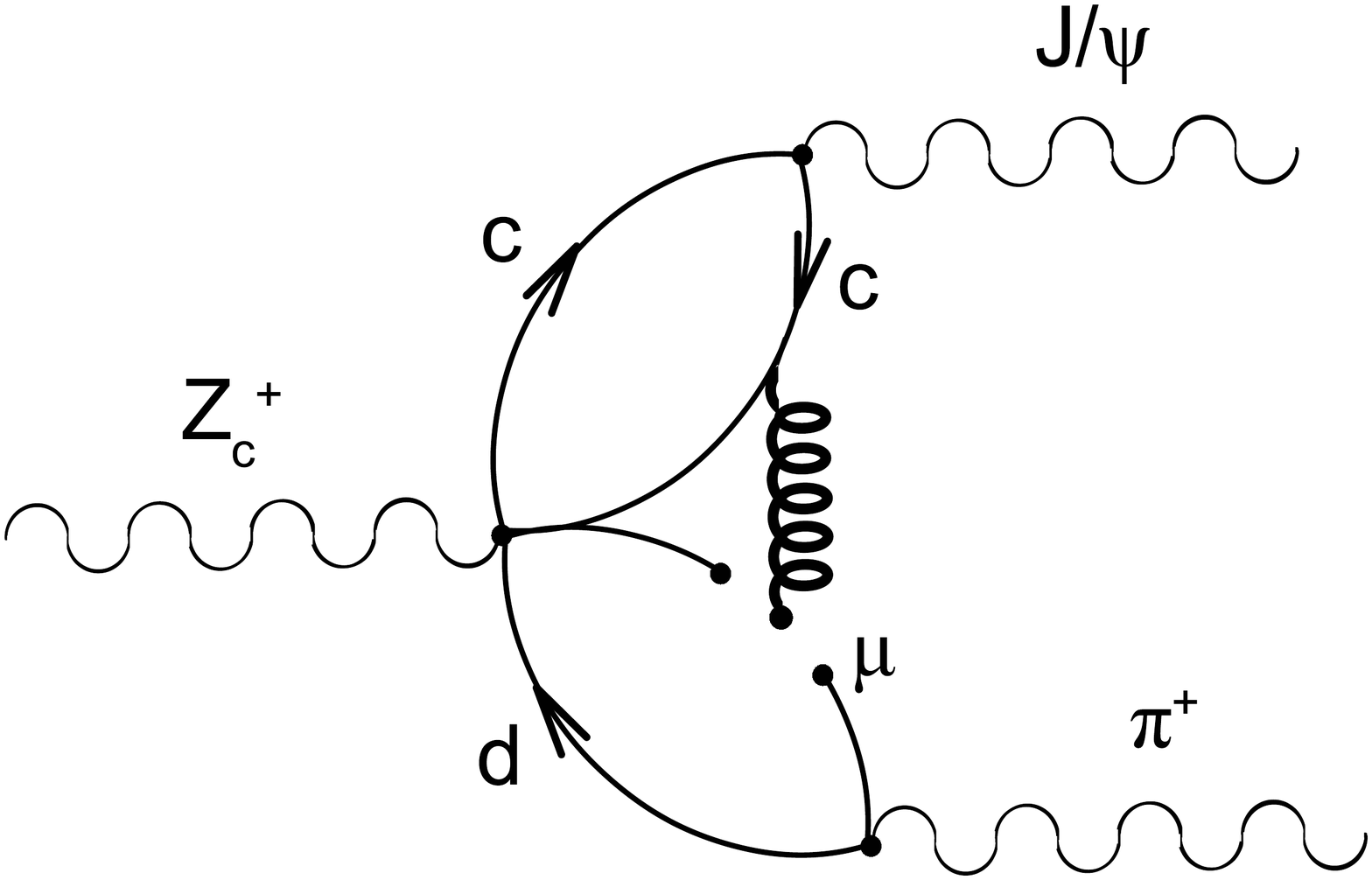,height=40mm}}
\caption{CC diagram which contributes to the OPE side of the sum rule. From Ref.~\cite{Dias:2013xfa}.}
\label{fig1}
\end{figure} 

The diagram in Fig. \ref{fig1} contributes only to the structures $q_\nu g_{\mu\alpha}$
and $q_\nu\pli_\mu\pli_\alpha$ appearing on the phenomenological side. On the OPE side  we choose to work with the $q_\nu\pli_\mu\pli_\alpha$ structure,  since 
structures with more momenta are supposed to give better results. We obtain:
\beq
\Pi^{(OPE)}= {\mix\over12\sqrt{2}\pi^2}{1\over q^2}
\int_0^1 d\alpha{\alpha(1-\al)\over m_c^2-\al(1-\al){\pli}^2}.
\label{ope}
\enq
Making a single Borel transformation to both $P^2={P^\prime}^2\rightarrow M^2$,
we finally get the sum rule:
\beqa
A\left(e^{-m_\psi^2/M^2}-e^{-m_{Z_c}^2/M^2}\right)+B~e^{-s_0/M^2}
={\mix\over12\sqrt{2}\pi^2}
\int_0^1 d\alpha \,  e^{- m_c^2\over \al(1-\al)M^2},
\label{sr}
\enqa
where $s_0$ is the continuum threshold parameter for $Z_c$,
\beq
A={g_{Z_c\psi \pi}\lambda_{Z_c} f_{\psi}F_{\pi}~(m_{Z_c}^2+m_\psi^2)
\over 2m_{Z_c}^2m_{\psi}(m_{Z_c}^2-m_{\psi}^2)},
\label{a}
\enq
and the parameter $B$ is introduced to take into account pole-continuum transitions (see Sec.~\ref{s-3p}).
To determine the coupling constant $g_{Z_c\psi \pi}$ we fit the QCDSR results 
with the analytical expression on the left-hand side (LHS) of Eq.(\ref{sr}),
and find: $A=1.46\times10^{-4}~\GeV^5$ and $B=-8.44\times10^{-4}~\GeV^5$. 
Using the definition of $A$ in Eq.(\ref{a}), the value obtained for the
coupling constant is~\cite{Dias:2013xfa}
\beq
g_{Z_c\psi \pi}=(3.89\pm0.56)~\GeV,
\label{coupling}
\enq
which is in excellent agreement with the estimate made in \cite{Faccini:2013lda}, based on  dimensional arguments. The corresponding decay width is~\cite{Dias:2013xfa}:
\beq
\Gamma(Z_c^+(3900)\to J/\psi\pi^+)=(29.1\pm8.2)~\MeV.
\label{width}
\enq

In Ref.~\cite{Dias:2013xfa} the three-point QCDSR was also used to evaluate 
the coupling constants at the vertices $Z_c^+(3900)\eta_c\rho^+$,   $Z_c^+(3900) D^+ \bar{D^*}^0 $ and   $Z_c^+(3900) \bar{D^0} {D^*}^+ $.  In all cases  only CC diagrams were considered.

To illustrate let us consider the  $Z_c^+(3900)\eta_c\rho^+$ case. In this case the phenomenological side is
\beqa
&&\Pi_{\mu\al}^{(phen)} (p,\pli,q)={-i\lambda_{Z_c} m_{\rho}f_{\rho}f_{\eta_c}
m^2_{\eta_c}~g_{Z_c\eta_c \rho}(q^2)
\over2m_c(p^2-m_{Z_c}^2)({\pli}^2-m_{\eta_c}^2)(q^2-m_\rho^2)}
\left(-g_{\mu\lambda}+{q_\mu q_\lambda\over m_{\rho}^2}\right)
\left(-g_\alpha^\lambda+{p_\alpha p^\lambda\over m_{Z_c}^2}\right)
+\cdots.
\lb{phen2}
\enqa

On the OPE side, for the $\pli_\alpha q_\mu$ structure we have:
\beq
\Pi^{(OPE)}= {-i m_c\mix\over48\sqrt{2}\pi^2}{1\over q^2}
\int_0^1 d\alpha{1\over m_c^2-\al(1-\al){\pli}^2}.
\label{ope2}
\enq
Isolating the $q_\al\pli_\mu$ structure in Eq.~(\ref{phen2}) and 
making a single Borel transformation on  both $P^2={P^\prime}^2\rightarrow M^2$,
we finally get the sum rule:
\beqa
C\left(e^{-m_{\eta_c}^2/M^2}-e^{-m_{Z_c}^2/M^2}\right)+D~e^{-s_0/M^2}=
{Q^2+m_\rho^2\over Q^2}{m_c\mix\over48\sqrt{2}\pi^2}
\int_0^1 d\alpha {e^{- m_c^2\over \al(1-\al)M^2}\over\al(1-\al)},
\label{sr2}
\enqa
with $Q^2=-q^2$ and
\beq
C={g_{Z_c\eta_c \rho}(Q^2)\lambda_{Z_c} m_\rho f_{\rho}f_{\eta_c}m_{\eta_c}^2
\over2m_c m_{Z_c}^2(m_{Z_c}^2-m_{\eta_c}^2)}.
\label{c}
\enq

One can use Eq.~(\ref{sr2}) and its derivative with respect to $M^2$ to 
eliminate $D$ from Eq.~(\ref{sr2}) and to isolate $ g_{Z_c\eta_c \rho}(Q^2)$. 
In Sec.~\ref{exproc} the QCDSR results for the $ g_{Z_c\eta_c \rho}(Q^2)$ form
factor are  illustrated together with the extrapolation procedure used to extract
the coupling constant. The QCDSR results are shown in Figs.~\ref{gexample} and
\ref{gQ2rho}. The squares in Fig. \ref{gQ2rho} show the $Q^2$ dependence 
of $g_{Z_c\eta_c \rho}(Q^2)$, obtained for $M^2=5.0$ GeV$^2$. For 
other values of the Borel mass, in  the range  
$4.0 \leq M^2 \leq 10.0$ GeV$^2$, the results are equivalent.
Using the parametrization in Eq.~(\ref{exprho}), also shown in Fig. \ref{gQ2rho}
as a  line, the  coupling constant is obtained as~\cite{Dias:2013xfa}:

\beq
g_{Z_c\eta_c\rho}=(4.85 \pm 0.81)~~\mbox{GeV}.
\label{couprho}
\enq

 The couplings, with the respective decay widths, for all studied decays in~\cite{Dias:2013xfa} are given in Table~\ref{ta3}. A total width of $\Gamma = (63.0 \pm 18.1)$ MeV was found for the  $Z_c(3900)$,  
in good agreement with the two experimental values:
$\Gamma=(46\pm 22)$ MeV from BESIII \cite{Ablikim:2013mio}, and
$\Gamma=(63\pm35)$ MeV from BELLE \cite{Liu:2013dau}.

\begin{table}[h]\begin{center}
\caption{Coupling constants and decay widths in  different channels}
{\begin{tabular} {ccc}  
\hline\hline
Vertex & coupling constant (GeV) & decay width (MeV)\\
\hline
$Z_c^+(3900)J/\psi\pi^+$ & $3.89\pm0.56$ & $29.1\pm8.2$ \\
 $Z_c^+(3900)\eta_c\rho^+$ & $4.85 \pm 0.81$ & $27.5\pm8.5$ \\
 $Z_c^+(3900) D^+ \bar{D^*}^0 $ & $2.5 \pm 0.3$ & $3.2 \pm 0.7$ \\
 $Z_c^+(3900) \bar{D^0} {D^*}^+ $ & $2.5 \pm 0.3$ & $3.2 \pm 0.7$ \\ 
\hline\hline
\end{tabular}\label{ta3}}
  \end{center}
\end{table}

From the results in Table~\ref{ta3} it is possible to evaluate the ratio
\begin{equation}
{\Gamma(Z_c(3900) \to D\bar{D}^*)\over
\Gamma(Z_c(3900) \to\pi J/\psi)}=0.22 \pm 0.12, 
\label{ratio-te}
\end{equation}
which is not compatible with the result in Eq.~(\ref{ratiozc}).
In Ref.~\cite{Agaev:2016dev}, also with a QCDSR calculation, it was obtained that just the decays  $Z_c^+(3900)J/\psi\pi^+$ and $Z_c^+(3900)\eta_c\rho^+$ lead to a total $Z_c$ decay width of $\Gamma = (65.7 \pm 10.3)$ MeV. This means that the ratio in Eq.~(\ref{ratio-te}) would be even smaller than the quoted one. From the results presented in Ref.~\cite{Faccini:2013lda}, using the diquark model, one arrives at
\begin{equation}
{\Gamma(Z_c(3900) \to D\bar{D}^*)\over
\Gamma(Z_c(3900) \to\pi J/\psi)}\sim0.14. 
\end{equation}
Therefore, from these results  one should conclude that the states $Z_c(3885)$ \cite{Ablikim:2013xfr} and $Z_c(3900)$ \cite{Ablikim:2013mio} are not the same.

\subsection{$Z_c^+(4020)$ (former $Z_c^+(4025)$)}\label{Zc4025}

The charged $Z^\pm_c(4025)$ state was first discovered by the BESIII collaboration~\cite{Ablikim:2013wzq} in the $\pi^\pm h_c(1P)$ invariant mass distribution of the process $e^+ e^-\to \pi^+\pi^- h_c(1P)$,
and it was also observed by the same collaboration~\cite{Ablikim:2013emm} as a peak in the $(D^*\bar D^*)^\pm$ invariant mass distribution from the reaction $e^+e^-\to (D^*\bar D^*)^\pm\pi^\mp$. The BESIII collaboration also found a signal for a neutral $Z^0_c(4020)$ in the corresponding $\pi^0h_c(1P)$ and $(D^*\bar D^*)^0$ invariant masses of the reactions $e^+ e^-\to \pi^0\pi^0 h_c(1P), (D^*\bar D^*)^0\pi^0$~\cite{Ablikim:2014dxl,Ablikim:2015vvn}. Production rates and mass values support putting together the manifestation of these charged and neutral states as an evidence for the existence of an isospin 1 particle with mass $(4024.1\pm 1.9)$ MeV and width $(13\pm 5)$ MeV, which nowadays is named as $X(4020)$. With the exception of parity, the quantum numbers of this isospin 1 state have not been determined, but all the above mentioned experimental analysis assume $s$-wave production and thus the quantum number assignment $I^G(J^{PC})=1^+(1^{+-})$.

\subsubsection{Theoretical explanations for $Z_c^+(4020)$}

The proximity of the mass of $Z_c(4020)$  to the $D^* \bar D^*$ threshold  motivated the association of a molecule-like structure to it.  In such a case, it should be possible to determine the origin of  $Z_c(4020)$ purely from the dynamics of the open charm vector mesons.  The idea of forming molecular $D^*\bar D^*$ resonances close to the threshold is not new and was foreseen in Refs.~\cite{DeRujula:1976zlg,Tornqvist:1993vu,Tornqvist:1993ng,Dubynskiy:2006sg,Voloshin:1976ap} much before the experimental findings in Refs.~\cite{Ablikim:2013wzq,Ablikim:2013emm,Ablikim:2014dxl,Ablikim:2015vvn}.

 The same motivation, that is, the proximity of the $(D^*\bar D^*)^\pm$ threshold (4017 MeV) to the resonance mass, has encouraged the search of a state arising from the dynamics involved in the $D^*\bar D^*$ system and coupled channels within other formalisms too. Within the context of effective field theories, in Ref.~\cite{Guo:2013sya}, using arguments of heavy quark symmetry and solving the Lippmann-Schwinger equation for the $D^*\bar D^*$ system, a (virtual) state with mass in the range 3950-4017 MeV is found with isospin 1 and $J^{PC}=1^{+-}$. In Ref.~\cite{Molina:2009ct}, using effective field theories based on the local hidden symmetry, it was found that a $I^G(J^{PC})=1^-(2^{++})$ state arises from the coupled channel dynamics involved in the $D^*\bar D^*$, $J/\psi \rho$ coupled system with a mass around 3920 MeV. This calculation was updated in Ref.~\cite{Aceti:2014kja} and the binding energy was reduced, finding the state with a mass around 3990 MeV and width arou!
 nd 160 Me
 V. 

Using the model of Refs.~\cite{Molina:2009ct,Aceti:2014kja}, in Ref.~\cite{Torres:2013lka}, the cross section for the $e^+e^-\to (D^*\bar D^*)^\pm\pi^\mp$ reaction was calculated and it was shown that the experimental result is compatible with a $J^P=1^+$ state with mass around 4025 MeV and small width, $\sim30$ MeV, and, thus, in line with the result in Ref.~\cite{Ablikim:2013emm}. But it was also found to be consistent with the existence of a $J^P=2^+$ state below the $D^*\bar D^*$ threshold with mass around 3990 MeV and width of 160 MeV, as found in Ref.~\cite{Aceti:2014kja}. Note that even if the state found in Refs.~\cite{Molina:2009ct,Aceti:2014kja} is below the $D^*\bar D^*$ threshold, because of its large width, when considering the phase space for the $e^+e^-\to (D^*\bar D^*)^\pm\pi^\mp$ process, a narrow peak will be produced slightly above the threshold in the $D^*\bar D^*$ invariant mass distribution of the reaction (for more details see Ref.~\cite{Torres:2013lka!
 }). In th
 is sense, the signal observed in Ref.~\cite{Ablikim:2013emm} could be due to the presence of a state below the $D^*\bar D^*$ threshold. There are some facts in favor of such an interpretation. For example, if, as assumed in Ref.~\cite{Ablikim:2013emm}, the signal found in the $D^*\bar D^*$ invariant mass distribution of $e^+e^-\to (D^*\bar D^*)^\pm\pi^\mp$ would correspond to a $J^P=1^+$ state produced in $s$-wave, such a state could easily decay to $J/\psi \pi$, implying that a clear signal around 4025 MeV (which in this case is quite far from the $J/\psi\pi$ threshold) should be seen in the $J/\psi \pi$ invariant mass distribution of the reaction $e^+e^-\to \pi^+\pi^- J/\psi$. This reaction was precisely studied by the BESIII collaboration~\cite{Ablikim:2013mio} and they found a $Z_c$ state around 3900 MeV, but no signal for $Z_c(4025)$ was found.

Additionally, if $Z_c(4025)$ should be interpreted as a state generated from the $D^*\bar D^*$ interaction it should be expected to be below the threshold and not above: the dominant contribution to the wave function of the $Z_c$ state would  come from the $D^*\bar D^*$ component. In effective field theories, the lowest order amplitude describing the $D^*\bar D^*$ interaction has a weak energy dependence, and as shown in Ref.~\cite{YamagataSekihara:2010pj}, in a single channel system with an energy-independent amplitude, when used as kernel in the Bethe-Salpeter equation to determine the scattering $T$-matrix, a state below the threshold, and not above, is produced. In this sense, if a $Z_c$ state could be generated from the $D^*\bar D^*$ system, it is expected to be below the $D^*\bar D^*$ threshold. This fact, together with the results of Refs.~\cite{Molina:2009ct,Aceti:2014kja,Torres:2013lka} could hint to a possible misinterpretation of the signal observed in Ref.~\cite{!
 Ablikim:2
 013emm}.

In Refs.~\cite{Deng:2014gqa, Deng:2015lca}, a study of tetraquark states, within a diquark-antidiquark configuration, was performed in the context of a color flux tube with a multibody confinement potential and it was found that the nearest state to $Z^+_c (4025)$ obtained with the model was the one with quantum numbers $J^P=2^+$. 
However, several works support the $1^{+-}$ quantum numbers too. Using a framework of non-relativistic quark model and Cornell-type potentials, a $Q\bar q$-$\bar Qq$ molecular-like four quark state with $J^{PC}=1^{+-}$ and mass around 4036 MeV was found~\cite{Patel:2014vua}, among others with a similar mass but other quantum numbers,  and identified with $Z_c(4025)$.

Alternative explanations for the experimental findings have also been brought forward. In Refs.~\cite{Swanson:2014tra,Swanson:2015bsa} a simple model inspired by effective field theories is used to show that the signal observed in the experiments could be an artifact arising from a coupled channel cusp effect. Studies based on a finite volume seem not to support the existence of a $Z_c$ state arising from the $D^*\bar D^*$ interaction~\cite{Prelovsek:2014swa,Chen:2015jwa}, considering interpolating currents with $I^G (J^{PC}) = 1^+ (1^{+-})$. Though it is unclear if these findings imply the absence of $Z_c$'s with $I^G (J^{PC}) = 1^+ (1^{+-})$ or if they are due to the limitations of the simulations (for example, the use of unphysical masses for the quarks $u$ and $d$).

\subsubsection{QCDSR calculations for $Z_c^+(4020)$}
\lb{spin-pro}

To study the $Z_c^+(4020)$ within the scheme of the QCD sum rules, the authors in Ref.~\cite{Khemchandani:2013iwa} considered the current:
\begin{equation}\label{j}
j_{\mu \nu} (x) = \left[ \bar{c}_a(x) \gamma_\mu u_a(x)\right]\left[\bar{d}_b(x) \gamma_\nu c_b(x)\right].
\end{equation}
With this current, which does not have a defined spin-parity, 
 one can construct the two-point  correlation function
\begin{equation}
\Pi_{\mu \nu \alpha \beta} (q^2) = i \int  d^4x e^{iqx} \langle 0 \mid T \left[ j_{\mu \nu} (x) j^\dagger_{\alpha \beta} (0) \right] \mid 0 \rangle, \label{Pi}
\end{equation}
from which it is necessary to extract the contributions to different well defined spin-parity combinations, in order to interpret the results. Considering an effective field theory point of view, bound states of two mesons are expected to be formed, most likely, when the constituent mesons interact at low energies. At such energies, the interactions can be well described by taking the lowest relative angular momentum in the system, in other words by considering interactions in the $s$-partial wave. In such a picture, the $D^* \bar D^*$ system can have total spin 0, 1 or 2. In the QCD sum rules approach, as described in the previous sections, the correlation function is written in terms of quark propagators and is calculated within the OPE. To do such calculations for a defined total spin 0, 1 or 2, we need to project Eq.~(\ref{Pi}) on a particular spin, which can be done using  the spin projectors given in Ref.~\cite{Torres:2013saa}.  These projectors were obtained in Ref.~\!
 cite{Torr
 es:2013saa} to study the $D^* \rho$ system, inspired by a study of the same system within an effective field approach done in Ref.~\cite{Molina:2009eb}. They are:
\begin{align} 
\mathcal{P}^{(0)}&=\frac{1}{3}\Delta^{\mu\nu}\Delta^{\alpha\beta},\nonumber\\
\mathcal{P}^{(1)}&=\frac{1}{2}\left(\Delta^{\mu\alpha}\Delta^{\nu\beta}-\Delta^{\mu\beta}\Delta^{\nu\alpha}\right),\label{proj}\\
\mathcal{P}^{(2)}&=\frac{1}{2}\left(\Delta^{\mu\alpha}\Delta^{\nu\beta}+\Delta^{\mu\beta}\Delta^{\nu\alpha}\right)-\frac{1}{3}\Delta^{\mu\nu}\Delta^{\alpha\beta},\nonumber
\end{align}
where $\Delta_{\mu\nu}$ is defined in terms of the metric tensor, $g^{\mu\nu}$, and the four momentum $q$ of the correlation function as
 \begin{align}
\Delta_{\mu\nu}\equiv -g_{\mu\nu}+\frac{q_\mu q_\nu}{q^2}.\label{Delta}
 \end{align}
 The spin projected spectral density was calculated by going in the OPE series up to dimension six in Ref.~\cite{Khemchandani:2013iwa}:
\begin{equation}
\rho^{S}_{\textrm{OPE}}=\rho^{S}_{\textrm{pert}}+\rho^{S}_{\langle\bar q q\rangle}+\rho^{S}_{\langle g^2 G^2\rangle}+\rho^{S}_{\langle\bar q g\sigma G q\rangle}+\rho^{S}_{{\langle\bar q q\rangle}^2}+\rho^{S}_{\langle g^3 G^3\rangle},\label{rho}
\end{equation}
where the superscript $S$ denotes the spin of the states in the spectrum. The spectral density on the phenomenological side is written as a sum of a narrow, sharp state and a smooth continuum 
\begin{align}
\rho^{S}_{\textrm{phenom}}(s)={\lambda^2_{S}}\delta(s-m_{S}^2)+\rho^{S}_{\textrm{cont}}(s), \label{rhopheno}
\end{align}
where $s = q^2$ is the squared four-momentum flowing in the correlation function,  $\lambda_{S}$ is the coupling of the current to the state we are looking for and $m_{S}$ denotes its mass.
As explained in the previous sections, the continuum is assumed to be separated from the ground state by about 500 MeV, although this separation energy is considered as an unknown parameter and is varied in the calculations to estimate errors in the results. As the standard procedure, a Borel transform of both the OPE and phenomenological sides is taken, which introduces a dependence of the results on the Borel mass. The criteria to establish a range of Borel mass, within which the results are considered to be reliable, are: (1) to ensure the dominance of the first term of Eq.~(\ref{rhopheno}), called as the pole term, over the continuum contribution, in agreement with the ansatz chosen for the phenomenological description of the spectral density and (2) to have a converging OPE series. It can be seen from the left panel of Fig.~\ref{fig1:4025}, 
\begin{figure}[h!]
\centering
\includegraphics[width=0.45\textwidth]{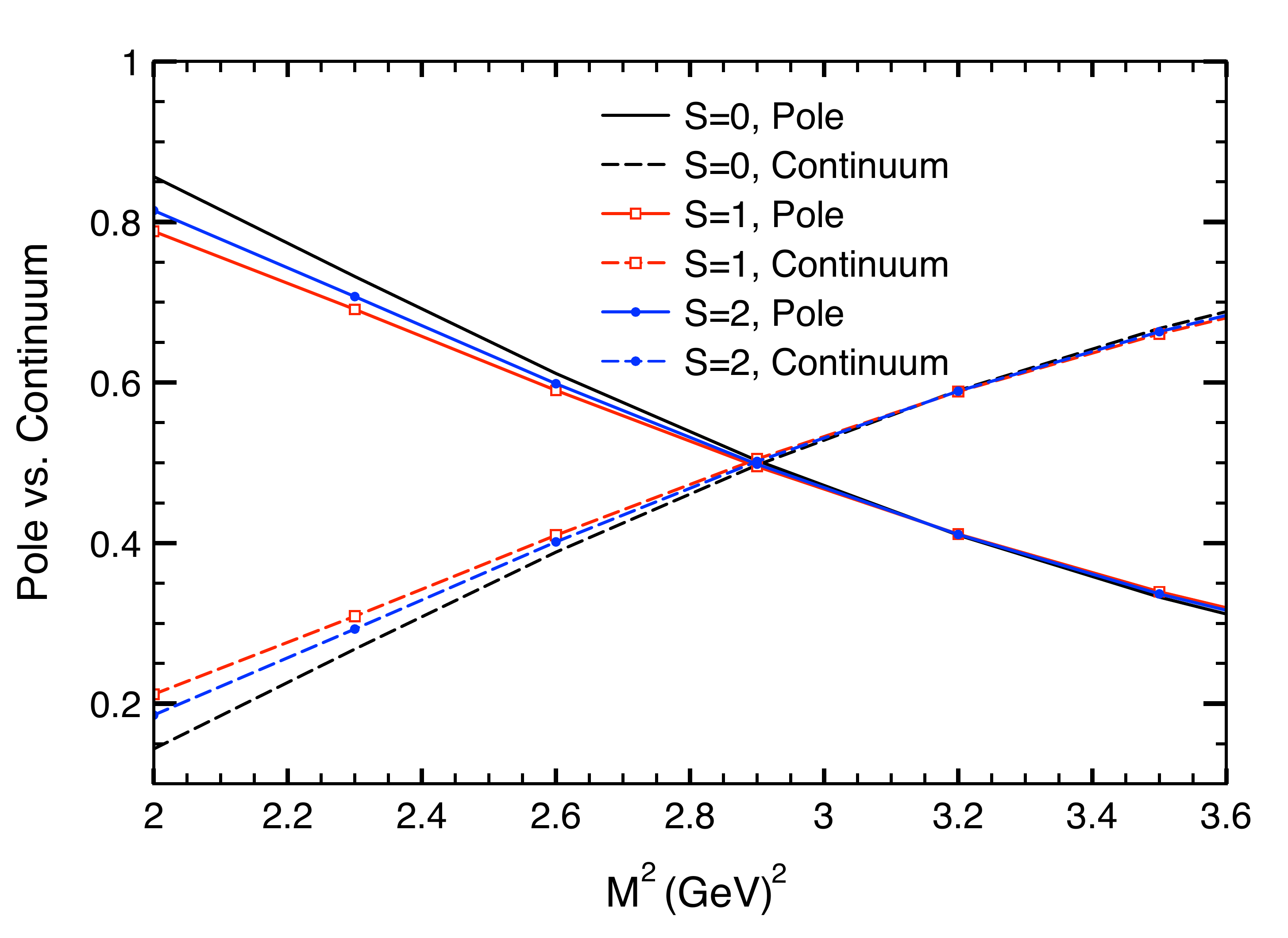}
\includegraphics[width=0.45\textwidth]{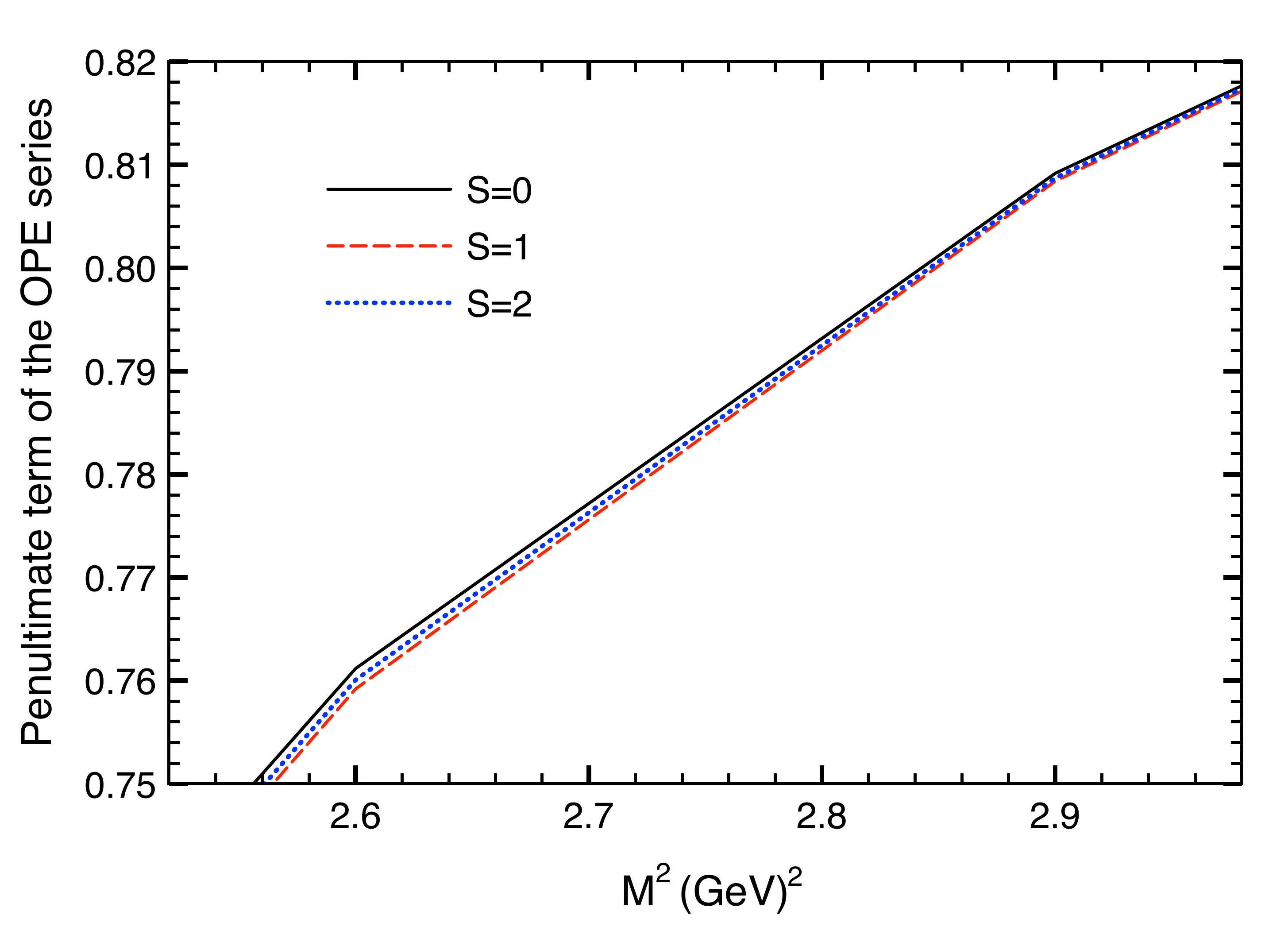}
\caption{Figures depicting the pole dominance on the phenomenological side and the convergence of the OPE series for the correlation function of Eq.~(\ref{Pi}).}\label{fig1:4025}
\end{figure}
for a fixed value of $\sqrt{s_0} = 4.55$ GeV, that the pole term dominates for all the three spin configurations until the squared Borel Mass value of about 2.9 GeV$^2$. The right panel of the same figure shows that 
the  contribution of the penultimate term reduces from $\sim25\%$ to $\sim20\%$ in the range of Borel mass 2.56 GeV$^2\leq M^2 \leq 2.9$ GeV$^2$. This range can be considered as a Borel window within which the results are reliable and thus we can look at the mass values obtained in this range for the three cases of spin. As can be seen from Fig.~\ref{fig2:4025}, 
\begin{figure}[h!]
\centering
\includegraphics[width=0.45\textwidth]{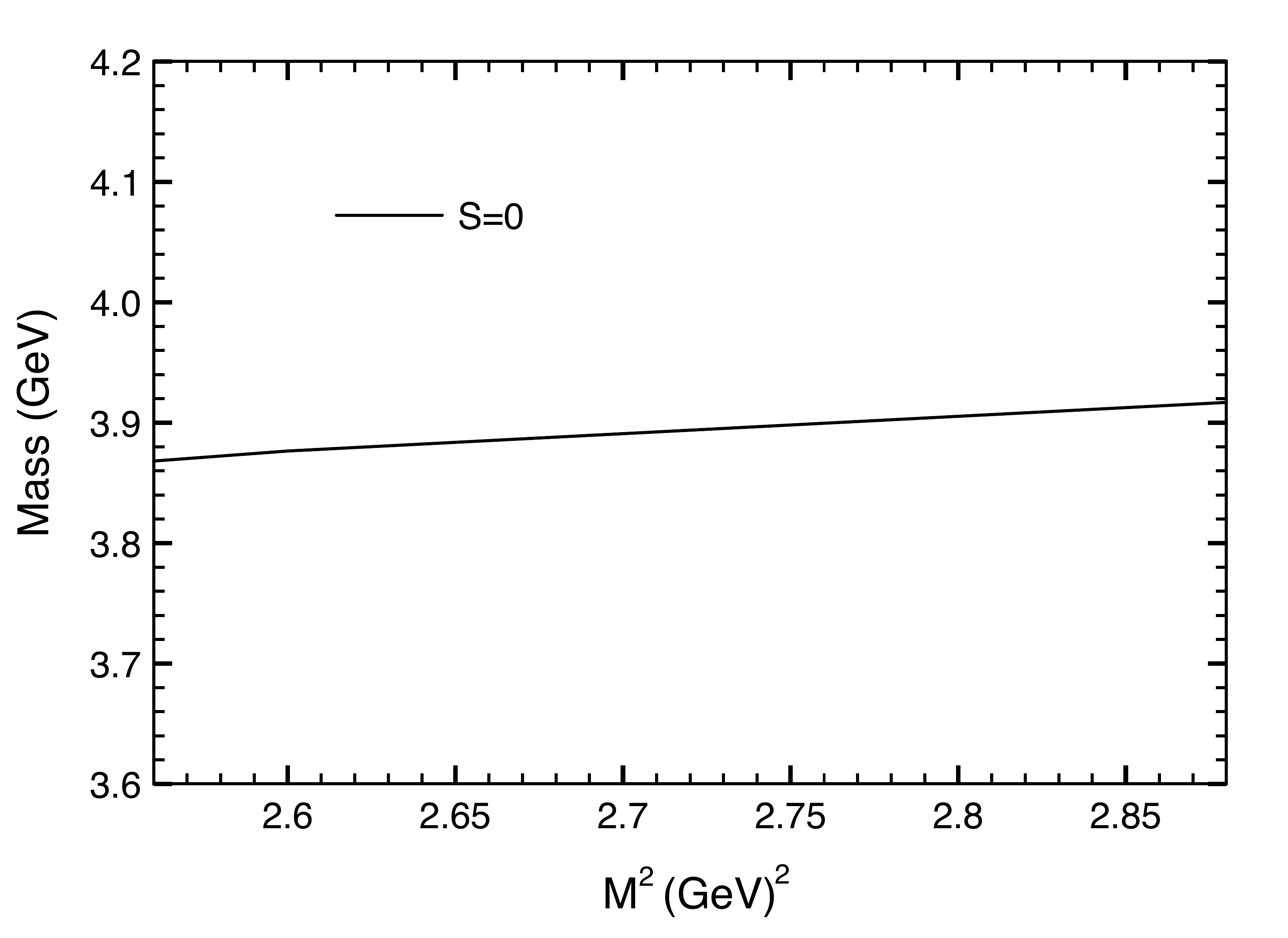}
\caption{Mass of a state with spin-parity $0^+$ as a function of Borel mass,  varying in the reliable range of the validity of results.}\label{fig2:4025}
\end{figure}
 the mass value is reasonably stable and it ranges from 3.87-3.91 GeV. In Ref.~\cite{Khemchandani:2013iwa} the authors have varied $\sqrt{s_0}$ by $\pm$0.5GeV and have better estimated the errors by taking into account the uncertainties involved in the values of other parameters, such as the condensates, the quark mass, etc.. The obtained values for the masses of the states with spin 0, 1 and 2, are respectively,
\begin{eqnarray}\label{res1}
M^{S=0} &=& \left(3943 \pm 104 \right)\,\, {\rm MeV},       \\\nonumber
M^{S=1} &=& \left(3950 \pm 105  \right)\,\,  {\rm MeV},        \\\nonumber
M^{S=2} &=& \left(3946 \pm 104 \right)\,\,{\rm MeV}.
\end{eqnarray}
In Ref.~\cite{Khemchandani:2013iwa}, three states with different spin but with almost similar mass are found, all with a $D^* \bar D^*$ molecular nature. The fact that the $s$-wave interaction must dominate in the formation of a molecular state implies that all these states have a positive parity. With the discovery of the neutral member of the isospin 1 triplet, in the $D^{*0} \bar D^{*0}$ system, it is also possible to define the $C$- and $G$-parity of the a state with spin 0, 1, and 2 formed in such a system. The spin 0 and  2 states formed in the $D^{*} \bar D^{*}$ should have  positive $C$-parity and negative $G$-parity, while the state with spin 1 has negative $C$-parity and positive $G$-parity. As discussed in ~\cite{Khemchandani:2013iwa}, the formation of a $0^{+}$ state in the $D^* \bar D^*$ system,  in the process $e^+ e^- \to (D^* \bar D^*)^{\pm,0} \pi^{\mp,0}$ is not possible, due to the conservation of parity and angular momentum. However, both $I^G (J^{PC}) = 1!
 ^+(1^{+-}
 )$ and $1^-(2^{++})$ can be assigned to $Z_c(4020)$. As can be seen from Table~\ref{zlist}, the particle data group assigns a positive $G$-parity and negative $C$-parity to this state, which seems to be motivated by the assumption of spin 1 for the state  $Z_c(4020)$. From Ref.~\cite{Khemchandani:2013iwa}, either of the possible quantum numbers $1^+(1^{+-})$ or $1^-(2^{++})$ can be associated to $Z_c(4020)$.

The central values of the results, given in Eq.~(\ref{res1}), are in line with the findings of Refs.~\cite{Molina:2009ct,Aceti:2014kja,Torres:2013lka}, although the mass values can be above the threshold when including error bars.

Other interpretations for the internal structure of this state, like a tetraquark nature, have also been suggested. Using QCD sum rules, in Ref.~\cite{Wang:2013llv}, a diquark-antidiquark structure was studied, giving rise to two possible spin-parity assignments for the state, $J^{PC}=1^{+-}$ or $2^{++}$. Using the same technique, in Ref.~\cite{Qiao:2013dda}, considering $J^P=1^-$ and $2^+$ tetraquark currents, it was found that the mass obtained with the $2^+$ current was consistent with the experimental data of $Z^+_c(4025)$, while the mass determined with the $1^-$ current was not compatible within the error-bars, suggesting then, a $2^+$ assignment for the state.

The possibility of this $Z_c$ state being a $D^*\bar D^*$ molecule kind has also been investigated within  QCD sum rules, in Ref.~\cite{Wang:2015nwa}, using a color octet-octet axial-vector current and values for the mass and width were found to be compatible with the experimental data, supporting the association of quantum numbers $J^{PC}=1^{+-}$ to $Z_c(4020)$. In Ref.~\cite{Chen:2013omd} a $J^P=1^{+}$ molecular current was used and the mass extracted was $(4.04\pm 0.24)$ GeV. A similar conclusion was found in Ref.~\cite{Cui:2013vfa} with a different current and with calculations done up to leading order in $\alpha_s$.  In Ref.~\cite{Chen:2015ata},  charmonium-like molecular interpolating currents with quantum numbers $J^{PC}=1^{+-}$ were constructed including both color singlet-singlet and color octet-octet structures and the authors arrived to the conclusion that $Z_c(4025)$ could be a $D^*\bar D^*$ or a $D_1\bar D_1$ molecular state.

Evidence for a state with spin-parity $0^{++}$, and a mass around 4000 MeV, also comes from other formalisms; within the constituent quark model, in Ref.~\cite{Yang:2017rmm} a tetraquark with mass $4005.7$ MeV and quantum numbers $I(J^{PC})=1(0^{++})$ was found with a mass similar to $Z_c(4025)$. This result is in agreement with the $0^{++}$ state found in~\cite{Khemchandani:2013iwa} (as summarized in Eq.~\ref{res1}).

There exist other studies based on QCDSR, which find  states with different spins, but almost same mass, as in~\cite{Khemchandani:2013iwa}.  Such is the case of Ref.~\cite{Wang:2014gwa},  where using vector $\times$ vector interpolating currents, the formation of $J^{PC}=0^{++},~1^{+-},~2^{++}$ molecular states was studied and  the existence of three states was found, one for each spin, with basically the same mass, within the uncertainties, with a central value of 4.01-4.04 GeV.

\subsection{$Z_1^+(4050)$, $Z_c^+(4055)$ and $Z_2^+(4250)$}\label{Zc4050}

 In the energy region around 4050 MeV, two states with basically the same mass have been claimed,  $Z_1(4050)$ and  $Z_c(4055)$, nowadays named as $X(4050)$ and $X(4055)$. In spite of their similar masses, the quantum numbers assignments seem to differ, with the former being a state with positive $C$-parity, isospin 1 and negative $G$-parity, while the latter has isospin 1 but opposite $C$- and $G$-parities. The other quantum numbers are unknown. 

The experimental evidence for $Z_1(4050)$ is controversial, as in case of the state $Z_2(4250)$ (or $X(4250)$ in Ref.~\cite{pdg} notation): they were observed by the Belle collaboration~\cite{Mizuk:2008me} in the $\pi^+\chi_{c1}(1P)$ invariant mass distribution of the reaction $\bar B^0\to K^-\pi^+\chi_{c1}(1P)$.  The fit to the data performed in Ref.~\cite{Mizuk:2008me} indicates that the consideration of two resonances is preferred by $13.2~\sigma$, and that the inclusion of two resonances is preferable to the consideration of one by $5.7\sigma$. These fits were done assuming  $J=0$ and 1 and the lowest possible orbital angular momentum for the system, but the $\chi^2$ result of the fit is not significantly altered  by changing $J$. However, the BaBar collaboration~\cite{Lees:2011ik} could reproduce the data on the $\pi^+\chi_{1c}(1P)$ invariant mass distributions of the processes $\bar B^0\to K^-\pi^+\chi_{c1}(1P)$ and $B^+\to K^0_S\pi^+\chi_{c1}(1P)$ with the $Z$ resonan!
 ce contri
 bution consistent with zero. 

In case of $Z_c(4055)$, the Belle collaboration claimed its existence based on an excess of events in the $\pi^\pm\psi(2S)$ invariant mass of the process $Y(4360)\to \pi^+\pi^-\psi(2S)$ with a $3.5\sigma$ significance~\cite{Wang:2014hta}.

\subsubsection{Theoretical explanations for $Z_1^+(4050)$, $Z_c^+(4055)$ and $Z_2^+(4250)$}

The possibility of $Z_1(4050)$ and $Z_2(4250)$ being tetraquarks has been studied with different models. In Ref.~\cite{Patel:2014vua}, using a Cornell potential the four-quark configuration of $Z_1(4050)$ as a cluster of $Q\bar q$ and $\bar Q q$ with some residual color forces that bind the two clusters is investigated and  the existence of a state with mass 4046 MeV is found with  quantum numbers $J^{PC}=2^{+-}$ together with a state with mass 4054 MeV and quantum number $3^{++}$. Both results are associated with $Z_1(4050)$. In Ref.~\cite{Deng:2015lca}, using a color flux-tube model, the authors arrived to the conclusion that $Z_1(4050)$ has a tetraquark $[cu][\bar c\bar d]$ nature, with a compact three-dimensional spatial configurations, with spin-parity $J^P=1^-$, while $Z_2(4250)$ can be interpreted as a $[cu][\bar c\bar d]$ tetraquark with $J^P=1^+$. No tetraquark candidate was found in Ref.~\cite{Ebert:2008kb} for $Z_1(4050)$ by using a relativistic quark model. On th!
 e other h
 and, $Z_2(4250)$ could be interpreted as a tetraquark~\cite{Ebert:2008kb}.

In Ref.~\cite{Liu:2008tn}, possible molecular states composed by a pair of heavy mesons, as $D^*$ and $\bar D^*$,
were studied by means of a meson exchange potential obtained from an effective Lagrangian based on the chiral and heavy quark symmetries. The authors of Ref.~\cite{Liu:2008tn} found two states, one with $J^P=0^+$ and other with $J^P=1^+$, which could correspond to $Z_1(4050)$. However, as the authors mention, the values of the cut-off used seem too large (4GeV for the $0^+$ state and 10 GeV for the $1^+$ state) and it is too naive to exclude, in such a situation, other components in the wave function of such a state. By using a chiral SU(3) quark model, the authors of Ref.~\cite{Liu:2008mi} reached the conclusion that $Z_1(4050)$ is unlikely to be an $s$-wave isospin 1 $D^*\bar D^*$ molecule.
In Ref.~\cite{Ding:2008gr}, using heavy meson chiral perturbation and an effective meson exchange potential, the authors found that the consideration of $Z_2(4250)$ as a $D_1\bar D$ molecule with quantum numbers $I^G(J^P)=1^-(1^-)$ is disfavored, since it requires a cut-off in their model of around $\Lambda\sim3$ GeV.

\subsubsection{QCDSR calculations for $Z_1^+(4050)$, $Z_c^+(4055)$ and $Z_2^+(4250)$}

Within the context of QCD sum rules, the tetraquark and molecular pictures have also been investigated. In Ref.~\cite{Wang:2013llv} it was concluded that the association of $Z_1(4050)$  and $Z_2(4250)$ with a $J^{PC}=0^{++}$ diquark-antidiquark tetraquark was disfavored. In Ref.~\cite{Wang:2008af}, by considering  a superposition of  $C\gamma_5-C\gamma_5$ and  $C-C$ currents a state with mass $(4.36\pm 0.18)$ GeV was found and associated with $Z_2(4250)$.   In Ref.~\cite{Chen:2015ata}, the authors constructed charmonium-like molecular interpolating currents with quantum numbers $J^{PC}=1^{+-}$ and found that $Z_1(4050)$ could be described by a $D^*\bar D^*$ or a $D_1\bar D_1$ molecular current, and $Z_2(4250)$ with a $D\bar D^*$ molecular current.
In Ref.~\cite{Lee:2008gn} the $I^G(J^P)=1^-(0^+)$ current
\begin{equation}
(\bar d\gamma_\mu c)(\bar c\gamma^\mu u)
\end{equation}
  was used to study the $D^{*}\bar D^{*}$ system and the $I^G(J^P)=1^-(1^-)$ current
  \begin{equation}
    \frac{i}{\sqrt{2}}[(\bar d\gamma_\mu\gamma_5c)(\bar c\gamma_5 u)+(\bar d\gamma_5 c)(\bar c\gamma_\mu\gamma_5 u)]
    \label{curz2}
  \end{equation}
  was used to investigate the $D_1\bar D$ system. In the case of the $D^*\bar D^*$ system, a state with a mass around 130 MeV above the threshold and 100 MeV above the nominal mass of $Z_1(4050)$ was found, while for the $D_1\bar D$ system a state with a mass around 100 MeV below the threshold and 60 MeV below the mass of $Z_2(4250)$ was obtained. In the former case, it was concluded that the signal found could be probably a virtual state which is not related to $Z_1(4050)$, while in the latter, the mass found is consistent with both $Z_1(4050)$ and $Z_2(4250)$ and thus no definite conclusions could be drawn.

  \begin{figure}[h]
\centerline{\scalebox{0.5}{\includegraphics[angle=0]{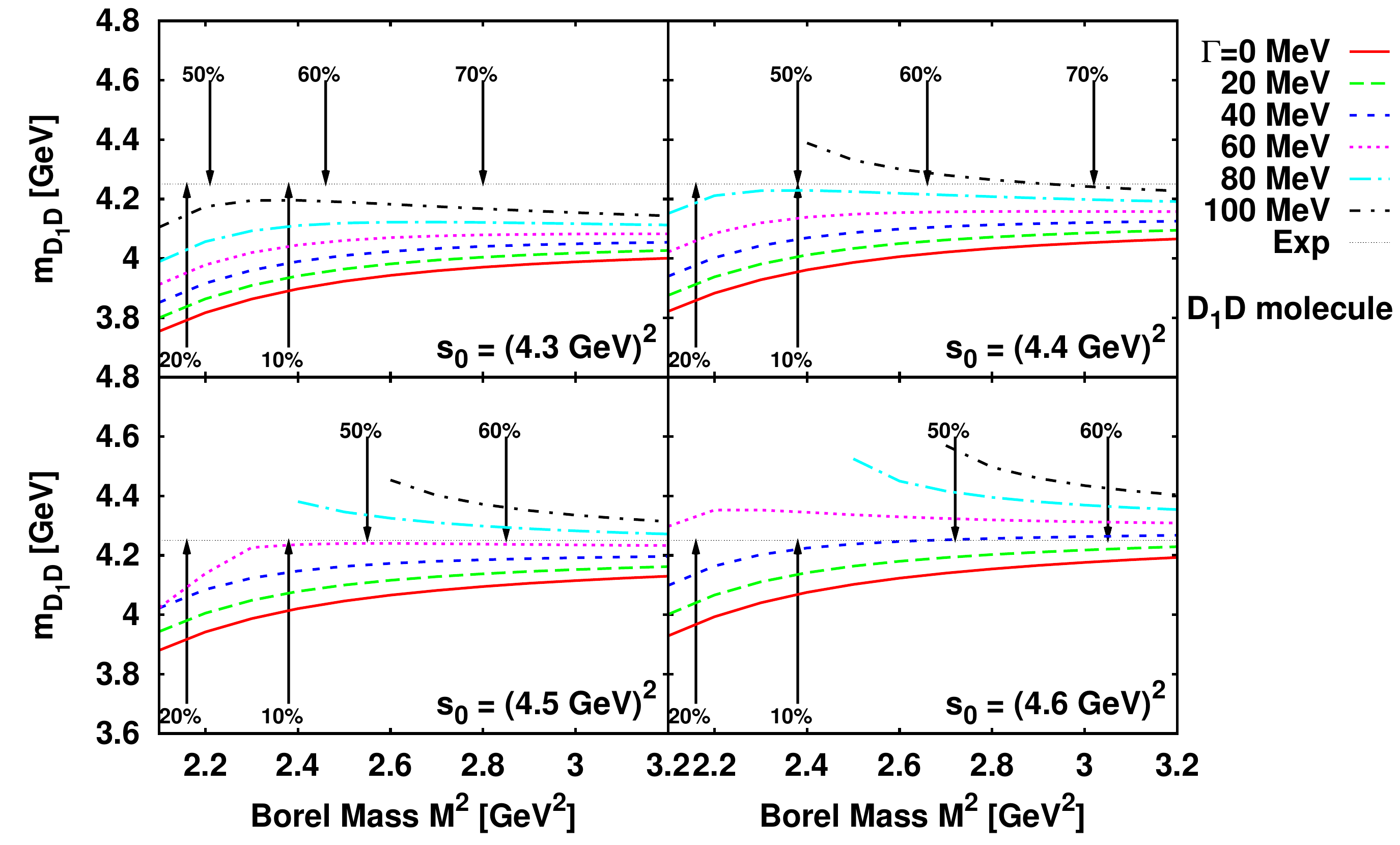}}}
\caption{Results for the $D_1 D$ molecule from Ref.~\cite{Lee:2008tz}. Each panel shows a  different choice of the continuum threshold. Upward and downward arrows  indicate the region of the Borel window $M^2_{\text{min}}$ and
  $M^2_{\text{max}}$, respectively. Associated numbers in \% denote the
  dimension eight condensate  contribution for upward arrows and
  continuum contribution for downward ones. Taken from ~\cite{Lee:2008tz}.
\label{figz2}}
\end{figure}
These conclusions were reached by ignoring the width of the $Z$ states in the spectral density of the phenomenological side of the sum rule. In view of this, in Ref.~\cite{Lee:2008tz} the same authors incorporated the width in the phenomenological spectral density and studied its effect on the mass, finding that using the current in Eq.~(\ref{curz2}), it is possible to obtain a mass $m_{D_1D} = 4.25~\GeV$ with a width
  $40\leq\Gamma\leq60~\MeV$, in agreement with the mass and width of $Z_2(4250)$,  as can be seen in Fig.~\ref{figz2}.

 Therefore, the authors of Ref.~\cite{Lee:2008tz} conclude that it
is possible to describe the $Z_2^+(4250)$ resonance  structure with a
$D_1\bar{D}$ molecular current with $I^GJ^P=1^-1^-$ quantum numbers, and that
the $D^{*}\bar{D}^{*}$ current is probably not related with the $Z_1^+(4050)$ resonance-like structure.

\subsection{$Z_c^+(4200)$}

The existence of $Z_c(4200)$ is based on the study of the $\psi \pi^+$ mass spectrum, obtained in the decay process $\bar B^0 \to J/\psi K^- \pi^+$, by the Belle Collaboration~\cite{Chilikin:2014bkk}. The  $\chi^2$-fit to the experimental data with the spin-parity assignment of $1^+$ for the $Z_c(4200)$ state leads to the highest statistical significance (8.2$\sigma$)~\cite{Chilikin:2014bkk}. The same data set also shows a signal for the better known $Z_c(4430)$. The mass and width for  $Z_c(4200)$ are found to be $4196^{+31+17}_{-70-132}$ MeV and $370^{+70+70}_{-70-132}$~MeV, respectively, in Ref.~\cite{Chilikin:2014bkk}. The inclusion of a $Z_c$ state with mass around 4200 MeV, apart from $Z_c(4430)$, in the experimental data on $B^0 \to J/\psi^\prime K^- \pi^+$, was also considered by the LHCb collaboration \cite{Aaij:2014jqa} and the quality of the fit was found to improve with the consideration of two $Z_c$ states rather than $Z_c(4430)$ alone.  Although the mass and wi!
 dth of th
 e state near 4200 MeV found by the LHCb collaboration, $4239\pm18^{+45}_{-10}$ MeV and $220 \pm 47^{+108}_{-74}$ MeV, are compatible with those determined in Ref.~\cite{Chilikin:2014bkk}, the preferred spin-parity was found to be different, $0^-$. The LHCb collaboration, though, does not claim the observation of any state with mass around 4200. The particle data group, thus, lists its spin-parity to be $1^{+}$. In a recent article \cite{Abazov:2018cyu}, the $D0$ collaboration confirms finding a resonant structure above 4 GeV which is similar to the one found in Ref.~\cite{Chilikin:2014bkk}. An analysis of the decay $\Lambda_b^0 \to J/\psi p \pi^-$ was made by the LHCb collaboration and it was found that the inclusion of either  the two exotic pentaquarks, $P_c^+$, or that of $Z_c(4200)$, was important to describe the data \cite{Aaij:2016ymb}. 

\subsubsection{Theoretical explanations  and QCDSR calculations for $Z_c^+(4200)$}

From a theoretical point of view, a tetraquark nature of $Z_c(4200)$ has been studied within different model calculations.  For example, the authors of Ref.~\cite{Deng:2017xlb} find a [$cu$][$\bar c \bar d]$ state, with quantum numbers $n \left({}^{2S+1} L_J \right) = 1\left({}^3D_1\right)$ and spin-parity $1^+$ and associate it with $Z_c(4200)$, in a model treating quark-quark interactions through one gluon exchange,  one boson exchange and $\sigma$ exchange. Evidence for $Z_c(4200)$ has also been found in a light-front holographic QCD model with a generic dilaton profile \cite{Guo:2016uaf}.  Using a formalism based on color magnetic interactions, in Ref.~\cite{Zhao:2014qva}, $Z_c(4200)$ is described as an axial vector tetraquark state.

More information is available from other works.  A study of the cross sections of the process $p \bar p \to Z_c^0 (4200) \pi^0$ has been done in Ref.~\cite{Wang:2015uua} and it has been suggested that proton-antiproton annihilation is an ideal process to investigate the neutral $Z_c(4200)$. Further, an estimate of the upper limit of the partial width $\Gamma(Z_c(4200) \to J/\psi \pi)$ has been done in Ref.~\cite{Wang:2015lwa} to be $\sim 37$ MeV. Based on a tetraquark nature of $Z_c(4200)$ with $J^{PC} = 1^{+-}$, it is also argued  in Ref.~\cite{Ma:2015nmy} that  the decay $Z_c(4200) \to h_c \pi $ should be suppressed.

Within QCD sum rules, using  the current
\begin{equation}
J_\mu (x) = \frac{\bar u i \gamma_5 \lambda^a c(x) \bar c(x)\gamma_\mu \lambda^a d(x) + \bar u \gamma_\mu\lambda^a c(x)\bar c(x) i \gamma_5 \lambda^a d(x)}{\sqrt{2}},
\end{equation}
$Z_c(4200)$ is described as an octet-octet type axial vector molecule-like state \cite{Wang:2015nwa}. The decay widths of $Z_c(4200) \to J/\psi \pi^+$, $\eta_c \rho^+$ $D^+ \bar D^{*0}$, $\bar D^0 D^{*+}$ were obtained in Ref.~\cite{Chen:2015fsa}, respectively, as  $87.3 \pm 47.1$ MeV, $334.4 \pm 119.8$ MeV and $6.6 \pm 6.4$ MeV, using the following current for $Z_c(4200)$
\begin{equation}
J_\nu = u_a^T C\gamma_5c_b\left(\bar d_a \gamma_\nu C \bar c_b^T + \bar d_b \gamma_\nu C \bar c_a^T\right) -  u_a^T C\gamma_\nu c_b\left(\bar d_a \gamma_5 C \bar c_b^T + \bar d_b \gamma_5 C \bar c_a^T\right).
\end{equation}
The obtained total width is consistent with the experimental value  $\Gamma=(370\pm70^{+70}_{-132})$ MeV \cite{Chilikin:2014bkk}.

Another study, based on  QCD sum rules, has been done in Ref.~\cite{Chen:2015ata}, where a $D \bar D^*$ current has been suggested for $Z_c(4200)$. However, this suggestion is not consistent with the usual interpretations where the $D \bar D^*$ current is associated with the $X(3872)$ state.

\subsection{$Z_c^-(4100)$}

The $Z_c^-(4100)$ state is the newest acquisition to the list of charged exotic charmonium states. The evidence for the existence of $Z_c(4100)$ was reported by the LHCb Collaboration \cite{Aaij:2018bla} and is based on the study of the $\eta_c(1S) \pi^-$ mass spectrum, obtained from the decay process $B^0 \to \eta_c(1S) K^+ \pi^-$. The reported significance of this exotic resonance is more than three standard deviations. Its mass and width are $M=(4096\pm20^{+18}_{-22})$ MeV and $\Gamma=(152\pm58^{+60}_{-35})$ MeV respectively,  and the spin-parity assignments $J^P=0^+$ and $1^-$ are both consistent with data \cite{Aaij:2018bla}. 
Since  $Z_c^-(4100)$ decays into $\eta_c(1S) \pi^-$ its $G$-parity is $-$. Therefore, the possible quantum numbers for its neutral partner are $I^G(J^{PC})=1^-(0^{++})$ or $1^-(1^{-+})$. While $J^{PC}=0^{++}$ are quantum numbers that can be also associated with quark-antiquark states,  $J^{PC}=1^{-+}$ is not consistent with the constituent quark model for a quark-antiquark system and it is considered exotic. Up to now only one state is known with such quantum numbers:  $\pi_1(1600)$ \cite{pdg}.

\subsubsection{Theoretical explanations and QCDSR calculations for $Z_c^-(4100)$}

In Ref.~\cite{Aaij:2018bla} it was suggested that this state could be the $J^P=0^+$ tetraquark  predicted by the diquark model in \cite{Maiani:2004vq}. However, the predicted masses for the $0^{++}$ tetraquark states were 3723 MeV and 3823 MeV, well below the observed mass \cite{Maiani:2004vq}. Up to now there are only a few theoretical calculations for $Z_c^-(4100)$. In Ref.~\cite{Wu:2018xdi} the authors used a simple chromomagnetic model to study the mass splitting among tetraquark states, including $Z_c^-(4100)$. The model is based on the description that the mass splitting among hadron states, with the same quark content, are mainly due to the chromomagnetic interaction term in the one-gluon-exchange potential. Based on these findings the authors concluded that $Z_c^-(4100)$ seems to be a $0^{++}$ $(cq)(\bar{c}\bar{q})$ tetraquark state. In Ref.~\cite{Voloshin:2018vym} it was argued that $Z_c^-(4100)$ is (dominantly) a hadrocharmonium with the $\eta_c$ embedded in a ligh!
 t-quark e
 xcitation, with quantum numbers of the pion, in the same way that $Z_c(4200)$ is a similar four-quark state containing $J/\psi$ instead of $\eta_c$. In this approach the natural quantum numbers for $Z_c^-(4100)$ would be $0^+$. In Ref.~\cite{Zhao:2018xrd} the  $Z_c^-(4100)$ can either be  interpreted as a $P$-wave resonance state arising from the $D^*\bar{D}^*$ interaction, or be caused by  final state interaction effects.
In Ref.~\cite{Cao:2018vmv} the author conjecture that the $Z_c^-(4100)$ observed in the $\eta_c\pi^-$ decay channel is the charge conjugate of the $Z_c^+(4050)$ observed in the  $\chi_{c1}\pi^+$ decay channel.

In Ref.~\cite{Wang:2018ntv}  the author used a $P$-wave  diquark-antidiquark  $J^{PC}=1^{--}$ interpolating field, in a QCDSR calculation, to study the mass of possible $J^{PC}=1^{--}$ tetraquark charmonium states. The obtained mass disfavors the assignment of $Z_c^-(4100)$ as a vector tetraquark state. Besides, as pointed out above, $J^{PC}=1^{--}$ quantum numbers are not allowed for $Z_c^-(4100)$.  A QCDSR calculation for a $J^{PC}=1^{-+}$ four-quark state, using a molecular $D^*D_0^*$ current, was done in Ref.~\cite{Albuquerque:2010fm}. The obtained mass was $m_{D^*D_0^*}=(4.92\pm0.08)~\GeV$ \cite{Albuquerque:2010fm}. In another QCDSR calculation, using a tetraquark current with $J^{PC}=1^{-+}$, the obtained mass was around 4.6 GeV \cite{Chen:2010ze}. In both cases the obtained masses are not consistent with the  $Z_c^-(4100)$ observed mass. On the other hand a mass $m_{X,0^{++}}=(3.81\pm0.19)$ GeV was obtained in a QCDSR calculation for a scalar $0^{++}$ tetraquark state!
  \cite{Ch
 en:2017dpy}, in excellent agreement with the mass obtained using a spin 0 projection of a $\bar{D}^*D^*$ interpolating current as shown in Fig.~\ref{fig2:4025} and given in Eq.~(\ref{res1}): $M^{S=0}=(3.94\pm0.10)$ MeV ~\cite{Khemchandani:2013iwa}, and with a  $0^{++}$ $\bar{D}_0^*D_0^*$ interpolating current in a N2LO QCDSR calculation: $m_{X,0^{++}}=(3.95\pm0.22)$ GeV \cite{Albuquerque:2016znh}. Therefore, the theoretical calculations seem to indicate that the $0^{++}$ quantum numbers are more compatible with the observed mass.

\subsection{Summary for the isovector $Z$ states}

The discovery of several manifestly exotic states, the  $Z^+$ states, may be considered as one of the most exciting findings of the last years. The description of such states  unavoidably requires  (at least) four valence quarks in the wave function. For the first of them, $Z^+(4430)$, there is now basically a consensus that it is the first radial excitation of $Z_c^+(3900)$. This interpretation favors a   diquark-antidiquark tetraquark assignment for the $Z_c^+(3900)$. Using a four-quark current in the QCDSR approach, one can explain not only the mass, but also the decay width of $Z_c^+(3900)$. The ratio in Eq.~(\ref{ratiozc}) still remains to be explained. Regarding the state $Z_c^+(4020)$, it is possible to explain its mass either as a $J^{PC}=1^{+-}$ or a $J^{PC}=2^{++}$ state, from a $D^*\bar{D}^*$ current. More experimental information is needed to confirm the existence of the states $Z_1^+(4050)$ and  $Z_2^+(4250)$, as well as the new state just observed by LHCb, $Z^-!
 (4100)$.

\section{Controversial $Y$ states}

\subsection{$Y(3940)$ or $X(3915)$ state}

The $Y(3940)$ was first observed by the Belle Collaboration  
in the decay $B\to KY(3940)\to K\omega J/\psi$ \cite{Abe:2004zs}. 
It was later confirmed by  
BaBar  in two channels $B^+\to K^+\omega J/\psi$ and 
$B^0\to K^0\omega J/\psi$~\cite{Aubert:2007vj,delAmoSanchez:2010jr}. 
The measured mass from these two Collaborations are: 
$(3943\pm11)$ MeV from Belle and $(3919.1^{+3.8}_{-3.4}\pm2)$  
MeV from BaBar, which gives an average mass of  $(3929\pm7)$ MeV and the total
width $(31^{+10}_{-8}\pm5)$ MeV.  
This state has positive $C$ and $G$ parities  and its possible spin-parity
is $J^P=0^+$ or $2^+$.
A similar state was observed also by Belle
Collaboration~\cite{Uehara:2009tx} in $\gamma\gamma\to \omega
J/\psi$ process. The measured mass and decay width are
$M=(3915\pm 3\pm 2)$ MeV and $\Gamma=(17\pm 10\pm 3)$ MeV,
respectively and the state was called $X(3915)$. It also carries positive
$C$ and $G$ parities. From the $\gamma\gamma$
fusion process, the possible spin-parity for the $X(3915)$ is also
$J^P=0^+$ or $2^+$. The BaBar Collaboration confirmed the existence of the
$X(3915)$ decaying into $\omega J/\psi$ in $\gamma\gamma\to \omega J/\psi$
process, with the
mass $(3919.4 \pm 2.2 \pm 1.6)$ MeV and width $(13 \pm 6 \pm 3)$ MeV
\cite{Lees:2012me}. Their analysis favored the $J^{P}=0^{+}$ assignment.

Due to the recent smaller mass observed by the BaBar Collaboration for the
$Y(3940)$ \cite{delAmoSanchez:2010jr}, in PDG~\cite{pdg} both states are considered as the same and are called $X(3915)$. In this review we keep the original name $Y(3940)$. The decay  
$Y\to J/\psi\omega$ is OZI suppressed for a charmonium state and hence   
the  $Y(3940)$  is a candidate to be an exotic,  a hybrid,  a molecular or a tetraquark
state.

\subsection{$Y(4140)$ state}

The first observation of the $Y(4140)$ structure 
was reported by the CDF Collaboration in the exclusive 
$B^+\to J/\psi\phi K^+$ decays, in 2009 \cite{Aaltonen:2009tz}. 
A prominent peak was observed in the 
$J/\psi\phi$ mass spectrum with mass 
$M=4143.4^{+2.9}_{-3.0}\,(\textrm{stat})
\pm 0.6 \,(\textrm{syst})$, and decay width 
$\Gamma=15.3^{10.4}_{-6.1}\,(\textrm{stat})\pm 2.5\, 
(\textrm{syst})$. In 2010, however, the Belle 
Collaboration performed an analysis of the 
data for events on the two-photon process and, 
as a result, no signal of the $Y(4140)$ state 
was found \cite{Shen:2009vs}. Soon after, in 2011, 
the LHCb Collaboration corroborated  Belle's 
data, also reporting negative results in the 
search for the $Y$ structure \cite{Aaij:2012pz}. 
Later on, the CMS Collaboration confirmed the  
observation of the $Y(4140)$ and another state, 
called $Y(4274)$, in exclusive $B^+\to J/\psi\phi K^+$ 
decays \cite{Chatrchyan:2013dma}.
Recently, the LHCb Collaboration  performed a 
full amplitude analysis of $B^+\to J/\psi\phi k^+$ 
decays \cite{Aaij:2016iza,Aaij:2016nsc} and claimed 
that four structures were required to fit the data, and 
the $Y(4140)$ was one of them. However, according 
to the LHCb results, the $Y(4140)$ is most likely a 
$J^{PC}=1^{++}$ state and has a width 
$\Gamma\approx 83\pm24\pm^{+21}_{-14}$, which 
is significantly larger than the one reported 
by the former experiments. Finally, the BESIII Collaboration searched for
$Y(4140)$ via $e^+~e^- \to \gamma \phi\jpsi$ at
$\sqrt{s}=$4.23, 4.26, 4.36, and 4.60~GeV, but no significant
$Y(4140)$ signal is observed in any of the data
samples~\cite{Ablikim:2014atq,Ablikim:2017cbv}.

In spite of these controversial experimental 
results, over the last decade there have been 
many experimental studies of the $J/\psi\phi$ mass
spectrum, and we expect the $Y(4140)$ to be confirmed in the future analysis.

\subsection{Theoretical interpretations for the $Y(3940)$ and $Y(4140)$ }

On the theoretical side, many efforts to understand the $Y(3940)$ and
$Y(4140)$ nature have been made. The $Y(4140)$ is the first 
state to be observed decaying into two heavy mesons 
containing both $c\bar{c}$ and $s\bar{s}$ content. 
Since it is far above the open charm  
threshold, it would be expected to decay 
into  open charm states with a large decay width. 
However, this feature is not observed by the 
experimental collaborations. In addition, since 
both $J/\psi$ and $\phi$ mesons have 
$J^{PC}=1^{--}$ quantum numbers, the states 
observed in the $J/\psi\phi$ mass spectra must have 
positive $C$-parity such that the 
exotic $1^{-+}$ quantum numbers are accessible. 
This set of quantum numbers is not allowed for any 
conventional charmonium state. Thus, these features 
put the $Y(4140)$ into the list of candidates 
that require  exotic quark configurations to be 
understood.

In Ref.~\cite{Mahajan:2009pj} the exotic 
$1^{-+}$ quantum numbers were assigned to $Y(4140)$ 
assuming it to be a hybrid charmonium state, although 
it was  argued that the $D_s^*\bar{D}_s^*$ molecular 
interpretation could also  be applied to  the $Y$ state.

A conventional charmonium interpretation was adopted in 
Ref.~\cite{Liu:2009iw}, where the $Y(4140)$ 
was assumed to be the second radial excitation of the 
$P$-wave charmonium $\chi^{\prime\prime}_{cJ}(J=0,\,1)$. 
With this assumption, the hidden charm decay mode of $Y$ was estimated 
in terms of the rescattering mechanism and, as a result, 
the value obtained was much smaller than the one reported 
by the CDF Collaboration in Ref.~\cite{Aaltonen:2009tz}. 
This result indicates that the conventional charmonium picture 
cannot be ascribed to the $Y$ structure.

A tetraquark model was used in Ref.~\cite{Stancu:2009ka}. 
However, this model provides a decay width around $100$ MeV for the 
$Y$ with $J^{PC}=0^{++}$. According to the experimental 
results, this interpretation cannot be supported since 
the decay width measured for the $Y(4140)$ 
is around $11$-$20$ MeV. On the other hand, the $J^{PC}=1^{++}$ 
quantum numbers seem to be favored by the tetraquark model 
since the decay width estimated, in this case, provides a 
result smaller than the one obtained assigning $J^{PC}=0^{++}$ 
for $Y$.

Since the $Y(3940)$ and  $Y(4140)$ masses are close to $D^*\bar{D}^*$ and
$D_s^*\bar{D}_s^*$ thresholds respectively, it seems natural to adopt a
molecular picture to understand their features. In fact, in 
Ref.~\cite{Liu:2009ei} the authors 
have investigated these meson molecules through a 
meson-exchange model and claimed that the  $Y(3940)$
must   be  the   molecular, $D^*\bar{D}^*$,   partner  of   the  $Y(4140)$,   a 
$D^*_s\bar{D}^*_s$ molecule.   This interpretation has been  tested in
several    approaches,    such   as    phenomenological    Lagrangians
\cite{Branz:2009yt} and vector-meson   dominance  models \cite{Branz:2010rj}.

On the other hand, within the QCDSR approach such quark 
configuration can be described, and an estimate for 
the $Y(4140)$ mass can be done for a current with 
$J^{PC}=0^{++}$. In addition, a tensorial $2^{++}$ 
state can also be studied with QCDSR with the help of 
spin projectors, which will be discussed later. In the following  
subsection we start to discuss the results obtained for $Y(3940)$ and $Y(4140)$
using the $D^*\bar{D}^*$ and $D_s^*\bar{D}_s^*$ multiquark configurations with $0^{++}$ 
quantum numbers, in the QCDSR approach.

\subsection{QCDSR calculations for the $Y(3940)$ and $Y(4140)$}

Considering $Y(4140)$ as a $D_s^*\bar{D}_s^*$ structure 
with $I^G(J^{PC})=0^+(0^{++})$, a suitable interpolating 
current describing such state, considered in 
Ref.~\cite{Albuquerque:2009ak}, is given by
\begin{equation}\label{jY4140}
j_{D_s^*\bar{D}_s^*}=(\bar{s}_a\gamma_{\mu}c_a)
(\bar{c}_a\gamma^{\mu}s_b)\, .
\end{equation}

The authors  considered contributions 
from the condensates up to dimension eight on the 
OPE side. The QCDSR analysis done showed a good Borel 
stability for $M^2$ values in the interval 
$2.3\leq M^2\leq 3.0$ GeV$^2$, with $4.4\leq \sqrt{s_0}\leq 4.7$. 
GeV. The effects related to the violation of the factorization hypothesis 
were also included in the error estimates. 
The result obtained for the mass was
\begin{equation}\label{mY4140}
m_{D_s^*\bar{D}_s^*}=(4.14\pm0.09) \, \textrm{GeV}\,,
\end{equation}
in good agreement with the experimental mass of the narrow 
structure $Y(4140)$.

A similar analysis based on the QCDSR technique was done in 
Ref.~\cite{Wang:2009ue} in which the 
author  also adopted a $D_s^*\bar{D}_s^*$ molecular 
interpolating current with $0^{++}$. However, the 
result found in this latter work was different, around $290$ MeV 
higher than the one obtained in Eq.~\eqref{mY4140}, and equal to 
\begin{equation}
m_{D_s^*\bar{D}_s^*}=(4.43\pm0.16)\, \textrm{GeV}\, .
\end{equation}
In order to get this result, in Ref.~\cite{Wang:2009ue} 
the condensates up to dimension eight were also included in 
the OPE. But, in this case, the convergence  seems to be too slow 
in comparison to the one in Ref.~\cite{Albuquerque:2009ak}. 
In fact, the minimum for the Borel mass in the former 
work is $2.6$ GeV$^2$ against $2.3$ GeV$^2$ in the latter one. 
In addition, the Borel window in the work of 
Ref.~\cite{Wang:2009ue} is smaller, 
implying a smaller $M^2$ region of stability.

From the QCDSR calculations done in 
Ref.~\cite{Albuquerque:2009ak},
it is possible to get a sum rule for a $D^*\bar{D}^*$ 
structure, by simply replacing the strange/anti-strange quark 
fields by the corresponding up/anti-up and down/anti-down quark in the 
interpolating current definition in Eq.~\eqref{jY4140}. 
In this case, the interpolating current for $D^*\bar{D}^*$ is 
\begin{equation}\label{DesDes}
j_{D^*\bar{D}^*}=(\bar{q}_a\gamma_{\mu}c_a)
(\bar{c}_a\gamma^{\mu}q_b)\, .
\end{equation}
It was expected that the structure described by this current 
could be used to explain the $Y(3940)$ state as suggested in
Ref.~\cite{Liu:2009ei}. The mass obtained in \cite{Albuquerque:2009ak} 
for such a structure is consistent, considering the errors, with the result in Eq.~(\ref{res1}) and also consistent with the result in Eq.~\eqref{mY4140}.
The degeneracy between the QCDSR results for these two currents in Eqs.~\eqref{DesDes} and \eqref{jY4140} is not surprising. In fact, from 
the point of view of a standard mass sum rule, like 
the one used here, the difference in studying 
the $D^*\bar{D}^*$ or the $D_s^*\bar{D}_s^*$ system 
arises from the fact that the latter one involves 
diagrams with strange quarks instead of  the 
light ones. In the $D^*\bar{D}^*$ case, the mass of the 
light quarks ($u$ and $d$) is taken to be negligible, 
thus, the corrections to the free propagator are 
related to condensates involving light quarks and 
gluons, while in the $D_s^*\bar{D}_s^*$ case we have corrections 
associated with strange quark condensates, gluon 
condensates and the mass of the $s$ quark. Since, 
as can be seen from Table~\ref{QCDParam}, 
the value for the light quark condensate is larger than
the one of the strange quark, this difference is compensated by
the $m_s$ corrections for the case of the $D_s^*\bar{D}_s^*$ 
system.

In Ref.~\cite{Zhang:2009vs} a mass of   
$3.9$ GeV was obtained from the application 
of the QCDSR approach to the $D^*\bar{D}^*$ current. 
In this case, the OPE was calculated up to dimension 
six, and the OPE convergence was found to be slower. 
It can be shown that the 
OPE convergence could be improved if the contributions 
from the condensates up to dimension eight would  
be included. Furthermore, the QCDSR analysis 
of Ref.~\cite{Zhang:2009vs} does not 
present any pole dominance, which means that a 
reliable Borel window cannot be established, 
compromising the results for the mass.

Although in Ref.~\cite{Liu:2009ei} it 
has been argued that the $D_s^*\bar{D}_s^*$ 
and $D^*\bar{D}^*$ molecules could be assigned  
to the $Y(4140)$ and $Y(3940)$, respectively, 
within QCDSR both currents provide similar masses. 
On the other hand, a molecular-charmonium
interpretation for the $Y(3940)$ is also possible. Indeed, in Ref.~\cite{Albuquerque:2013owa}, a mixed
$(\chi_{c0})\!-\!(D^\ast  \!\bar{D}^\ast)$   current,  with   $J^{PC}  =
0^{++}$: 
\beqa j  &=& - \:\mbox{cos}\:\theta\frac{\comq}{\sqrt{2}}
   ~j_{_{\chi_{c0}}}   ~+~   \mbox{sin}   \:\theta
~j_{_{D^\ast \!D^\ast}},
  \label{mixjy}
  \enqa
was used to study the $Y(3940)$ state. The current in Eq.~(\ref{mixjy}) is
similar to the ones used in Secs.\ref{sec-mixing} and \ref{sec-Ymass}
to study the states
$X(3872)$ and $Y(4260)$ respectively, with $\theta$  being the  mixing angle.
The molecular $D^*\bar{D}^*$ current is given in Eq.~(\ref{DesDes}) and the
$\chi_{c0}$ current is: $ j_{_{\chi_{c0}}}=\bar{c}_k  c_k$.

Fixing the mixing angle  in the range $\theta=(76.0\pm  5.0)^0$,
the mass obtained in~\cite{Albuquerque:2013owa} for the mixed state is:
\begin{equation}
M_{Y} = (3.95 \pm 0.11) ~ \mbox{GeV} ~,
\end{equation}
which  is in agreement, within the errors, with  the experimental mass for  the $Y(3940)$
state  observed by   the {Belle} Collaboration: $(3943\pm11)$ MeV\cite{Abe:2004zs}, 
and by BaBar: $(3919.1^{+3.8}_{-3.4}\pm2)$~\cite{delAmoSanchez:2010jr}.

Since it is possible to explain the mass of the $Y(3940)$ with the mixed current, in Ref.~\cite{Albuquerque:2013owa} the decay width $Y(3940) \to J/\psi \:\omega$ was also considered. The three-point function for the vertex $Y J/\psi \:\omega$ is defined as:
\beq \Pi_{\mu \nu}(p,\pli,  q) = \int
d^4x \:d^4y ~e^{i\pli \cdot x} \:e^{iq\cdot y} ~\lag 0|T\{j_{\mu}^{\psi}(x)
j_{\nu}^{\omega}(y)j^{\dagger}(0)\}|0\rag,
\label{3po}
\enq
where the  mixed $(\chi_{c0})\!-\!(D^\ast     \!\bar{D}^\ast)$  current, given
in Eq.~(\ref{mixjy}), was considered, and
\beq
j_{\mu}^{\psi}=(\bar{c}_a \gamma_\mu c_a)
,\;\;j^{\omega}_\nu = \frac{1}{6}\Big(\bar{u}_a \gamma_\nu u_a +
\bar{d}_a \gamma_\nu d_a \Big).
\label{jomega}
\enq

The OPE  side of the  correlation function was evaluated  at  the leading
order in  $\alpha_s$, considering condensates  up to  dimension 7, in the $q_{\mu}p^\prime_\nu$ structure~\cite{Albuquerque:2013owa}. After extrapolating the form factor to the $\omega$ pole, following the procedure discussed in Sec.~\ref{exproc} the coupling constant, using the same mixing angle $\theta=(76.0\pm  5.0)^0$, was obtained as,
$g_{_{\!Y \!\psi\omega}}=(0.58~ \pm~0.14 )~~\mbox{GeV}^{-1}$,

The decay width for the process $Y(3940) \rightarrow J/\psi\: \omega$ is
given by
\beqa
\Gamma_{{Y(3940) \to J/\psi\:\omega}}= \frac{g^2_{_{\!Y \!\psi \omega}}}{3}
{p(M_Y,M_\omega ,M_\psi)\over8\pi M_{Y}^2} \left(M_\psi^2 M_\omega^2 + 
{1\over2}(M_Y^2- M^2_{\psi}-M_\omega^2)^2\right),\label{gamma3940}
\enqa
where
\beq
p(a,b,c) \equiv {\sqrt{a^4+b^4+c^4-2a^2b^2-2a^2c^2-2b^2c^2}\over 2a}.
\enq
Inserting the  value obtained  for  the
coupling constant in Eq.(\ref{gamma3940}) one gets~\cite{Albuquerque:2013owa}:
\beq
\Gamma_{Y(3940)\to J/\psi \:\omega}=(1.7~\pm ~0.6)~\MeV.
\label{width-3940}
\enq
This result is consistent with the experimental width of the state and
the   lower    limit   for    the   process    $Y\to   J/\psi\:\omega$
\cite{Abe:2004zs,Aubert:2007vj}.  It is also of the same
order as other available theoretical evaluations \cite{Branz:2009yt,Branz:2010rj}.

As we have seen, many efforts have been done in 
order to understand the quark configuration of the $Y(3940)$ and
$Y(4140)$ structures as well as their $J^{PC}$ 
quantum numbers. From the theoretical side, 
many models such as the conventional charmonium, 
hybrid, tetraquark,  meson molecule and mixed states
have been employed to investigate these systems.

As mentioned before, the new LHCb analysis on 
the $J/\psi\phi$ data brought surprises to 
the spectrum around $4.1$ GeV since 
other structures were seen in this channel. 
According to the LHCb results, in order 
to fit the data, the $Y(4140)$ is required to have 
a large decay width. This new value is close, 
within the error bars, to the one reported for 
the $Y(4140)$ state, measured by the Belle 
Collaboration \cite{Abe:2007sya} 
in the $D^*\bar{D}^*$ mass spectra in the process 
$e^+e^-\to J/\psi D^*\bar{D}^*$. 
The mass and width obtained for this state 
are $M=4156\pm29$ MeV and $\Gamma=139^{+113}_{-65}$ MeV, 
respectively \cite{Abe:2007sya}. Its 
quantum numbers, such as, spin, isospin, and parity are 
not known yet. The decay of $X(4160)$ to $D^*\bar{D}^*$, 
in a $D_s^*\bar{D}_s^*$ molecular model, could be understood 
in terms of triangular loop diagrams involving a strange meson. 
In this perspective, the $X(4160)$ state could be a 
$D_s^*\bar{D}_s^*$ molecule-like structure. 
In fact, a dynamically generated state that could be 
assigned to $X(4160)$ was found in many coupled channel
investigations \cite{Branz:2010rj,Liang:2015twa,Molina:2009ct}. 
If we compare its mass and width with the 
corresponding ones known for $Y(4140)$, and also 
have in mind that both are close to the 
$D^*_s\bar{D}_s^*$ threshold, a question arises: 
are they the same state?

With the aim to shed light on this discussion, 
in Ref.~\cite{Torres:2016oyz}, a more 
general QCDSR analysis was done, considering a 
tensorial current that allows exploring spin 
contributions other than the $0^{++}$, 
considered by the QCDSR works discussed above. 
We are going to discuss the details of this 
analysis in the next subsection.

\subsection{The $Y(4140)$ and $X(4160)$ with a tensorial current}

The simplest vector times vector type interpolating 
current coupling to the $D_s^{*+}D_s^{*-}$ meson-meson 
system,  used in Ref.~\cite{Torres:2016oyz}, 
is given by
\begin{equation}\label{jtensor}
j_{\mu\nu}(x)=\Big(\bar{c}_a\gamma_{\mu}s_a\Big)
\Big(\bar{s}_b \gamma_{\nu} c_b\Big) \, .
\end{equation}
As discussed in Sec.~\ref{spin-pro}, since the current in Eq.~\eqref{jtensor} is tensorial, it 
provides contributions for spin $0$, $1$, and $2$ particles. From calculations
in the QCDSR approach for both, mass and decay width,  
the contributions from different spins can be studied separately by 
projecting a particular spin using the projectors defined in Eqs.~(\ref{proj}),
which separate  the $J^P=0^+, 1^+, 2^+$ 
contributions to the correlation function. If a state 
of isospin $I$, and spin $S$ is formed as a consequence 
of the interaction between a $\bar D^*_s$ and a  $D^*_s$ 
meson with  angular momentum $L$, the parity $P$, charge conjugation $C$, 
and $G$-parity, associated with a particular state can be 
determined through: $P=(-1)^L$, $C=(-1)^{L+S}$, and 
$G=(-1)^{L+S+I}$. In the case considered here, since $I=0$ and $L=0$, 
the possible states must have positive parity and 
$C=G=(-1)^S$. This means that the states with quantum numbers 
$I^G (J^{PC})=0^+ (0^{++}),\,0^- (1^{+-}),\,0^+ (2^{++})$.

In Ref.~\cite{Torres:2016oyz} a QCDSR calculation, based on the current
in Eq.~(\ref{jtensor}), was done considering contributions from the 
QCD condensates up to dimension eight, for each 
spin projection. As a result,  
three states were found with isospin $0$, nearly degenerate, with masses around
$4.1$ GeV: 
\begin{align}
M_0&=(4.114\pm 0.130)~\text{GeV}\label{M0final}\, ,\\
M_1&=(4.120\pm0.127)~\text{GeV}\label{M1final}\,,\\
M_2&=(4.117\pm 0.123)~\text{GeV}\label{M2final}\, .
\end{align}
This result suggests the presence of a 
$I^{G}(J^{PC})=0^{++}$, a $0^-(1^{+-})$ and a $0^+(2^{++})$ 
states with masses given by the values in 
Eqs.~\eqref{M0final}, \eqref{M1final} and \eqref{M2final}, respectively.

It is worth mentioning that the $\bar D^* D^*$ system, described in
Sec.~\ref{spin-pro}, revealed the possibility of having spin $0$, $1$, and 
$2$ states with masses around $3950 \pm 100$ MeV (see Eq.~(\ref{res1})) with isospin 
$0$ and $1$. Comparing the masses obtained in the 
study of the $\bar D^* D^*$ (isoscalar) 
current~\cite{Khemchandani:2013iwa} with  
Eqs.~\eqref{M0final}, \eqref{M1final}, \eqref{M2final}, 
we clearly see that, within the errors, the two results 
are compatible. Within the QCDSR technique the 
reason for this degeneracy can be 
understood recalling that the additional
 $m_s$ contributions on the OPE side of the $D_s^*\bar{D}_s^*$ 
sum rule, for each spin 
case, are compensated by the smaller value of the 
strange quark condensate as compared with the light quark 
condensate. Therefore,  one ends up with no significant 
difference in the sum rules results for both, 
$D_s^*\bar{D}_s^*$ and $D^*\bar{D}^*$ currents. 
Furthermore, this degeneracy can also be interpreted 
as a manifestation of heavy quark symmetry. In fact, 
although the strange quarks are heavier than the light quarks 
when compared a $D^*\bar{D}^*$ state with a $D_s^*\bar{D}_s^*$ state, there are
two charm quarks in both systems.  Since the charm quarks are much heavier than 
both, strange and light quarks, their  presence  overshadows the 
mass difference between light and strange quarks. 
This can be compared to the use of heavy 
quark symmetry in the calculations based on effective 
field theories, where $D^*\bar{D}^*$ and $D_s^*\bar{D}_s^*$ 
are considered as coupled channels. The  
results obtained in these calculations 
\cite{HidalgoDuque:2012pq} seem to be very 
similar to the ones discussed in this subsection.

Even arguing that this degeneracy can be 
understood as a  manifestation of heavy quark symmetry or 
by the QCDSR point of view, an intriguing question still 
remains: are the states found in Ref.~\cite{Khemchandani:2013iwa} 
the same as the ones in Ref.~\cite{Torres:2016oyz}? 
There are two possibilities. One is that 
the states found in the QCDSR studies can couple to both 
currents, since they have the same quantum numbers. Therefore, they
can be  the same states and hence, we conclude that there are only
three states in the mass interval $3.8$-$4.2$ GeV,  
with $0^{++}$, $1^{+-}$ and $2^{++}$ quantum numbers. 
Another possibility is that 
both QCDSR studies describe different states, due to their quark content and, in this case, 
we have six different states in the mass region $3.8$-$4.2$ GeV. 
In either case, just the mass analysis is not enough to discriminate
between them. Since 
the $D^*\bar{D}^*$ states 
cannot decay into $J/\psi\phi$ because this channel is 
Okubo-Zweig-Iizuka (OZI) suppressed, a QCDSR calculation 
of the decay width is useful to distinguish the states 
coupling to $D^*\bar{D}^*$ and $D_s^*\bar{D}_s^*$.

The decay width sum rule analysis was done in 
Ref.~\cite{Torres:2016oyz} for each 
spin projection of the $D_s^*\bar{D}_s^*$ current in 
Eq.~\eqref{jtensor}. The 
three-point function of the  $X J/\psi\phi$ vertex, 
with $X$ denoting each spin component of the 
$D_s^*\bar{D}_s^*$ current, is given by:
\begin{equation}\label{3point-pro}
\Pi_{\mu \nu \alpha \beta} (p^2)= \int d^4x\, d^4y e^{i\,p^\prime \cdot x} e^{i\, q \cdot y} \langle 0 | T\{j_{\mu}^{\psi}(x)\,j_{\nu}^{\phi}(y)\, j_{\alpha \beta}^{\dagger X}(0) \} | 0 \rangle \, .
\end{equation}
Since a state with $J^{PC}=1^{+-}$ 
cannot decay into $J/\psi\phi$,  only 
the scalar and tensorial components of the $j_{\alpha \beta}$ 
current were considered. The interpolating currents used for $\omega$ and
$J/\psi$ are given in Eq.~(\ref{jomega}) and the $\phi$ meson current is: 
$j^{\,\phi}_{\nu}=\bar{s}_b\gamma_{\nu}s_b$.

\begin{figure}
\includegraphics[width=0.49\textwidth]{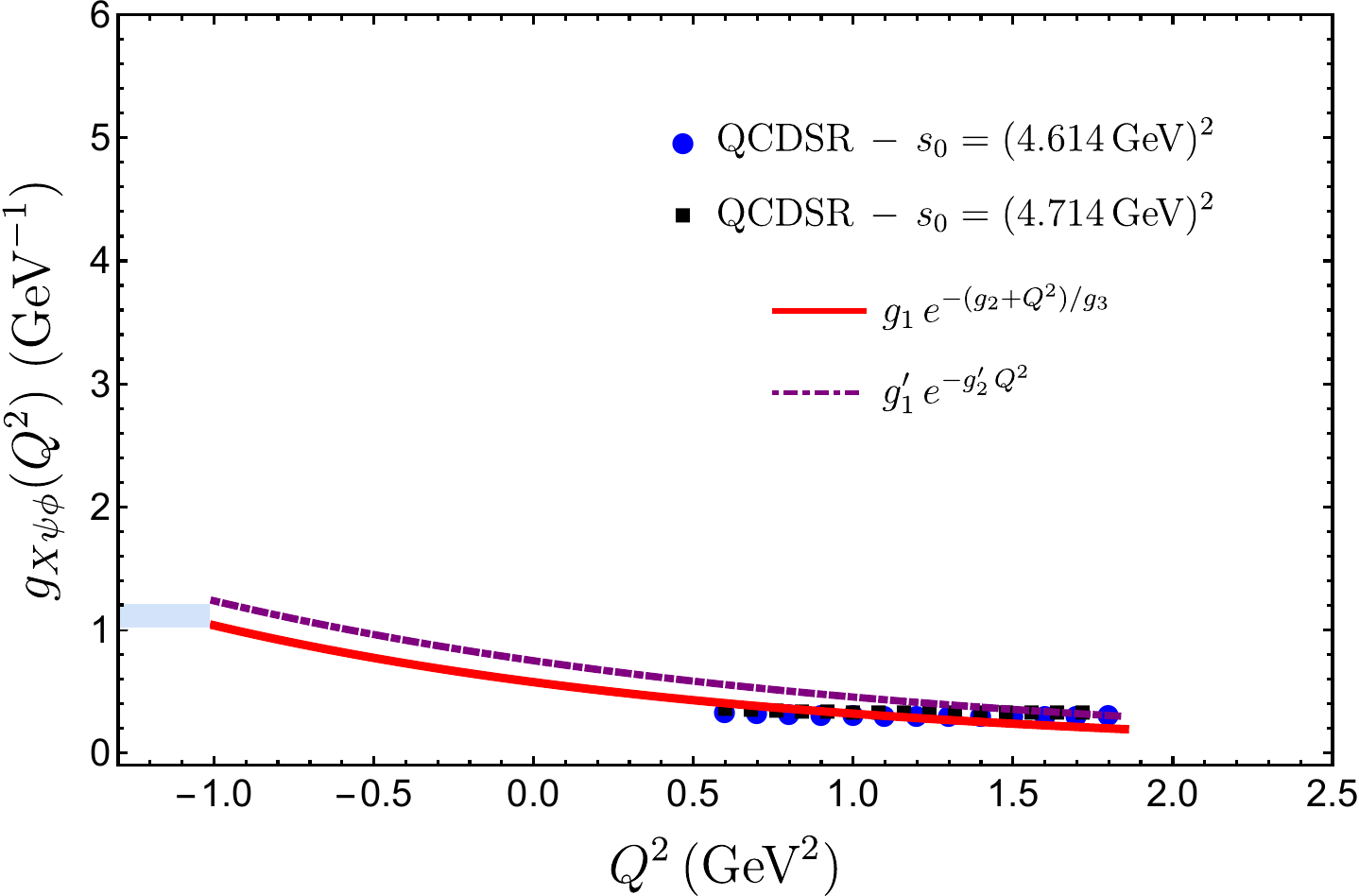}
\includegraphics[width=0.49\textwidth]{FQ2Spin2}
\caption{Solutions for the form factors for spin $0$ (left panel) 
and spin 2 (right panel) cases for a fixed $M^2$. The shaded 
region in the figures represents the range of the values 
for the corresponding coupling constants due to the different choices 
for the parametrization of the form factor. Taken from~\cite{Torres:2016oyz}.}\label{coupFG}
\end{figure}

The three-point function QCDSR  was evaluated in Ref.~\cite{Torres:2016oyz}
at leading order in $\alpha_s$, considering contributions from 
condensates up to dimension five. 
The details of the calculations can be found in ~\cite{Torres:2016oyz}.
In Fig.~\ref{coupFG} the solutions for 
both $0^{++}$ and $2^{++}$ cases are shown, with the corresponding 
form factors used to extrapolate the QCDSR results to the 
time-like region, as discussed in Sec.~\ref{exproc}. 
The resulting coupling constants found for spin $0$ and $2$ cases  
denoted by $g^{S=0}_{X\psi\phi}$ and $F^{S=2}_{X\psi\phi}$, are~\cite{Torres:2016oyz}:
\begin{eqnarray}
g^{S=0}_{X\psi\phi}(-m_{\phi}^2)&\approx& 1.115 \pm 0.085~\text{GeV}^{-1}\, ,\nonumber\\
F^{S=2}_{X\psi\phi}(-m_{\phi}^2)&\approx& 5.0\pm 0.6 ~\text{GeV}\, .
\end{eqnarray}

The decay width is given by:
\begin{align}
\Gamma=\frac{1}{8\pi}\frac{p(m^2_X,m^2_{\psi},m^2_{\phi})}{m^2_X}\frac{1}{2S_X+1}\sum_\text{pol}|\mathcal{M}|^2\, ,\label{dw}
\end{align}
where $p(m^2_X,m^2_{\psi},m^2_{\phi})$ is the center of mass momentum, $S_X$ is the spin of $X$ state. The matrix elements for the $0^{++}$ 
and $2^{++}$  cases are given by 
\begin{align}\label{gamma}
\sum_\text{pol}|\mathcal{M}|^2&=\Big(g^{S=0}_{X\psi\phi}\Big)^2
\Big[ m^2_{\psi}m^2_{\phi}+\frac{1}{2}(m^2_X-m^2_{\psi}-m^2_{\phi})^2\Big]\,,
\end{align}
and
\begin{align}
\sum_\text{pol}|\mathcal{M}|^2&=\frac{\Big(F^{S=2}_{X\psi\phi}\Big)^2}{24m^4_Xm^2_{\psi}m^2_{\phi}}\Big\{ m^8_X+6m^6_X(m^2_{\psi}+m^2_{\phi})-14m^4_X(m^4_{\psi}+m^4_{\phi}-6m^2_{\psi}m^2_{\phi})\nonumber\\
&+6m^2_X(m^2_{\phi}-m^2_{\psi})^2(m^2_{\psi}+m^2_{\phi})+(m^2_{\phi}-m^2_{\psi})^4\Big\}\, .
\end{align}

Taking into account the error associated 
with the mass found within QCD sum rules for the spin 
$0$ and $2$ states, given by Eqs.~\eqref{M0final} and 
\eqref{M2final}, the obtained values of the decay width are~\cite{Torres:2016oyz}:
\begin{equation}
\Gamma_{S=0}\approx (34\pm 14) ~\text{MeV}\,
\end{equation}
for the spin $0$ state, and 
\begin{equation}\lb{decay2}
\Gamma_{S=2}\approx (20\pm 7) ~\text{MeV}\, ,
\end{equation}
for the spin $2$ state.

According to the above results, both $0^{++}$ and 
$2^{++}$ can be associated with the $X(4160)$ state.
In Ref.~\cite{Molina:2009ct} the 
authors have studied the $D_s^*\bar{D}_s^*$ system 
using effective field theories, and have concluded that 
the interaction between $D_s^*$ and $\bar{D}_s^*$ mesons 
generates a state with mass $4170$ MeV and full 
width equal to $130$ MeV, which is compatible with  
the corresponding experimental values for the  $X(4160)$ 
state. From the couplings provided in 
Ref.~\cite{Molina:2009ct}, it is
possible to estimate the partial decay width of 
the $2^{++}$ state to be $20$-$30$ MeV. Thus, the result for the 
decay width associated with the $2^{++}$ state 
found in Ref.~\cite{Torres:2016oyz} within QCDSR, given in Eq.~(\ref{decay2})
is compatible with the one obtained in Ref.~\cite{Molina:2009ct}.

The findings in Ref.~\cite{Torres:2016oyz} presented above
suggest that the $D_s^*\bar{D}_s^*$ molecular current
with $2^{++}$ is most likely the charmonium-like 
state $X(4160)$, and not the $Y(4140)$ . It is 
worth  pointing out that although the $1^{++}$ quantum 
numbers for $Y(4140)$ are favored by the LHCb 
analysis \cite{Aaij:2016iza,Aaij:2016nsc}, 
the set of $0^{++}$ and $2^{++}$ cannot be excluded 
since fits to the data have a good statistical significance. 
This is what the results of Ref.~\cite{Torres:2016oyz} 
seem to indicate, i. e., that more than one resonance may 
contribute around $4.1$ GeV in the data of 
Ref.~\cite{Aaij:2016iza,Aaij:2016nsc}. Recently, 
this conclusion was reinforced by the study 
conducted in Ref.~\cite{Wang:2017mrt}. So far  
no fit to the data of Ref.~\cite{Aaij:2016iza,Aaij:2016nsc} 
with more than one resonance around $4.1$ GeV has been done. 
According to the reasoning made in Ref.~\cite{Wang:2017mrt} 
the peak seen by LHCb in the $J/\psi\phi$ spectrum can be fitted 
using more than one resonance in the $4.1$ GeV region. 
More concretely, the authors claim that in the LHCb data there is a lack 
of information related to the $X(4160)$ contribution that, if 
included in the analysis, would provide a narrow width 
 for  $Y(4140)$.

 \subsection{Summary for the Controversial $Y$ states}
 From the QCDSR studies presented in this section it is possible to explain not only the mass, but also the decay width of the $Y(3940)$ (or $X(3915)$)
 considered as a mixed
 $(\chi_{c0})\!-\!(D^\ast  \!\bar{D}^\ast)$ state, with a mixing angle:
 $\theta=(76.0\pm  5.0)^0$.

 Regarding the $Y(4140)$,
 considering the observations and non-confirmations of its existence,
 from the experimental point of view it is very important to make
 an effort to determine if the state $Y(4140)$ really exists. Supposing it
 exists, with a QCDSR calculation  it is possible to explain its mass either
 as a $D^*\bar{D}^*$ or a $D_s^*\bar{D}_s^*$, with quantum numbers $J^{PC}=0^{++}$ or $2^{++}$. However, the analysis done for the decay width presented above
 suggest that the $D_s^*\bar{D}_s^*$ molecular current
with $2^{++}$ is more likely related with  the charmonium-like 
state $X(4160)$, than to  the $Y(4140)$ one.

\section{Higher Order Perturbative Corrections in Sum Rules}

In previous sections, all the works done with the QCDSR method take into account approximations 
at Leading Order (LO) of perturbative QCD and include non-perturbative condensates in the OPE, to study the masses and coupling constants of the $XYZ$-states. 
In the literature we have many successful examples of using this approach at LO to explain the 
properties of such states.
A natural improvement of such a method would be including, whenever possible, up to 
Next-to-Leading Order (NLO) corrections in perturbative QCD and estimate their relevance 
for the available LO sum rule calculations. 

In this section, we explain how to include up to order $\alpha_s^2$ (N2LO) perturbative corrections in the chiral limit and, 
adding to these, the SU(3) NLO corrections to the heavy-light exotic (molecules and 
tetraquarks) correlators. In doing so, we assume the factorization of the four-quark operator into a 
convolution of two-quark operators, built from bilinear quark currents. The 
contributions of the unknown order $\alpha_s^3$ (N3LO) correction are estimated from a geometric 
growth of the perturbative series\,\cite{Narison:2009ag} and are added as a source of systematic 
uncertainties in the truncation of the perturbative series. For a complete description on these NLO 
techniques in QCDSR, see~\cite{Albuquerque:2016znh,Albuquerque:2017vfq} and references therein.

\subsection{NLO Sum Rules}
The conventional use of QCDSR obviously suffers from the ill-defined heavy quark mass definition 
used at LO. The favored numerical input values for the charm quark mass, 
$m_c (m_c) \approx (1.23-1.26)$~GeV, used in the current literature correspond numerically to the one 
of the running masses. However, there is no reason to discard values like $m_c\approx 1.5$ GeV of the 
on-shell (pole) quark masses, which are more natural because the spectral functions have been evaluated 
using the on-shell heavy quark propagator. For this reason, the perturbative expression of the spectral 
function obtained using on-shell renormalization must be transformed into the $\overline{MS}$-scheme 
by using the relation between the $\overline{MS}$ running mass, $\overline{m}_c(\mu)$, and the on-shell 
mass (pole), $M_c$, to order $\alpha_s^2$
\cite{Tarrach:1980up, Coquereaux:1979eq, Binetruy:1979hc, Narison:1987qh, Narison:1988xi, 
Gray:1990yh, Fleischer:1998dw, Chetyrkin:1999qi}:
\begin{eqnarray}
M_c &=& \overline{m}_c(\mu)\bigg{[}
1+\frac{4}{3} a_s+ (16.2163 -1.0414 n_l)a_s^2
+\ln{\left( \frac{\mu}{ M_c} \right)^2} \left( a_s+(8.8472 -0.3611 n_l) a_s^2 \right) \nnb\\
&&+\ln^2{\left( \frac{\mu}{ M_c} \right)^2} \left( 1.7917 -0.0833 n_l\right) a_s^2 + ...\bigg{]},
\label{eq:pole}
\end{eqnarray}
for $n_l$ light flavors where $\mu$ is the arbitrary subtraction point and $a_s \equiv \alpha_s/\pi$.

To extract the perturbative $\alpha_s^n$ corrections to the correlator of
a four-quark current, and due to the technical 
complexity of the calculations, we shall assume that these radiative corrections are dominated by 
the ones from the factorized diagrams shown in Figs.\,\ref{fig:factor}a,b, while we neglect the ones from 
non-factorized diagrams in Figs.\,\ref{fig:factor}c-f. This fact has been proven explicitly in
Refs.~\cite{Narison:1994zt,Hagiwara:2002hf}, in the case of the $\bar B^0B^0$ systems (very similar 
correlator as the ones discussed in the following)~, where the non-factorized $\alpha_s$ corrections 
do not exceed 10\% of the total $\alpha_s$ contributions. As can be seen in 
Refs.~\cite{Albuquerque:2016znh,Albuquerque:2017vfq}, the effect of factorization in a sum rule 
calculation at LO is about 2.2\% for the decay constant, and 0.5\% for the mass which is quite tiny. 
However, to avoid this (small) effect, we shall work in the following with the full non-factorized 
perturbative $\oplus$ condensates of the LO expressions. For the NLO expressions, we work with 
the spectral function as a convolution of the ones associated to quark bilinear current, as illustrated 
by the Feynman diagrams in Fig.\,\ref{fig:factor}. In this way, we use the low-energy representation 
of the effective spectral function, as suggested in Ref.~\cite{Pich:1985ab}, to obtain the expression:
\begin{eqnarray}
  \frac{1}{\pi} {\rm Im} \:\Pi^{(1)}(s) &=& \theta (s - 4 M_c^2) \left( \frac{s}{4\pi} \right)^2
  \int\limits_{M_c^2}^{(\sqrt{s}-M_c)^2} \!\!\!\!\!\!ds_1 ~\frac{1}{\pi}{\rm Im} \:\Pi^{(1)}(s_1) 
  \!\!\int\limits_{M_c^2}^{(\sqrt{s}-\sqrt{s_1})^2} \!\!\!\!\!\!ds_2
  ~\frac{1}{\pi}{\rm Im} \:\Pi^{(0)}(s_2) \cdot \lambda^{3/2},
\label{eq:convolution}
\end{eqnarray}
for the $J=1$ states, and
\begin{eqnarray}
  \frac{1}{\pi}{\rm Im} \:\Pi_A^{(0)}(s) &=& \theta(s-4M_Q^2) \left( \frac{s}{4\pi} \right)^2 
  \int\limits_{M_c^2}^{(\sqrt{s}-M_c)^2} \!\!\!\!\!\!ds_1 ~\frac{1}{\pi}{\rm Im} \:\Pi^{(0)}(s_1)
  \!\!\int\limits_{M_c^2}^{(\sqrt{s}-\sqrt{s_1})^2} \!\!\!\!\!\!ds_2  
  ~\frac{1}{\pi}{\rm Im} \:\Pi^{(0)}(s_2) \cdot \lambda^{1/2} \left( \frac{s_1}{s} \!+\! \frac{s_2}{s} \!-\! 1 \right)^2, \\
  && \nnb\\
  \frac{1}{\pi}{\rm Im} \:\Pi_B^{(0)}(s) &=& \theta(s-4M_Q^2) \left( \frac{s}{4\pi} \right)^2 
  \int\limits_{M_c^2}^{(\sqrt{s}-M_c)^2} \!\!\!\!\!\!ds_1 ~\frac{1}{\pi}{\rm Im} \:\Pi^{(1)}(s_1)
  \!\!\int\limits_{M_c^2}^{(\sqrt{s}-\sqrt{s_1})^2} \!\!\!\!\!\!ds_2  
  ~\frac{1}{\pi}{\rm Im} \:\Pi^{(1)}(s_2) \cdot 
  \lambda^{1/2} \left[ \left( \frac{s_1}{s} \!+\! \frac{s_2}{s} \!-\! 1 \right)^2  + \frac{8s_1 s_2}{s^2} \right], 
  ~~~~~~
\end{eqnarray}
for the $J=0$ states. We use the definition
\begin{eqnarray}
  \lambda &\equiv& \left( 1 - \frac{\left( \sqrt{s_1} - \sqrt{s_2} \right)^2}{s} \right) 
  \left( 1 - \frac{\left( \sqrt{s_1} + \sqrt{s_2} \right)^2}{s} \right),
\end{eqnarray}
which is the phase space factor. The invariant spectral function, Im $\Pi^{(1)}(s)$, is associated to 
the bilinear $\bar{c} \gamma_\mu q$ vector or $\bar{c} \gamma_\mu \gamma_5 q$ axial-vector current, while 
Im $\Pi^{(0)}(s)$ is associated to the $\bar{c} q$ scalar or  $\bar{c} \gamma_5 q$ pseudoscalar current. 
Notice that, in the limit where the light quark mass $m_q \to 0$, the perturbative expressions of the vector 
(scalar) and axial-vector (pseudoscalar) spectral functions are the same.
These representations allow us to evaluate the perturbative $\alpha_s^n$-corrections of 
the spectral functions of heavy-light bilinear currents, since we can use the expressions that are already known to order $\alpha_s$ 
(NLO) from Ref.~\cite{Broadhurst:1981jk}. N2LO corrections are known in the chiral limit, 
$m_q=0$, from Ref.~\cite{Chetyrkin:2000mq,Chetyrkin:2001je}. We use
the NLO SU(3) breaking perturbative corrections obtained in\,\cite{Gelhausen:2013wia} 
from the two-point function formed by bilinear currents. From the above representation, the anomalous 
dimensions of the molecular correlators come from the (pseudo)scalar current. Therefore, the corresponding 
renormalization group invariant interpolating current reads to NLO as:
\begin{eqnarray}\lb{ano-mol}
  \bar {\cal O}^{(0)}_{mol}(\mu) ~=~ a_s(\mu)^{4/\beta_1} {\cal O}^{(0)}_{mol}~, &~~~~~~~~&
  \bar {\cal O}^{(1)}_{mol}(\mu) ~=~ a_s(\mu)^{2/\beta_1} {\cal O}^{(1)}_{mol}~,  
\end{eqnarray}
where $\beta_1=-(1/2)(11-2n_f/3)$ is the first coefficient of the QCD $\beta$-function for $n_f$ flavors, and 
$a_s\equiv (\alpha_s/\pi)$. 
The spin-0 currents built from two (axial)-vector currents have no anomalous dimension. We have 
introduced the super-index $(1)$ for denoting the vector and axial-vector spin-1 channels.

\begin{figure}[t] 
\begin{center}
{\includegraphics[width=9cm]{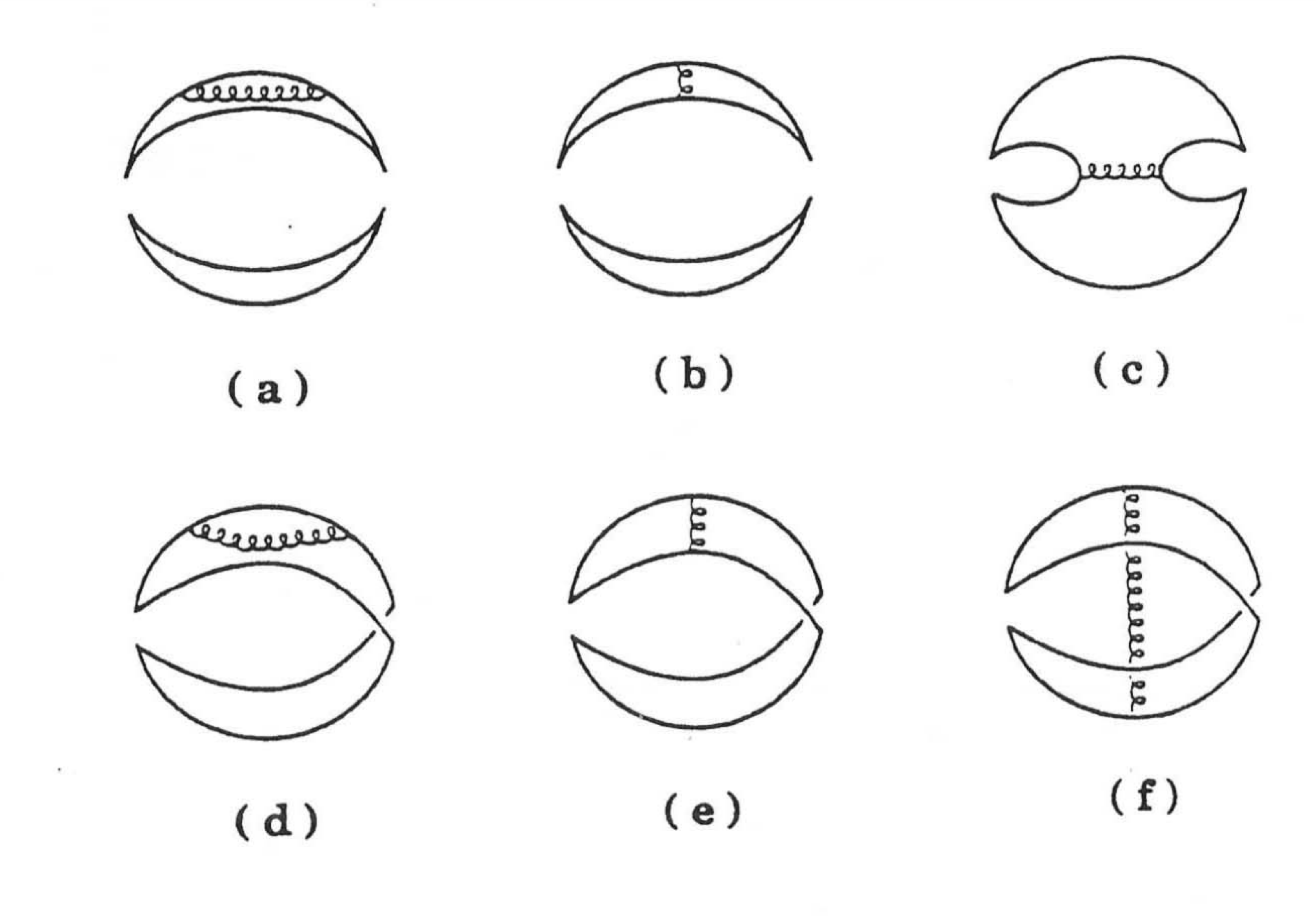}}
\caption{\scriptsize 
{\bf (a)} Factorized contributions to the four-quark correlator at NLO of perturbative contribution; 
{\bf (b)} Non-factorized diagrams at NLO of perturbative contribution.
Taken from\,\cite{Pich:1985ab}.
}
\label{fig:factor}
\end{center}
\end{figure} 

Another aspect of including higher order perturbative corrections is related with the $1/q^2$ corrections due to a 
tachyonic gluon mass discussed in \cite{Chetyrkin:1998yr, Narison:2001ix}. However, such corrections will 
not be included here because they are dual to the sum of the large order 
perturbative series\,\cite{Narison:2009ag}. Therefore, we shall consider
the inclusion of the N3LO terms, estimated from 
the geometric growth of the QCD perturbative series \cite{Narison:2009ag}, as a source of the errors. The estimate of these errors 
is given in Refs.~\cite{Albuquerque:2016znh,Albuquerque:2017vfq}.
We are still using the ansatz of pole plus continuum, in Eq.~(\ref{eq:duality}), for parametrizing the spectral function (generic notation):
\begin{eqnarray}
  \frac{1}{\pi}\mbox{Im} \:\Pi(s) &\simeq& f_H^2 M_H^8 \:\delta(s - M_H^2) \:+\: 
  ``\mbox{QCD continuum}" ~\theta(s - s_0),
  \label{eq:duality}  
\end{eqnarray}
where $f_H$ is the hadronic decay constant defined as:
\begin{eqnarray}
  \langle 0| \:{\cal O}^{(0)}\: | H \rangle ~=~ f_H^{(0)} M_H^4 ~, &~~~~~~~~~~~~&
  \langle 0| \:{\cal O}_{\mu}^{(1)}\: | H \rangle ~=~ f_H^{(1)} M_H^5 \,\epsilon_{\mu}~,
\label{eq:coupling}  
\end{eqnarray}
respectively, for spin 0 and 1 hadronic states, with  $\epsilon^{\,\mu}$ being the vector polarization.
The higher state contributions are given by the ``QCD continuum" coming from the discontinuity 
of the QCD diagrams and starting from a constant threshold $s_0$. Finite width corrections to this simple model 
have been studied in, e.g.,\,\cite{Lee:2008tz,Narison:1988ts, Narison:1996fm} and have been found to be negligible. 

Noting that, due to Eq.~(\ref{ano-mol}), the bilinear (pseudo)scalar currents, 
in Table \ref{tab:current}, 
acquires an anomalous dimension due to its normalization, thus the decay constants run to order 
$\alpha_s^2$ as:
\begin{eqnarray}
  f^{(0)}_{mol}(\mu) ~=~ \hat f^{(0)}_{mol} \left( -\beta_1 a_s \right)^{4/\beta_1}/r_m^2~, &~~~~&
  f^{(1)}_{mol}(\mu) ~=~ \hat f^{(1)}_{mol} \left( -\beta_1 a_s \right)^{2/\beta_1}/r_m~,
\label{eq:fhat}  
\end{eqnarray}
where  the 
QCD corrections ($r_m$) numerically read to N2LO as \cite{Narison:2007spa, Gray:1990yh}:
\begin{eqnarray}
  r_m(n_f = 4) &=& 1+1.014 \,a_s +1.389 \,a_s^2~.
\end{eqnarray}
In Eq.~(\ref{eq:fhat}) we have introduced the renormalization group invariant coupling $\hat f_{mol}$ and the first 
coefficient of the QCD $\beta$-function for $n_f$ flavors, $a_s \equiv \alpha_s / \pi$.
Notice that the coupling of the (pseudo)scalar molecule built from two (axial)-vector currents 
has no anomalous dimension and does not run.

As usual in a QCDSR calculation, to obtain the hadronic masses we  use the expression:
\begin{eqnarray}\label{eq:ratioLSR}
  M_H^2 &\simeq& \frac{\int_{4m_c^2}^{s_0} ds ~s ~e^{-s/M^2} 
  ~\mbox{Im} \:\Pi(s,\mu)}
  {\int_{4m_c^2}^{s_0} ds ~e^{-s/M^2} ~\mbox{Im} \:\Pi(s,\mu)},  
\end{eqnarray}
where $\mu$ is the subtraction point which appears in the approximate QCD series when radiative 
corrections are included.  Since our studies 
are concentrated on charmonium-like states, the lower limit integral in Eq.\,(\ref{eq:ratioLSR}) is given by 
the square of constituent quark masses of these hadrons.

In this section, we shall add to the previous well-known $M^2$- and $s_0$-stability criteria, discussed in Sec.~\ref{sr-sta}, the one 
associated with the requirement of stability versus the arbitrary subtraction constant $\mu$. The $\mu$-stability procedure has been applied recently in\,
Refs.~\cite{Narison:2012xy, Narison:2014vka, Narison:2014ska, Narison:2018dcr}. It gives 
a much better meaning of the choice of $\mu$-value at which the observable is extracted. The errors 
in the determination of the results have been reduced due to a better control of the $\mu$ region of variation.

\subsection{QCD Input Parameters at NLO}
The QCD parameters appearing in the following analysis will be the charm quark mass $m_{c}$, 
the strange quark mass $m_s$ (we shall neglect the light quark masses $m_{u,d}$),
the light quark condensates $\qq[q]$ ($q\equiv u,d,s$), the gluon condensates $\GG$ and 
$\GGG$, the mixed quark condensate $\qGq[q]$ and the four-quark condensate $\rho\alpha_s \qq[q]^2$, 
where $\rho \simeq 3-4$ indicates the deviation from the four-quark vacuum saturation. Their values are 
given in Tables \ref{QCDParam} and  \ref{tab:param}. We shall work with the running light quark condensates which, in leading 
order in $\alpha_s$, is given by: 
\begin{equation}
  {\qq[q]}(M_\tau) = -{\qq[q]} \left( -\beta_1 \,a_s(M_\tau) \right)^{2/{\beta_1}},~~~~~~~~
  {\qGq[q]}(M_\tau) = -{m_0^2 \:{{\qq[q]}} \left( -\beta_1 \,a_s(M_\tau) \right)^{1/{3\beta_1}}}~,
  \label{d4g}
\end{equation}
and the running strange quark mass to NLO (for the number of flavours $n_f=3$):
\begin{equation}
  \overline{m}_s(M_\tau)={\hat m_s} \left( -\beta_1 \,a_s(M_\tau) \right)^{-2/{\beta_1}}(1+0.8951\,a_s(M_\tau))~,
  \label{ms}
\end{equation}
where  ${\qq[q]}$ (given in Table~\ref{QCDParam}) and $ \hat m_s$ are the spontaneous RGI light quark 
condensate and strange quark mass\,\cite{Floratos:1978jb}.
For the heavy quarks, we shall use the running mass and the corresponding value of $\alpha_s$ evaluated 
at the scale $\mu$, whose value used here corresponds to the optimal one obtained 
in\, Ref.~\cite{Albuquerque:2016znh}. 

For the $\langle \alpha_s G^2 \rangle$ condensate, we have enlarged the original error by a factor about 3 
in order to have a conservative result for recovering the original SVZ estimate and the alternative extraction 
in Ref.~\cite{Ioffe:2002be,Ioffe:2005ym} from charmonium sum rules. However, a direct naive comparison of this range of 
values obtained within short QCD series (few terms) with the one from lattice calculations \cite{Bali:2014sja} 
obtained within a long QCD series\,\cite{Lee:2010hd} can be misleading. 
We shall see later on that the effects of the gluon and four-quark condensates on the values of the decay 
constants and masses are relatively small even though they play an important role in the stability analysis. 

{\scriptsize
\begin{table}[t]
 \begin{center}
 \setlength{\tabcolsep}{1.25pc}
 \caption{QCD input parameters used in sum rules at NLO.}
 {\small
 \def\arraystretch{1.25}
 \begin{tabular}{lll}
 &\\
 \hline
 Parameters & Values & Ref. \\
 \hline
 $\alpha_s(M_\tau)$& $0.325 \pm 0.008$&\cite{Narison:2009vy,Braaten:1991qm, Narison:1988ni}\\
 $\hat m_s$&$(0.114\pm0.006)$ GeV &\cite{Narison:2007spa, Narison:1988xi,Narison:2014vka, 
 Narison:2009vy,Narison:2005ny, Narison:1999mv,Dosch:1997wb}\\
 $\overline{m}_c({m}_c)$&$(1261 \pm 12)$ MeV 
 &average \cite{pdg, Ioffe:2002be, Ioffe:2005ym, Narison:2010cg, Narison:2011xe, Narison:2011rn}\\
 $\langle \alpha_s G^2 \rangle$& $(7\pm 3)\times 10^{-2}$ GeV$^4$&
 \cite{Narison:2004vz, Bertlmann:1983pf, Bertlmann:1984rs, Narison:2009vy, Narison:2010cg, 
 Narison:2011xe, Narison:2011rn, Launer:1983ib, Narison:1992ru, Narison:1995jr, Yndurain:1999pb, 
 Narison:1995tw}\\
 $\GGG$& $(8.2\pm 2.0)$ GeV$^2\times\langle \alpha_s G^2 \rangle$& 
 \cite{Narison:2010cg, Narison:2011xe, Narison:2011rn}\\
 $\rho \alpha_s \qq[q]^2$&$(5.8\pm 1.8)\times 10^{-4}$ GeV$^6$&\cite{Narison:2009vy, 
 Chung:1984gr, Dosch:1988vv, Launer:1983ib}\\
 \hline
 \hline
\end{tabular}}
\label{tab:param}
\end{center}
\end{table}
}

\subsection{Heavy-Light Molecular States}

For describing these molecular states, we consider the usual lowest dimension local interpolating 
currents where each bilinear current has the quantum number and quark content of the open-charm mesons:
\begin{equation}
  D(0^-)\:, ~~D^\ast_0(0^+)\:, ~~D^*(1^-)\:, ~~D_1(1^+)
\end{equation}
and their respective extension to the strange sector:
\begin{equation}
  D_s(0^-)\:, ~~D_{s0}^\ast(0^+)\:, ~~D_s^*(1^-)\:, ~~D_{s1}(1^+) \:.
\end{equation}
For simplicity, we do not consider colored and more general combinations of interpolating operators 
discussed e.g in Ref.~\cite{Dias:2013xfa, Chen:2015ata}, as well as the higher dimension ones involving derivatives.  This choice justifies the approximate 
use (up to order $1/N_c$) of the factorization of the four-quark currents as a convolution of two bilinear 
quark-antiquark currents when estimating higher order perturbative corrections. These states and the 
corresponding interpolating currents, for the states studied in this review, are given in Table \ref{tab:current}. A more general study can be found in Refs.~\cite{Albuquerque:2016znh,Albuquerque:2017vfq}.
\begin{figure}[b] 
\begin{center}
{\includegraphics[width=6.5cm  ]{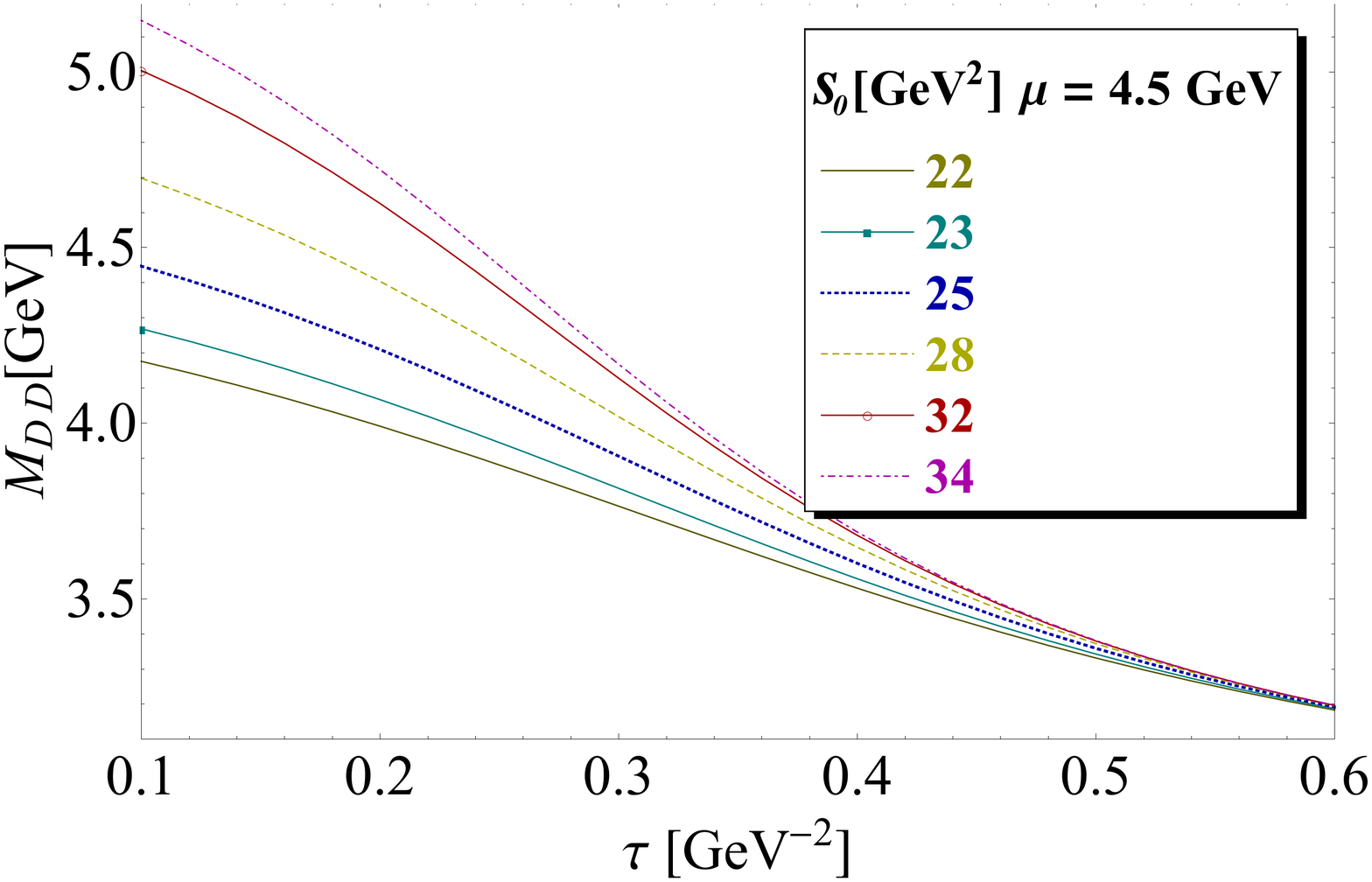}}
\caption{
\scriptsize 
 $M_{DD}$ at LO as function of $1/M^2$ for different values of $s_0$, for $\mu=4.5$~GeV 
and for the QCD parameters in Table\,\ref{tab:param}. Figure taken from \cite{Albuquerque:2016znh}.}
\label{fig:d-lo} 
\end{center}
\end{figure} 
\noindent

{\scriptsize
\begin{table}[t]
\begin{center}
\setlength{\tabcolsep}{1.8pc}
\newlength{\digitwidth} \settowidth{\digitwidth}{\rm 0}
\catcode`?=\active \def?{\kern\digitwidth}

\caption{ Interpolating currents with a definite $C$-parity describing the $\bar DD$- and $\bar D_s D_s$-like 
molecular states for $J^{PC}=0^{++}, ~1^{++}$ and $1^{--}$. $q\equiv u,d,s$.}
{\small
\def\arraystretch{1.25}
\begin{tabular}{lll}
&\\
\hline
$J^{PC}$& States & Molecule Currents  $\equiv{\cal O}_{mol}(x)$  \\
\hline
\\
$\bf 0^{++}$& $\bar DD\:, ~~~\bar D_s D_s$ 
&$( \bar{q} \gamma_5 c ) (\bar{c} \gamma_5 q)$ \\
&$\bar D^\ast D^\ast \:, ~~~\bar D_s^\ast D_s^\ast$
&$( \bar{q} \gamma_\mu c ) (\bar{c} \gamma^\mu q)$ \\ 
&$\bar D^\ast_0 D^\ast_0\:, ~~~\bar D_{s0}^\ast D_{s0}^\ast$ 
& $( \bar{q} c ) (\bar{c} q)$\\ 
&$\bar D_{1} D_{1}$\:, ~~~$\bar D_{s1} D_{s1}$ 
&$( \bar{q} \gamma_\mu \gamma_5 c ) (\bar{c} \gamma^\mu \gamma_5 q)$ \\ 
\\
$\bf 1^{++}$& $\bar D^*D\:, ~~~\bar D_s^\ast D_s$ 
&$ \frac{i}{\sqrt{2}} \Big[ (\bar{c} \gamma_\mu q) ( \bar{q} \gamma_5 c ) 
     - (\bar{q} \gamma_\mu c) ( \bar{c} \gamma_5 q ) \Big]$ \\ 
&$\bar D_{0}^\ast D_{1}\:, ~~~\bar D_{s0}^\ast D_{s1}$
&$ \frac{1}{\sqrt{2}} \Big[ ( \bar{q} c ) (\bar{c} \gamma_\mu \gamma_5 q)
     + ( \bar{c} q ) (\bar{q} \gamma_\mu \gamma_5 c) \Big]$\\ 
\\
%
$\bf 1^{--}$& $\bar D_0^\ast D^\ast\:, ~~~\bar D_{s0}^\ast D_s^\ast$
&$ \frac{1}{\sqrt{2}} \Big[ ( \bar{q} c ) (\bar{c} \gamma_\mu q) 
     + ( \bar{c} q ) (\bar{q} \gamma_\mu c) \Big]$\\
& $\bar D D_1\:, ~~~\bar D_s D_{s1}$ &
$ \frac{i}{\sqrt{2}} \Big[ ( \bar{c} \gamma_\mu \gamma_5 q ) (\bar{q} \gamma_5 c)
     - ( \bar{q} \gamma_\mu \gamma_5 c ) (\bar{c} \gamma_5 q) \Big]$\\
     \\
\hline
\hline
\end{tabular}}
\label{tab:current}
\end{center}
\end{table}
}

\begin{figure}[t] 
\begin{center}
{\includegraphics[width=6.5cm  ]{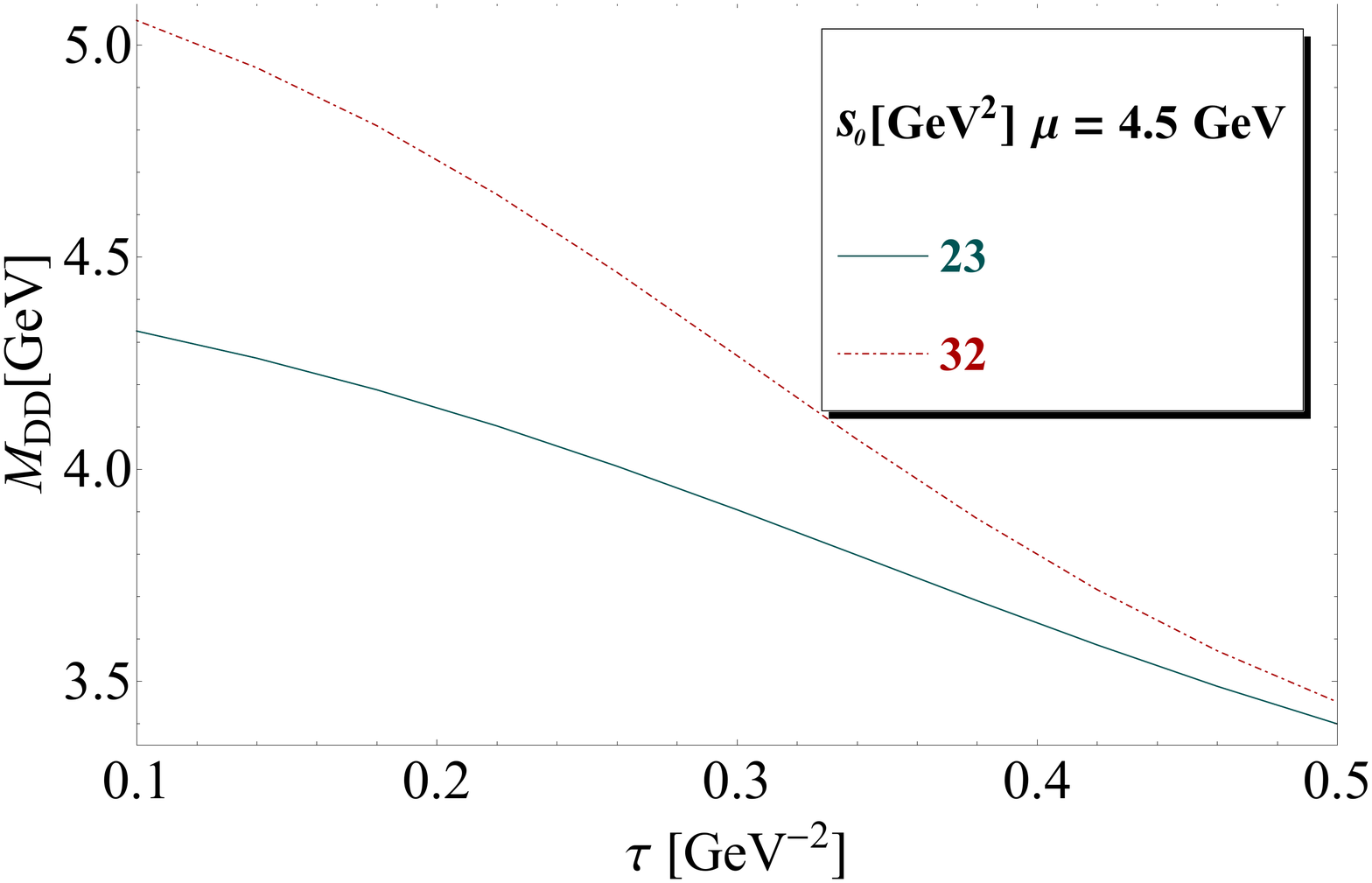}} \hspace{1cm}
{\includegraphics[width=6.2cm  ]{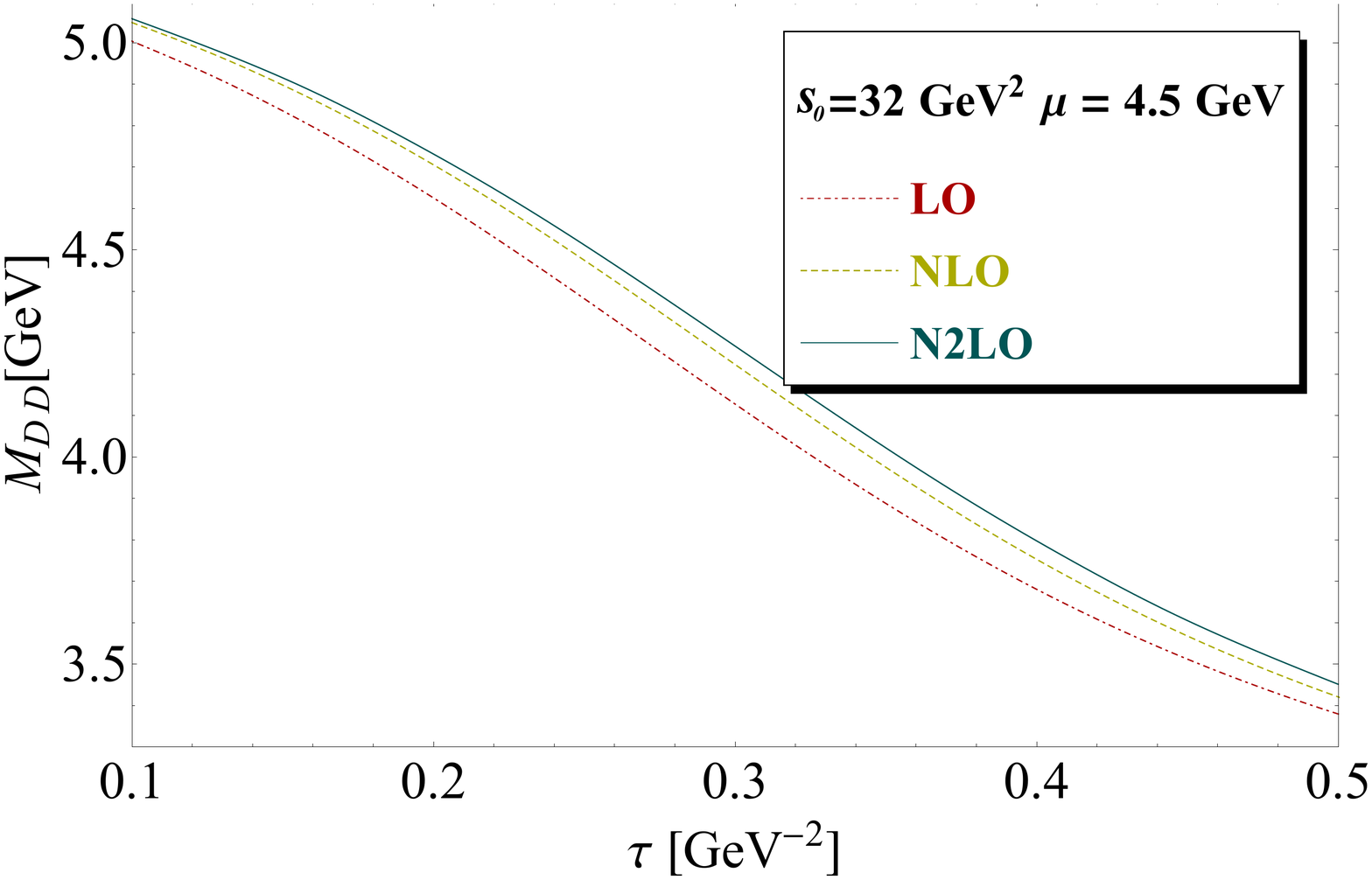}}
\centerline{\hspace*{-5cm} a)  \hspace*{7.5cm} b)}
\caption{
\scriptsize 
{\bf a)} $M_{DD}$ at N2LO as function of $\tau=1/M^2$ for different values of $s_0$, for $\mu=4.5$~GeV
and for the QCD parameters in Table\,\ref{tab:param}; 
{\bf b)} The same as a) but for a given value of $s_0=32$~GeV$^2$, and for 
different truncation of the perturbative series. Figures taken from \cite{Albuquerque:2016znh}.
}
\label{fig:d-n2lo} 
\end{center}
\end{figure} 
\noindent

Next, we show the techniques and strategies for evaluating the mass and decay constants of the 
$\bar DD$ molecular state. Essentially, the sum rule analysis for the other interpolating 
currents are very similar. Therefore, we only quote the results for the remaining states. 
We start with the mass, $M_{DD}$, obtained with the $\bar DD$ molecular current, at 
LO approximation. We show in Fig.\ref{fig:d-lo} the results in terms of the Borel mass variable 
$\tau= 1/M^2$ at different values of continuum threshold $s_0$. Then we implement the 
higher order perturbative corrections to this sum rule and the results are shown in 
Fig.\,\ref{fig:d-n2lo}a  at N2LO.

We consider as an optimal choice the mean value of the coupling and mass obtained at the 
minimum or inflection point for the common range of $s_0$-values corresponding to the starting of 
the $\tau$-stability and the one where $s_0$-stability is reached. In these stability regions, the 
requirement that the pole contribution is larger than the one of the continuum is automatically 
satisfied (see e.g.\,\cite{Nielsen:2009uh}). Therefore, from the Fig.\,\ref{fig:d-n2lo}a at N2LO, we obtain the 
range for the continuum threshold as $s_0 \simeq  (27.5 \pm 4.5)$~GeV$^2$, for 
$1/M^2= (0.30 \pm 0.05)$~GeV$^{-2}$.
Another interesting point discussed in Refs.~\cite{Albuquerque:2016znh,Albuquerque:2017vfq} is about 
the running versus the pole quark mass definitions in QCDSR at LO. It was pointed out that the effect of 
the definitions (running and pole) of the heavy quark mass should be added as errors in the LO analysis.

As we can see in Fig. {\ref{fig:d-n2lo}}b, using $s_0 = 32$~GeV$^2$, the convergence of the perturbative 
series is obtained for an optimal choice: $\mu = (4.5 \pm 0.5) ~\mbox{GeV}$. We observe that from NLO 
to N2LO the mass of the state decreases by about only 0.1~\%, as another indication of the good 
convergence of the perturbative series. Using the fact that the final result must be independent of 
the arbitrary parameter $\mu$ (plateau / inflection point for the coupling and minimum for the mass), 
we consider as an optimal result the one at $\mu\simeq 4.5$ GeV, where we deduce the result at N2LO, 
\begin{eqnarray}
  M_{DD} ~=~ (3898 \pm 36) ~\mbox{MeV}, &~~~~\mbox{and}~~~~&
  f_{DD} ~=~ (170 \pm 15), ~\mbox{keV} 
\end{eqnarray}
which is included in Table\,\ref{tab:resultc}. Notice that the mass obtained for the $\bar DD$ molecular 
state is above the $\bar DD$-threshold ($\sim 3729$~MeV) and, for this reason, such a molecule would not 
be consistent with a bound state. Therefore, from a QCDSR point of view, we can not use the 
$\bar DD$ molecular state to describe any new $XYZ$ states observed so far.

We proceed with our 
investigation for other molecular currents in Table~\ref{tab:current}, where we apply the same 
analysis for extracting the optimal values for the masses and couplings at N2LO approximation. 
Then we check if the molecular currents could describe some of these $XYZ$ states. The final results 
are summarized in Table\,\ref{tab:resultc}.
{\scriptsize
\begin{table}[t]
\setlength{\tabcolsep}{0.8pc}
\caption{The masses and couplings of the $\bar DD$- and $\bar D_s D_s$-like Molecular states from 
QCDSR within stability criteria at N2LO of perturbative series. The value of the continuum threshold, $s_0$, is also included for completeness.}
\vspace{0.2cm}
\begin{center}
{\small
\def\arraystretch{1.25}
 \begin{tabular}{lccc}
 \hline
 States & $\sqrt{s_0}$ & Mass & Coupling \\
 $J^{PC}$& (GeV) & (MeV) & (keV) \\
 \hline
 $\bf 0^{++}$ \\
 $\bar DD$& $4.7 \sim 5.7$ & $3898 \pm 36$ & $170 \pm 15$\\
 $\bar D^\ast D^\ast$& $4.7 \sim 5.7$ & $3903 \pm 179$ & $302 \pm 47$ \\
 $\bar D_0^\ast D_0^\ast$& $4.7 \sim 5.7$ & $3954 \pm 224$ & $114 \pm 18$ \\
 $\bar D_1 D_1$& $4.7 \sim 5.7$ & $3784 \pm 56$ & $274 \pm 37$ \\
\\
 $\bf 1^{+\pm}$ \\
 $\bar D^\ast D$ & $4.7 \sim 5.7$ & $3903 \pm 62$ & $161 \pm 17$ \\
 $\bar D_0^\ast D_1$ & $6.5 \sim 6.9$ & $3854 \pm 182$ & $112 \pm 17$ \\
\\
 $\bf 1^{--}$ \\
 $\bar D_0^\ast D^\ast$ & $6.5 \sim 6.9$ & $5748 \pm 101$ & $261 \pm 17$ \\
 $\bar D D_1$ & $6.5 \sim 6.9$ & $5544 \pm162$ & $231 \pm 21$ \\
\hline
\hline
\end{tabular}}
\quad \quad
{\small
\def\arraystretch{1.25}
 \begin{tabular}{lcccc}
 \hline
 States & $\sqrt{s_0}$ & Mass & Coupling \\
 $J^{PC}$& (GeV) & (MeV) & (keV) \\
 \hline
 $\bf 0^{++}$ \\
 $\bar D_s D_s$ & $5.0 \sim 6.9$ & $4169 \pm 48$ & $167 \pm 18$ \\
 $\bar D_s^\ast D_s^\ast$ & $5.0 \sim 6.9$ & $4196 \pm 200$ & $284 \pm 34$ \\
 $\bar D_{s0}^\ast D_{s0}^\ast$ & $5.5 \sim 7.0$ & $4225 \pm 132$ & $102 \pm 14$ \\
 $\bar D_{s1} D_{s1}$ & $5.3 \sim 6.9$ & $4124 \pm 61$ & $229 \pm 31$ \\
\\
 $\bf 1^{+\pm}$ \\
 $\bar D_s^\ast D_s$ & $5.0 \sim 6.9$ & $4188 \pm 67$ & $156 \pm 17$ \\
 $\bar D_{s0}^\ast D_{s1}$ & $5.5 \sim 6.9$ & $4275 \pm 206$ & $110 \pm 18$ \\
\\
 $\bf 1^{--}$ \\
 $\bar D_{s0}^\ast D_s^\ast$ & $6.0 \sim 7.5$ & $5571 \pm 180$ & $216 \pm 11$ \\
 $\bar D_s D_{s1}$ & $6.2 \sim 7.5$ & $5272 \pm 120$ & $213 \pm 13$ \\
%
\hline
\hline
\end{tabular}}
\end{center}
\label{tab:resultc}
\end{table}
}

\subsection{Heavy-Light Tetraquark States}

The tetraquark states were first introduced in Ref.~\cite{Jaffe:1976ig} for interpreting the complex 
spectra of light scalar mesons. Recent analysis based on 
$1/N_c$ expansion has shown that the tetraquark states should be narrow\,\cite{Weinberg:2013cfa,
Knecht:2013yqa, Rossi:2016szw}, which do not then favor the tetraquark interpretation of the light scalar 
meson $f_0(500)$. In the following, we also study the tetraquark currents to investigate possible hadronic structures to 
describe the $XYZ$ states in the charmonia spectra. The tetraquark states $\big[ \,c q \bar{c} \bar{q} \,\big]$ 
will be described by the interpolating currents given in Table\,\ref{tab:4qcurrent}. A more general study can be found in Refs.~\ref{Albuquerque:2016znh,Albuquerque:2017vfq}}.    We use the following 
naming scheme for the scalar, pseudoscalar, axial-vector and vector states for the tetraquark currents:
\begin{equation}
  S_c\,(0^+)\:, ~~A_c\,(1^-)\:, ~~V_c\,(1^+)
\end{equation}
and their respective extension to the strange sector:
\begin{equation}
  S_{cs}\,(0^+)\:, ~~A_{cs}\,(1^+)\:, ~~V_{cs}\,(1^-) ~~.
\end{equation}
{\scriptsize
\begin{table}[t]
\setlength{\tabcolsep}{2.2pc}
\caption{ Interpolating currents with a definite $P$-parity describing the tetraquark states. $q\equiv u,d,s$. }
\vspace{0.2cm}
\begin{center}
{\small
\def\arraystretch{1.25}
\begin{tabular}{cll}
\hline
$J^{P}$ & States & Tetraquark Currents  $\equiv{\cal O}_{4q}(x)$  \\
\hline
\\
$\bf 0^{+}$ & $S_c\:, ~~~S_{cs}$ & $ \epsilon_{abc}\epsilon_{dec}  \bigg[
		\big( q^T_a \: C\gamma_5 \:c_b \big) \big( \bar{q}_d \: \gamma_5 C \: \bar{c}^T_e \big) 
		+ k \big( q^T_a \: C \:c_b \big) \big( \bar{q}_d \: C \: \bar{c}^T_e \big) \bigg] $\\		
$\bf 1^{+}$ & $A_c\:, ~~~A_{cs}$ & $\epsilon_{abc}\epsilon_{dec}  \bigg[
		\big( q^T_a \: C\gamma_5 \:c_b \big) \big( \bar{q}_d \: \gamma_\mu C \: \bar{c}^T_e \big) 
		+ k \big( q^T_a \: C \:c_b \big) \big( \bar{q}_d \: \gamma_\mu\gamma_5 C \: \bar{c}^T_e \big) \bigg]  $ \\
%
$\bf 1^{-}$ & $V_c\:, ~~~V_{cs}$ & $  \epsilon_{abc}\epsilon_{dec}  \bigg[
		\big( q^T_a \: C\gamma_5 \:c_b \big) \big( \bar{q}_d \: \gamma_\mu\gamma_5 C \: \bar{c}^T_e \big) 
		+ k \big( q^T_a \: C \:c_b \big) \big( \bar{q}_d \: \gamma_\mu C \: \bar{c}^T_e \big) \bigg]$\\
		\\
\hline
\hline
\end{tabular}
}
\label{tab:4qcurrent}
\end{center}
\end{table}
}
The analysis of the masses and couplings of tetraquark states is very similar to the one of the 
molecules and present analogous features: presence of minima or/and inflection points, good 
convergence of the perturbative series and the OPE. The results for the tetraquark states are 
summarized in Table\,\ref{tab:4q-resultc} and for a complete discussion see 
Refs.~\cite{Albuquerque:2016znh,Albuquerque:2017vfq}.

Just to show an example of the results on the use of the sum rules for the tetraquarks, we state the mass and coupling 
of the scalar $S_c \,(0^+)$ tetraquark state. At N2LO, the corresponding set of parameters are:
\begin{equation}
\tau\simeq (0.35 \pm 0.05)~{\rm GeV}^{-2}, ~~~~s_0 \simeq (27.5 \pm 4.5)~{\rm GeV}^2~~~~ 
{\rm and}~~~ ~\mu \simeq  4.5~\rm{ GeV}, 
\end{equation}
and taking into account the uncertainties, as indicated in Table~\ref{tab:param}, we get:
\begin{eqnarray}
  M_{S_c} ~=~ 3898 \pm 54 ~{\rm MeV} &\hspace{0.6cm}{\rm and}\hspace{0.6cm}&  
  f_{S_c} ~=~ 191 \pm 20 ~{\rm keV} ~~.
\end{eqnarray}
This value for the mass is compatible with the $Y(3940)$ mass and we could naively describe it 
as a non-strange scalar tetraquark state. It is important to notice that this mass value, calculated 
at N2LO approximation, has almost the same magnitude as the one calculated at LO, and the 
difference is in order of $\sim 0.1\%$. This weak impact on the results from LO to N2LO is verified 
for the other tetraquark states too (see Refs.~\cite{Albuquerque:2016znh,Albuquerque:2017vfq}).
{\scriptsize
\begin{table}[ht]
\setlength{\tabcolsep}{0.8pc}
\caption{The masses and couplings of the $[\bar{c} \bar{q} cq]$ and $[\bar{c}\bar{s}cs]$ tetraquark 
states from QCDSR within stability criteria at N2LO of perturbative series.}
\vspace{0.2cm}
\begin{center}
{\small
\def\arraystretch{1.25}
 \begin{tabular}{cccc}
 \hline
 States & $\sqrt{s_0}$ & Mass & Coupling \\
 ($J^P$) & (Gev) & (MeV) & (MeV) \\
 \hline
   $S_c\,(0^{+})$ & $4.7 \sim 5.7$ & $3898 \pm 54$ & $191 \pm 20$ \\
   $A_c\,(1^{+})$ & $4.7 \sim 5.7$ & $3888 \pm 130$ & $184 \pm 30$ \\
   $P_c\,(0^{-})$ & $6.0 \sim 6.4$ & $5750 \pm 127$ & $310 \pm 13$ \\
   $V_c\,(1^{-})$ & $6.0 \sim 6.4$ & $5793 \pm 122$ & $296 \pm 19$ \\
\hline
\hline
\end{tabular}}
\quad \quad
{\small
\def\arraystretch{1.25}
 \begin{tabular}{lccc}
 \hline
 States & $\sqrt{s_0}$ & Mass & Coupling \\
 ($J^P$) & (Gev) & (MeV) & (MeV) \\
 \hline
   $S_{cs}\,(0^{+})$ & $5.2 \sim 6.7$ & $4233 \pm 61$ & $187 \pm 19$ \\
   $A_{cs}\,(1^{+})$ & $5.3 \sim 6.7$ & $4209 \pm 112$ & $160 \pm 17$ \\
   $P_{cs}\,(0^{-})$ & $6.2 \sim 7.5$ & $5524 \pm 176$ & $267 \pm 30$ \\
   $V_{cs}\,(1^{-})$ & $6.2 \sim 7.5$ & $5539 \pm 234$ & $258 \pm 33$ \\
\hline
\hline
\end{tabular}}
\end{center}
\label{tab:4q-resultc}
\end{table}
}

\subsection{Summary for the higher order corrections in QCDSR}

The N2LO predictions for the masses differ only slightly from the LO ones when the value of the 
running mass is used for the latter. However, the magnitude of the meson couplings is strongly affected by the 
radiative corrections in some channels, which consequently may modify the existing estimates of the meson 
hadronic widths based on vertex functions.
The $0^{++}$ $\bar DD$-like molecular states are almost degenerated with the $1^{+\pm}$ states. They have masses around $3900$\,MeV, which is consistent, within the errors, with the mass of the $Y(3940)$ state. It is also consistent with the scalar $S_c\,(0^+)$ tetraquark state.

As mentioned in the introduction, there are several $1^{++}$ observed states. In addition to the 
well-established $X(3872)$, we have  $X(4140)$ and $X(4274)$. From the results presented here it is possible to describe the mass of the 
 $X(3872)$ as a $\bar{D}^\ast D$ molecule or a $A_c$ tetraquark state, which is consistent with the results presented in Sec.~\ref{X(3872)}. For the remaining $1^{++}$ states, 
$X(4140)$ and $X(4274)$, one could interpret them as a $\bar{D}_s^\ast D_s$ molecule and a $A_{cs}$ tetraquark state, respectively.
From a QCDSR calculation at N2LO, and considering the relevant uncertainties, there are 
some molecule and/or tetraquark states which could be consistent with some of the observed charged states. 
The $\bar{D}^\ast D ~(1^{+-})$ and $\bar{D}_0^\ast D_1 ~(1^{+-})$ molecules would be  good 
candidates to explain the $Z_c^+(3900)$. In particular, the $\bar{D}_0^\ast D_1 ~(1^{+-})$ molecule 
could also be compatible with the $Z_c^+(4020)$. The $Z_c^+(4200)$ state might be identified as a 
charmed-strange axial $A_{cs}$ tetraquark state and the $\bar{D}_0^\ast D_0^\ast ~(0^{++})$ molecule could be naively associated with the recently observed charged state $Z_c^-(4100)$.

\section{\label{sum} Summary}

In  this review we have discussed the exotic charmonium states, observed by 
BaBar, Belle, CDF, \DZero, LHCb and BESIII Collaborations, from the perspective of QCD sum rules. We have computed the masses of several $X$, $Y$ 
and $Z$ states and, as it was seen case by case, the method of QCDSR, in spite of its limitations, contributes a great deal to the understanding of the structure of these new states. In some cases a tetraquark current gives a better agreement with the observed mass and in some other cases the agreement is better with a molecular current. However, as a general result, the two kinds of currents, molecular or tetraquark, lead to almost the same result for the mass of the state. Besides, since the used currents are local, a molecular current does not represent a true molecule. It is just the color combination between the quarks that is similar to a molecular state. For the non-charged states, QCDSR results favor a mixing between two and four-quark currents.

The limitations in statements made with QCDSR estimates come mostly from uncertainties in the method. However these statements can be made progressively more precise as we know more experimental information about the state in question. One good example is the $X(3872)$, from which, besides the mass, several decay modes were measured. Combining all the available information and using QCDSR to calculate the observed decay widths, we were able to say that the $X(3872)$ is a mixed state, where the most important component is a $c \bar{c}$ pair, which is mixed with a small four-quark component, out of which a large fraction is composed by neutral combinations of $D$-like and $D^*$-like two-quark states with  only a tiny fraction of  charged $D$-like and $D^*$-like  states. This conclusion is very specific and precise and it is more elaborated than the other results presented addressing only the masses of the new charmonia. This improvement was a consequence of studying simultaneo!
 usly the 
 mass and the decay width. This type of combined calculation should be extended to all states.

One of the drawbacks of previous QCDSR calculations was the absence of $\alpha_s$ corrections. In a series of recent works it was shown that these corrections are tiny. Another problem is the use of the factorization hypothesis, according to which $\langle \bar{q} q \bar{q} q \rangle = \rho \,\langle \bar{q} q\rangle^2$, where $\rho \simeq 1$.  Also here, there has been some progress on the theoretical side showing the violation of this hypothesis, its origin and the best value of $\rho$. 

The next generation of experimental data from $e^+ e^-$ colliders will increase a lot the statistics and will yield invariant mass spectra with high precision. Consequently, what was previously seen as a single bump, a single state, will reveal itself as a series of different peaks at different masses. The beginning of this ``unfolding'' has already been observed in the case of the $Y(4360)$ and more recently in the case of the $Y(4260)$.  For QCDSR this is challenging and asks for improvement of the pole-continuum model of the spectral function. Efforts along this direction should become a priority in the work of our community.

New data on hadronic production of exotic charmonium are also expected to appear at the LHC. In central collisions exotic  charmonium production is essentially a high energy process, calculable either with perturbative QCD \cite{Cho:2017dcy,Meng:2013gga} or  with effective theories \cite{Torres:2014fxa}.  On the other hand in ultraperipheral collisions these states are produced by photon-photon fusion \cite{Moreira:2016ciu,Goncalves:2018hiw} and the cross section is proportional to the 2-photon decay width of the states. Here again QCDSR is relevant since this decay receives large non-perturbative contributions. The decay width of some states has been calculated but there is room for improvement and also other states to be considered. 

The above mentioned uncertainties put some limits on the predictive power of  QCDSR. In QCDSR, as in quark model calculations \cite{Debastiani:2017msn}, one may find more states than those really observed. On the other hand, exactly because of this feature, QCDSR calculations have  ``veto power'', i.e.: {\it If it is not true in QCDSR it is not true in QCD}. This veto power has been already used to say that the existence of tetraquarks made only of light quarks is disfavored~\cite{Matheus:2007ta}. 

We close this review with some conclusions from the results presented in the previous sections. They are contained in Table ~\ref{tabfinal} where we present a summary of the most plausible interpretations for some of the states presented in Table~\ref{tab_summary}. It is important to remember that a molecular or tetraquark structure assignment in Table~\ref{tabfinal} is just the indication of the current used in the calculations, and that they are equivalent, from a QCDSR perspective. 


\begin{table}[h]

  \begin{center}
    \caption{Structure and quantum numbers from  QCDSR studies. In the case of isovector states, the quoted charge conjugation, $C$, is for the neutral state in the multiplet.}
  \label{tabfinal}
    \begin{tabular}{|c|c|c|} \hline
      state    & structure & $J^{PC}$    \\ \hline
  $X(3872)$ & mixed $\chi_{c1}-D\bar{D}^*$ & $1^{++}$    \\
  $Z_c(3900)$ & $D\bar{D}^*$ & $1^{+-}$    \\
  $Y(3940)$ & mixed $\chi_{c0}-D^*\bar{D}^*$   & $0^{++}$ \\
  $Z_c(4020)$ & $D^*\bar{D}^*$ & $1^{+-}$ or $2^{++}$    \\
  $Z_c(4100)$ & $D_0^*\bar{D}_0^*$ & $0^{++}$     \\
  $Y(4140)$ & mixed $D^*\bar{D}^*-D_s^*\bar{D}_s^*$  & $0^{++}$  \\
  $X(4160)$ & $D_s^*\bar{D}_s^*$  & $2^{++}$ \\
  $Z_c(4200)$ & $[cs][\bar{c}\bar{s}]$ & $1^{+}$     \\
  $Z_2(4250)$ & $D\bar{D}_1$  & $1^-$  \\
  $Y(4260)$ & mixed $J/\psi-[cq][\bar{c}\bar{q}]$  & $1^{--}$  \\
  $Y(4360)$ & $[cq][\bar{c}\bar{q}]$ & $1^{--}$  \\
  $Y(4660)$ & $[cs][\bar{c}\bar{s}]$  & $1^{--}$ \\
\hline
     \end{tabular}
  \end{center}
\end{table}


Table~\ref{tabfinal} represents the final result of a comprehensive effort and a careful analysis of several theoretical possibilities in the light of existing data. It is an encouraging example of what QCDSR can do. This Table contains a short summary of what we have learned about the new charmonium states in the recent past. In particular, similar to the $X(3872)$ state, mixed charmonium-tetraquark currents give a better description for all neutral exotic states, like the  $Y(3940)$ (or $X(3915)$), $Y(4140)$ and the $Y(4260)$.

The discovery of several manifestly exotic states, the  $Z^+$ states, may be considered as one of the most exciting findings of the last years. The description of such states  unavoidably requires  (at least) four valence quarks in the wave function. 


\vskip 1cm

\noindent
{\bf Acknowledgements:} The authors would like to thank M.~E. Bracco, S.~H. Lee, R.D. Matheus, K. Morita, S. Narison J.-M.~Richard, R. Rodrigues da Silva with whom they have collaborated in one or more of the works described in this review. The  authors  are indebted to the brazilian funding agency CNPq.

\markboth{\sl QCD Sum Rules Approach to the $X,~Y$ and $Z$ States } {\sl Bibliography  }

\bibliographystyle{elsarticle-num.bst}
\bibliography{biblio}

\begin{thebibliography}{100}
\expandafter\ifx\csname url\endcsname\relax
  \def\url#1{\texttt{#1}}\fi
\expandafter\ifx\csname urlprefix\endcsname\relax\def\urlprefix{URL }\fi
\expandafter\ifx\csname href\endcsname\relax
  \def\href#1#2{#2} \def\path#1{#1}\fi

\bibitem{Jaffe:2004ph}
R.~L. Jaffe, {Exotica}, Phys. Rept. 409 (2005) 1--45.
\newblock \href {http://arxiv.org/abs/hep-ph/0409065}
  {\path{arXiv:hep-ph/0409065}}, \href
  {http://dx.doi.org/10.1016/j.physrep.2004.11.005}
  {\path{doi:10.1016/j.physrep.2004.11.005}}.

\bibitem{Swanson:2006st}
E.~S. Swanson, {The New heavy mesons: A Status report}, Phys. Rept. 429 (2006)
  243--305.
\newblock \href {http://arxiv.org/abs/hep-ph/0601110}
  {\path{arXiv:hep-ph/0601110}}, \href
  {http://dx.doi.org/10.1016/j.physrep.2006.04.003}
  {\path{doi:10.1016/j.physrep.2006.04.003}}.

\bibitem{Zhu:2007wz}
S.-L. Zhu, {New hadron states}, Int. J. Mod. Phys. E17 (2008) 283--322.
\newblock \href {http://arxiv.org/abs/hep-ph/0703225}
  {\path{arXiv:hep-ph/0703225}}, \href
  {http://dx.doi.org/10.1142/S0218301308009446}
  {\path{doi:10.1142/S0218301308009446}}.

\bibitem{Klempt:2007cp}
E.~Klempt, A.~Zaitsev, {Glueballs, Hybrids, Multiquarks. Experimental facts
  versus QCD inspired concepts}, Phys. Rept. 454 (2007) 1--202.
\newblock \href {http://arxiv.org/abs/0708.4016} {\path{arXiv:0708.4016}},
  \href {http://dx.doi.org/10.1016/j.physrep.2007.07.006}
  {\path{doi:10.1016/j.physrep.2007.07.006}}.

\bibitem{Voloshin:2007dx}
M.~B. Voloshin, {Charmonium}, Prog. Part. Nucl. Phys. 61 (2008) 455--511.
\newblock \href {http://arxiv.org/abs/0711.4556} {\path{arXiv:0711.4556}},
  \href {http://dx.doi.org/10.1016/j.ppnp.2008.02.001}
  {\path{doi:10.1016/j.ppnp.2008.02.001}}.

\bibitem{Godfrey:2008nc}
S.~Godfrey, S.~L. Olsen, {The Exotic XYZ Charmonium-like Mesons}, Ann. Rev.
  Nucl. Part. Sci. 58 (2008) 51--73.
\newblock \href {http://arxiv.org/abs/0801.3867} {\path{arXiv:0801.3867}},
  \href {http://dx.doi.org/10.1146/annurev.nucl.58.110707.171145}
  {\path{doi:10.1146/annurev.nucl.58.110707.171145}}.

\bibitem{Nielsen:2009uh}
M.~Nielsen, F.~S. Navarra, S.~H. Lee, {New Charmonium States in QCD Sum Rules:
  A Concise Review}, Phys. Rept. 497 (2010) 41--83.
\newblock \href {http://arxiv.org/abs/0911.1958} {\path{arXiv:0911.1958}},
  \href {http://dx.doi.org/10.1016/j.physrep.2010.07.005}
  {\path{doi:10.1016/j.physrep.2010.07.005}}.

\bibitem{Brambilla:2010cs}
N.~Brambilla, et~al., {Heavy quarkonium: progress, puzzles, and opportunities},
  Eur. Phys. J. C71 (2011) 1534.
\newblock \href {http://arxiv.org/abs/1010.5827} {\path{arXiv:1010.5827}},
  \href {http://dx.doi.org/10.1140/epjc/s10052-010-1534-9}
  {\path{doi:10.1140/epjc/s10052-010-1534-9}}.

\bibitem{Druzhinin:2011qd}
V.~P. Druzhinin, S.~I. Eidelman, S.~I. Serednyakov, E.~P. Solodov, {Hadron
  Production via e+e- Collisions with Initial State Radiation}, Rev. Mod. Phys.
  83 (2011) 1545.
\newblock \href {http://arxiv.org/abs/1105.4975} {\path{arXiv:1105.4975}},
  \href {http://dx.doi.org/10.1103/RevModPhys.83.1545}
  {\path{doi:10.1103/RevModPhys.83.1545}}.

\bibitem{Li:2012me}
N.~Li, Z.-F. Sun, J.~He, X.~Liu, Z.-G. Luo, S.-L. Zhu, {Few-Body Systems
  Composed of Heavy Quarks}, Few Body Syst. 54 (2013) 807--812.
\newblock \href {http://arxiv.org/abs/1208.6347} {\path{arXiv:1208.6347}},
  \href {http://dx.doi.org/10.1007/s00601-012-0564-2}
  {\path{doi:10.1007/s00601-012-0564-2}}.

\bibitem{Liu:2013waa}
X.~Liu, {An overview of $XYZ$ new particles}, Chin. Sci. Bull. 59 (2014)
  3815--3830.
\newblock \href {http://arxiv.org/abs/1312.7408} {\path{arXiv:1312.7408}},
  \href {http://dx.doi.org/10.1007/s11434-014-0407-2}
  {\path{doi:10.1007/s11434-014-0407-2}}.

\bibitem{Brambilla:2014jmp}
N.~Brambilla, et~al., {QCD and Strongly Coupled Gauge Theories: Challenges and
  Perspectives}, Eur. Phys. J. C74~(10) (2014) 2981.
\newblock \href {http://arxiv.org/abs/1404.3723} {\path{arXiv:1404.3723}},
  \href {http://dx.doi.org/10.1140/epjc/s10052-014-2981-5}
  {\path{doi:10.1140/epjc/s10052-014-2981-5}}.

\bibitem{Olsen:2014qna}
S.~L. Olsen, {A New Hadron Spectroscopy}, Front. Phys.(Beijing) 10~(2) (2015)
  121--154.
\newblock \href {http://arxiv.org/abs/1411.7738} {\path{arXiv:1411.7738}},
  \href {http://dx.doi.org/10.1007/S11467-014-0449-6}
  {\path{doi:10.1007/S11467-014-0449-6}}.

\bibitem{Nielsen:2014mva}
M.~Nielsen, F.~S. Navarra, {Charged Exotic Charmonium States}, Mod. Phys. Lett.
  A29 (2014) 1430005.
\newblock \href {http://arxiv.org/abs/1401.2913} {\path{arXiv:1401.2913}},
  \href {http://dx.doi.org/10.1142/S0217732314300055}
  {\path{doi:10.1142/S0217732314300055}}.

\bibitem{Esposito:2014rxa}
A.~Esposito, A.~L. Guerrieri, F.~Piccinini, A.~Pilloni, A.~D. Polosa,
  {Four-Quark Hadrons: an Updated Review}, Int. J. Mod. Phys. A30 (2015)
  1530002.
\newblock \href {http://arxiv.org/abs/1411.5997} {\path{arXiv:1411.5997}},
  \href {http://dx.doi.org/10.1142/S0217751X15300021}
  {\path{doi:10.1142/S0217751X15300021}}.

\bibitem{Briceno:2015rlt}
R.~A. Briceno, et~al., {Issues and Opportunities in Exotic Hadrons}, Chin.
  Phys. C40~(4) (2016) 042001.
\newblock \href {http://arxiv.org/abs/1511.06779} {\path{arXiv:1511.06779}},
  \href {http://dx.doi.org/10.1088/1674-1137/40/4/042001}
  {\path{doi:10.1088/1674-1137/40/4/042001}}.

\bibitem{Hosaka:2016pey}
A.~Hosaka, T.~Iijima, K.~Miyabayashi, Y.~Sakai, S.~Yasui, {Exotic hadrons with
  heavy flavors: X, Y, Z, and related states}, PTEP 2016~(6) (2016) 062C01.
\newblock \href {http://arxiv.org/abs/1603.09229} {\path{arXiv:1603.09229}},
  \href {http://dx.doi.org/10.1093/ptep/ptw045}
  {\path{doi:10.1093/ptep/ptw045}}.

\bibitem{Chen:2016qju}
H.-X. Chen, W.~Chen, X.~Liu, S.-L. Zhu, {The hidden-charm pentaquark and
  tetraquark states}, Phys. Rept. 639 (2016) 1--121.
\newblock \href {http://arxiv.org/abs/1601.02092} {\path{arXiv:1601.02092}},
  \href {http://dx.doi.org/10.1016/j.physrep.2016.05.004}
  {\path{doi:10.1016/j.physrep.2016.05.004}}.

\bibitem{Esposito:2016noz}
A.~Esposito, A.~Pilloni, A.~D. Polosa, {Multiquark Resonances}, Phys. Rept. 668
  (2016) 1--97.
\newblock \href {http://arxiv.org/abs/1611.07920} {\path{arXiv:1611.07920}},
  \href {http://dx.doi.org/10.1016/j.physrep.2016.11.002}
  {\path{doi:10.1016/j.physrep.2016.11.002}}.

\bibitem{Guo:2017jvc}
F.-K. Guo, C.~Hanhart, U.-G. Meißner, Q.~Wang, Q.~Zhao, B.-S. Zou, {Hadronic
  molecules}, Rev. Mod. Phys. 90~(1) (2018) 015004.
\newblock \href {http://arxiv.org/abs/1705.00141} {\path{arXiv:1705.00141}},
  \href {http://dx.doi.org/10.1103/RevModPhys.90.015004}
  {\path{doi:10.1103/RevModPhys.90.015004}}.

\bibitem{Yuan:2018inv}
C.-Z. Yuan, {The XYZ states revisited}, Int. J. Mod. Phys. A33~(21) (2018)
  1830018.
\newblock \href {http://arxiv.org/abs/1808.01570} {\path{arXiv:1808.01570}},
  \href {http://dx.doi.org/10.1142/S0217751X18300181}
  {\path{doi:10.1142/S0217751X18300181}}.

\bibitem{pdg}
M.~Tanabashi, et~al., {Review of Particle Physics}, Phys. Rev. D98~(3) (2018)
  030001.
\newblock \href {http://dx.doi.org/10.1103/PhysRevD.98.030001}
  {\path{doi:10.1103/PhysRevD.98.030001}}.

\bibitem{Choi:2003ue}
S.~K. Choi, et~al., {Observation of a narrow charmonium - like state in
  exclusive $B^\pm \to K^\pm \pi^+ \pi^- J/\psi$ decays}, Phys. Rev. Lett. 91
  (2003) 262001.
\newblock \href {http://arxiv.org/abs/hep-ex/0309032}
  {\path{arXiv:hep-ex/0309032}}, \href
  {http://dx.doi.org/10.1103/PhysRevLett.91.262001}
  {\path{doi:10.1103/PhysRevLett.91.262001}}.

\bibitem{Adachi:2008te}
I.~Adachi, et~al.,
  \href{http://inspirehep.net/record/795806/files/arXiv:0809.1224.pdf}{{Study
  of $X(3872) $ in $B$ meson decays}}, in: {Proceedings, 34th International
  Conference on High Energy Physics (ICHEP 2008): Philadelphia, Pennsylvania,
  July 30-August 5, 2008}, 2008.
\newblock \href {http://arxiv.org/abs/0809.1224} {\path{arXiv:0809.1224}}.
\newline\urlprefix\url{http://inspirehep.net/record/795806/files/arXiv:0809.1224.pdf}

\bibitem{Choi:2011fc}
S.~K. Choi, et~al., {Bounds on the width, mass difference and other properties
  of X(3872) --> pi+pi-J/psi decays}, Phys. Rev. D84 (2011) 052004.
\newblock \href {http://arxiv.org/abs/1107.0163} {\path{arXiv:1107.0163}},
  \href {http://dx.doi.org/10.1103/PhysRevD.84.052004}
  {\path{doi:10.1103/PhysRevD.84.052004}}.

\bibitem{Aubert:2008gu}
B.~Aubert, et~al., {A Study of $B \to X(3872) K$, with $X_{3872} \to J/\Psi
  \pi^{+} \pi^{-}$}, Phys. Rev. D77 (2008) 111101.
\newblock \href {http://arxiv.org/abs/0803.2838} {\path{arXiv:0803.2838}},
  \href {http://dx.doi.org/10.1103/PhysRevD.77.111101}
  {\path{doi:10.1103/PhysRevD.77.111101}}.

\bibitem{Acosta:2003zx}
D.~Acosta, et~al., {Observation of the narrow state $X(3872) \to J/\psi \pi^+
  \pi^-$ in $\bar{p}p$ collisions at $\sqrt{s} = 1.96$ TeV}, Phys. Rev. Lett.
  93 (2004) 072001.
\newblock \href {http://arxiv.org/abs/hep-ex/0312021}
  {\path{arXiv:hep-ex/0312021}}, \href
  {http://dx.doi.org/10.1103/PhysRevLett.93.072001}
  {\path{doi:10.1103/PhysRevLett.93.072001}}.

\bibitem{Abulencia:2006ma}
A.~Abulencia, et~al., {Analysis of the quantum Numbers $J^{PC}$ of the
  $X(3872)$}, Phys. Rev. Lett. 98 (2007) 132002.
\newblock \href {http://arxiv.org/abs/hep-ex/0612053}
  {\path{arXiv:hep-ex/0612053}}, \href
  {http://dx.doi.org/10.1103/PhysRevLett.98.132002}
  {\path{doi:10.1103/PhysRevLett.98.132002}}.

\bibitem{Aaltonen:2009vj}
T.~Aaltonen, et~al., {Precision Measurement of the $X(3872)$ Mass in $J/\psi
  \pi^+ \pi^-$ Decays}, Phys. Rev. Lett. 103 (2009) 152001.
\newblock \href {http://arxiv.org/abs/0906.5218} {\path{arXiv:0906.5218}},
  \href {http://dx.doi.org/10.1103/PhysRevLett.103.152001}
  {\path{doi:10.1103/PhysRevLett.103.152001}}.

\bibitem{Abazov:2004kp}
V.~M. Abazov, et~al., {Observation and properties of the $X(3872)$ decaying to
  $J/\psi \pi^+ \pi^-$ in $p\bar{p}$ collisions at $\sqrt{s} = 1.96$ TeV},
  Phys. Rev. Lett. 93 (2004) 162002.
\newblock \href {http://arxiv.org/abs/hep-ex/0405004}
  {\path{arXiv:hep-ex/0405004}}, \href
  {http://dx.doi.org/10.1103/PhysRevLett.93.162002}
  {\path{doi:10.1103/PhysRevLett.93.162002}}.

\bibitem{Abe:2005ix}
K.~Abe, et~al., {Evidence for $X(3872) \to \gamma J/\psi$ and the sub-threshold
  decay $X(3872) \to \omega J/\psi$}, in: {Lepton and photon interactions at
  high energies. Proceedings, 22nd International Symposium, LP 2005, Uppsala,
  Sweden, June 30-July 5, 2005}, 2005.
\newblock \href {http://arxiv.org/abs/hep-ex/0505037}
  {\path{arXiv:hep-ex/0505037}}.

\bibitem{delAmoSanchez:2010jr}
P.~del Amo~Sanchez, et~al., {Evidence for the decay $X(3872) \to J/\psi
  \omega$}, Phys. Rev. D82 (2010) 011101.
\newblock \href {http://arxiv.org/abs/1005.5190} {\path{arXiv:1005.5190}},
  \href {http://dx.doi.org/10.1103/PhysRevD.82.011101}
  {\path{doi:10.1103/PhysRevD.82.011101}}.

\bibitem{Gokhroo:2006bt}
G.~Gokhroo, et~al., {Observation of a Near-threshold D0 anti-D0 pi0 Enhancement
  in B $\to$ D0 anti-D0 pi0 K Decay}, Phys. Rev. Lett. 97 (2006) 162002.
\newblock \href {http://arxiv.org/abs/hep-ex/0606055}
  {\path{arXiv:hep-ex/0606055}}, \href
  {http://dx.doi.org/10.1103/PhysRevLett.97.162002}
  {\path{doi:10.1103/PhysRevLett.97.162002}}.

\bibitem{Adachi:2008sua}
T.~Aushev, et~al., {Study of the B $\to$ X(3872)(D*0 anti-D0) K decay}, Phys.
  Rev. D81 (2010) 031103.
\newblock \href {http://arxiv.org/abs/0810.0358} {\path{arXiv:0810.0358}},
  \href {http://dx.doi.org/10.1103/PhysRevD.81.031103}
  {\path{doi:10.1103/PhysRevD.81.031103}}.

\bibitem{Aubert:2007rva}
B.~Aubert, et~al., {Study of Resonances in Exclusive B Decays to anti-D(*) D(*)
  K}, Phys. Rev. D77 (2008) 011102.
\newblock \href {http://arxiv.org/abs/0708.1565} {\path{arXiv:0708.1565}},
  \href {http://dx.doi.org/10.1103/PhysRevD.77.011102}
  {\path{doi:10.1103/PhysRevD.77.011102}}.

\bibitem{Aubert:2006aj}
B.~Aubert, et~al., {Search for $B^{+} \to X(3872) K^{+}$, $X_{3872} \to J/\psi
  \gamma$}, Phys. Rev. D74 (2006) 071101.
\newblock \href {http://arxiv.org/abs/hep-ex/0607050}
  {\path{arXiv:hep-ex/0607050}}, \href
  {http://dx.doi.org/10.1103/PhysRevD.74.071101}
  {\path{doi:10.1103/PhysRevD.74.071101}}.

\bibitem{Aubert:2008ae}
B.~Aubert, et~al., {Evidence for $X(3872) \to \psi_{2S} \gamma$ in $B^\pm \to
  X_{3872} K^\pm$ decays, and a study of $B \to c \bar{c} \gamma K$}, Phys.
  Rev. Lett. 102 (2009) 132001.
\newblock \href {http://arxiv.org/abs/0809.0042} {\path{arXiv:0809.0042}},
  \href {http://dx.doi.org/10.1103/PhysRevLett.102.132001}
  {\path{doi:10.1103/PhysRevLett.102.132001}}.

\bibitem{Aaij:2014ala}
R.~Aaij, et~al., {Evidence for the decay $X(3872)\rightarrow\psi(2S)\gamma$},
  Nucl. Phys. B886 (2014) 665--680.
\newblock \href {http://arxiv.org/abs/1404.0275} {\path{arXiv:1404.0275}},
  \href {http://dx.doi.org/10.1016/j.nuclphysb.2014.06.011}
  {\path{doi:10.1016/j.nuclphysb.2014.06.011}}.

\bibitem{Ablikim:2013dyn}
M.~Ablikim, et~al., {Observation of $e^+e^− → γX$(3872) at BESIII}, Phys.
  Rev. Lett. 112~(9) (2014) 092001.
\newblock \href {http://arxiv.org/abs/1310.4101} {\path{arXiv:1310.4101}},
  \href {http://dx.doi.org/10.1103/PhysRevLett.112.092001}
  {\path{doi:10.1103/PhysRevLett.112.092001}}.

\bibitem{Aaij:2011sn}
R.~Aaij, et~al., {Observation of $X(3872) $ production in $pp$ collisions at
  $\sqrt{s}=7$ TeV}, Eur. Phys. J. C72 (2012) 1972.
\newblock \href {http://arxiv.org/abs/1112.5310} {\path{arXiv:1112.5310}},
  \href {http://dx.doi.org/10.1140/epjc/s10052-012-1972-7}
  {\path{doi:10.1140/epjc/s10052-012-1972-7}}.

\bibitem{Aaij:2013zoa}
R.~Aaij, et~al., {Determination of the X(3872) meson quantum Numbers}, Phys.
  Rev. Lett. 110 (2013) 222001.
\newblock \href {http://arxiv.org/abs/1302.6269} {\path{arXiv:1302.6269}},
  \href {http://dx.doi.org/10.1103/PhysRevLett.110.222001}
  {\path{doi:10.1103/PhysRevLett.110.222001}}.

\bibitem{Chatrchyan:2013cld}
S.~Chatrchyan, et~al., {Measurement of the X(3872) production cross section via
  decays to J/psi pi pi in pp collisions at sqrt(s) = 7 TeV}, JHEP 04 (2013)
  154.
\newblock \href {http://arxiv.org/abs/1302.3968} {\path{arXiv:1302.3968}},
  \href {http://dx.doi.org/10.1007/JHEP04(2013)154}
  {\path{doi:10.1007/JHEP04(2013)154}}.

\bibitem{Ablikim:2013mio}
M.~Ablikim, et~al., {Observation of a Charged Charmoniumlike Structure in
  $e^+e^- \to \pi^+\pi^- J/\psi$ at $\sqrt{s}$ =4.26 GeV}, Phys. Rev. Lett. 110
  (2013) 252001.
\newblock \href {http://arxiv.org/abs/1303.5949} {\path{arXiv:1303.5949}},
  \href {http://dx.doi.org/10.1103/PhysRevLett.110.252001}
  {\path{doi:10.1103/PhysRevLett.110.252001}}.

\bibitem{Liu:2013dau}
Z.~Q. Liu, et~al., {Study of $e^+e^- \to \pi^+ \pi^- J/\psi$ and Observation of
  a Charged Charmoniumlike State at Belle}, Phys. Rev. Lett. 110 (2013) 252002.
\newblock \href {http://arxiv.org/abs/1304.0121} {\path{arXiv:1304.0121}},
  \href {http://dx.doi.org/10.1103/PhysRevLett.110.252002}
  {\path{doi:10.1103/PhysRevLett.110.252002}}.

\bibitem{Xiao:2013iha}
T.~Xiao, S.~Dobbs, A.~Tomaradze, K.~K. Seth, {Observation of the Charged Hadron
  $Z_c^{\pm}(3900)$ and Evidence for the Neutral $Z_c^0(3900)$ in $e^+e^-\to
  \pi\pi J/\psi$ at $\sqrt{s}=4170$ MeV}, Phys. Lett. B727 (2013) 366--370.
\newblock \href {http://arxiv.org/abs/1304.3036} {\path{arXiv:1304.3036}},
  \href {http://dx.doi.org/10.1016/j.physletb.2013.10.041}
  {\path{doi:10.1016/j.physletb.2013.10.041}}.

\bibitem{Ablikim:2013xfr}
M.~Ablikim, et~al., {Observation of a charged $(D\bar{D}^{*})^\pm$ mass peak in
  $e^{+}e^{-} \to \pi D\bar{D}^{*}$ at $\sqrt{s} =$ 4.26 GeV}, Phys. Rev. Lett.
  112~(2) (2014) 022001.
\newblock \href {http://arxiv.org/abs/1310.1163} {\path{arXiv:1310.1163}},
  \href {http://dx.doi.org/10.1103/PhysRevLett.112.022001}
  {\path{doi:10.1103/PhysRevLett.112.022001}}.

\bibitem{Abe:2004zs}
K.~Abe, et~al., {Observation of a near-threshold omega J/psi mass enhancement
  in exclusive B $\to$ K omega J/psi decays}, Phys. Rev. Lett. 94 (2005)
  182002.
\newblock \href {http://arxiv.org/abs/hep-ex/0408126}
  {\path{arXiv:hep-ex/0408126}}, \href
  {http://dx.doi.org/10.1103/PhysRevLett.94.182002}
  {\path{doi:10.1103/PhysRevLett.94.182002}}.

\bibitem{Aubert:2007vj}
B.~Aubert, et~al., {Observation of Y(3940) $\to J/\psi \omega$ in $B \to J/\psi
  \omega K$ at BABAR}, Phys. Rev. Lett. 101 (2008) 082001.
\newblock \href {http://arxiv.org/abs/0711.2047} {\path{arXiv:0711.2047}},
  \href {http://dx.doi.org/10.1103/PhysRevLett.101.082001}
  {\path{doi:10.1103/PhysRevLett.101.082001}}.

\bibitem{Uehara:2009tx}
S.~Uehara, et~al., {Observation of a charmonium-like enhancement in the $\gamma
  \gamma \to \omega J/\psi$ process}, Phys. Rev. Lett. 104 (2010) 092001.
\newblock \href {http://arxiv.org/abs/0912.4451} {\path{arXiv:0912.4451}},
  \href {http://dx.doi.org/10.1103/PhysRevLett.104.092001}
  {\path{doi:10.1103/PhysRevLett.104.092001}}.

\bibitem{Lees:2012me}
J.~P. Lees, et~al., {Search for resonances decaying to $\eta_c \pi^+ \pi^-$ in
  two-photon interactions}, Phys. Rev. D86 (2012) 092005.
\newblock \href {http://arxiv.org/abs/1206.2008} {\path{arXiv:1206.2008}},
  \href {http://dx.doi.org/10.1103/PhysRevD.86.092005}
  {\path{doi:10.1103/PhysRevD.86.092005}}.

\bibitem{Abe:2007jna}
K.~Abe, et~al., {Observation of a new charmonium state in double charmonium
  production in e+ e- annihilation at s**(1/2) ~ 10.6-GeV}, Phys. Rev. Lett. 98
  (2007) 082001.
\newblock \href {http://arxiv.org/abs/hep-ex/0507019}
  {\path{arXiv:hep-ex/0507019}}, \href
  {http://dx.doi.org/10.1103/PhysRevLett.98.082001}
  {\path{doi:10.1103/PhysRevLett.98.082001}}.

\bibitem{Abe:2007sya}
P.~Pakhlov, et~al., {Production of New Charmoniumlike States in e+ e- --> J/psi
  D(*) anti-D(*) at s**(1/2) ~ 10. GeV}, Phys. Rev. Lett. 100 (2008) 202001.
\newblock \href {http://arxiv.org/abs/0708.3812} {\path{arXiv:0708.3812}},
  \href {http://dx.doi.org/10.1103/PhysRevLett.100.202001}
  {\path{doi:10.1103/PhysRevLett.100.202001}}.

\bibitem{Yuan:2007sj}
C.~Z. Yuan, et~al., {Measurement of $e^+ e^- \to \pi^+ \pi^- J/\psi$
  cross-section via initial state radiation at Belle}, Phys. Rev. Lett. 99
  (2007) 182004.
\newblock \href {http://arxiv.org/abs/0707.2541} {\path{arXiv:0707.2541}},
  \href {http://dx.doi.org/10.1103/PhysRevLett.99.182004}
  {\path{doi:10.1103/PhysRevLett.99.182004}}.

\bibitem{Ablikim:2016qzw}
M.~Ablikim, et~al., {Precise measurement of the $e^+e^-\to \pi^+\pi^-J/\psi$
  cross section at center-of-mass energies from 3.77 to 4.60 GeV}, Phys. Rev.
  Lett. 118~(9) (2017) 092001.
\newblock \href {http://arxiv.org/abs/1611.01317} {\path{arXiv:1611.01317}},
  \href {http://dx.doi.org/10.1103/PhysRevLett.118.092001}
  {\path{doi:10.1103/PhysRevLett.118.092001}}.

\bibitem{Ablikim:2013wzq}
M.~Ablikim, et~al., {Observation of a Charged Charmoniumlike Structure
  $Z_c$(4020) and Search for the $Z_c$(3900) in $e^+ e^- \to \pi^+\pi^- h_c$},
  Phys. Rev. Lett. 111~(24) (2013) 242001.
\newblock \href {http://arxiv.org/abs/1309.1896} {\path{arXiv:1309.1896}},
  \href {http://dx.doi.org/10.1103/PhysRevLett.111.242001}
  {\path{doi:10.1103/PhysRevLett.111.242001}}.

\bibitem{Ablikim:2013emm}
M.~Ablikim, et~al., {Observation of a charged charmoniumlike structure in
  $e^+e^- \to (D^{*} \bar{D}^{*})^{\pm} \pi^\mp$ at $\sqrt{s}=4.26$GeV}, Phys.
  Rev. Lett. 112~(13) (2014) 132001.
\newblock \href {http://arxiv.org/abs/1308.2760} {\path{arXiv:1308.2760}},
  \href {http://dx.doi.org/10.1103/PhysRevLett.112.132001}
  {\path{doi:10.1103/PhysRevLett.112.132001}}.

\bibitem{Mizuk:2008me}
R.~Mizuk, et~al., {Observation of two resonance-like structures in the pi+
  chi(c1) mass distribution in exclusive anti-B0 $\to$ K- pi+ chi(c1) decays},
  Phys. Rev. D78 (2008) 072004.
\newblock \href {http://arxiv.org/abs/0806.4098} {\path{arXiv:0806.4098}},
  \href {http://dx.doi.org/10.1103/PhysRevD.78.072004}
  {\path{doi:10.1103/PhysRevD.78.072004}}.

\bibitem{Lees:2011ik}
J.~P. Lees, et~al., {Search for the $Z_1(4050)^+$ and $Z_2(4250)^+$ states in
  $\bar B^0 \to \chi_{c1} K^- \pi^+$ and $B^+ \to \chi_{c1} K^0_S \pi^+$},
  Phys. Rev. D85 (2012) 052003.
\newblock \href {http://arxiv.org/abs/1111.5919} {\path{arXiv:1111.5919}},
  \href {http://dx.doi.org/10.1103/PhysRevD.85.052003}
  {\path{doi:10.1103/PhysRevD.85.052003}}.

\bibitem{Wang:2014hta}
X.~L. Wang, et~al., {Measurement of $e^+e^- \to \pi^+\pi^-\psi(2S)$ via Initial
  State Radiation at Belle}, Phys. Rev. D91 (2015) 112007.
\newblock \href {http://arxiv.org/abs/1410.7641} {\path{arXiv:1410.7641}},
  \href {http://dx.doi.org/10.1103/PhysRevD.91.112007}
  {\path{doi:10.1103/PhysRevD.91.112007}}.

\bibitem{Aaij:2018bla}
R.~Aaij, et~al., {Evidence for an $\eta_c(1S) \pi^-$ resonance in $B^0 \to
  \eta_c(1S) K^+\pi^-$ decays}, Submitted to: Eur. Phys. J.\href
  {http://arxiv.org/abs/1809.07416} {\path{arXiv:1809.07416}}.

\bibitem{Aaltonen:2009tz}
T.~Aaltonen, et~al., {Evidence for a Narrow Near-Threshold Structure in the
  $J/\psi\phi$ Mass Spectrum in $B^+\to J/\psi\phi K^+$ Decays}, Phys. Rev.
  Lett. 102 (2009) 242002.
\newblock \href {http://arxiv.org/abs/0903.2229} {\path{arXiv:0903.2229}},
  \href {http://dx.doi.org/10.1103/PhysRevLett.102.242002}
  {\path{doi:10.1103/PhysRevLett.102.242002}}.

\bibitem{Aaltonen:2011at}
T.~Aaltonen, et~al., {Observation of the $Y(4140)$ structure in the
  $J/\psi\phi$ mass spectrum in $B^\pm\to J/\psi\phi K^\pm$ decays}, Mod. Phys.
  Lett. A32~(26) (2017) 1750139.
\newblock \href {http://arxiv.org/abs/1101.6058} {\path{arXiv:1101.6058}},
  \href {http://dx.doi.org/10.1142/S0217732317501395}
  {\path{doi:10.1142/S0217732317501395}}.

\bibitem{Abazov:2013xda}
V.~M. Abazov, et~al., {Search for the $X$(4140) state in $B^+ \to
  $J$_{\psi,phi}K^+$ decays with the D0 Detector}, Phys. Rev. D89~(1) (2014)
  012004.
\newblock \href {http://arxiv.org/abs/1309.6580} {\path{arXiv:1309.6580}},
  \href {http://dx.doi.org/10.1103/PhysRevD.89.012004}
  {\path{doi:10.1103/PhysRevD.89.012004}}.

\bibitem{Aaij:2016iza}
R.~Aaij, et~al., {Observation of $J/\psi\phi$ structures consistent with exotic
  states from amplitude analysis of $B^+\to J/\psi \phi K^+$ decays}, Phys.
  Rev. Lett. 118~(2) (2017) 022003.
\newblock \href {http://arxiv.org/abs/1606.07895} {\path{arXiv:1606.07895}},
  \href {http://dx.doi.org/10.1103/PhysRevLett.118.022003}
  {\path{doi:10.1103/PhysRevLett.118.022003}}.

\bibitem{Ablikim:2014atq}
M.~Ablikim, et~al., {Search for the Y(4140) via $e^+e^- → γϕJ/ψ$ at
  $\sqrt{s}$=4.23 , 4.26 and 4.36 GeV}, Phys. Rev. D91~(3) (2015) 032002.
\newblock \href {http://arxiv.org/abs/1412.1867} {\path{arXiv:1412.1867}},
  \href {http://dx.doi.org/10.1103/PhysRevD.91.032002}
  {\path{doi:10.1103/PhysRevD.91.032002}}.

\bibitem{Ablikim:2017cbv}
M.~Ablikim, et~al., {Observation of $e^{+}e^{-} \to \phi\chi_{c1}$ and
  $\phi\chi_{c2}$ at $\sqrt{s}$=4.600 GeV}, Phys. Rev. D97~(3) (2018) 032008.
\newblock \href {http://arxiv.org/abs/1712.09240} {\path{arXiv:1712.09240}},
  \href {http://dx.doi.org/10.1103/PhysRevD.97.032008}
  {\path{doi:10.1103/PhysRevD.97.032008}}.

\bibitem{Chilikin:2014bkk}
K.~Chilikin, et~al., {Observation of a new charged charmoniumlike state in
  $\bar{B}^0 \to J/\psi K^- \pi^+$ decays}, Phys. Rev. D90~(11) (2014) 112009.
\newblock \href {http://arxiv.org/abs/1408.6457} {\path{arXiv:1408.6457}},
  \href {http://dx.doi.org/10.1103/PhysRevD.90.112009}
  {\path{doi:10.1103/PhysRevD.90.112009}}.

\bibitem{Ablikim:2014qwy}
M.~Ablikim, et~al., {Study of $e^+e^-\to\omega\chi_{cJ}$ at center-of-mass
  energies from 4.21 to 4.42 GeV}, Phys. Rev. Lett. 114~(9) (2015) 092003.
\newblock \href {http://arxiv.org/abs/1410.6538} {\path{arXiv:1410.6538}},
  \href {http://dx.doi.org/10.1103/PhysRevLett.114.092003}
  {\path{doi:10.1103/PhysRevLett.114.092003}}.

\bibitem{BESIII:2016adj}
M.~Ablikim, et~al., {Evidence of Two Resonant Structures in $e^+ e^- \to \pi^+
  \pi^- h_c$}, Phys. Rev. Lett. 118~(9) (2017) 092002.
\newblock \href {http://arxiv.org/abs/1610.07044} {\path{arXiv:1610.07044}},
  \href {http://dx.doi.org/10.1103/PhysRevLett.118.092002}
  {\path{doi:10.1103/PhysRevLett.118.092002}}.

\bibitem{Ablikim:2017oaf}
M.~Ablikim, et~al., {Measurement of $e^{+}e^{-}\rightarrow
  \pi^{+}\pi^{-}\psi(3686)$ from 4.008 to 4.600~GeV and observation of a
  charged structure in the $\pi^{\pm}\psi(3686)$ mass spectrum}, Phys. Rev.
  D96~(3) (2017) 032004.
\newblock \href {http://arxiv.org/abs/1703.08787} {\path{arXiv:1703.08787}},
  \href {http://dx.doi.org/10.1103/PhysRevD.96.032004}
  {\path{doi:10.1103/PhysRevD.96.032004}}.

\bibitem{Ablikim:2018vxx}
M.~Ablikim, et~al., {Evidence of a resonant structure in the $e^+e^-\to
  \pi^+D^0D^{*-}$ cross section between 4.05 and 4.60 GeV}\href
  {http://arxiv.org/abs/1808.02847} {\path{arXiv:1808.02847}}.

\bibitem{Aubert:2005rm}
B.~Aubert, et~al., {Observation of a broad structure in the $\pi^+ \pi^-
  J/\psi$ mass spectrum around 4.26-GeV/c$^2$}, Phys. Rev. Lett. 95 (2005)
  142001.
\newblock \href {http://arxiv.org/abs/hep-ex/0506081}
  {\path{arXiv:hep-ex/0506081}}, \href
  {http://dx.doi.org/10.1103/PhysRevLett.95.142001}
  {\path{doi:10.1103/PhysRevLett.95.142001}}.

\bibitem{Lees:2012cn}
J.~P. Lees, et~al., {Study of the reaction $e^{+}e^{-} \to
  J/\psi\pi^{+}\pi^{-}$ via initial-state radiation at BaBar}, Phys. Rev. D86
  (2012) 051102.
\newblock \href {http://arxiv.org/abs/1204.2158} {\path{arXiv:1204.2158}},
  \href {http://dx.doi.org/10.1103/PhysRevD.86.051102}
  {\path{doi:10.1103/PhysRevD.86.051102}}.

\bibitem{He:2006kg}
Q.~He, et~al., {Confirmation of the Y(4260) resonance production in ISR}, Phys.
  Rev. D74 (2006) 091104.
\newblock \href {http://arxiv.org/abs/hep-ex/0611021}
  {\path{arXiv:hep-ex/0611021}}, \href
  {http://dx.doi.org/10.1103/PhysRevD.74.091104}
  {\path{doi:10.1103/PhysRevD.74.091104}}.

\bibitem{Coan:2006rv}
T.~E. Coan, et~al., {Charmonium decays of Y(4260), psi(4160) and psi(4040)},
  Phys. Rev. Lett. 96 (2006) 162003.
\newblock \href {http://arxiv.org/abs/hep-ex/0602034}
  {\path{arXiv:hep-ex/0602034}}, \href
  {http://dx.doi.org/10.1103/PhysRevLett.96.162003}
  {\path{doi:10.1103/PhysRevLett.96.162003}}.

\bibitem{Shen:2009vs}
C.~P. Shen, et~al., {Evidence for a new resonance and search for the Y(4140) in
  the gamma gamma $\to$ phi J/psi process}, Phys. Rev. Lett. 104 (2010) 112004.
\newblock \href {http://arxiv.org/abs/0912.2383} {\path{arXiv:0912.2383}},
  \href {http://dx.doi.org/10.1103/PhysRevLett.104.112004}
  {\path{doi:10.1103/PhysRevLett.104.112004}}.

\bibitem{Aubert:2007zz}
B.~Aubert, et~al., {Evidence of a broad structure at an invariant mass of 4.32-
  $GeV/c^{2}$ in the reaction $e^{+} e^{-} \to \pi^{+} \pi^{-} \psi_{2S}$
  measured at BaBar}, Phys. Rev. Lett. 98 (2007) 212001.
\newblock \href {http://arxiv.org/abs/hep-ex/0610057}
  {\path{arXiv:hep-ex/0610057}}, \href
  {http://dx.doi.org/10.1103/PhysRevLett.98.212001}
  {\path{doi:10.1103/PhysRevLett.98.212001}}.

\bibitem{Lees:2012pv}
J.~P. Lees, et~al., {Study of the reaction $e^{+}e^{-}\to
  \psi(2S)\pi^{+}\pi^{-}$ via initial-state radiation at BaBar}, Phys. Rev.
  D89~(11) (2014) 111103.
\newblock \href {http://arxiv.org/abs/1211.6271} {\path{arXiv:1211.6271}},
  \href {http://dx.doi.org/10.1103/PhysRevD.89.111103}
  {\path{doi:10.1103/PhysRevD.89.111103}}.

\bibitem{Wang:2007ea}
X.~L. Wang, et~al., {Observation of Two Resonant Structures in e+e- to pi+ pi-
  psi(2S) via Initial State Radiation at Belle}, Phys. Rev. Lett. 99 (2007)
  142002.
\newblock \href {http://arxiv.org/abs/0707.3699} {\path{arXiv:0707.3699}},
  \href {http://dx.doi.org/10.1103/PhysRevLett.99.142002}
  {\path{doi:10.1103/PhysRevLett.99.142002}}.

\bibitem{Choi:2007wga}
S.~K. Choi, et~al., {Observation of a resonance-like structure in the pi+-
  psi-prime mass distribution in exclusive $B \to K \pi^\pm \psi^prime$
  decays}, Phys. Rev. Lett. 100 (2008) 142001.
\newblock \href {http://arxiv.org/abs/0708.1790} {\path{arXiv:0708.1790}},
  \href {http://dx.doi.org/10.1103/PhysRevLett.100.142001}
  {\path{doi:10.1103/PhysRevLett.100.142001}}.

\bibitem{Mizuk:2009da}
R.~Mizuk, et~al., {Dalitz analysis of $B \to K \pi^+ \psi^\prime$ decays and
  the $Z(4430)^+$}, Phys. Rev. D80 (2009) 031104.
\newblock \href {http://arxiv.org/abs/0905.2869} {\path{arXiv:0905.2869}},
  \href {http://dx.doi.org/10.1103/PhysRevD.80.031104}
  {\path{doi:10.1103/PhysRevD.80.031104}}.

\bibitem{Chilikin:2013tch}
K.~Chilikin, et~al., {Experimental constraints on the spin and parity of the
  $Z$(4430)$^+$}, Phys. Rev. D88~(7) (2013) 074026.
\newblock \href {http://arxiv.org/abs/1306.4894} {\path{arXiv:1306.4894}},
  \href {http://dx.doi.org/10.1103/PhysRevD.88.074026}
  {\path{doi:10.1103/PhysRevD.88.074026}}.

\bibitem{Aubert:2008aa}
B.~Aubert, et~al., {Search for the Z(4430)- at BABAR}, Phys. Rev. D79 (2009)
  112001.
\newblock \href {http://arxiv.org/abs/0811.0564} {\path{arXiv:0811.0564}},
  \href {http://dx.doi.org/10.1103/PhysRevD.79.112001}
  {\path{doi:10.1103/PhysRevD.79.112001}}.

\bibitem{Aaij:2014jqa}
R.~Aaij, et~al., {Observation of the resonant character of the $Z(4430)^-$
  state}, Phys. Rev. Lett. 112~(22) (2014) 222002.
\newblock \href {http://arxiv.org/abs/1404.1903} {\path{arXiv:1404.1903}},
  \href {http://dx.doi.org/10.1103/PhysRevLett.112.222002}
  {\path{doi:10.1103/PhysRevLett.112.222002}}.

\bibitem{Pakhlova:2008vn}
G.~Pakhlova, et~al., {Observation of a near-threshold enhancement in the
  $e^+e^- \to \Lambda^+_c \Lambda^-_c$ cross section using initial-state
  radiation}, Phys. Rev. Lett. 101 (2008) 172001.
\newblock \href {http://arxiv.org/abs/0807.4458} {\path{arXiv:0807.4458}},
  \href {http://dx.doi.org/10.1103/PhysRevLett.101.172001}
  {\path{doi:10.1103/PhysRevLett.101.172001}}.

\bibitem{Matheus:2004gx}
R.~D. Matheus, F.~S. Navarra, M.~Nielsen, R.~R. da~Silva, {Comparative study of
  pentaquark interpolating currents}, Phys. Lett. B602 (2004) 185--196.
\newblock \href {http://arxiv.org/abs/hep-ph/0406246}
  {\path{arXiv:hep-ph/0406246}}, \href
  {http://dx.doi.org/10.1016/j.physletb.2004.09.072}
  {\path{doi:10.1016/j.physletb.2004.09.072}}.

\bibitem{Navarra:2006nd}
F.~S. Navarra, M.~Nielsen, {X(3872) $\to$ J / psi pi+ pi- and X(3872) $\to$ J /
  psi pi+ pi- pi0 decay widths from QCD sum rules}, Phys. Lett. B639 (2006)
  272--277.
\newblock \href {http://arxiv.org/abs/hep-ph/0605038}
  {\path{arXiv:hep-ph/0605038}}, \href
  {http://dx.doi.org/10.1016/j.physletb.2006.06.054}
  {\path{doi:10.1016/j.physletb.2006.06.054}}.

\bibitem{Matheus:2006xi}
R.~D. Matheus, S.~Narison, M.~Nielsen, J.~M. Richard, {Can the X(3872) be a 1++
  four-quark state?}, Phys. Rev. D75 (2007) 014005.
\newblock \href {http://arxiv.org/abs/hep-ph/0608297}
  {\path{arXiv:hep-ph/0608297}}, \href
  {http://dx.doi.org/10.1103/PhysRevD.75.014005}
  {\path{doi:10.1103/PhysRevD.75.014005}}.

\bibitem{Lee:2007gs}
S.~H. Lee, A.~Mihara, F.~S. Navarra, M.~Nielsen, {QCD sum rules study of the
  meson Z+(4430)}, Phys. Lett. B661 (2008) 28--32.
\newblock \href {http://arxiv.org/abs/0710.1029} {\path{arXiv:0710.1029}},
  \href {http://dx.doi.org/10.1016/j.physletb.2008.01.062}
  {\path{doi:10.1016/j.physletb.2008.01.062}}.

\bibitem{Lee:2008tz}
S.~H. Lee, K.~Morita, M.~Nielsen, {Width of exotics from QCD sum rules:
  Tetraquarks or molecules?}, Phys. Rev. D78 (2008) 076001.
\newblock \href {http://arxiv.org/abs/0808.3168} {\path{arXiv:0808.3168}},
  \href {http://dx.doi.org/10.1103/PhysRevD.78.076001}
  {\path{doi:10.1103/PhysRevD.78.076001}}.

\bibitem{Lee:2008gn}
S.~H. Lee, K.~Morita, M.~Nielsen, {Can the pi+ chi(c1) resonance structures be
  D* anti-D* and D(1) anti-D molecules?}, Nucl. Phys. A815 (2009) 29--39.
\newblock \href {http://arxiv.org/abs/0808.0690} {\path{arXiv:0808.0690}},
  \href {http://dx.doi.org/10.1016/j.nuclphysa.2008.10.012}
  {\path{doi:10.1016/j.nuclphysa.2008.10.012}}.

\bibitem{Albuquerque:2008up}
R.~M. Albuquerque, M.~Nielsen, {QCD sum rules study of the J(PC) = 1--
  charmonium Y mesons}, Nucl. Phys. A815 (2009) 53--66, [Erratum: Nucl.
  Phys.A857,48(2011)].
\newblock \href {http://arxiv.org/abs/0804.4817} {\path{arXiv:0804.4817}},
  \href {http://dx.doi.org/10.1016/j.nuclphysa.2011.04.001,
  10.1016/j.nuclphysa.2008.10.015} {\path{doi:10.1016/j.nuclphysa.2011.04.001,
  10.1016/j.nuclphysa.2008.10.015}}.

\bibitem{Bracco:2008jj}
M.~E. Bracco, S.~H. Lee, M.~Nielsen, R.~Rodrigues~da Silva, {The Meson Z+(4430)
  as a tetraquark state}, Phys. Lett. B671 (2009) 240--244.
\newblock \href {http://arxiv.org/abs/0807.3275} {\path{arXiv:0807.3275}},
  \href {http://dx.doi.org/10.1016/j.physletb.2008.12.021}
  {\path{doi:10.1016/j.physletb.2008.12.021}}.

\bibitem{Lee:2008uy}
S.~H. Lee, M.~Nielsen, U.~Wiedner, {D(s)D* molecule as an axial meson}, J.
  Korean Phys. Soc. 55 (2009) 424.
\newblock \href {http://arxiv.org/abs/0803.1168} {\path{arXiv:0803.1168}},
  \href {http://dx.doi.org/10.3938/jkps.55.424}
  {\path{doi:10.3938/jkps.55.424}}.

\bibitem{Albuquerque:2009ak}
R.~M. Albuquerque, M.~E. Bracco, M.~Nielsen, {A QCD sum rule calculation for
  the Y(4140) narrow structure}, Phys. Lett. B678 (2009) 186--190.
\newblock \href {http://arxiv.org/abs/0903.5540} {\path{arXiv:0903.5540}},
  \href {http://dx.doi.org/10.1016/j.physletb.2009.06.022}
  {\path{doi:10.1016/j.physletb.2009.06.022}}.

\bibitem{Matheus:2009vq}
R.~D. Matheus, F.~S. Navarra, M.~Nielsen, C.~M. Zanetti, {QCD Sum Rules for the
  X(3872) as a mixed molecule-charmoniun state}, Phys. Rev. D80 (2009) 056002.
\newblock \href {http://arxiv.org/abs/0907.2683} {\path{arXiv:0907.2683}},
  \href {http://dx.doi.org/10.1103/PhysRevD.80.056002}
  {\path{doi:10.1103/PhysRevD.80.056002}}.

\bibitem{Albuquerque:2010fm}
R.~M. Albuquerque, J.~M. Dias, M.~Nielsen, {Can the X(4350) narrow structure be
  a $1^{-+}$ exotic state?}, Phys. Lett. B690 (2010) 141--144.
\newblock \href {http://arxiv.org/abs/1001.3092} {\path{arXiv:1001.3092}},
  \href {http://dx.doi.org/10.1016/j.physletb.2010.05.024}
  {\path{doi:10.1016/j.physletb.2010.05.024}}.

\bibitem{Nielsen:2010ij}
M.~Nielsen, C.~M. Zanetti, {Radiative decay of the X(3872) as a mixed
  molecule-charmonium state in QCD Sum Rules}, Phys. Rev. D82 (2010) 116002.
\newblock \href {http://arxiv.org/abs/1006.0467} {\path{arXiv:1006.0467}},
  \href {http://dx.doi.org/10.1103/PhysRevD.82.116002}
  {\path{doi:10.1103/PhysRevD.82.116002}}.

\bibitem{Narison:2010pd}
S.~Narison, F.~S. Navarra, M.~Nielsen, {On the nature of the X(3872) from QCD},
  Phys. Rev. D83 (2011) 016004.
\newblock \href {http://arxiv.org/abs/1006.4802} {\path{arXiv:1006.4802}},
  \href {http://dx.doi.org/10.1103/PhysRevD.83.016004}
  {\path{doi:10.1103/PhysRevD.83.016004}}.

\bibitem{Finazzo:2011he}
S.~I. Finazzo, M.~Nielsen, X.~Liu, {QCD sum rule calculation for the
  charmonium-like structures in the $J/\psi \phi$ and $J/\psi \omega$ invariant
  mass spectra}, Phys. Lett. B701 (2011) 101--106.
\newblock \href {http://arxiv.org/abs/1102.2347} {\path{arXiv:1102.2347}},
  \href {http://dx.doi.org/10.1016/j.physletb.2011.05.042}
  {\path{doi:10.1016/j.physletb.2011.05.042}}.

\bibitem{Zanetti:2011ju}
C.~M. Zanetti, M.~Nielsen, R.~D. Matheus, {QCD Sum Rules for the production of
  the X(3872) as a mixed molecule-charmonium state in B meson decay}, Phys.
  Lett. B702 (2011) 359--363.
\newblock \href {http://arxiv.org/abs/1105.1343} {\path{arXiv:1105.1343}},
  \href {http://dx.doi.org/10.1016/j.physletb.2011.07.018}
  {\path{doi:10.1016/j.physletb.2011.07.018}}.

\bibitem{Dias:2011mi}
J.~M. Dias, S.~Narison, F.~S. Navarra, M.~Nielsen, J.~M. Richard, {Relation
  between $T_{cc,bb}$ and $X_{c,b}$ from QCD}, Phys. Lett. B703 (2011)
  274--280.
\newblock \href {http://arxiv.org/abs/1105.5630} {\path{arXiv:1105.5630}},
  \href {http://dx.doi.org/10.1016/j.physletb.2011.07.082}
  {\path{doi:10.1016/j.physletb.2011.07.082}}.

\bibitem{Albuquerque:2011ix}
R.~M. Albuquerque, M.~Nielsen, R.~Rodrigues~da Silva, {Exotic $1^{--}$ States
  in QCD Sum Rules}, Phys. Rev. D84 (2011) 116004.
\newblock \href {http://arxiv.org/abs/1110.2113} {\path{arXiv:1110.2113}},
  \href {http://dx.doi.org/10.1103/PhysRevD.84.116004}
  {\path{doi:10.1103/PhysRevD.84.116004}}.

\bibitem{Sun:2012sy}
Z.-F. Sun, X.~Liu, M.~Nielsen, S.-L. Zhu, {Hadronic molecules with both open
  charm and bottom}, Phys. Rev. D85 (2012) 094008.
\newblock \href {http://arxiv.org/abs/1203.1090} {\path{arXiv:1203.1090}},
  \href {http://dx.doi.org/10.1103/PhysRevD.85.094008}
  {\path{doi:10.1103/PhysRevD.85.094008}}.

\bibitem{Dias:2012ek}
J.~M. Dias, R.~M. Albuquerque, M.~Nielsen, C.~M. Zanetti, {Y(4260) as a mixed
  charmonium-tetraquark state}, Phys. Rev. D86 (2012) 116012.
\newblock \href {http://arxiv.org/abs/1209.6592} {\path{arXiv:1209.6592}},
  \href {http://dx.doi.org/10.1103/PhysRevD.86.116012}
  {\path{doi:10.1103/PhysRevD.86.116012}}.

\bibitem{Dias:2013xfa}
J.~M. Dias, F.~S. Navarra, M.~Nielsen, C.~M. Zanetti, {$Z^+_c$(3900) decay
  width in QCD sum rules}, Phys. Rev. D88~(1) (2013) 016004.
\newblock \href {http://arxiv.org/abs/1304.6433} {\path{arXiv:1304.6433}},
  \href {http://dx.doi.org/10.1103/PhysRevD.88.016004}
  {\path{doi:10.1103/PhysRevD.88.016004}}.

\bibitem{Torres:2013saa}
A.~Martinez~Torres, K.~P. Khemchandani, M.~Nielsen, F.~S. Navarra, E.~Oset,
  {Exploring the $D^*\rho$ system within QCD sum rules}, Phys. Rev. D88~(7)
  (2013) 074033.
\newblock \href {http://arxiv.org/abs/1307.1724} {\path{arXiv:1307.1724}},
  \href {http://dx.doi.org/10.1103/PhysRevD.88.074033}
  {\path{doi:10.1103/PhysRevD.88.074033}}.

\bibitem{Dias:2013qga}
J.~M. Dias, X.~Liu, M.~Nielsen, {Predicition for the decay width of a charged
  state near the $D_s\bar{D}^*/D^*_s\bar{D}$ threshold}, Phys. Rev. D88~(9)
  (2013) 096014.
\newblock \href {http://arxiv.org/abs/1307.7100} {\path{arXiv:1307.7100}},
  \href {http://dx.doi.org/10.1103/PhysRevD.88.096014}
  {\path{doi:10.1103/PhysRevD.88.096014}}.

\bibitem{Khemchandani:2013iwa}
K.~P. Khemchandani, A.~Martinez~Torres, M.~Nielsen, F.~S. Navarra, {Relating
  $D^* \bar{D}^*$ currents with $J^\pi= 0^+,1^+$ and $2^+$ to $Z_c$ states},
  Phys. Rev. D89~(1) (2014) 014029.
\newblock \href {http://arxiv.org/abs/1310.0862} {\path{arXiv:1310.0862}},
  \href {http://dx.doi.org/10.1103/PhysRevD.89.014029}
  {\path{doi:10.1103/PhysRevD.89.014029}}.

\bibitem{Torres:2013lka}
A.~Martinez~Torres, K.~P. Khemchandani, F.~S. Navarra, M.~Nielsen, E.~Oset,
  {Reanalysis of the $e^+ e^- \to (D^* \bar D^*)^{\pm} \pi^{\mp}$ reaction and
  the claim for the $Z_c (4025)$ resonance}, Phys. Rev. D89~(1) (2014) 014025.
\newblock \href {http://arxiv.org/abs/1310.1119} {\path{arXiv:1310.1119}},
  \href {http://dx.doi.org/10.1103/PhysRevD.89.014025}
  {\path{doi:10.1103/PhysRevD.89.014025}}.

\bibitem{Albuquerque:2013owa}
R.~M. Albuquerque, J.~M. Dias, M.~Nielsen, C.~M. Zanetti, {Y(3940) as a Mixed
  Charmonium-Molecule State}, Phys. Rev. D89~(7) (2014) 076007.
\newblock \href {http://arxiv.org/abs/1311.6411} {\path{arXiv:1311.6411}},
  \href {http://dx.doi.org/10.1103/PhysRevD.89.076007}
  {\path{doi:10.1103/PhysRevD.89.076007}}.

\bibitem{Albuquerque:2015nwa}
R.~M. Albuquerque, M.~Nielsen, C.~M. Zanetti, {Production of the Y (4260) state
  in B meson decay}, Phys. Lett. B747 (2015) 83--87.
\newblock \href {http://arxiv.org/abs/1502.00119} {\path{arXiv:1502.00119}},
  \href {http://dx.doi.org/10.1016/j.physletb.2015.05.022}
  {\path{doi:10.1016/j.physletb.2015.05.022}}.

\bibitem{Albuquerque:2015kia}
R.~M. Albuquerque, R.~D. Matheus, {The $J/\Psi\Phi$ decay channel of the
  Y(4140) molecular state}, Nucl. Part. Phys. Proc. 258-259 (2015) 148--151.
\newblock \href {http://dx.doi.org/10.1016/j.nuclphysbps.2015.01.032}
  {\path{doi:10.1016/j.nuclphysbps.2015.01.032}}.

\bibitem{Torres:2016oyz}
A.~Mart{\'\i}nez~Torres, K.~P. Khemchandani, J.~M. Dias, F.~S. Navarra,
  M.~Nielsen, {Understanding close-lying exotic charmonia states within QCD sum
  rules}, Nucl. Phys. A966 (2017) 135--157.
\newblock \href {http://arxiv.org/abs/1606.07505} {\path{arXiv:1606.07505}},
  \href {http://dx.doi.org/10.1016/j.nuclphysa.2017.06.022}
  {\path{doi:10.1016/j.nuclphysa.2017.06.022}}.

\bibitem{Albuquerque:2016znh}
R.~Albuquerque, S.~Narison, F.~Fanomezana, A.~Rabemananjara, D.~Rabetiarivony,
  G.~Randriamanatrika, {XYZ-like Spectra from Laplace Sum Rule at N2LO in the
  Chiral Limit}, Int. J. Mod. Phys. A31~(36) (2016) 1650196.
\newblock \href {http://arxiv.org/abs/1609.03351} {\path{arXiv:1609.03351}},
  \href {http://dx.doi.org/10.1142/S0217751X16501967}
  {\path{doi:10.1142/S0217751X16501967}}.

\bibitem{Albuquerque:2017vfq}
R.~Albuquerque, S.~Narison, D.~Rabetiarivony, G.~Randriamanatrika, {XYZ-SU3
  Breakings from Laplace Sum Rules at Higher Orders}, Int. J. Mod. Phys.
  A33~(16) (2018) 1850082.
\newblock \href {http://arxiv.org/abs/1709.09023} {\path{arXiv:1709.09023}},
  \href {http://dx.doi.org/10.1142/S0217751X18500823}
  {\path{doi:10.1142/S0217751X18500823}}.

\bibitem{Albuquerque:2018jss}
R.~M. Albuquerque, F.~Fanomezana, S.~Narison, A.~Rabemananjara,
  D.~Rabetiarivony, G.~Randriamanatrika,
  \href{http://inspirehep.net/record/1650836/files/arXiv:1801.09110.pdf}{{0$^+$
  and 1$^+$ heavy-light exotic mesons at N2LO in the chiral limit}}, in: {20th
  High-Energy Physics International Conference in Quantum Chromodynamics (QCD
  17) Montpellier, France, July 3-7, 2017}, 2018.
\newblock \href {http://arxiv.org/abs/1801.09110} {\path{arXiv:1801.09110}}.
\newline\urlprefix\url{http://inspirehep.net/record/1650836/files/arXiv:1801.09110.pdf}

\bibitem{Shifman:1978bx}
M.~A. Shifman, A.~I. Vainshtein, V.~I. Zakharov, {QCD and Resonance Physics.
  Theoretical Foundations}, Nucl. Phys. B147 (1979) 385--447.
\newblock \href {http://dx.doi.org/10.1016/0550-3213(79)90022-1}
  {\path{doi:10.1016/0550-3213(79)90022-1}}.

\bibitem{Shifman:1978by}
M.~A. Shifman, A.~I. Vainshtein, V.~I. Zakharov, {QCD and Resonance Physics:
  Applications}, Nucl. Phys. B147 (1979) 448--518.
\newblock \href {http://dx.doi.org/10.1016/0550-3213(79)90023-3}
  {\path{doi:10.1016/0550-3213(79)90023-3}}.

\bibitem{Reinders:1984sr}
L.~J. Reinders, H.~Rubinstein, S.~Yazaki, {Hadron Properties from QCD Sum
  Rules}, Phys. Rept. 127 (1985) 1.
\newblock \href {http://dx.doi.org/10.1016/0370-1573(85)90065-1}
  {\path{doi:10.1016/0370-1573(85)90065-1}}.

\bibitem{Narison:2007spa}
S.~Narison,
  \href{http://www.cambridge.org/zw/academic/subjects/physics/particle-physics-and-nuclear-physics/qcd-theory-hadrons-partons-confinement?format=PB}{{QCD
  as a Theory of Hadrons}}, Vol.~17, Cambridge University Press, 2007.
\newline\urlprefix\url{http://www.cambridge.org/zw/academic/subjects/physics/particle-physics-and-nuclear-physics/qcd-theory-hadrons-partons-confinement?format=PB}

\bibitem{Narison:1989aq}
S.~Narison, {QCD spectral sum rules}, World Sci. Lect. Notes Phys. 26 (1989)
  1--527.

\bibitem{Narison:1980ti}
S.~Narison, {Techniques of Dimensional Renormalization and Applications to the
  Two Point Functions of QCD and QED}, Phys. Rept. 84 (1982) 263--399.
\newblock \href {http://dx.doi.org/10.1016/0370-1573(82)90023-0}
  {\path{doi:10.1016/0370-1573(82)90023-0}}.

\bibitem{Narison:2002pw}
S.~Narison, {Withdrawn: QCD as a theory of hadrons from partons to
  confinement}\href {http://arxiv.org/abs/hep-ph/0205006}
  {\path{arXiv:hep-ph/0205006}}.

\bibitem{Aaij:2015tga}
R.~Aaij, et~al., {Observation of $J/\psi p$ Resonances Consistent with
  Pentaquark States in $\Lambda_b^0 \to J/\psi K^- p$ Decays}, Phys. Rev. Lett.
  115 (2015) 072001.
\newblock \href {http://arxiv.org/abs/1507.03414} {\path{arXiv:1507.03414}},
  \href {http://dx.doi.org/10.1103/PhysRevLett.115.072001}
  {\path{doi:10.1103/PhysRevLett.115.072001}}.

\bibitem{Aaij:2016phn}
R.~Aaij, et~al., {Model-independent evidence for $J/\psi p$ contributions to
  $\Lambda_b^0\to J/\psi p K^-$ decays}, Phys. Rev. Lett. 117~(8) (2016)
  082002.
\newblock \href {http://arxiv.org/abs/1604.05708} {\path{arXiv:1604.05708}},
  \href {http://dx.doi.org/10.1103/PhysRevLett.117.082002}
  {\path{doi:10.1103/PhysRevLett.117.082002}}.

\bibitem{Aaij:2016ymb}
R.~Aaij, et~al., {Evidence for exotic hadron contributions to $\Lambda_b^0 \to
  J/\psi p \pi^-$ decays}, Phys. Rev. Lett. 117~(8) (2016) 082003, [Addendum:
  Phys. Rev. Lett.118,119901(2017)].
\newblock \href {http://arxiv.org/abs/1606.06999} {\path{arXiv:1606.06999}},
  \href {http://dx.doi.org/10.1103/PhysRevLett.118.119901,
  10.1103/PhysRevLett.117.082003, 10.1103/PhysRevLett.117.109902}
  {\path{doi:10.1103/PhysRevLett.118.119901, 10.1103/PhysRevLett.117.082003,
  10.1103/PhysRevLett.117.109902}}.

\bibitem{Ioffe:1981kw}
B.~L. Ioffe, {Calculation of Baryon Masses in Quantum Chromodynamics}, Nucl.
  Phys. B188 (1981) 317--341, [Erratum: Nucl. Phys.B191,591(1981)].
\newblock \href {http://dx.doi.org/10.1016/0550-3213(81)90315-1,
  10.1016/0550-3213(81)90259-5} {\path{doi:10.1016/0550-3213(81)90315-1,
  10.1016/0550-3213(81)90259-5}}.

\bibitem{Shifman:2001ck}
M.~Shifman, B.~Ioffe (Eds.), {At the frontier of particle physics. Handbook of
  QCD. Vol. 1-3}, World Scientific, Singapore, Singapore, 2001.
\newblock \href {http://dx.doi.org/10.1142/4544} {\path{doi:10.1142/4544}}.

\bibitem{Radyushkin:1998du}
A.~V. Radyushkin, {Introduction into QCD sum rule approach}, in: {Strong
  interactions at low and intermediate energies. Proceedings, 13th Annual
  Hampton University Graduate Studies, HUGS'98, Newport News, USA, May 26-June
  12, 1998}, 1998.
\newblock \href {http://arxiv.org/abs/hep-ph/0101227}
  {\path{arXiv:hep-ph/0101227}}.

\bibitem{Novikov:1977dq}
V.~A. Novikov, L.~B. Okun, M.~A. Shifman, A.~I. Vainshtein, M.~B. Voloshin,
  V.~I. Zakharov, {Charmonium and Gluons: Basic Experimental Facts and
  Theoretical Introduction}, Phys. Rept. 41 (1978) 1--133.
\newblock \href {http://dx.doi.org/10.1016/0370-1573(78)90120-5}
  {\path{doi:10.1016/0370-1573(78)90120-5}}.

\bibitem{Shifman:1998rb}
M.~A. Shifman, {Snapshots of hadrons or the story of how the vacuum medium
  determines the properties of the classical mesons which are produced, live
  and die in the QCD vacuum}, Prog. Theor. Phys. Suppl. 131 (1998) 1--71,
  [,111(1998)].
\newblock \href {http://arxiv.org/abs/hep-ph/9802214}
  {\path{arXiv:hep-ph/9802214}}, \href {http://dx.doi.org/10.1143/PTPS.131.1}
  {\path{doi:10.1143/PTPS.131.1}}.

\bibitem{Cohen:1994wm}
T.~D. Cohen, R.~J. Furnstahl, D.~K. Griegel, X.-m. Jin, {QCD sum rules and
  applications to nuclear physics}, Prog. Part. Nucl. Phys. 35 (1995) 221--298,
  [,221(1994)].
\newblock \href {http://arxiv.org/abs/hep-ph/9503315}
  {\path{arXiv:hep-ph/9503315}}, \href
  {http://dx.doi.org/10.1016/0146-6410(95)00043-I}
  {\path{doi:10.1016/0146-6410(95)00043-I}}.

\bibitem{Shifman:2000jv}
M.~A. Shifman, {Quark hadron duality}, in: {At the frontier of particle
  physics. Handbook of QCD. Vol. 1-3}, World Scientific, World Scientific,
  Singapore, 2001, pp. 1447--1494, [3,1447(2000)].
\newblock \href {http://arxiv.org/abs/hep-ph/0009131}
  {\path{arXiv:hep-ph/0009131}}.

\bibitem{Shifman:2001qm}
M.~Shifman, {Lectures on quark hadron duality}, Czech. J. Phys. 52 (2002)
  B102--B135.
\newblock \href {http://dx.doi.org/10.1007/s10582-002-0080-6}
  {\path{doi:10.1007/s10582-002-0080-6}}.

\bibitem{Wilson:1969zs}
K.~G. Wilson, {Nonlagrangian models of current algebra}, Phys. Rev. 179 (1969)
  1499--1512.
\newblock \href {http://dx.doi.org/10.1103/PhysRev.179.1499}
  {\path{doi:10.1103/PhysRev.179.1499}}.

\bibitem{Leinweber:1995fn}
D.~B. Leinweber, {QCD sum rules for skeptics}, Annals Phys. 254 (1997)
  328--396.
\newblock \href {http://arxiv.org/abs/nucl-th/9510051}
  {\path{arXiv:nucl-th/9510051}}, \href
  {http://dx.doi.org/10.1006/aphy.1996.5641}
  {\path{doi:10.1006/aphy.1996.5641}}.

\bibitem{Dominguez:2013ata}
C.~A. Dominguez, {Introduction to QCD sum rules}, Mod. Phys. Lett. A28 (2013)
  1360002.
\newblock \href {http://arxiv.org/abs/1305.7047} {\path{arXiv:1305.7047}},
  \href {http://dx.doi.org/10.1142/S021773231360002X}
  {\path{doi:10.1142/S021773231360002X}}.

\bibitem{Reinders:1980wy}
L.~J. Reinders, H.~R. Rubinstein, S.~Yazaki, {QCD CONTRIBUTIONS TO VACUUM
  POLARIZATION}, Phys. Lett. 94B (1980) 203--206.
\newblock \href {http://dx.doi.org/10.1016/0370-2693(80)90859-X}
  {\path{doi:10.1016/0370-2693(80)90859-X}}.

\bibitem{Reinders:1981si}
L.~J. Reinders, H.~R. Rubinstein, S.~Yazaki, {QCD Sum Rules for Heavy Quark
  Systems}, Nucl. Phys. B186 (1981) 109--146.
\newblock \href {http://dx.doi.org/10.1016/0550-3213(81)90095-X}
  {\path{doi:10.1016/0550-3213(81)90095-X}}.

\bibitem{Narison:2010cg}
S.~Narison, {Gluon condensates and c, b quark masses from quarkonia ratios of
  moments}, Phys. Lett. B693 (2010) 559--566, [Erratum: Phys.
  Lett.B705,544(2011)].
\newblock \href {http://arxiv.org/abs/1004.5333} {\path{arXiv:1004.5333}},
  \href {http://dx.doi.org/10.1016/j.physletb.2011.09.116,
  10.1016/j.physletb.2010.09.007} {\path{doi:10.1016/j.physletb.2011.09.116,
  10.1016/j.physletb.2010.09.007}}.

\bibitem{GellMann:1968rz}
M.~Gell-Mann, R.~J. Oakes, B.~Renner, {Behavior of current divergences under
  SU(3) x SU(3)}, Phys. Rev. 175 (1968) 2195--2199.
\newblock \href {http://dx.doi.org/10.1103/PhysRev.175.2195}
  {\path{doi:10.1103/PhysRev.175.2195}}.

\bibitem{Tanabashi:2018oca}
M.~Tanabashi, et~al., {Review of Particle Physics}, Phys. Rev. D98~(3) (2018)
  030001.
\newblock \href {http://dx.doi.org/10.1103/PhysRevD.98.030001}
  {\path{doi:10.1103/PhysRevD.98.030001}}.

\bibitem{Bagan:1984zt}
E.~Bagan, J.~I. Latorre, P.~Pascual, R.~Tarrach, {Heavy Quark Expansion,
  Factorization and Eight-dimensional Gluon Condensates}, Nucl. Phys. B254
  (1985) 555--568.
\newblock \href {http://dx.doi.org/10.1016/0550-3213(85)90233-0}
  {\path{doi:10.1016/0550-3213(85)90233-0}}.

\bibitem{Launer:1983ib}
G.~Launer, S.~Narison, R.~Tarrach, {Nonperturbative {QCD} Vacuum From $e^+ e^-
  \to$ I = 1 Hadron Data}, Z. Phys. C26 (1984) 433--439.
\newblock \href {http://dx.doi.org/10.1007/BF01452571}
  {\path{doi:10.1007/BF01452571}}.

\bibitem{Narison:2009vy}
S.~Narison, {Power corrections to alpha(s)(M(tau)),|V(us)| and anti-m(s)},
  Phys. Lett. B673 (2009) 30--36.
\newblock \href {http://arxiv.org/abs/0901.3823} {\path{arXiv:0901.3823}},
  \href {http://dx.doi.org/10.1016/j.physletb.2009.01.062}
  {\path{doi:10.1016/j.physletb.2009.01.062}}.

\bibitem{Braghin:2014nva}
F.~L. Braghin, F.~S. Navarra, {Factorization breaking of four-quark condensates
  in the Nambu–Jona-Lasinio model}, Phys. Rev. D91~(7) (2015) 074008.
\newblock \href {http://arxiv.org/abs/1404.4094} {\path{arXiv:1404.4094}},
  \href {http://dx.doi.org/10.1103/PhysRevD.91.074008}
  {\path{doi:10.1103/PhysRevD.91.074008}}.

\bibitem{Pascual:1984zb}
P.~Pascual, R.~Tarrach, {QCD: RENORMALIZATION FOR THE PRACTITIONER}, Lect.
  Notes Phys. 194 (1984) 1--277.

\bibitem{Lucha:2007pz}
W.~Lucha, D.~Melikhov, S.~Simula, {Systematic uncertainties of hadron
  parameters obtained with QCD sum rules}, Phys. Rev. D76 (2007) 036002.
\newblock \href {http://arxiv.org/abs/0705.0470} {\path{arXiv:0705.0470}},
  \href {http://dx.doi.org/10.1103/PhysRevD.76.036002}
  {\path{doi:10.1103/PhysRevD.76.036002}}.

\bibitem{Narison:2004vz}
S.~Narison, {V-A hadronic tau decays: A Laboratory for the QCD vacuum}, Phys.
  Lett. B624 (2005) 223--232.
\newblock \href {http://arxiv.org/abs/hep-ph/0412152}
  {\path{arXiv:hep-ph/0412152}}, \href
  {http://dx.doi.org/10.1016/j.physletb.2005.08.007}
  {\path{doi:10.1016/j.physletb.2005.08.007}}.

\bibitem{Kallen:1958ifa}
G.~Kallen, A.~S. Wightman, {The Analytic Properties of the Vacuum Expectation
  Value\ of a Product of Three Scalar Local Fields}, Vol.~1, 1958.

\bibitem{Kallen:1959kza}
G.~Kallen, H.~Wilhelmsson, {Generalized singular functions}, Vol.~1, 1959.

\bibitem{Martin:1999cr}
A.~Martin, {The Rigorous analyticity unitarity program and its\ successes},
  Lect. Notes Phys. 558 (2000) 127--135, [,127(1999)].
\newblock \href {http://arxiv.org/abs/hep-ph/9906393}
  {\path{arXiv:hep-ph/9906393}}.

\bibitem{Zwicky:2016lka}
R.~Zwicky, {A brief Introduction to Dispersion Relations and\ Analyticity}, in:
  {Proceedings, Quantum Field Theory at the Limits: from\ Strong Fields to
  Heavy Quarks (HQ 2016): Dubna, Russia, July 18-30, 2016}, 2017, pp. 93--120.
\newblock \href {http://arxiv.org/abs/1610.06090} {\path{arXiv:1610.06090}},
  \href {http://dx.doi.org/10.3204/DESY-PROC-2016-04/Zwicky}
  {\path{doi:10.3204/DESY-PROC-2016-04/Zwicky}}.

\bibitem{Ioffe:1982qb}
B.~L. Ioffe, A.~V. Smilga, {Meson Widths and Form-Factors at Intermediate
  Momentum Transfer in Nonperturbative QCD}, Nucl. Phys. B216 (1983) 373--407.
\newblock \href {http://dx.doi.org/10.1016/0550-3213(83)90291-2}
  {\path{doi:10.1016/0550-3213(83)90291-2}}.

\bibitem{Reinders:1983wi}
L.~J. Reinders, {{QCD} Sum Rules: An Introduction and Some Applications}, Acta
  Phys. Polon. B15 (1984) 329.

\bibitem{Nesterenko:1982gc}
V.~A. Nesterenko, A.~V. Radyushkin, {Sum Rules and Pion Form-Factor in QCD},
  Phys. Lett. 115B (1982) 410.
\newblock \href {http://dx.doi.org/10.1016/0370-2693(82)90528-7}
  {\path{doi:10.1016/0370-2693(82)90528-7}}.

\bibitem{Bracco:2011pg}
M.~E. Bracco, M.~Chiapparini, M.~Navarra, F. S. and\~Nielsen, {Charm couplings
  and form factors in QCD sum rules}, Prog. Part. Nucl. Phys. 67 (2012)
  1019--1052.
\newblock \href {http://arxiv.org/abs/1104.2864} {\path{arXiv:1104.2864}},
  \href {http://dx.doi.org/10.1016/j.ppnp.2012.03.002}
  {\path{doi:10.1016/j.ppnp.2012.03.002}}.

\bibitem{Cutkosky:1960sp}
R.~E. Cutkosky, {Singularities and discontinuities of Feynman\ amplitudes}, J.
  Math. Phys. 1 (1960) 429--433.
\newblock \href {http://dx.doi.org/10.1063/1.1703676}
  {\path{doi:10.1063/1.1703676}}.

\bibitem{Dias:2016dme}
J.~M. Dias, K.~P. Khemchandani, A.~Mart{\'\i}nez~Torres, M.~Nielsen, C.~M.
  Zanetti, {A QCD sum rule calculation of the $X^\pm(5568) \to B_{s}^0\pi^\pm$
  decay width}, Phys. Lett. B758 (2016) 235--238.
\newblock \href {http://arxiv.org/abs/1603.02249} {\path{arXiv:1603.02249}},
  \href {http://dx.doi.org/10.1016/j.physletb.2016.05.015}
  {\path{doi:10.1016/j.physletb.2016.05.015}}.

\bibitem{Duraes:2002px}
F.~O. Duraes, S.~H. Lee, F.~S. Navarra, M.~Nielsen, {J / psi dissociation by
  pions in QCD}, Phys. Lett. B564 (2003) 97--103.
\newblock \href {http://arxiv.org/abs/nucl-th/0210075}
  {\path{arXiv:nucl-th/0210075}}, \href
  {http://dx.doi.org/10.1016/S0370-2693(03)00709-3}
  {\path{doi:10.1016/S0370-2693(03)00709-3}}.

\bibitem{Bracco:1999xe}
M.~E. Bracco, F.~S. Navarra, M.~Nielsen, {$g_{N K \Lambda}$ and $g_{N K
  \Sigma}$ from QCD sum rules in the $\gamma_5 \sigma_{\mu \nu}$ structure},
  Phys. Lett. B454 (1999) 346--352.
\newblock \href {http://arxiv.org/abs/nucl-th/9902007}
  {\path{arXiv:nucl-th/9902007}}, \href
  {http://dx.doi.org/10.1016/S0370-2693(99)00354-8}
  {\path{doi:10.1016/S0370-2693(99)00354-8}}.

\bibitem{Ioffe:1983ju}
B.~L. Ioffe, A.~V. Smilga, {Nucleon Magnetic Moments and Magnetic Properties
  of\ Vacuum in QCD}, Nucl. Phys. B232 (1984) 109--142.
\newblock \href {http://dx.doi.org/10.1016/0550-3213(84)90364-X}
  {\path{doi:10.1016/0550-3213(84)90364-X}}.

\bibitem{Eidemuller:2005jm}
M.~Eidemuller, F.~S. Navarra, R.~Nielsen, M. and\ Rodrigues da~Silva,
  {Pentaquark decay width in QCD sum rules}, Phys. Rev. D72 (2005) 034003.
\newblock \href {http://arxiv.org/abs/hep-ph/0503193}
  {\path{arXiv:hep-ph/0503193}}, \href
  {http://dx.doi.org/10.1103/PhysRevD.72.034003}
  {\path{doi:10.1103/PhysRevD.72.034003}}.

\bibitem{Navarra:2001ju}
F.~S. Navarra, M.~Nielsen, M.~E. Bracco, {D* D pi form-factor revisited}, Phys.
  Rev. D65 (2002) 037502.
\newblock \href {http://arxiv.org/abs/hep-ph/0109188}
  {\path{arXiv:hep-ph/0109188}}, \href
  {http://dx.doi.org/10.1103/PhysRevD.65.037502}
  {\path{doi:10.1103/PhysRevD.65.037502}}.

\bibitem{Bracco:2004rx}
M.~E. Bracco, M.~Chiapparini, F.~S. Navarra, M.~Nielsen, {J/psi D*D* vertex
  from QCD sum rules}, Phys. Lett. B605 (2005) 326--334.
\newblock \href {http://arxiv.org/abs/hep-ph/0410071}
  {\path{arXiv:hep-ph/0410071}}, \href
  {http://dx.doi.org/10.1016/j.physletb.2004.11.024}
  {\path{doi:10.1016/j.physletb.2004.11.024}}.

\bibitem{Duraes:2004uc}
F.~O. Duraes, F.~S. Navarra, M.~Nielsen, M.~R. Robilotta, {Meson loops and the
  g(D*D pi) coupling}, Braz. J. Phys. 36 (2006) 1232--1237.
\newblock \href {http://arxiv.org/abs/hep-ph/0403064}
  {\path{arXiv:hep-ph/0403064}}, \href
  {http://dx.doi.org/10.1590/S0103-97332006000700021}
  {\path{doi:10.1590/S0103-97332006000700021}}.

\bibitem{Dosch:1994wj}
H.~G. Dosch, {Nonperturbative methods in quantum chromodynamics}, Prog. Part.
  Nucl. Phys. 33 (1994) 121--200.
\newblock \href {http://dx.doi.org/10.1016/0146-6410(94)90044-2}
  {\path{doi:10.1016/0146-6410(94)90044-2}}.

\bibitem{Goerke:2016hxf}
F.~Goerke, T.~Gutsche, M.~A. Ivanov, J.~G. Korner, V.~E. Lyubovitskij,
  P.~Santorelli, {Four-quark structure of Zc(3900), Z(4430) and Xb(5568)
  states}, Phys. Rev. D94~(9) (2016) 094017.
\newblock \href {http://arxiv.org/abs/1608.04656} {\path{arXiv:1608.04656}},
  \href {http://dx.doi.org/10.1103/PhysRevD.94.094017}
  {\path{doi:10.1103/PhysRevD.94.094017}}.

\bibitem{Kondo:2004cr}
Y.~Kondo, O.~Morimatsu, T.~Nishikawa, {Two-hadron-irreducible QCD sum rule for
  pentaquark baryon}, Phys. Lett. B611 (2005) 93--101.
\newblock \href {http://arxiv.org/abs/hep-ph/0404285}
  {\path{arXiv:hep-ph/0404285}}, \href
  {http://dx.doi.org/10.1016/j.physletb.2005.01.070}
  {\path{doi:10.1016/j.physletb.2005.01.070}}.

\bibitem{Lee:2004xk}
S.~H. Lee, H.~Kim, Y.~Kwon, {Parity of Theta+(1540) from QCD sum rules}, Phys.
  Lett. B609 (2005) 252--258.
\newblock \href {http://arxiv.org/abs/hep-ph/0411104}
  {\path{arXiv:hep-ph/0411104}}, \href
  {http://dx.doi.org/10.1016/j.physletb.2005.01.029}
  {\path{doi:10.1016/j.physletb.2005.01.029}}.

\bibitem{Lee:2007mva}
H.-J. Lee, N.~I. Kochelev, {On the pi pi contribution to the QCD sum rules for
  the light tetraquark}, Phys. Rev. D78 (2008) 076005.
\newblock \href {http://arxiv.org/abs/hep-ph/0702225}
  {\path{arXiv:hep-ph/0702225}}, \href
  {http://dx.doi.org/10.1103/PhysRevD.78.076005}
  {\path{doi:10.1103/PhysRevD.78.076005}}.

\bibitem{Chen:2009gs}
H.-X. Chen, A.~Hosaka, H.~Toki, S.-L. Zhu, {Light Scalar Meson sigma(600) in
  QCD Sum Rule with Continuum}, Phys. Rev. D81 (2010) 114034.
\newblock \href {http://arxiv.org/abs/0912.5138} {\path{arXiv:0912.5138}},
  \href {http://dx.doi.org/10.1103/PhysRevD.81.114034}
  {\path{doi:10.1103/PhysRevD.81.114034}}.

\bibitem{Swanson:2014tra}
E.~S. Swanson, {$Z_b$ and $Z_c$ Exotic States as Coupled Channel Cusps}, Phys.
  Rev. D91~(3) (2015) 034009.
\newblock \href {http://arxiv.org/abs/1409.3291} {\path{arXiv:1409.3291}},
  \href {http://dx.doi.org/10.1103/PhysRevD.91.034009}
  {\path{doi:10.1103/PhysRevD.91.034009}}.

\bibitem{Bugg:2011jr}
D.~V. Bugg, {An Explanation of Belle states $Z_b(10610)$ and $Z_b(10650)$}, EPL
  96~(1) (2011) 11002.
\newblock \href {http://arxiv.org/abs/1105.5492} {\path{arXiv:1105.5492}},
  \href {http://dx.doi.org/10.1209/0295-5075/96/11002}
  {\path{doi:10.1209/0295-5075/96/11002}}.

\bibitem{Guo:2014iya}
F.-K. Guo, C.~Hanhart, Q.~Wang, Q.~Zhao, {Could the near-threshold $XYZ$ states
  be simply kinematic effects?}, Phys. Rev. D91~(5) (2015) 051504.
\newblock \href {http://arxiv.org/abs/1411.5584} {\path{arXiv:1411.5584}},
  \href {http://dx.doi.org/10.1103/PhysRevD.91.051504}
  {\path{doi:10.1103/PhysRevD.91.051504}}.

\bibitem{Aubert:2004ns}
B.~Aubert, et~al., {Study of the $B \to J/\psi K^- \pi^+ \pi^-$ decay and
  measurement of the $B \to X(3872) K^-$ branching fraction}, Phys. Rev. D71
  (2005) 071103.
\newblock \href {http://arxiv.org/abs/hep-ex/0406022}
  {\path{arXiv:hep-ex/0406022}}, \href
  {http://dx.doi.org/10.1103/PhysRevD.71.071103}
  {\path{doi:10.1103/PhysRevD.71.071103}}.

\bibitem{Aaij:2015eva}
R.~Aaij, et~al., {Quantum Numbers of the $X(3872)$ state and orbital angular
  momentum in its $\rho^0 J\psi$ decay}, Phys. Rev. D92~(1) (2015) 011102.
\newblock \href {http://arxiv.org/abs/1504.06339} {\path{arXiv:1504.06339}},
  \href {http://dx.doi.org/10.1103/PhysRevD.92.011102}
  {\path{doi:10.1103/PhysRevD.92.011102}}.

\bibitem{Barnes:2003vb}
T.~Barnes, S.~Godfrey, {Charmonium options for the X(3872)}, Phys. Rev. D69
  (2004) 054008.
\newblock \href {http://arxiv.org/abs/hep-ph/0311162}
  {\path{arXiv:hep-ph/0311162}}, \href
  {http://dx.doi.org/10.1103/PhysRevD.69.054008}
  {\path{doi:10.1103/PhysRevD.69.054008}}.

\bibitem{Barnes:2005pb}
T.~Barnes, S.~Godfrey, E.~S. Swanson, {Higher charmonia}, Phys. Rev. D72 (2005)
  054026.
\newblock \href {http://arxiv.org/abs/hep-ph/0505002}
  {\path{arXiv:hep-ph/0505002}}, \href
  {http://dx.doi.org/10.1103/PhysRevD.72.054026}
  {\path{doi:10.1103/PhysRevD.72.054026}}.

\bibitem{Okamoto:2001jb}
M.~Okamoto, et~al., {Charmonium spectrum from quenched anisotropic lattice
  QCD}, Phys. Rev. D65 (2002) 094508.
\newblock \href {http://arxiv.org/abs/hep-lat/0112020}
  {\path{arXiv:hep-lat/0112020}}, \href
  {http://dx.doi.org/10.1103/PhysRevD.65.094508}
  {\path{doi:10.1103/PhysRevD.65.094508}}.

\bibitem{Tornqvist:1993ng}
N.~A. Tornqvist, {From the deuteron to deusons, an analysis of deuteron - like
  meson meson bound states}, Z. Phys. C61 (1994) 525--537.
\newblock \href {http://arxiv.org/abs/hep-ph/9310247}
  {\path{arXiv:hep-ph/9310247}}, \href {http://dx.doi.org/10.1007/BF01413192}
  {\path{doi:10.1007/BF01413192}}.

\bibitem{Swanson:2004pp}
E.~S. Swanson, {Diagnostic decays of the X(3872)}, Phys. Lett. B598 (2004)
  197--202.
\newblock \href {http://arxiv.org/abs/hep-ph/0406080}
  {\path{arXiv:hep-ph/0406080}}, \href
  {http://dx.doi.org/10.1016/j.physletb.2004.07.059}
  {\path{doi:10.1016/j.physletb.2004.07.059}}.

\bibitem{Swanson:2003tb}
E.~S. Swanson, {Short range structure in the X(3872)}, Phys. Lett. B588 (2004)
  189--195.
\newblock \href {http://arxiv.org/abs/hep-ph/0311229}
  {\path{arXiv:hep-ph/0311229}}, \href
  {http://dx.doi.org/10.1016/j.physletb.2004.03.033}
  {\path{doi:10.1016/j.physletb.2004.03.033}}.

\bibitem{Aceti:2012cb}
F.~Aceti, R.~Molina, E.~Oset, {The $X(3872) \to J/\psi \gamma$ decay in the $D
  \bar D^*$ molecular picture}, Phys. Rev. D86 (2012) 113007.
\newblock \href {http://arxiv.org/abs/1207.2832} {\path{arXiv:1207.2832}},
  \href {http://dx.doi.org/10.1103/PhysRevD.86.113007}
  {\path{doi:10.1103/PhysRevD.86.113007}}.

\bibitem{Maiani:2004vq}
L.~Maiani, F.~Piccinini, A.~D. Polosa, V.~Riquer, {Diquark-antidiquarks with
  hidden or open charm and the nature of X(3872)}, Phys. Rev. D71 (2005)
  014028.
\newblock \href {http://arxiv.org/abs/hep-ph/0412098}
  {\path{arXiv:hep-ph/0412098}}, \href
  {http://dx.doi.org/10.1103/PhysRevD.71.014028}
  {\path{doi:10.1103/PhysRevD.71.014028}}.

\bibitem{Eichten:2005ga}
E.~J. Eichten, K.~Lane, C.~Quigg, {New states above charm threshold}, Phys.
  Rev. D73 (2006) 014014, [Erratum: Phys. Rev.D73,079903(2006)].
\newblock \href {http://arxiv.org/abs/hep-ph/0511179}
  {\path{arXiv:hep-ph/0511179}}, \href
  {http://dx.doi.org/10.1103/PhysRevD.73.014014, 10.1103/PhysRevD.73.079903}
  {\path{doi:10.1103/PhysRevD.73.014014, 10.1103/PhysRevD.73.079903}}.

\bibitem{Suzuki:2005ha}
M.~Suzuki, {The X(3872) boson: Molecule or charmonium}, Phys. Rev. D72 (2005)
  114013.
\newblock \href {http://arxiv.org/abs/hep-ph/0508258}
  {\path{arXiv:hep-ph/0508258}}, \href
  {http://dx.doi.org/10.1103/PhysRevD.72.114013}
  {\path{doi:10.1103/PhysRevD.72.114013}}.

\bibitem{Meng:2005er}
C.~Meng, Y.-J. Gao, K.-T. Chao, {B → $χ_{c1}$(1P,2P)K decays in QCD
  factorization and X(3872)}, Phys. Rev. D87~(7) (2013) 074035.
\newblock \href {http://arxiv.org/abs/hep-ph/0506222}
  {\path{arXiv:hep-ph/0506222}}, \href
  {http://dx.doi.org/10.1103/PhysRevD.87.074035}
  {\path{doi:10.1103/PhysRevD.87.074035}}.

\bibitem{Bugg:2004rk}
D.~V. Bugg, {Reinterpreting several narrow `resonances' as threshold cusps},
  Phys. Lett. B598 (2004) 8--14.
\newblock \href {http://arxiv.org/abs/hep-ph/0406293}
  {\path{arXiv:hep-ph/0406293}}, \href
  {http://dx.doi.org/10.1016/j.physletb.2004.07.047}
  {\path{doi:10.1016/j.physletb.2004.07.047}}.

\bibitem{Li:2004sta}
B.~A. Li, {Is X(3872) a possible candidate of hybrid meson}, Phys. Lett. B605
  (2005) 306--310.
\newblock \href {http://arxiv.org/abs/hep-ph/0410264}
  {\path{arXiv:hep-ph/0410264}}, \href
  {http://dx.doi.org/10.1016/j.physletb.2004.11.062}
  {\path{doi:10.1016/j.physletb.2004.11.062}}.

\bibitem{Close:2003mb}
F.~E. Close, S.~Godfrey, {Charmonium hybrid production in exclusive B meson
  decays}, Phys. Lett. B574 (2003) 210--216.
\newblock \href {http://arxiv.org/abs/hep-ph/0305285}
  {\path{arXiv:hep-ph/0305285}}, \href
  {http://dx.doi.org/10.1016/j.physletb.2003.09.011}
  {\path{doi:10.1016/j.physletb.2003.09.011}}.

\bibitem{Seth:2004zb}
K.~K. Seth, {An Alternative Interpretation of X(3872)}, Phys. Lett. B612 (2005)
  1--4.
\newblock \href {http://arxiv.org/abs/hep-ph/0411122}
  {\path{arXiv:hep-ph/0411122}}, \href
  {http://dx.doi.org/10.1016/j.physletb.2005.02.057}
  {\path{doi:10.1016/j.physletb.2005.02.057}}.

\bibitem{Zhang:2009em}
J.-R. Zhang, M.-Q. Huang, {{Q anti-s}{anti-Q-(prime)s} molecular states in QCD
  sum rules}, Commun. Theor. Phys. 54 (2010) 1075--1090.
\newblock \href {http://arxiv.org/abs/0905.4672} {\path{arXiv:0905.4672}},
  \href {http://dx.doi.org/10.1088/0253-6102/54/6/22}
  {\path{doi:10.1088/0253-6102/54/6/22}}.

\bibitem{Wang:2013daa}
Z.-G. Wang, T.~Huang, {Possible assignments of the $X(3872)$, $Z_c(3900)$ and
  $Z_b(10610)$ as axial-vector molecular states}, Eur. Phys. J. C74~(5) (2014)
  2891.
\newblock \href {http://arxiv.org/abs/1312.7489} {\path{arXiv:1312.7489}},
  \href {http://dx.doi.org/10.1140/epjc/s10052-014-2891-6}
  {\path{doi:10.1140/epjc/s10052-014-2891-6}}.

\bibitem{Mutuk:2018zxs}
H.~Mutuk, Y.~Saraç, H.~Gümüs, A.~Ozpineci, {X(3872) and Its Heavy Quark Spin
  Symmetry Partners in QCD Sum Rules}, Eur. Phys. J. C78~(11) (2018) 904.
\newblock \href {http://arxiv.org/abs/1807.04091} {\path{arXiv:1807.04091}},
  \href {http://dx.doi.org/10.1140/epjc/s10052-018-6382-z}
  {\path{doi:10.1140/epjc/s10052-018-6382-z}}.

\bibitem{Chen:2010ze}
W.~Chen, S.-L. Zhu, {The Vector and Axial-Vector Charmonium-like States}, Phys.
  Rev. D83 (2011) 034010.
\newblock \href {http://arxiv.org/abs/1010.3397} {\path{arXiv:1010.3397}},
  \href {http://dx.doi.org/10.1103/PhysRevD.83.034010}
  {\path{doi:10.1103/PhysRevD.83.034010}}.

\bibitem{Wang:2013vex}
Z.-G. Wang, T.~Huang, {Analysis of the $X(3872)$, $Z_c(3900)$ and $Z_c(3885)$
  as axial-vector tetraquark states with QCD sum rules}, Phys. Rev. D89~(5)
  (2014) 054019.
\newblock \href {http://arxiv.org/abs/1310.2422} {\path{arXiv:1310.2422}},
  \href {http://dx.doi.org/10.1103/PhysRevD.89.054019}
  {\path{doi:10.1103/PhysRevD.89.054019}}.

\bibitem{Harnett:2012gs}
D.~Harnett, R.~T. Kleiv, T.~G. Steele, H.-y. Jin, {Axial Vector $J^{PC}=1^{++}$
  Charmonium and Bottomonium Hybrid Mass Predictions with QCD Sum-Rules}, J.
  Phys. G39 (2012) 125003.
\newblock \href {http://arxiv.org/abs/1206.6776} {\path{arXiv:1206.6776}},
  \href {http://dx.doi.org/10.1088/0954-3899/39/12/125003}
  {\path{doi:10.1088/0954-3899/39/12/125003}}.

\bibitem{Chen:2013zia}
W.~Chen, R.~T. Kleiv, T.~G. Steele, B.~Bulthuis, D.~Harnett, J.~Ho,
  T.~Richards, S.-L. Zhu, {Mass Spectrum of Heavy Quarkonium Hybrids}, JHEP 09
  (2013) 019.
\newblock \href {http://arxiv.org/abs/1304.4522} {\path{arXiv:1304.4522}},
  \href {http://dx.doi.org/10.1007/JHEP09(2013)019}
  {\path{doi:10.1007/JHEP09(2013)019}}.

\bibitem{Chen:2013pya}
W.~Chen, H.-y. Jin, R.~T. Kleiv, T.~G. Steele, M.~Wang, Q.~Xu, {QCD sum-rule
  interpretation of X(3872) with $J^{PC}=1^{++}$ mixtures of hybrid charmonium
  and $\overline{D}D^*$ molecular currents}, Phys. Rev. D88~(4) (2013) 045027.
\newblock \href {http://arxiv.org/abs/1305.0244} {\path{arXiv:1305.0244}},
  \href {http://dx.doi.org/10.1103/PhysRevD.88.045027}
  {\path{doi:10.1103/PhysRevD.88.045027}}.

\bibitem{Palameta:2018yce}
A.~Palameta, D.~Harnett, T.~G. Steele, {Meson-Hybrid Mixing in $J^{PC}=1^{++}$
  Heavy Quarkonium from QCD Sum-Rules}, Phys. Rev. D98~(7) (2018) 074014.
\newblock \href {http://arxiv.org/abs/1805.04230} {\path{arXiv:1805.04230}},
  \href {http://dx.doi.org/10.1103/PhysRevD.98.074014}
  {\path{doi:10.1103/PhysRevD.98.074014}}.

\bibitem{Azizi:2017ubq}
K.~Azizi, N.~Er, {X (3872): propagating in a dense medium}, Nucl. Phys. B936
  (2018) 151--168.
\newblock \href {http://arxiv.org/abs/1710.02806} {\path{arXiv:1710.02806}},
  \href {http://dx.doi.org/10.1016/j.nuclphysb.2018.09.014}
  {\path{doi:10.1016/j.nuclphysb.2018.09.014}}.

\bibitem{Sugiyama:2007sg}
J.~Sugiyama, T.~Nakamura, N.~Ishii, T.~Nishikawa, M.~Oka, {Mixings of 4-quark
  components in light non-singlet scalar mesons in QCD sum rules}, Phys. Rev.
  D76 (2007) 114010.
\newblock \href {http://arxiv.org/abs/0707.2533} {\path{arXiv:0707.2533}},
  \href {http://dx.doi.org/10.1103/PhysRevD.76.114010}
  {\path{doi:10.1103/PhysRevD.76.114010}}.

\bibitem{Dong:2008gb}
Y.-b. Dong, A.~Faessler, T.~Gutsche, V.~E. Lyubovitskij, {Estimate for the
  $X(3872) \to \gamma J/\psi$ decay width}, Phys. Rev. D77 (2008) 094013.
\newblock \href {http://arxiv.org/abs/0802.3610} {\path{arXiv:0802.3610}},
  \href {http://dx.doi.org/10.1103/PhysRevD.77.094013}
  {\path{doi:10.1103/PhysRevD.77.094013}}.

\bibitem{Aubert:2005vi}
B.~Aubert, et~al., {Measurements of the absolute branching fractions of $B^\pm
  \to K^\pm$ X($c \bar{c}$)}, Phys. Rev. Lett. 96 (2006) 052002.
\newblock \href {http://arxiv.org/abs/hep-ex/0510070}
  {\path{arXiv:hep-ex/0510070}}, \href
  {http://dx.doi.org/10.1103/PhysRevLett.96.052002}
  {\path{doi:10.1103/PhysRevLett.96.052002}}.

\bibitem{Mo:2006ss}
X.~H. Mo, G.~Li, C.~Z. Yuan, K.~L. He, H.~M. Hu, J.~H. Hu, P.~Wang, Z.~Y. Wang,
  {Determining the upper limit of Gamma(ee) for the Y(4260)}, Phys. Lett. B640
  (2006) 182--187.
\newblock \href {http://arxiv.org/abs/hep-ex/0603024}
  {\path{arXiv:hep-ex/0603024}}, \href
  {http://dx.doi.org/10.1016/j.physletb.2006.07.060}
  {\path{doi:10.1016/j.physletb.2006.07.060}}.

\bibitem{Eichten:1978tg}
E.~Eichten, K.~Gottfried, T.~Kinoshita, K.~D. Lane, T.-M. Yan, {Charmonium: The
  Model}, Phys. Rev. D17 (1978) 3090, [Erratum: Phys. Rev.D21,313(1980)].
\newblock \href {http://dx.doi.org/10.1103/PhysRevD.17.3090,
  10.1103/physrevd.21.313.2} {\path{doi:10.1103/PhysRevD.17.3090,
  10.1103/physrevd.21.313.2}}.

\bibitem{Eichten:1979ms}
E.~Eichten, K.~Gottfried, T.~Kinoshita, K.~D. Lane, T.-M. Yan, {Charmonium:
  Comparison with Experiment}, Phys. Rev. D21 (1980) 203.
\newblock \href {http://dx.doi.org/10.1103/PhysRevD.21.203}
  {\path{doi:10.1103/PhysRevD.21.203}}.

\bibitem{Godfrey:1985xj}
S.~Godfrey, N.~Isgur, {Mesons in a Relativized Quark Model with
  Chromodynamics}, Phys. Rev. D32 (1985) 189--231.
\newblock \href {http://dx.doi.org/10.1103/PhysRevD.32.189}
  {\path{doi:10.1103/PhysRevD.32.189}}.

\bibitem{Olsen:2009gi}
S.~L. Olsen, {Hadronic Spectrum - Multiquark States}, Nucl. Phys. A827 (2009)
  53C--60C, [,53(2009)].
\newblock \href {http://arxiv.org/abs/0901.2371} {\path{arXiv:0901.2371}},
  \href {http://dx.doi.org/10.1016/j.nuclphysa.2009.05.018}
  {\path{doi:10.1016/j.nuclphysa.2009.05.018}}.

\bibitem{Yuan:2007bt}
C.~Z. Yuan, et~al., {Observation of $e^+e^- \to K^+ K^- J/\psi$ via initial
  state radiation at Belle}, Phys. Rev. D77 (2008) 011105.
\newblock \href {http://arxiv.org/abs/0709.2565} {\path{arXiv:0709.2565}},
  \href {http://dx.doi.org/10.1103/PhysRevD.77.011105}
  {\path{doi:10.1103/PhysRevD.77.011105}}.

\bibitem{Shen:2014gdm}
C.~P. Shen, et~al., {Updated cross section measurement of $e^+ e^- \to K^+ K^-
  J/\psi$ and $K_S^0K_S^0J/\psi$ via initial state radiation at Belle}, Phys.
  Rev. D89~(7) (2014) 072015.
\newblock \href {http://arxiv.org/abs/1402.6578} {\path{arXiv:1402.6578}},
  \href {http://dx.doi.org/10.1103/PhysRevD.89.072015}
  {\path{doi:10.1103/PhysRevD.89.072015}}.

\bibitem{Abe:2006fj}
K.~Abe, et~al., {Measurement of the near-threshold $e^+ e^- \to D^{*\pm}
  D^{*\mp}$ cross section using initial-state radiation}, Phys. Rev. Lett. 98
  (2007) 092001.
\newblock \href {http://arxiv.org/abs/hep-ex/0608018}
  {\path{arXiv:hep-ex/0608018}}, \href
  {http://dx.doi.org/10.1103/PhysRevLett.98.092001}
  {\path{doi:10.1103/PhysRevLett.98.092001}}.

\bibitem{Aubert:2009aq}
B.~Aubert, et~al., {Exclusive Initial-State-Radiation Production of the D
  anti-D, D* anti-D*, and D* anti-D* Systems}, Phys. Rev. D79 (2009) 092001.
\newblock \href {http://arxiv.org/abs/0903.1597} {\path{arXiv:0903.1597}},
  \href {http://dx.doi.org/10.1103/PhysRevD.79.092001}
  {\path{doi:10.1103/PhysRevD.79.092001}}.

\bibitem{Pakhlova:2008zza}
G.~Pakhlova, et~al., {Measurement of the near-threshold $e^+ e^- \to$ D anti-D
  cross section using initial-state radiation}, Phys. Rev. D77 (2008) 011103.
\newblock \href {http://arxiv.org/abs/0708.0082} {\path{arXiv:0708.0082}},
  \href {http://dx.doi.org/10.1103/PhysRevD.77.011103}
  {\path{doi:10.1103/PhysRevD.77.011103}}.

\bibitem{Ding:2008gr}
G.-J. Ding, {Are Y(4260) and Z+(2) are D(1) D or D(0) D* Hadronic Molecules?},
  Phys. Rev. D79 (2009) 014001.
\newblock \href {http://arxiv.org/abs/0809.4818} {\path{arXiv:0809.4818}},
  \href {http://dx.doi.org/10.1103/PhysRevD.79.014001}
  {\path{doi:10.1103/PhysRevD.79.014001}}.

\bibitem{Wang:2013cya}
Q.~Wang, C.~Hanhart, Q.~Zhao, {Decoding the riddle of $Y(4260)$ and
  $Z_c(3900)$}, Phys. Rev. Lett. 111~(13) (2013) 132003.
\newblock \href {http://arxiv.org/abs/1303.6355} {\path{arXiv:1303.6355}},
  \href {http://dx.doi.org/10.1103/PhysRevLett.111.132003}
  {\path{doi:10.1103/PhysRevLett.111.132003}}.

\bibitem{Yuan:2005dr}
C.~Z. Yuan, P.~Wang, X.~H. Mo, {The Y(4260) as an omega chi(c1) molecular
  state}, Phys. Lett. B634 (2006) 399--402.
\newblock \href {http://arxiv.org/abs/hep-ph/0511107}
  {\path{arXiv:hep-ph/0511107}}, \href
  {http://dx.doi.org/10.1016/j.physletb.2006.01.031}
  {\path{doi:10.1016/j.physletb.2006.01.031}}.

\bibitem{Liu:2005ay}
X.~Liu, X.-Q. Zeng, X.-Q. Li, {Possible molecular structure of the newly
  observed Y(4260)}, Phys. Rev. D72 (2005) 054023.
\newblock \href {http://arxiv.org/abs/hep-ph/0507177}
  {\path{arXiv:hep-ph/0507177}}, \href
  {http://dx.doi.org/10.1103/PhysRevD.72.054023}
  {\path{doi:10.1103/PhysRevD.72.054023}}.

\bibitem{MartinezTorres:2009xb}
A.~Martinez~Torres, K.~P. Khemchandani, D.~Gamermann, E.~Oset, {The Y(4260) as
  a J/psi K anti-K system}, Phys. Rev. D80 (2009) 094012.
\newblock \href {http://arxiv.org/abs/0906.5333} {\path{arXiv:0906.5333}},
  \href {http://dx.doi.org/10.1103/PhysRevD.80.094012}
  {\path{doi:10.1103/PhysRevD.80.094012}}.

\bibitem{Zhu:2005hp}
S.-L. Zhu, {The Possible interpretations of Y(4260)}, Phys. Lett. B625 (2005)
  212.
\newblock \href {http://arxiv.org/abs/hep-ph/0507025}
  {\path{arXiv:hep-ph/0507025}}, \href
  {http://dx.doi.org/10.1016/j.physletb.2005.08.068}
  {\path{doi:10.1016/j.physletb.2005.08.068}}.

\bibitem{Qiao:2005av}
C.-F. Qiao, {One explanation for the exotic state Y(4260)}, Phys. Lett. B639
  (2006) 263--265.
\newblock \href {http://arxiv.org/abs/hep-ph/0510228}
  {\path{arXiv:hep-ph/0510228}}, \href
  {http://dx.doi.org/10.1016/j.physletb.2006.06.038}
  {\path{doi:10.1016/j.physletb.2006.06.038}}.

\bibitem{vanBeveren:2006ih}
E.~van Beveren, G.~Rupp, {Is the Y(4260) just a coupled-channel signal?}\href
  {http://arxiv.org/abs/hep-ph/0605317} {\path{arXiv:hep-ph/0605317}}.

\bibitem{vanBeveren:2009fb}
E.~van Beveren, G.~Rupp, {The X(4260) and possible confirmation of psi(3D),
  psi(5S), psi(4D), psi(6S) and psi(5D) in J/psi pi+ pi-}\href
  {http://arxiv.org/abs/0904.4351} {\path{arXiv:0904.4351}}.

\bibitem{vanBeveren:2009jk}
E.~van Beveren, G.~Rupp, {Interference effects in the X(4260) signal}, Phys.
  Rev. D79 (2009) 111501.
\newblock \href {http://arxiv.org/abs/0905.1595} {\path{arXiv:0905.1595}},
  \href {http://dx.doi.org/10.1103/PhysRevD.79.111501}
  {\path{doi:10.1103/PhysRevD.79.111501}}.

\bibitem{Wang:2018ntv}
Z.-G. Wang, {Lowest vector tetraquark states: $Y(4260/4220)$ or $Z_c(4100)$},
  Eur. Phys. J. C78~(11) (2018) 933.
\newblock \href {http://arxiv.org/abs/1809.10299} {\path{arXiv:1809.10299}},
  \href {http://dx.doi.org/10.1140/epjc/s10052-018-6417-5}
  {\path{doi:10.1140/epjc/s10052-018-6417-5}}.

\bibitem{Aitala:2000xu}
E.~M. Aitala, et~al., {Experimental evidence for a light and broad scalar
  resonance in $D^+ \to \pi^- \pi^+ \pi^+$ decay}, Phys. Rev. Lett. 86 (2001)
  770--774.
\newblock \href {http://arxiv.org/abs/hep-ex/0007028}
  {\path{arXiv:hep-ex/0007028}}, \href
  {http://dx.doi.org/10.1103/PhysRevLett.86.770}
  {\path{doi:10.1103/PhysRevLett.86.770}}.

\bibitem{Dosch:2002rh}
H.~G. Dosch, E.~M. Ferreira, F.~S. Navarra, M.~Nielsen, {Semileptonic D decay
  into scalar mesons: A QCD sum rule approach}, Phys. Rev. D65 (2002) 114002.
\newblock \href {http://arxiv.org/abs/hep-ph/0203225}
  {\path{arXiv:hep-ph/0203225}}, \href
  {http://dx.doi.org/10.1103/PhysRevD.65.114002}
  {\path{doi:10.1103/PhysRevD.65.114002}}.

\bibitem{Aubert:2005zh}
B.~Aubert, et~al., {Study of $J/\psi \pi^+ \pi^-$ states produced in $B^0 \to
  J/\psi \pi^+ \pi^- K^0$ and $B^- \to J/\psi \pi^+ \pi^- K^-$}, Phys. Rev. D73
  (2006) 011101.
\newblock \href {http://arxiv.org/abs/hep-ex/0507090}
  {\path{arXiv:hep-ex/0507090}}, \href
  {http://dx.doi.org/10.1103/PhysRevD.73.011101}
  {\path{doi:10.1103/PhysRevD.73.011101}}.

\bibitem{Buchalla:1995vs}
G.~Buchalla, A.~J. Buras, M.~E. Lautenbacher, {Weak decays beyond leading
  logarithms}, Rev. Mod. Phys. 68 (1996) 1125--1144.
\newblock \href {http://arxiv.org/abs/hep-ph/9512380}
  {\path{arXiv:hep-ph/9512380}}, \href
  {http://dx.doi.org/10.1103/RevModPhys.68.1125}
  {\path{doi:10.1103/RevModPhys.68.1125}}.

\bibitem{Cleven:2013mka}
M.~Cleven, Q.~Wang, F.-K. Guo, C.~Hanhart, U.-G. Mei{\ss}ner, Q.~Zhao,
  {$Y(4260)$ as the first $S$-wave open charm vector molecular state?}, Phys.
  Rev. D90~(7) (2014) 074039.
\newblock \href {http://arxiv.org/abs/1310.2190} {\path{arXiv:1310.2190}},
  \href {http://dx.doi.org/10.1103/PhysRevD.90.074039}
  {\path{doi:10.1103/PhysRevD.90.074039}}.

\bibitem{Pakhlova:2009jv}
G.~Pakhlova, et~al., {Measurement of the e+ e- $\to$ D0 D*- pi+ cross section
  using initial-state radiation}, Phys. Rev. D80 (2009) 091101.
\newblock \href {http://arxiv.org/abs/0908.0231} {\path{arXiv:0908.0231}},
  \href {http://dx.doi.org/10.1103/PhysRevD.80.091101}
  {\path{doi:10.1103/PhysRevD.80.091101}}.

\bibitem{Ding:2007rg}
G.-J. Ding, J.-J. Zhu, M.-L. Yan, {Canonical Charmonium Interpretation for
  Y(4360) and Y(4660)}, Phys. Rev. D77 (2008) 014033.
\newblock \href {http://arxiv.org/abs/0708.3712} {\path{arXiv:0708.3712}},
  \href {http://dx.doi.org/10.1103/PhysRevD.77.014033}
  {\path{doi:10.1103/PhysRevD.77.014033}}.

\bibitem{Li:2009zu}
B.-Q. Li, K.-T. Chao, {Higher Charmonia and X,Y,Z states with Screened
  Potential}, Phys. Rev. D79 (2009) 094004.
\newblock \href {http://arxiv.org/abs/0903.5506} {\path{arXiv:0903.5506}},
  \href {http://dx.doi.org/10.1103/PhysRevD.79.094004}
  {\path{doi:10.1103/PhysRevD.79.094004}}.

\bibitem{Qiao:2007ce}
C.-F. Qiao, {A Uniform description of the states recently observed at
  B-factories}, J. Phys. G35 (2008) 075008.
\newblock \href {http://arxiv.org/abs/0709.4066} {\path{arXiv:0709.4066}},
  \href {http://dx.doi.org/10.1088/0954-3899/35/7/075008}
  {\path{doi:10.1088/0954-3899/35/7/075008}}.

\bibitem{Cotugno:2009ys}
G.~Cotugno, R.~Faccini, A.~D. Polosa, C.~Sabelli, {Charmed Baryonium}, Phys.
  Rev. Lett. 104 (2010) 132005.
\newblock \href {http://arxiv.org/abs/0911.2178} {\path{arXiv:0911.2178}},
  \href {http://dx.doi.org/10.1103/PhysRevLett.104.132005}
  {\path{doi:10.1103/PhysRevLett.104.132005}}.

\bibitem{Kalashnikova:2008qr}
{\relax Yu}.~S. Kalashnikova, A.~V. Nefediev, {Spectra and decays of hybrid
  charmonia}, Phys. Rev. D77 (2008) 054025.
\newblock \href {http://arxiv.org/abs/0801.2036} {\path{arXiv:0801.2036}},
  \href {http://dx.doi.org/10.1103/PhysRevD.77.054025}
  {\path{doi:10.1103/PhysRevD.77.054025}}.

\bibitem{Close:2010wq}
F.~Close, C.~Downum, C.~E. Thomas, {Novel Charmonium and Bottomonium
  Spectroscopies due to Deeply Bound Hadronic Molecules from Single Pion
  Exchange}, Phys. Rev. D81 (2010) 074033.
\newblock \href {http://arxiv.org/abs/1001.2553} {\path{arXiv:1001.2553}},
  \href {http://dx.doi.org/10.1103/PhysRevD.81.074033}
  {\path{doi:10.1103/PhysRevD.81.074033}}.

\bibitem{Guo:2008zg}
F.-K. Guo, C.~Hanhart, U.-G. Meissner, {Evidence that the Y(4660) is a
  f(0)(980)psi-prime bound state}, Phys. Lett. B665 (2008) 26--29.
\newblock \href {http://arxiv.org/abs/0803.1392} {\path{arXiv:0803.1392}},
  \href {http://dx.doi.org/10.1016/j.physletb.2008.05.057}
  {\path{doi:10.1016/j.physletb.2008.05.057}}.

\bibitem{Dubynskiy:2008mq}
S.~Dubynskiy, M.~B. Voloshin, {Hadro-Charmonium}, Phys. Lett. B666 (2008)
  344--346.
\newblock \href {http://arxiv.org/abs/0803.2224} {\path{arXiv:0803.2224}},
  \href {http://dx.doi.org/10.1016/j.physletb.2008.07.086}
  {\path{doi:10.1016/j.physletb.2008.07.086}}.

\bibitem{Ebert:2008kb}
D.~Ebert, R.~N. Faustov, V.~O. Galkin, {Excited heavy tetraquarks with hidden
  charm}, Eur. Phys. J. C58 (2008) 399--405.
\newblock \href {http://arxiv.org/abs/0808.3912} {\path{arXiv:0808.3912}},
  \href {http://dx.doi.org/10.1140/epjc/s10052-008-0754-8}
  {\path{doi:10.1140/epjc/s10052-008-0754-8}}.

\bibitem{Zhang:2010mw}
J.-R. Zhang, M.-Q. Huang, {The $P$-wave $[cs][\bar{c}\bar{s}]$ tetraquark
  state: $Y(4260)$ or $Y(4660)$?}, Phys. Rev. D83 (2011) 036005.
\newblock \href {http://arxiv.org/abs/1011.2818} {\path{arXiv:1011.2818}},
  \href {http://dx.doi.org/10.1103/PhysRevD.83.036005}
  {\path{doi:10.1103/PhysRevD.83.036005}}.

\bibitem{Sundu:2018toi}
H.~Sundu, S.~S. Agaev, K.~Azizi, {Resonance $Y(4660)$ as a vector tetraquark
  and its strong decay channels}, Phys. Rev. D98~(5) (2018) 054021.
\newblock \href {http://arxiv.org/abs/1805.04705} {\path{arXiv:1805.04705}},
  \href {http://dx.doi.org/10.1103/PhysRevD.98.054021}
  {\path{doi:10.1103/PhysRevD.98.054021}}.

\bibitem{Hooft:2008we}
G.~'t~Hooft, G.~Isidori, L.~Maiani, A.~D. Polosa, V.~Riquer, {A Theory of
  Scalar Mesons}, Phys. Lett. B662 (2008) 424--430.
\newblock \href {http://arxiv.org/abs/0801.2288} {\path{arXiv:0801.2288}},
  \href {http://dx.doi.org/10.1016/j.physletb.2008.03.036}
  {\path{doi:10.1016/j.physletb.2008.03.036}}.

\bibitem{Matheus:2007ta}
R.~D. Matheus, F.~S. Navarra, M.~Nielsen, R.~Rodrigues~da Silva, {Do the QCD
  sum rules support four-quark states?}, Phys. Rev. D76 (2007) 056005.
\newblock \href {http://arxiv.org/abs/0705.1357} {\path{arXiv:0705.1357}},
  \href {http://dx.doi.org/10.1103/PhysRevD.76.056005}
  {\path{doi:10.1103/PhysRevD.76.056005}}.

\bibitem{Rosner:2007mu}
J.~L. Rosner, {Threshold effect and pi+- psi(2S) peak}, Phys. Rev. D76 (2007)
  114002.
\newblock \href {http://arxiv.org/abs/0708.3496} {\path{arXiv:0708.3496}},
  \href {http://dx.doi.org/10.1103/PhysRevD.76.114002}
  {\path{doi:10.1103/PhysRevD.76.114002}}.

\bibitem{Bugg:2007vp}
D.~V. Bugg, {Z+(4430) as a cusp in D*(2010)D(1)(2420)}\href
  {http://arxiv.org/abs/0709.1254} {\path{arXiv:0709.1254}}.

\bibitem{Meng:2007fu}
C.~Meng, K.-T. Chao, {Z+(4430) as a resonance in the D(1)(D(1)-prime)D*
  channel}\href {http://arxiv.org/abs/0708.4222} {\path{arXiv:0708.4222}}.

\bibitem{Liu:2007bf}
X.~Liu, Y.-R. Liu, W.-Z. Deng, S.-L. Zhu, {Is Z+(4430) a loosely bound
  molecular state?}, Phys. Rev. D77 (2008) 034003.
\newblock \href {http://arxiv.org/abs/0711.0494} {\path{arXiv:0711.0494}},
  \href {http://dx.doi.org/10.1103/PhysRevD.77.034003}
  {\path{doi:10.1103/PhysRevD.77.034003}}.

\bibitem{Liu:2008xz}
X.~Liu, Y.-R. Liu, W.-Z. Deng, S.-L. Zhu, {Z+(4430) as a D(1)-prime D* (D(1)
  D*) molecular state}, Phys. Rev. D77 (2008) 094015.
\newblock \href {http://arxiv.org/abs/0803.1295} {\path{arXiv:0803.1295}},
  \href {http://dx.doi.org/10.1103/PhysRevD.77.094015}
  {\path{doi:10.1103/PhysRevD.77.094015}}.

\bibitem{Ding:2008mp}
G.-J. Ding, W.~Huang, J.-F. Liu, M.-L. Yan, {Z+(4430) and analogous heavy
  flavor molecules}, Phys. Rev. D79 (2009) 034026.
\newblock \href {http://arxiv.org/abs/0805.3822} {\path{arXiv:0805.3822}},
  \href {http://dx.doi.org/10.1103/PhysRevD.79.034026}
  {\path{doi:10.1103/PhysRevD.79.034026}}.

\bibitem{Zhang:2009vs}
J.-R. Zhang, M.-Q. Huang, {{Q anti-q}{anti-Q-(prime)q} molecular states}, Phys.
  Rev. D80 (2009) 056004.
\newblock \href {http://arxiv.org/abs/0906.0090} {\path{arXiv:0906.0090}},
  \href {http://dx.doi.org/10.1103/PhysRevD.80.056004}
  {\path{doi:10.1103/PhysRevD.80.056004}}.

\bibitem{Meng:2009qt}
G.-Z. Meng, et~al., {Low-energy D*+ D0(1) Scattering and the Resonance-like
  Structure Z+ (4430)}, Phys. Rev. D80 (2009) 034503.
\newblock \href {http://arxiv.org/abs/0905.0752} {\path{arXiv:0905.0752}},
  \href {http://dx.doi.org/10.1103/PhysRevD.80.034503}
  {\path{doi:10.1103/PhysRevD.80.034503}}.

\bibitem{Maiani:2007wz}
L.~Maiani, A.~D. Polosa, V.~Riquer, {The Charged Z(4433): Towards a new
  spectroscopy}\href {http://arxiv.org/abs/0708.3997} {\path{arXiv:0708.3997}}.

\bibitem{Faccini:2013lda}
L.~Maiani, V.~Riquer, R.~Faccini, F.~Piccinini, A.~Pilloni, A.~D. Polosa, {A
  $J^{PG}=1^{++}$ Charged Resonance in the $Y(4260) \to \pi^+ \pi^- J/\psi$
  Decay?}, Phys. Rev. D87~(11) (2013) 111102.
\newblock \href {http://arxiv.org/abs/1303.6857} {\path{arXiv:1303.6857}},
  \href {http://dx.doi.org/10.1103/PhysRevD.87.111102}
  {\path{doi:10.1103/PhysRevD.87.111102}}.

\bibitem{Maiani:2014aja}
L.~Maiani, F.~Piccinini, A.~D. Polosa, V.~Riquer, {The Z(4430) and a New
  Paradigm for Spin Interactions in Tetraquarks}, Phys. Rev. D89 (2014) 114010.
\newblock \href {http://arxiv.org/abs/1405.1551} {\path{arXiv:1405.1551}},
  \href {http://dx.doi.org/10.1103/PhysRevD.89.114010}
  {\path{doi:10.1103/PhysRevD.89.114010}}.

\bibitem{Navarra:2011xa}
F.~S. Navarra, M.~Nielsen, J.-M. Richard, {Exotic Charmonium and
  Bottomonium-like Resonances}, J. Phys. Conf. Ser. 348 (2012) 012007.
\newblock \href {http://arxiv.org/abs/1108.1230} {\path{arXiv:1108.1230}},
  \href {http://dx.doi.org/10.1088/1742-6596/348/1/012007}
  {\path{doi:10.1088/1742-6596/348/1/012007}}.

\bibitem{Patel:2014vua}
S.~Patel, M.~Shah, P.~C. Vinodkumar, {Mass spectra of four-quark states in the
  hidden charm sector}, Eur. Phys. J. A50 (2014) 131.
\newblock \href {http://arxiv.org/abs/1402.3974} {\path{arXiv:1402.3974}},
  \href {http://dx.doi.org/10.1140/epja/i2014-14131-9}
  {\path{doi:10.1140/epja/i2014-14131-9}}.

\bibitem{Hadizadeh:2015cvx}
M.~R. Hadizadeh, A.~Khaledi-Nasab, {Heavy tetraquarks in the
  diquark--antidiquark picture}, Phys. Lett. B753 (2016) 8--12.
\newblock \href {http://arxiv.org/abs/1511.08542} {\path{arXiv:1511.08542}},
  \href {http://dx.doi.org/10.1016/j.physletb.2015.11.072}
  {\path{doi:10.1016/j.physletb.2015.11.072}}.

\bibitem{Wang:2014vha}
Z.-G. Wang, {Analysis of the $Z(4430)$ as the first radial excitation of the
  $Z_c(3900)$}, Commun. Theor. Phys. 63~(3) (2015) 325--330.
\newblock \href {http://arxiv.org/abs/1405.3581} {\path{arXiv:1405.3581}},
  \href {http://dx.doi.org/10.1088/0253-6102/63/3/325}
  {\path{doi:10.1088/0253-6102/63/3/325}}.

\bibitem{Agaev:2017tzv}
S.~S. Agaev, K.~Azizi, H.~Sundu, {Treating $Z_c(3900)$ and $Z(4430)$ as the
  ground-state and first radially excited tetraquarks}, Phys. Rev. D96~(3)
  (2017) 034026.
\newblock \href {http://arxiv.org/abs/1706.01216} {\path{arXiv:1706.01216}},
  \href {http://dx.doi.org/10.1103/PhysRevD.96.034026}
  {\path{doi:10.1103/PhysRevD.96.034026}}.

\bibitem{Deng:2015lca}
C.~Deng, J.~Ping, H.~Huang, F.~Wang, {Systematic study of Z$_c^+$ family from a
  multiquark color flux-tube model}, Phys. Rev. D92~(3) (2015) 034027.
\newblock \href {http://arxiv.org/abs/1507.06408} {\path{arXiv:1507.06408}},
  \href {http://dx.doi.org/10.1103/PhysRevD.92.034027}
  {\path{doi:10.1103/PhysRevD.92.034027}}.

\bibitem{Ma:2014zua}
L.~Ma, X.-H. Liu, X.~Liu, S.-L. Zhu, {Exotic Four Quark Matter: $Z_1(4475)$},
  Phys. Rev. D90~(3) (2014) 037502.
\newblock \href {http://arxiv.org/abs/1404.3450} {\path{arXiv:1404.3450}},
  \href {http://dx.doi.org/10.1103/PhysRevD.90.037502}
  {\path{doi:10.1103/PhysRevD.90.037502}}.

\bibitem{Barnes:2014csa}
T.~Barnes, F.~E. Close, E.~S. Swanson, {Molecular Interpretation of the
  Supercharmonium State Z(4475)}, Phys. Rev. D91~(1) (2015) 014004.
\newblock \href {http://arxiv.org/abs/1409.6651} {\path{arXiv:1409.6651}},
  \href {http://dx.doi.org/10.1103/PhysRevD.91.014004}
  {\path{doi:10.1103/PhysRevD.91.014004}}.

\bibitem{Liu:2014eka}
X.-H. Liu, L.~Ma, L.-P. Sun, X.~Liu, S.-L. Zhu, {Resolving the puzzling decay
  patterns of charged $Z_c$ and $Z_b$ states}, Phys. Rev. D90~(7) (2014)
  074020.
\newblock \href {http://arxiv.org/abs/1407.3684} {\path{arXiv:1407.3684}},
  \href {http://dx.doi.org/10.1103/PhysRevD.90.074020}
  {\path{doi:10.1103/PhysRevD.90.074020}}.

\bibitem{Dosch:2015nwa}
H.~G. Dosch, G.~F. de~Teramond, S.~J. Brodsky, {Superconformal Baryon-Meson
  Symmetry and Light-Front Holographic QCD}, Phys. Rev. D91~(8) (2015) 085016.
\newblock \href {http://arxiv.org/abs/1501.00959} {\path{arXiv:1501.00959}},
  \href {http://dx.doi.org/10.1103/PhysRevD.91.085016}
  {\path{doi:10.1103/PhysRevD.91.085016}}.

\bibitem{Dosch:2015bca}
H.~G. Dosch, G.~F. de~Teramond, S.~J. Brodsky, {Supersymmetry Across the Light
  and Heavy-Light Hadronic Spectrum}, Phys. Rev. D92~(7) (2015) 074010.
\newblock \href {http://arxiv.org/abs/1504.05112} {\path{arXiv:1504.05112}},
  \href {http://dx.doi.org/10.1103/PhysRevD.92.074010}
  {\path{doi:10.1103/PhysRevD.92.074010}}.

\bibitem{Brodsky:2016yod}
S.~J. Brodsky, G.~F. de~T{\'e}ramond, H.~G. Dosch, C.~Lorc{\'e}, {Universal
  Effective Hadron Dynamics from Superconformal Algebra}, Phys. Lett. B759
  (2016) 171--177.
\newblock \href {http://arxiv.org/abs/1604.06746} {\path{arXiv:1604.06746}},
  \href {http://dx.doi.org/10.1016/j.physletb.2016.05.068}
  {\path{doi:10.1016/j.physletb.2016.05.068}}.

\bibitem{Dosch:2016zdv}
H.~G. Dosch, G.~F. de~Teramond, S.~J. Brodsky, {Supersymmetry Across the Light
  and Heavy-Light Hadronic Spectrum II}, Phys. Rev. D95~(3) (2017) 034016.
\newblock \href {http://arxiv.org/abs/1612.02370} {\path{arXiv:1612.02370}},
  \href {http://dx.doi.org/10.1103/PhysRevD.95.034016}
  {\path{doi:10.1103/PhysRevD.95.034016}}.

\bibitem{Nielsen:2018uyn}
M.~Nielsen, S.~J. Brodsky, {Hadronic superpartners from a superconformal and
  supersymmetric algebra}, Phys. Rev. D97~(11) (2018) 114001.
\newblock \href {http://arxiv.org/abs/1802.09652} {\path{arXiv:1802.09652}},
  \href {http://dx.doi.org/10.1103/PhysRevD.97.114001}
  {\path{doi:10.1103/PhysRevD.97.114001}}.

\bibitem{Nielsen:2018ytt}
M.~Nielsen, S.~J. Brodsky, G.~F. de~T{\'e}ramond, H.~G. Dosch, F.~S. Navarra,
  L.~Zou, {Supersymmetry in the Double-Heavy Hadronic Spectrum}, Phys. Rev.
  D98~(3) (2018) 034002.
\newblock \href {http://arxiv.org/abs/1805.11567} {\path{arXiv:1805.11567}},
  \href {http://dx.doi.org/10.1103/PhysRevD.98.034002}
  {\path{doi:10.1103/PhysRevD.98.034002}}.

\bibitem{Collaboration:2017njt}
M.~Ablikim, et~al., {Determination of the Spin and Parity of the $Z_c(3900)$},
  Phys. Rev. Lett. 119~(7) (2017) 072001.
\newblock \href {http://arxiv.org/abs/1706.04100} {\path{arXiv:1706.04100}},
  \href {http://dx.doi.org/10.1103/PhysRevLett.119.072001}
  {\path{doi:10.1103/PhysRevLett.119.072001}}.

\bibitem{Ablikim:2015tbp}
M.~Ablikim, et~al., {Observation of $Z_c(3900)^{0}$ in $e^+e^-\to\pi^0\pi^0
  J/\psi$}, Phys. Rev. Lett. 115~(11) (2015) 112003.
\newblock \href {http://arxiv.org/abs/1506.06018} {\path{arXiv:1506.06018}},
  \href {http://dx.doi.org/10.1103/PhysRevLett.115.112003}
  {\path{doi:10.1103/PhysRevLett.115.112003}}.

\bibitem{Ablikim:2015swa}
M.~Ablikim, et~al., {Confirmation of a charged charmoniumlike state
  $Z_c(3885)^{\mp}$ in $e^+e^-\to\pi^{\pm}(D\bar{D}^*)^\mp$ with double $D$
  tag}, Phys. Rev. D92~(9) (2015) 092006.
\newblock \href {http://arxiv.org/abs/1509.01398} {\path{arXiv:1509.01398}},
  \href {http://dx.doi.org/10.1103/PhysRevD.92.092006}
  {\path{doi:10.1103/PhysRevD.92.092006}}.

\bibitem{Ablikim:2015gda}
M.~Ablikim, et~al., {Observation of a Neutral Structure near the $D\bar{D}^{*}$
  Mass Threshold in $e^{+}e^{-}\to (D \bar{D}^*)^0\pi^0$ at $\sqrt{s}$ = 4.226
  and 4.257 GeV}, Phys. Rev. Lett. 115~(22) (2015) 222002.
\newblock \href {http://arxiv.org/abs/1509.05620} {\path{arXiv:1509.05620}},
  \href {http://dx.doi.org/10.1103/PhysRevLett.115.222002}
  {\path{doi:10.1103/PhysRevLett.115.222002}}.

\bibitem{Zhao:2014gqa}
L.~Zhao, L.~Ma, S.-L. Zhu, {Spin-orbit force, recoil corrections, and possible
  $B \bar{B}^{*}$ and $D \bar{D}^{*}$ molecular states}, Phys. Rev. D89~(9)
  (2014) 094026.
\newblock \href {http://arxiv.org/abs/1403.4043} {\path{arXiv:1403.4043}},
  \href {http://dx.doi.org/10.1103/PhysRevD.89.094026}
  {\path{doi:10.1103/PhysRevD.89.094026}}.

\bibitem{He:2014nya}
J.~He, {Study of the $B\bar{B}^*/D\bar{D}^*$ bound states in a Bethe-Salpeter
  approach}, Phys. Rev. D90~(7) (2014) 076008.
\newblock \href {http://arxiv.org/abs/1409.8506} {\path{arXiv:1409.8506}},
  \href {http://dx.doi.org/10.1103/PhysRevD.90.076008}
  {\path{doi:10.1103/PhysRevD.90.076008}}.

\bibitem{Prelovsek:2013xba}
S.~Prelovsek, L.~Leskovec, {Search for $Z^{+}_{c}$(3900) in the $1^{+-}$
  Channel on the Lattice}, Phys. Lett. B727 (2013) 172--176.
\newblock \href {http://arxiv.org/abs/1308.2097} {\path{arXiv:1308.2097}},
  \href {http://dx.doi.org/10.1016/j.physletb.2013.10.009}
  {\path{doi:10.1016/j.physletb.2013.10.009}}.

\bibitem{Prelovsek:2014swa}
S.~Prelovsek, C.~B. Lang, L.~Leskovec, D.~Mohler, {Study of the $Z_c^+$ channel
  using lattice QCD}, Phys. Rev. D91~(1) (2015) 014504.
\newblock \href {http://arxiv.org/abs/1405.7623} {\path{arXiv:1405.7623}},
  \href {http://dx.doi.org/10.1103/PhysRevD.91.014504}
  {\path{doi:10.1103/PhysRevD.91.014504}}.

\bibitem{Chen:2014afa}
Y.~Chen, et~al., {Low-energy scattering of the $(D\bar{D}^*)^\pm$ system and
  the resonance-like structure $Z_c(3900)$}, Phys. Rev. D89~(9) (2014) 094506.
\newblock \href {http://arxiv.org/abs/1403.1318} {\path{arXiv:1403.1318}},
  \href {http://dx.doi.org/10.1103/PhysRevD.89.094506}
  {\path{doi:10.1103/PhysRevD.89.094506}}.

\bibitem{Aceti:2014uea}
F.~Aceti, M.~Bayar, E.~Oset, A.~Martinez~Torres, K.~P. Khemchandani, J.~M.
  Dias, F.~S. Navarra, M.~Nielsen, {Prediction of an $I=1$ $D \bar D^*$ state
  and relationship to the claimed $Z_c(3900)$, $Z_c(3885)$}, Phys. Rev. D90~(1)
  (2014) 016003.
\newblock \href {http://arxiv.org/abs/1401.8216} {\path{arXiv:1401.8216}},
  \href {http://dx.doi.org/10.1103/PhysRevD.90.016003}
  {\path{doi:10.1103/PhysRevD.90.016003}}.

\bibitem{Karliner:2015ina}
M.~Karliner, J.~L. Rosner, {New Exotic Meson and Baryon Resonances from
  Doubly-Heavy Hadronic Molecules}, Phys. Rev. Lett. 115~(12) (2015) 122001.
\newblock \href {http://arxiv.org/abs/1506.06386} {\path{arXiv:1506.06386}},
  \href {http://dx.doi.org/10.1103/PhysRevLett.115.122001}
  {\path{doi:10.1103/PhysRevLett.115.122001}}.

\bibitem{He:2015mja}
J.~He, {The $Z_c(3900)$ as a resonance from the $D\bar{D}^*$ interaction},
  Phys. Rev. D92~(3) (2015) 034004.
\newblock \href {http://arxiv.org/abs/1505.05379} {\path{arXiv:1505.05379}},
  \href {http://dx.doi.org/10.1103/PhysRevD.92.034004}
  {\path{doi:10.1103/PhysRevD.92.034004}}.

\bibitem{Wilbring:2013cha}
E.~Wilbring, H.~W. Hammer, U.~G. Mei{\ss}ner, {Electromagnetic Structure of the
  $Z_c(3900)$}, Phys. Lett. B726 (2013) 326--329.
\newblock \href {http://arxiv.org/abs/1304.2882} {\path{arXiv:1304.2882}},
  \href {http://dx.doi.org/10.1016/j.physletb.2013.08.059}
  {\path{doi:10.1016/j.physletb.2013.08.059}}.

\bibitem{Dong:2013iqa}
Y.~Dong, A.~Faessler, T.~Gutsche, V.~E. Lyubovitskij, {Strong decays of
  molecular states Z$_{c}^{+}$ and Z$_{c}^{'+}$}, Phys. Rev. D88~(1) (2013)
  014030.
\newblock \href {http://arxiv.org/abs/1306.0824} {\path{arXiv:1306.0824}},
  \href {http://dx.doi.org/10.1103/PhysRevD.88.014030}
  {\path{doi:10.1103/PhysRevD.88.014030}}.

\bibitem{Ke:2013gia}
H.-W. Ke, Z.-T. Wei, X.-Q. Li, {Is $Z_c(3900)$ a molecular state}, Eur. Phys.
  J. C73~(10) (2013) 2561.
\newblock \href {http://arxiv.org/abs/1307.2414} {\path{arXiv:1307.2414}},
  \href {http://dx.doi.org/10.1140/epjc/s10052-013-2561-0}
  {\path{doi:10.1140/epjc/s10052-013-2561-0}}.

\bibitem{Gutsche:2014zda}
T.~Gutsche, M.~Kesenheimer, V.~E. Lyubovitskij, {Radiative and dilepton decays
  of the hadronic molecule $Z_c^+$(3900)}, Phys. Rev. D90~(9) (2014) 094013.
\newblock \href {http://arxiv.org/abs/1410.0259} {\path{arXiv:1410.0259}},
  \href {http://dx.doi.org/10.1103/PhysRevD.90.094013}
  {\path{doi:10.1103/PhysRevD.90.094013}}.

\bibitem{Esposito:2014hsa}
A.~Esposito, A.~L. Guerrieri, A.~Pilloni, {Probing the nature of $Z_c^{(′)}$
  states via the $η_cρ$ decay}, Phys. Lett. B746 (2015) 194--201.
\newblock \href {http://arxiv.org/abs/1409.3551} {\path{arXiv:1409.3551}},
  \href {http://dx.doi.org/10.1016/j.physletb.2015.04.057}
  {\path{doi:10.1016/j.physletb.2015.04.057}}.

\bibitem{Chen:2015igx}
D.-Y. Chen, Y.-B. Dong, {Radiative decays of the neutral $Z_c(3900)$}, Phys.
  Rev. D93~(1) (2016) 014003.
\newblock \href {http://arxiv.org/abs/1510.00829} {\path{arXiv:1510.00829}},
  \href {http://dx.doi.org/10.1103/PhysRevD.93.014003}
  {\path{doi:10.1103/PhysRevD.93.014003}}.

\bibitem{Gong:2016hlt}
Q.-R. Gong, Z.-H. Guo, C.~Meng, G.-Y. Tang, Y.-F. Wang, H.-Q. Zheng,
  {$Z_c(3900)$ as a $D\bar{D}^*$ molecule from the pole counting rule}, Phys.
  Rev. D94~(11) (2016) 114019.
\newblock \href {http://arxiv.org/abs/1604.08836} {\path{arXiv:1604.08836}},
  \href {http://dx.doi.org/10.1103/PhysRevD.94.114019}
  {\path{doi:10.1103/PhysRevD.94.114019}}.

\bibitem{Ke:2016owt}
H.-W. Ke, X.-Q. Li, {Study on decays of $Z_c(4020)$ and $Z_c(3900)$ into
  $h_c+\pi $}, Eur. Phys. J. C76~(6) (2016) 334.
\newblock \href {http://arxiv.org/abs/1601.03575} {\path{arXiv:1601.03575}},
  \href {http://dx.doi.org/10.1140/epjc/s10052-016-4183-9}
  {\path{doi:10.1140/epjc/s10052-016-4183-9}}.

\bibitem{Chen:2015ata}
W.~Chen, T.~G. Steele, H.-X. Chen, S.-L. Zhu, {Mass spectra of Zc and Zb exotic
  states as hadron molecules}, Phys. Rev. D92~(5) (2015) 054002.
\newblock \href {http://arxiv.org/abs/1505.05619} {\path{arXiv:1505.05619}},
  \href {http://dx.doi.org/10.1103/PhysRevD.92.054002}
  {\path{doi:10.1103/PhysRevD.92.054002}}.

\bibitem{Zhang:2013aoa}
J.-R. Zhang, {Improved QCD sum rule study of $Z_{c}(3900)$ as a $\bar{D}D^{*}$
  molecular state}, Phys. Rev. D87~(11) (2013) 116004.
\newblock \href {http://arxiv.org/abs/1304.5748} {\path{arXiv:1304.5748}},
  \href {http://dx.doi.org/10.1103/PhysRevD.87.116004}
  {\path{doi:10.1103/PhysRevD.87.116004}}.

\bibitem{Cui:2013yva}
C.-Y. Cui, Y.-L. Liu, W.-B. Chen, M.-Q. Huang, {Could $Z_{c}(3900)$ be a
  $I^{G}J^{P}=1^{+}1^{+}$ $D^{*}\bar{D}$ molecular state?}, J. Phys. G41 (2014)
  075003.
\newblock \href {http://arxiv.org/abs/1304.1850} {\path{arXiv:1304.1850}},
  \href {http://dx.doi.org/10.1088/0954-3899/41/7/075003}
  {\path{doi:10.1088/0954-3899/41/7/075003}}.

\bibitem{Deng:2014gqa}
C.~Deng, J.~Ping, F.~Wang, {Interpreting $Z_c(3900)$ and $Z_c(4025)/Z_c(4020)$
  as charged tetraquark states}, Phys. Rev. D90 (2014) 054009.
\newblock \href {http://arxiv.org/abs/1402.0777} {\path{arXiv:1402.0777}},
  \href {http://dx.doi.org/10.1103/PhysRevD.90.054009}
  {\path{doi:10.1103/PhysRevD.90.054009}}.

\bibitem{Agaev:2016dev}
S.~S. Agaev, K.~Azizi, H.~Sundu, {Strong $Z_c^{+}(3900)\rightarrow J/\psi
  \pi^{+}; \eta_{c} \rho^{+}$ decays in QCD}, Phys. Rev. D93~(7) (2016) 074002.
\newblock \href {http://arxiv.org/abs/1601.03847} {\path{arXiv:1601.03847}},
  \href {http://dx.doi.org/10.1103/PhysRevD.93.074002}
  {\path{doi:10.1103/PhysRevD.93.074002}}.

\bibitem{Mahajan:2013qja}
N.~Mahajan, {Interpreting Z(3900)}\href {http://arxiv.org/abs/1304.1301}
  {\path{arXiv:1304.1301}}.

\bibitem{Ablikim:2014dxl}
M.~Ablikim, et~al., {Observation of $e^+e^- → π^0π^0h_c$ and a Neutral
  Charmoniumlike Structure $Z_c(4020)^0$}, Phys. Rev. Lett. 113~(21) (2014)
  212002.
\newblock \href {http://arxiv.org/abs/1409.6577} {\path{arXiv:1409.6577}},
  \href {http://dx.doi.org/10.1103/PhysRevLett.113.212002}
  {\path{doi:10.1103/PhysRevLett.113.212002}}.

\bibitem{Ablikim:2015vvn}
M.~Ablikim, et~al., {Observation of a neutral charmoniumlike state
  $Z_c(4025)^0$ in $e^{+} e^{-} \to (D^{*} \bar{D}^{*})^{0} \pi^0$}, Phys. Rev.
  Lett. 115~(18) (2015) 182002.
\newblock \href {http://arxiv.org/abs/1507.02404} {\path{arXiv:1507.02404}},
  \href {http://dx.doi.org/10.1103/PhysRevLett.115.182002}
  {\path{doi:10.1103/PhysRevLett.115.182002}}.

\bibitem{DeRujula:1976zlg}
A.~De~Rujula, H.~Georgi, S.~L. Glashow, {Molecular Charmonium: A New
  Spectroscopy?}, Phys. Rev. Lett. 38 (1977) 317.
\newblock \href {http://dx.doi.org/10.1103/PhysRevLett.38.317}
  {\path{doi:10.1103/PhysRevLett.38.317}}.

\bibitem{Tornqvist:1993vu}
N.~A. Tornqvist, {On deusons or deuteron - like meson meson bound states},
  Nuovo Cim. A107 (1994) 2471--2476.
\newblock \href {http://arxiv.org/abs/hep-ph/9310225}
  {\path{arXiv:hep-ph/9310225}}, \href {http://dx.doi.org/10.1007/BF02734018}
  {\path{doi:10.1007/BF02734018}}.

\bibitem{Dubynskiy:2006sg}
S.~Dubynskiy, M.~B. Voloshin, {Possible new resonance at the D* anti-D*
  threshold in e+ e- annihilation}, Mod. Phys. Lett. A21 (2006) 2779--2788.
\newblock \href {http://arxiv.org/abs/hep-ph/0608179}
  {\path{arXiv:hep-ph/0608179}}, \href
  {http://dx.doi.org/10.1142/S0217732306022195}
  {\path{doi:10.1142/S0217732306022195}}.

\bibitem{Voloshin:1976ap}
M.~B. Voloshin, L.~B. Okun, {Hadron Molecules and Charmonium Atom}, JETP Lett.
  23 (1976) 333--336, [Pisma Zh. Eksp. Teor. Fiz.23,369(1976)].

\bibitem{Guo:2013sya}
F.-K. Guo, C.~Hidalgo-Duque, J.~Nieves, M.~P. Valderrama, {Consequences of
  Heavy Quark Symmetries for Hadronic Molecules}, Phys. Rev. D88 (2013) 054007.
\newblock \href {http://arxiv.org/abs/1303.6608} {\path{arXiv:1303.6608}},
  \href {http://dx.doi.org/10.1103/PhysRevD.88.054007}
  {\path{doi:10.1103/PhysRevD.88.054007}}.

\bibitem{Molina:2009ct}
R.~Molina, E.~Oset, {The Y(3940), Z(3930) and the X(4160) as dynamically
  generated resonances from the vector-vector interaction}, Phys. Rev. D80
  (2009) 114013.
\newblock \href {http://arxiv.org/abs/0907.3043} {\path{arXiv:0907.3043}},
  \href {http://dx.doi.org/10.1103/PhysRevD.80.114013}
  {\path{doi:10.1103/PhysRevD.80.114013}}.

\bibitem{Aceti:2014kja}
F.~Aceti, M.~Bayar, J.~M. Dias, E.~Oset, {Prediction of a $Z_c(4000)$ $D^* \bar
  D^*$ state and relationship to the claimed $Z_c(4025)$}, Eur. Phys. J. A50
  (2014) 103.
\newblock \href {http://arxiv.org/abs/1401.2076} {\path{arXiv:1401.2076}},
  \href {http://dx.doi.org/10.1140/epja/i2014-14103-1}
  {\path{doi:10.1140/epja/i2014-14103-1}}.

\bibitem{YamagataSekihara:2010pj}
J.~Yamagata-Sekihara, J.~Nieves, E.~Oset, {Couplings in coupled channels versus
  wave functions in the case of resonances: application to the two
  $\Lambda(1405)$ states}, Phys. Rev. D83 (2011) 014003.
\newblock \href {http://arxiv.org/abs/1007.3923} {\path{arXiv:1007.3923}},
  \href {http://dx.doi.org/10.1103/PhysRevD.83.014003}
  {\path{doi:10.1103/PhysRevD.83.014003}}.

\bibitem{Swanson:2015bsa}
E.~S. Swanson, {Cusps and Exotic Charmonia}, Int. J. Mod. Phys. E25~(07) (2016)
  1642010.
\newblock \href {http://arxiv.org/abs/1504.07952} {\path{arXiv:1504.07952}},
  \href {http://dx.doi.org/10.1142/S0218301316420106}
  {\path{doi:10.1142/S0218301316420106}}.

\bibitem{Chen:2015jwa}
Y.~Chen, et~al., {Low-energy Scattering of $(D^{*}\bar{D}^{*})^\pm$ System and
  the Resonance-like Structure $Z_c(4025)$}, Phys. Rev. D92~(5) (2015) 054507.
\newblock \href {http://arxiv.org/abs/1503.02371} {\path{arXiv:1503.02371}},
  \href {http://dx.doi.org/10.1103/PhysRevD.92.054507}
  {\path{doi:10.1103/PhysRevD.92.054507}}.

\bibitem{Molina:2009eb}
R.~Molina, H.~Nagahiro, A.~Hosaka, E.~Oset, {Scalar, axial-vector and tensor
  resonances from the rho D*, omega D* interaction in the hidden gauge
  formalism}, Phys. Rev. D80 (2009) 014025.
\newblock \href {http://arxiv.org/abs/0903.3823} {\path{arXiv:0903.3823}},
  \href {http://dx.doi.org/10.1103/PhysRevD.80.014025}
  {\path{doi:10.1103/PhysRevD.80.014025}}.

\bibitem{Wang:2013llv}
Z.-G. Wang, {Reanalysis of the $Z_c(4020)$, $Z_c(4025)$, $Z(4050)$ and
  $Z(4250)$ as tetraquark states with QCD sum rules}, Commun. Theor. Phys.
  63~(4) (2015) 466--480.
\newblock \href {http://arxiv.org/abs/1312.1537} {\path{arXiv:1312.1537}},
  \href {http://dx.doi.org/10.1088/0253-6102/63/4/466}
  {\path{doi:10.1088/0253-6102/63/4/466}}.

\bibitem{Qiao:2013dda}
C.-F. Qiao, L.~Tang, {Interpretation of $Z_c(4025)$ as the hidden charm
  tetraquark states via QCD Sum Rules}, Eur. Phys. J. C74 (2014) 2810.
\newblock \href {http://arxiv.org/abs/1308.3439} {\path{arXiv:1308.3439}},
  \href {http://dx.doi.org/10.1140/epjc/s10052-014-2810-x}
  {\path{doi:10.1140/epjc/s10052-014-2810-x}}.

\bibitem{Wang:2015nwa}
Z.-G. Wang, {Analysis of the $Z_c(4200)$ as axial-vector molecule-like state},
  Int. J. Mod. Phys. A30~(30) (2015) 1550168.
\newblock \href {http://arxiv.org/abs/1502.01459} {\path{arXiv:1502.01459}},
  \href {http://dx.doi.org/10.1142/S0217751X15501687}
  {\path{doi:10.1142/S0217751X15501687}}.

\bibitem{Chen:2013omd}
W.~Chen, T.~G. Steele, M.-L. Du, S.-L. Zhu, {$D^*\bar D^*$ molecule
  interpretation of $Z_c(4025)$}, Eur. Phys. J. C74~(2) (2014) 2773.
\newblock \href {http://arxiv.org/abs/1308.5060} {\path{arXiv:1308.5060}},
  \href {http://dx.doi.org/10.1140/epjc/s10052-014-2773-y}
  {\path{doi:10.1140/epjc/s10052-014-2773-y}}.

\bibitem{Cui:2013vfa}
C.-Y. Cui, C.-Y. Cui, Y.-L. Liu, M.-Q. Huang, {Could $Z_{c}(4025)$ be a
  $J^{P}=1^{+}$ $D^{*}\bar{D^{*}}$ molecular state?Could $Z_c$(4025) be a $J^P$
  = $1^+ D^* \bar{D^*}$ molecular state?}, Eur. Phys. J. C73~(12) (2013) 2661.
\newblock \href {http://arxiv.org/abs/1308.3625} {\path{arXiv:1308.3625}},
  \href {http://dx.doi.org/10.1140/epjc/s10052-013-2661-x}
  {\path{doi:10.1140/epjc/s10052-013-2661-x}}.

\bibitem{Yang:2017rmm}
Y.-C. Yang, Z.-Y. Tan, H.-S. Zong, J.~Ping, {Dynamical study of $S$-wave
  $\bar{Q}Q\bar{q}q$ system}\href {http://arxiv.org/abs/1712.09285}
  {\path{arXiv:1712.09285}}.

\bibitem{Wang:2014gwa}
Z.-G. Wang, {Reanalysis of the $Y(3940)$, $Y(4140)$, $Z_c(4020)$, $Z_c(4025)$
  and $Z_b(10650)$ as molecular states with QCD sum rules}, Eur. Phys. J.
  C74~(7) (2014) 2963.
\newblock \href {http://arxiv.org/abs/1403.0810} {\path{arXiv:1403.0810}},
  \href {http://dx.doi.org/10.1140/epjc/s10052-014-2963-7}
  {\path{doi:10.1140/epjc/s10052-014-2963-7}}.

\bibitem{Liu:2008tn}
X.~Liu, Z.-G. Luo, Y.-R. Liu, S.-L. Zhu, {X(3872) and Other Possible Heavy
  Molecular States}, Eur. Phys. J. C61 (2009) 411--428.
\newblock \href {http://arxiv.org/abs/0808.0073} {\path{arXiv:0808.0073}},
  \href {http://dx.doi.org/10.1140/epjc/s10052-009-1020-4}
  {\path{doi:10.1140/epjc/s10052-009-1020-4}}.

\bibitem{Liu:2008mi}
Y.-R. Liu, Z.-Y. Zhang, {The Bound state problem of S-wave heavy quark
  meson-aitimeson systems}, Phys. Rev. C80 (2009) 015208.
\newblock \href {http://arxiv.org/abs/0810.1598} {\path{arXiv:0810.1598}},
  \href {http://dx.doi.org/10.1103/PhysRevC.80.015208}
  {\path{doi:10.1103/PhysRevC.80.015208}}.

\bibitem{Wang:2008af}
Z.-G. Wang, {Another tetraquark structure in the pi+ (chi(c1)) invariant mass
  distribution}, Eur. Phys. J. C62 (2009) 375--382.
\newblock \href {http://arxiv.org/abs/0807.4592} {\path{arXiv:0807.4592}},
  \href {http://dx.doi.org/10.1140/epjc/s10052-009-1043-x}
  {\path{doi:10.1140/epjc/s10052-009-1043-x}}.

\bibitem{Abazov:2018cyu}
V.~M. Abazov, et~al., {Evidence for $Z_c^{\pm}(3900)$ in semi-inclusive decays
  of $b$-flavored hadrons}\href {http://arxiv.org/abs/1807.00183}
  {\path{arXiv:1807.00183}}.

\bibitem{Deng:2017xlb}
C.~Deng, J.~Ping, H.~Huang, F.~Wang, {Hidden charmed states and multibody color
  flux-tube dynamics}, Phys. Rev. D98~(1) (2018) 014026.
\newblock \href {http://arxiv.org/abs/1801.00164} {\path{arXiv:1801.00164}},
  \href {http://dx.doi.org/10.1103/PhysRevD.98.014026}
  {\path{doi:10.1103/PhysRevD.98.014026}}.

\bibitem{Guo:2016uaf}
Z.~Guo, T.~Liu, B.-Q. Ma, {Light-front holographic QCD with generic dilaton
  profile}, Phys. Rev. D93~(7) (2016) 076010.
\newblock \href {http://arxiv.org/abs/1604.08463} {\path{arXiv:1604.08463}},
  \href {http://dx.doi.org/10.1103/PhysRevD.93.076010}
  {\path{doi:10.1103/PhysRevD.93.076010}}.

\bibitem{Zhao:2014qva}
L.~Zhao, W.-Z. Deng, S.-L. Zhu, {Hidden-Charm Tetraquarks and Charged $Z_c$
  States}, Phys. Rev. D90~(9) (2014) 094031.
\newblock \href {http://arxiv.org/abs/1408.3924} {\path{arXiv:1408.3924}},
  \href {http://dx.doi.org/10.1103/PhysRevD.90.094031}
  {\path{doi:10.1103/PhysRevD.90.094031}}.

\bibitem{Wang:2015uua}
X.-Y. Wang, X.-R. Chen, {Discovery Potential for the Neutral Charmonium-Like by
  Annihilation}, Adv. High Energy Phys. 2015 (2015) 918231.
\newblock \href {http://arxiv.org/abs/1509.08553} {\path{arXiv:1509.08553}},
  \href {http://dx.doi.org/10.1155/2015/918231}
  {\path{doi:10.1155/2015/918231}}.

\bibitem{Wang:2015lwa}
X.-Y. Wang, X.-R. Chen, A.~Guskov, {Photoproduction of the charged
  charmoniumlike $Z_{c}^{+}(4200)$}, Phys. Rev. D92~(9) (2015) 094017.
\newblock \href {http://arxiv.org/abs/1503.02125} {\path{arXiv:1503.02125}},
  \href {http://dx.doi.org/10.1103/PhysRevD.92.094017}
  {\path{doi:10.1103/PhysRevD.92.094017}}.

\bibitem{Ma:2015nmy}
L.~Ma, W.-Z. Deng, X.-L. Chen, S.-L. Zhu, {Strong decay patterns of the
  hidden-charm tetraquarks}\href {http://arxiv.org/abs/1512.01938}
  {\path{arXiv:1512.01938}}.

\bibitem{Chen:2015fsa}
W.~Chen, T.~G. Steele, H.-X. Chen, S.-L. Zhu, {$Z_c(4200)^+$ decay width as a
  charmonium-like tetraquark state}, Eur. Phys. J. C75~(8) (2015) 358.
\newblock \href {http://arxiv.org/abs/1501.03863} {\path{arXiv:1501.03863}},
  \href {http://dx.doi.org/10.1140/epjc/s10052-015-3578-3}
  {\path{doi:10.1140/epjc/s10052-015-3578-3}}.

\bibitem{Wu:2018xdi}
J.~Wu, X.~Liu, Y.-R. Liu, S.-L. Zhu, {Systematic studies of charmonium-,
  bottomonium-, and $B_c$-like tetraquark states}\href
  {http://arxiv.org/abs/1810.06886} {\path{arXiv:1810.06886}}.

\bibitem{Voloshin:2018vym}
M.~B. Voloshin, {$Z_c(4100)$ and $Z_c(4200)$ as hadrocharmonium}\href
  {http://arxiv.org/abs/1810.08146} {\path{arXiv:1810.08146}}.

\bibitem{Zhao:2018xrd}
Q.~Zhao, {Some insights into the newly observed $Z_c(4100)$ in $B^0\to \eta_c
  K^+ \pi^-$ by LHCb}\href {http://arxiv.org/abs/1811.05357}
  {\path{arXiv:1811.05357}}.

\bibitem{Cao:2018vmv}
X.~Cao, J.-P. Dai, {The spin parity of $Z_c^-$(4100), $Z_1^+$(4050) and
  $Z_2^+$(4250)}\href {http://arxiv.org/abs/1811.06434}
  {\path{arXiv:1811.06434}}.

\bibitem{Aaij:2012pz}
R.~Aaij, et~al., {Search for the $X(4140)$ state in $B^+ \to J/\psi \phi K^+$
  decays}, Phys. Rev. D85 (2012) 091103.
\newblock \href {http://arxiv.org/abs/1202.5087} {\path{arXiv:1202.5087}},
  \href {http://dx.doi.org/10.1103/PhysRevD.85.091103}
  {\path{doi:10.1103/PhysRevD.85.091103}}.

\bibitem{Chatrchyan:2013dma}
S.~Chatrchyan, et~al., {Observation of a peaking structure in the $J/\psi \phi$
  mass spectrum from $B^{\pm} \to J/\psi \phi K^{\pm}$ decays}, Phys. Lett.
  B734 (2014) 261--281.
\newblock \href {http://arxiv.org/abs/1309.6920} {\path{arXiv:1309.6920}},
  \href {http://dx.doi.org/10.1016/j.physletb.2014.05.055}
  {\path{doi:10.1016/j.physletb.2014.05.055}}.

\bibitem{Aaij:2016nsc}
R.~Aaij, et~al., {Amplitude analysis of $B^+\to J/\psi \phi K^+$ decays}, Phys.
  Rev. D95~(1) (2017) 012002.
\newblock \href {http://arxiv.org/abs/1606.07898} {\path{arXiv:1606.07898}},
  \href {http://dx.doi.org/10.1103/PhysRevD.95.012002}
  {\path{doi:10.1103/PhysRevD.95.012002}}.

\bibitem{Mahajan:2009pj}
N.~Mahajan, {Y(4140): Possible options}, Phys. Lett. B679 (2009) 228--230.
\newblock \href {http://arxiv.org/abs/0903.3107} {\path{arXiv:0903.3107}},
  \href {http://dx.doi.org/10.1016/j.physletb.2009.07.043}
  {\path{doi:10.1016/j.physletb.2009.07.043}}.

\bibitem{Liu:2009iw}
X.~Liu, {The Hidden charm decay of Y(4140) by the rescattering mechanism},
  Phys. Lett. B680 (2009) 137--140.
\newblock \href {http://arxiv.org/abs/0904.0136} {\path{arXiv:0904.0136}},
  \href {http://dx.doi.org/10.1016/j.physletb.2009.08.049}
  {\path{doi:10.1016/j.physletb.2009.08.049}}.

\bibitem{Stancu:2009ka}
F.~Stancu, {Can Y(4140) be a c anti-c s anti-s tetraquark?}, J. Phys. G37
  (2010) 075017.
\newblock \href {http://arxiv.org/abs/0906.2485} {\path{arXiv:0906.2485}},
  \href {http://dx.doi.org/10.1088/0954-3899/37/7/075017}
  {\path{doi:10.1088/0954-3899/37/7/075017}}.

\bibitem{Liu:2009ei}
X.~Liu, S.-L. Zhu, {Y(4143) is probably a molecular partner of Y(3930)}, Phys.
  Rev. D80 (2009) 017502, [Erratum: Phys. Rev.D85,019902(2012)].
\newblock \href {http://arxiv.org/abs/0903.2529} {\path{arXiv:0903.2529}},
  \href {http://dx.doi.org/10.1103/PhysRevD.85.019902,
  10.1103/PhysRevD.80.017502} {\path{doi:10.1103/PhysRevD.85.019902,
  10.1103/PhysRevD.80.017502}}.

\bibitem{Branz:2009yt}
T.~Branz, T.~Gutsche, V.~E. Lyubovitskij, {Hadronic molecule structure of the
  Y(3940) and Y(4140)}, Phys. Rev. D80 (2009) 054019.
\newblock \href {http://arxiv.org/abs/0903.5424} {\path{arXiv:0903.5424}},
  \href {http://dx.doi.org/10.1103/PhysRevD.80.054019}
  {\path{doi:10.1103/PhysRevD.80.054019}}.

\bibitem{Branz:2010rj}
T.~Branz, R.~Molina, E.~Oset, {Radiative decays of the Y(3940), Z(3930) and the
  X(4160) as dynamically generated resonances}, Phys. Rev. D83 (2011) 114015.
\newblock \href {http://arxiv.org/abs/1010.0587} {\path{arXiv:1010.0587}},
  \href {http://dx.doi.org/10.1103/PhysRevD.83.114015}
  {\path{doi:10.1103/PhysRevD.83.114015}}.

\bibitem{Wang:2009ue}
Z.-G. Wang, {Analysis of the Y(4140) with QCD sum rules}, Eur. Phys. J. C63
  (2009) 115--122.
\newblock \href {http://arxiv.org/abs/0903.5200} {\path{arXiv:0903.5200}},
  \href {http://dx.doi.org/10.1140/epjc/s10052-009-1097-9}
  {\path{doi:10.1140/epjc/s10052-009-1097-9}}.

\bibitem{Liang:2015twa}
W.-H. Liang, J.-J. Xie, E.~Oset, R.~Molina, M.~D{\"o}ring, {Predictions for the
  $\bar B^0 \to \bar K^{*0} X (YZ)$ and $\bar B^0_s \to \phi X (YZ)$ with
  $X(4160), Y(3940), Z(3930)$}, Eur. Phys. J. A51~(5) (2015) 58.
\newblock \href {http://arxiv.org/abs/1502.02932} {\path{arXiv:1502.02932}},
  \href {http://dx.doi.org/10.1140/epja/i2015-15058-3}
  {\path{doi:10.1140/epja/i2015-15058-3}}.

\bibitem{HidalgoDuque:2012pq}
C.~Hidalgo-Duque, J.~Nieves, M.~P. Valderrama, {Light flavor and heavy quark
  spin symmetry in heavy meson molecules}, Phys. Rev. D87~(7) (2013) 076006.
\newblock \href {http://arxiv.org/abs/1210.5431} {\path{arXiv:1210.5431}},
  \href {http://dx.doi.org/10.1103/PhysRevD.87.076006}
  {\path{doi:10.1103/PhysRevD.87.076006}}.

\bibitem{Wang:2017mrt}
E.~Wang, J.-J. Xie, L.-S. Geng, E.~Oset, {Analysis of the $B^+\to J/\psi \phi
  K^+$ data at low $J/\psi \phi$ invariant masses and the $X(4140)$ and
  $X(4160)$ resonances}, Phys. Rev. D97~(1) (2018) 014017.
\newblock \href {http://arxiv.org/abs/1710.02061} {\path{arXiv:1710.02061}},
  \href {http://dx.doi.org/10.1103/PhysRevD.97.014017}
  {\path{doi:10.1103/PhysRevD.97.014017}}.

\bibitem{Narison:2009ag}
S.~Narison, V.~I. Zakharov, {Duality between QCD Perturbative Series and Power
  Corrections}, Phys. Lett. B679 (2009) 355--361.
\newblock \href {http://arxiv.org/abs/0906.4312} {\path{arXiv:0906.4312}},
  \href {http://dx.doi.org/10.1016/j.physletb.2009.07.060}
  {\path{doi:10.1016/j.physletb.2009.07.060}}.

\bibitem{Tarrach:1980up}
R.~Tarrach, {The Pole Mass in Perturbative QCD}, Nucl. Phys. B183 (1981)
  384--396.
\newblock \href {http://dx.doi.org/10.1016/0550-3213(81)90140-1}
  {\path{doi:10.1016/0550-3213(81)90140-1}}.

\bibitem{Coquereaux:1979eq}
R.~Coquereaux, {Dimensional Renormalization and Comparison of Renormalization
  Schemes in Quantum Electrodynamics}, Annals Phys. 125 (1980) 401.
\newblock \href {http://dx.doi.org/10.1016/0003-4916(80)90139-6}
  {\path{doi:10.1016/0003-4916(80)90139-6}}.

\bibitem{Binetruy:1979hc}
P.~Binetruy, T.~Schucker, {Gauge and Renormalization Scheme Dependence in
  {GUTs}}, Nucl. Phys. B178 (1981) 293--306.
\newblock \href {http://dx.doi.org/10.1016/0550-3213(81)90410-7}
  {\path{doi:10.1016/0550-3213(81)90410-7}}.

\bibitem{Narison:1987qh}
S.~Narison, {Heavy Quark Mass in the MS Scheme: Revisited}, Phys. Lett. B197
  (1987) 405--408.
\newblock \href {http://dx.doi.org/10.1016/0370-2693(87)90410-2}
  {\path{doi:10.1016/0370-2693(87)90410-2}}.

\bibitem{Narison:1988xi}
S.~Narison, {Light and Heavy Quark Masses, Test of PCAC and Flavor Breakings of
  Condensates in QCD}, Phys. Lett. B216 (1989) 191--197.
\newblock \href {http://dx.doi.org/10.1016/0370-2693(89)91393-2}
  {\path{doi:10.1016/0370-2693(89)91393-2}}.

\bibitem{Gray:1990yh}
N.~Gray, D.~J. Broadhurst, W.~Grafe, K.~Schilcher, {Three Loop Relation of
  Quark (Modified) Ms and Pole Masses}, Z. Phys. C48 (1990) 673--680.
\newblock \href {http://dx.doi.org/10.1007/BF01614703}
  {\path{doi:10.1007/BF01614703}}.

\bibitem{Fleischer:1998dw}
J.~Fleischer, F.~Jegerlehner, O.~V. Tarasov, O.~L. Veretin, {Two loop QCD
  corrections of the massive fermion propagator}, Nucl. Phys. B539 (1999)
  671--690, [Erratum: Nucl. Phys.B571,511(2000)].
\newblock \href {http://arxiv.org/abs/hep-ph/9803493}
  {\path{arXiv:hep-ph/9803493}}, \href
  {http://dx.doi.org/10.1016/S0550-3213(99)00794-4,
  10.1016/S0550-3213(98)00705-6} {\path{doi:10.1016/S0550-3213(99)00794-4,
  10.1016/S0550-3213(98)00705-6}}.

\bibitem{Chetyrkin:1999qi}
K.~G. Chetyrkin, M.~Steinhauser, {The Relation between the MS-bar and the
  on-shell quark mass at order alpha(s)**3}, Nucl. Phys. B573 (2000) 617--651.
\newblock \href {http://arxiv.org/abs/hep-ph/9911434}
  {\path{arXiv:hep-ph/9911434}}, \href
  {http://dx.doi.org/10.1016/S0550-3213(99)00784-1}
  {\path{doi:10.1016/S0550-3213(99)00784-1}}.

\bibitem{Narison:1994zt}
S.~Narison, A.~A. Pivovarov, {QSSR estimate of the B(B) parameter at
  next-to-leading order}, Phys. Lett. B327 (1994) 341--346.
\newblock \href {http://arxiv.org/abs/hep-ph/9403225}
  {\path{arXiv:hep-ph/9403225}}, \href
  {http://dx.doi.org/10.1016/0370-2693(94)90739-0}
  {\path{doi:10.1016/0370-2693(94)90739-0}}.

\bibitem{Hagiwara:2002hf}
K.~Hagiwara, S.~Narison, D.~Nomura, {B0(d,s) - anti-B0(d,s) mass differences
  from QCD spectral sum rules}, Phys. Lett. B540 (2002) 233--240.
\newblock \href {http://arxiv.org/abs/hep-ph/0205092}
  {\path{arXiv:hep-ph/0205092}}, \href
  {http://dx.doi.org/10.1016/S0370-2693(02)02133-0}
  {\path{doi:10.1016/S0370-2693(02)02133-0}}.

\bibitem{Pich:1985ab}
A.~Pich, E.~De~Rafael, {K anti-K Mixing in the Standard Model}, Phys. Lett.
  158B (1985) 477--484.
\newblock \href {http://dx.doi.org/10.1016/0370-2693(85)90798-1}
  {\path{doi:10.1016/0370-2693(85)90798-1}}.

\bibitem{Broadhurst:1981jk}
D.~J. Broadhurst, {Chiral Symmetry Breaking and Perturbative QCD}, Phys. Lett.
  101B (1981) 423--426.
\newblock \href {http://dx.doi.org/10.1016/0370-2693(81)90167-2}
  {\path{doi:10.1016/0370-2693(81)90167-2}}.

\bibitem{Chetyrkin:2000mq}
K.~G. Chetyrkin, M.~Steinhauser, {Three loop nondiagonal current correlators in
  QCD and NLO corrections to single top quark production}, Phys. Lett. B502
  (2001) 104--114.
\newblock \href {http://arxiv.org/abs/hep-ph/0012002}
  {\path{arXiv:hep-ph/0012002}}, \href
  {http://dx.doi.org/10.1016/S0370-2693(01)00179-4}
  {\path{doi:10.1016/S0370-2693(01)00179-4}}.

\bibitem{Chetyrkin:2001je}
K.~G. Chetyrkin, M.~Steinhauser, {Heavy - light current correlators at order
  alpha-s**2 in QCD and HQET}, Eur. Phys. J. C21 (2001) 319--338.
\newblock \href {http://arxiv.org/abs/hep-ph/0108017}
  {\path{arXiv:hep-ph/0108017}}, \href
  {http://dx.doi.org/10.1007/s100520100744} {\path{doi:10.1007/s100520100744}}.

\bibitem{Gelhausen:2013wia}
P.~Gelhausen, A.~Khodjamirian, A.~A. Pivovarov, D.~Rosenthal, {Decay constants
  of heavy-light vector mesons from QCD sum rules}, Phys. Rev. D88 (2013)
  014015, [Erratum: Phys. Rev.D91,099901(2015)].
\newblock \href {http://arxiv.org/abs/1305.5432} {\path{arXiv:1305.5432}},
  \href {http://dx.doi.org/10.1103/PhysRevD.88.014015,
  10.1103/PhysRevD.91.099901, 10.1103/PhysRevD.89.099901}
  {\path{doi:10.1103/PhysRevD.88.014015, 10.1103/PhysRevD.91.099901,
  10.1103/PhysRevD.89.099901}}.

\bibitem{Chetyrkin:1998yr}
K.~G. Chetyrkin, S.~Narison, V.~I. Zakharov, {Short distance tachyonic gluon
  mass and 1 / Q**2 corrections}, Nucl. Phys. B550 (1999) 353--374.
\newblock \href {http://arxiv.org/abs/hep-ph/9811275}
  {\path{arXiv:hep-ph/9811275}}, \href
  {http://dx.doi.org/10.1016/S0550-3213(99)00167-4}
  {\path{doi:10.1016/S0550-3213(99)00167-4}}.

\bibitem{Narison:2001ix}
S.~Narison, V.~I. Zakharov, {Hints on the power corrections from current
  correlators in x space}, Phys. Lett. B522 (2001) 266--272.
\newblock \href {http://arxiv.org/abs/hep-ph/0110141}
  {\path{arXiv:hep-ph/0110141}}, \href
  {http://dx.doi.org/10.1016/S0370-2693(01)01274-6}
  {\path{doi:10.1016/S0370-2693(01)01274-6}}.

\bibitem{Narison:1988ts}
S.~Narison, G.~Veneziano, {{QCD} Tests of $G$ (1.6) = Glueball}, Int. J. Mod.
  Phys. A4 (1989) 2751.
\newblock \href {http://dx.doi.org/10.1142/S0217751X89001060}
  {\path{doi:10.1142/S0217751X89001060}}.

\bibitem{Narison:1996fm}
S.~Narison, {Masses, decays and mixings of gluonia in QCD}, Nucl. Phys. B509
  (1998) 312--356.
\newblock \href {http://arxiv.org/abs/hep-ph/9612457}
  {\path{arXiv:hep-ph/9612457}}, \href
  {http://dx.doi.org/10.1016/S0550-3213(97)00562-2}
  {\path{doi:10.1016/S0550-3213(97)00562-2}}.

\bibitem{Narison:2012xy}
S.~Narison, {A fresh look into $m_{c,b}$ and precise $f_{D_(s),B_(s)}$ from
  heavy-light QCD spectral sum rules}, Phys. Lett. B718 (2013) 1321--1333.
\newblock \href {http://arxiv.org/abs/1209.2023} {\path{arXiv:1209.2023}},
  \href {http://dx.doi.org/10.1016/j.physletb.2012.10.057}
  {\path{doi:10.1016/j.physletb.2012.10.057}}.

\bibitem{Narison:2014vka}
S.~Narison, {Improved light quark masses from pseudoscalar sum rules}, Phys.
  Lett. B738 (2014) 346--360.
\newblock \href {http://arxiv.org/abs/1401.3689} {\path{arXiv:1401.3689}},
  \href {http://dx.doi.org/10.1016/j.physletb.2014.09.056}
  {\path{doi:10.1016/j.physletb.2014.09.056}}.

\bibitem{Narison:2014ska}
S.~Narison, {Improved $f_{D*_{(s)}}, f_{{B*}_{(s)}}$ and $f_{B_{c}}$ from QCD
  Laplace sum rules}, Int. J. Mod. Phys. A30~(20) (2015) 1550116.
\newblock \href {http://arxiv.org/abs/1404.6642} {\path{arXiv:1404.6642}},
  \href {http://dx.doi.org/10.1142/S0217751X1550116X}
  {\path{doi:10.1142/S0217751X1550116X}}.

\bibitem{Narison:2018dcr}
S.~Narison, {QCD parameter correlations from heavy quarkonia}, Int. J. Mod.
  Phys. A33~(10) (2018) 1850045.
\newblock \href {http://arxiv.org/abs/1801.00592} {\path{arXiv:1801.00592}},
  \href {http://dx.doi.org/10.1142/S0217751X18500458}
  {\path{doi:10.1142/S0217751X18500458}}.

\bibitem{Floratos:1978jb}
E.~G. Floratos, S.~Narison, E.~de~Rafael, {Spectral Function Sum Rules in
  Quantum Chromodynamics. 1. Charged Currents Sector}, Nucl. Phys. B155 (1979)
  115--149.
\newblock \href {http://dx.doi.org/10.1016/0550-3213(79)90359-6}
  {\path{doi:10.1016/0550-3213(79)90359-6}}.

\bibitem{Ioffe:2002be}
B.~L. Ioffe, K.~N. Zyablyuk, {Gluon condensate in charmonium sum rules with
  three loop corrections}, Eur. Phys. J. C27 (2003) 229--241.
\newblock \href {http://arxiv.org/abs/hep-ph/0207183}
  {\path{arXiv:hep-ph/0207183}}, \href
  {http://dx.doi.org/10.1140/epjc/s2002-01099-8}
  {\path{doi:10.1140/epjc/s2002-01099-8}}.

\bibitem{Ioffe:2005ym}
B.~L. Ioffe, {QCD at low energies}, Prog. Part. Nucl. Phys. 56 (2006) 232--277.
\newblock \href {http://arxiv.org/abs/hep-ph/0502148}
  {\path{arXiv:hep-ph/0502148}}, \href
  {http://dx.doi.org/10.1016/j.ppnp.2005.05.001}
  {\path{doi:10.1016/j.ppnp.2005.05.001}}.

\bibitem{Bali:2014sja}
G.~S. Bali, C.~Bauer, A.~Pineda, {Model-independent determination of the gluon
  condensate in four-dimensional SU(3) gauge theory}, Phys. Rev. Lett. 113
  (2014) 092001.
\newblock \href {http://arxiv.org/abs/1403.6477} {\path{arXiv:1403.6477}},
  \href {http://dx.doi.org/10.1103/PhysRevLett.113.092001}
  {\path{doi:10.1103/PhysRevLett.113.092001}}.

\bibitem{Lee:2010hd}
T.~Lee, {Renormalon Subtraction from the Average Plaquette and the Gluon
  Condensate}, Phys. Rev. D82 (2010) 114021.
\newblock \href {http://arxiv.org/abs/1003.0231} {\path{arXiv:1003.0231}},
  \href {http://dx.doi.org/10.1103/PhysRevD.82.114021}
  {\path{doi:10.1103/PhysRevD.82.114021}}.

\bibitem{Braaten:1991qm}
E.~Braaten, S.~Narison, A.~Pich, {QCD analysis of the tau hadronic width},
  Nucl. Phys. B373 (1992) 581--612.
\newblock \href {http://dx.doi.org/10.1016/0550-3213(92)90267-F}
  {\path{doi:10.1016/0550-3213(92)90267-F}}.

\bibitem{Narison:1988ni}
S.~Narison, A.~Pich, {QCD Formulation of the tau Decay and Determination of
  Lambda (MS)}, Phys. Lett. B211 (1988) 183--188.
\newblock \href {http://dx.doi.org/10.1016/0370-2693(88)90830-1}
  {\path{doi:10.1016/0370-2693(88)90830-1}}.

\bibitem{Narison:2005ny}
S.~Narison, {Strange quark mass from e+ e- revisited and present status of
  light quark masses}, Phys. Rev. D74 (2006) 034013.
\newblock \href {http://arxiv.org/abs/hep-ph/0510108}
  {\path{arXiv:hep-ph/0510108}}, \href
  {http://dx.doi.org/10.1103/PhysRevD.74.034013}
  {\path{doi:10.1103/PhysRevD.74.034013}}.

\bibitem{Narison:1999mv}
S.~Narison, {On the strange quark mass from e+ e- and tau decay data, and test
  of the SU(2) isospin symmetry}, Phys. Lett. B466 (1999) 345--354.
\newblock \href {http://arxiv.org/abs/hep-ph/9905264}
  {\path{arXiv:hep-ph/9905264}}, \href
  {http://dx.doi.org/10.1016/S0370-2693(99)01093-X}
  {\path{doi:10.1016/S0370-2693(99)01093-X}}.

\bibitem{Dosch:1997wb}
H.~G. Dosch, S.~Narison, {Direct extraction of the chiral quark condensate and
  bounds on the light quark masses}, Phys. Lett. B417 (1998) 173--176.
\newblock \href {http://arxiv.org/abs/hep-ph/9709215}
  {\path{arXiv:hep-ph/9709215}}, \href
  {http://dx.doi.org/10.1016/S0370-2693(97)01370-1}
  {\path{doi:10.1016/S0370-2693(97)01370-1}}.

\bibitem{Narison:2011xe}
S.~Narison, {Gluon Condensates and precise $\overline{m}_{c,b}$ from
  QCD-Moments and their ratios to Order $\alpha_s^3$ and < G$^4$ >}, Phys.
  Lett. B706 (2012) 412--422.
\newblock \href {http://arxiv.org/abs/1105.2922} {\path{arXiv:1105.2922}},
  \href {http://dx.doi.org/10.1016/j.physletb.2011.11.058}
  {\path{doi:10.1016/j.physletb.2011.11.058}}.

\bibitem{Narison:2011rn}
S.~Narison, {Gluon Condensates and $m_b(m_b)$ from QCD-Exponential Moments at
  Higher Orders}, Phys. Lett. B707 (2012) 259--263.
\newblock \href {http://arxiv.org/abs/1105.5070} {\path{arXiv:1105.5070}},
  \href {http://dx.doi.org/10.1016/j.physletb.2011.12.047}
  {\path{doi:10.1016/j.physletb.2011.12.047}}.

\bibitem{Bertlmann:1983pf}
R.~A. Bertlmann, J.~S. Bell, {GLUON CONDENSATE POTENTIALS}, Nucl. Phys. B227
  (1983) 435, [,435(1983)].
\newblock \href {http://dx.doi.org/10.1016/0550-3213(83)90567-9}
  {\path{doi:10.1016/0550-3213(83)90567-9}}.

\bibitem{Bertlmann:1984rs}
R.~A. Bertlmann, H.~Neufeld, {Exponential {QCD} Moments for Charmonium Triplet
  $s$ Wave Up to Order $ < G^4 > $}, Z. Phys. C27 (1985) 437.
\newblock \href {http://dx.doi.org/10.1007/BF01548649}
  {\path{doi:10.1007/BF01548649}}.

\bibitem{Narison:1992ru}
S.~Narison, {Determination of the D = 2 'operator' from e+ e- data}, Phys.
  Lett. B300 (1993) 293--297.
\newblock \href {http://dx.doi.org/10.1016/0370-2693(93)90368-R}
  {\path{doi:10.1016/0370-2693(93)90368-R}}.

\bibitem{Narison:1995jr}
S.~Narison, {QCD tests from e+ e- $\to$ I = 1 hadrons data and implication on
  the value of alpha-s from tau decays}, Phys. Lett. B361 (1995) 121--130.
\newblock \href {http://arxiv.org/abs/hep-ph/9504334}
  {\path{arXiv:hep-ph/9504334}}, \href
  {http://dx.doi.org/10.1016/0370-2693(95)01125-A}
  {\path{doi:10.1016/0370-2693(95)01125-A}}.

\bibitem{Yndurain:1999pb}
F.~J. Yndurain, {Gluon condensate from superconvergent QCD sum rule}, Phys.
  Rept. 320 (1999) 287--293.
\newblock \href {http://arxiv.org/abs/hep-ph/9903457}
  {\path{arXiv:hep-ph/9903457}}, \href
  {http://dx.doi.org/10.1016/S0370-1573(99)00079-4}
  {\path{doi:10.1016/S0370-1573(99)00079-4}}.

\bibitem{Narison:1995tw}
S.~Narison, {Heavy quarkonia mass splittings in QCD: Gluon condensate, alpha-s
  and 1/m expansion}, Phys. Lett. B387 (1996) 162--172.
\newblock \href {http://arxiv.org/abs/hep-ph/9512348}
  {\path{arXiv:hep-ph/9512348}}, \href
  {http://dx.doi.org/10.1016/0370-2693(96)00954-9}
  {\path{doi:10.1016/0370-2693(96)00954-9}}.

\bibitem{Chung:1984gr}
Y.~Chung, H.~G. Dosch, M.~Kremer, D.~Schall, {Chiral Symmetry Breaking
  Condensates for Baryonic Sum Rules}, Z. Phys. C25 (1984) 151.
\newblock \href {http://dx.doi.org/10.1007/BF01557473}
  {\path{doi:10.1007/BF01557473}}.

\bibitem{Dosch:1988vv}
H.~G. Dosch, M.~Jamin, S.~Narison, {Baryon Masses and Flavor Symmetry Breaking
  of Chiral Condensates}, Phys. Lett. B220 (1989) 251--257.
\newblock \href {http://dx.doi.org/10.1016/0370-2693(89)90047-6}
  {\path{doi:10.1016/0370-2693(89)90047-6}}.

\bibitem{Jaffe:1976ig}
R.~L. Jaffe, {Multi-Quark Hadrons. 1. The Phenomenology of (2 Quark 2
  anti-Quark) Mesons}, Phys. Rev. D15 (1977) 267.
\newblock \href {http://dx.doi.org/10.1103/PhysRevD.15.267}
  {\path{doi:10.1103/PhysRevD.15.267}}.

\bibitem{Weinberg:2013cfa}
S.~Weinberg, {Tetraquark Mesons in Large $N$ Quantum Chromodynamics}, Phys.
  Rev. Lett. 110 (2013) 261601.
\newblock \href {http://arxiv.org/abs/1303.0342} {\path{arXiv:1303.0342}},
  \href {http://dx.doi.org/10.1103/PhysRevLett.110.261601}
  {\path{doi:10.1103/PhysRevLett.110.261601}}.

\bibitem{Knecht:2013yqa}
M.~Knecht, S.~Peris, {Narrow Tetraquarks at Large N}, Phys. Rev. D88 (2013)
  036016.
\newblock \href {http://arxiv.org/abs/1307.1273} {\path{arXiv:1307.1273}},
  \href {http://dx.doi.org/10.1103/PhysRevD.88.036016}
  {\path{doi:10.1103/PhysRevD.88.036016}}.

\bibitem{Rossi:2016szw}
G.~Rossi, G.~Veneziano, {The string-junction picture of multiquark states: an
  update}, JHEP 06 (2016) 041.
\newblock \href {http://arxiv.org/abs/1603.05830} {\path{arXiv:1603.05830}},
  \href {http://dx.doi.org/10.1007/JHEP06(2016)041}
  {\path{doi:10.1007/JHEP06(2016)041}}.

\bibitem{Cho:2017dcy}
S.~Cho, et~al., {Exotic Hadrons from Heavy Ion Collisions}, Prog. Part. Nucl.
  Phys. 95 (2017) 279--322.
\newblock \href {http://arxiv.org/abs/1702.00486} {\path{arXiv:1702.00486}},
  \href {http://dx.doi.org/10.1016/j.ppnp.2017.02.002}
  {\path{doi:10.1016/j.ppnp.2017.02.002}}.

\bibitem{Meng:2013gga}
C.~Meng, H.~Han, K.-T. Chao, {X(3872) and its production at hadron colliders},
  Phys. Rev. D96~(7) (2017) 074014.
\newblock \href {http://arxiv.org/abs/1304.6710} {\path{arXiv:1304.6710}},
  \href {http://dx.doi.org/10.1103/PhysRevD.96.074014}
  {\path{doi:10.1103/PhysRevD.96.074014}}.

\bibitem{Torres:2014fxa}
A.~Martinez~Torres, K.~P. Khemchandani, F.~S. Navarra, M.~Nielsen, L.~M. Abreu,
  {On $X(3872)$ production in high energy heavy ion collisions}, Phys. Rev.
  D90~(11) (2014) 114023, [Erratum: Phys. Rev.D93,no.5,059902(2016)].
\newblock \href {http://arxiv.org/abs/1405.7583} {\path{arXiv:1405.7583}},
  \href {http://dx.doi.org/10.1103/PhysRevD.93.059902,
  10.1103/PhysRevD.90.114023} {\path{doi:10.1103/PhysRevD.93.059902,
  10.1103/PhysRevD.90.114023}}.

\bibitem{Moreira:2016ciu}
B.~D. Moreira, C.~A. Bertulani, V.~P. Goncalves, F.~S. Navarra, {Production of
  exotic charmonium in $\gamma \gamma$ interactions at hadron colliders}, Phys.
  Rev. D94~(9) (2016) 094024.
\newblock \href {http://arxiv.org/abs/1610.06604} {\path{arXiv:1610.06604}},
  \href {http://dx.doi.org/10.1103/PhysRevD.94.094024}
  {\path{doi:10.1103/PhysRevD.94.094024}}.

\bibitem{Goncalves:2018hiw}
V.~P. Goncalves, B.~D. Moreira, {Probing the $X(4350)$ in $\gamma \gamma$
  interactions at the LHC}\href {http://arxiv.org/abs/1809.08125}
  {\path{arXiv:1809.08125}}.

\bibitem{Debastiani:2017msn}
V.~R. Debastiani, F.~S. Navarra, {A non-relativistic model for the
  $[cc][\bar{c}\bar{c}]$ tetraquark}\href {http://arxiv.org/abs/1706.07553}
  {\path{arXiv:1706.07553}}.

\end{thebibliography}


%

\end{document}